**Physics-Based Machine-Learning Approach for Modeling the Temperature-Dependent Yield Strengths of Medium- or High-Entropy Alloys**

B. Steingrimsson [*,1], X. Fan[2], R. Feng[3], and P. K. Liaw [4]

1. Imagars LLC, 2062 Thorncroft Drive Suite 1214, Hillsboro, OR 97124, USA. Email: baldur@imagars.com

2. Department of Materials Science and Engineering, The University of Tennessee, Knoxville, TN, 37996, USA.

3. Neutron Scattering Division, Oak Ridge National Laboratory, Oak Ridge, TN, 37830, USA.

4. Department of Materials Science and Engineering, The University of Tennessee, Knoxville, TN, 37996, USA.

[*] Corresponding authors: baldur@imagars.com.

**Abstract**

Machine learning is becoming a powerful tool to predict temperature-dependent yield strengths (YS) of structural materials, particularly for multi-principal-element systems. However, successful machine-learning predictions depend on the use of reasonable machine-learning models. Here, we present a comprehensive and up-to-date overview of a bilinear log model for predicting temperature-dependent YS of medium-entropy or high-entropy alloys (MEAs or HEAs). In this model, a break temperature, $T_{break}$, is introduced, which can guide the design of MEAs or HEAs with attractive high-temperature properties. Unlike assuming black-box structures, our model is based on the underlying physics, incorporated in form of a priori information. A technique of global optimization is employed to enable the concurrent optimization of model parameters over low- and high-temperature regimes, showing that the break temperature is consistent across YS and ultimate strength for a variety of HEA compositions. A high-level comparison between YS of MEAs/HEAs and those of nickel-based superalloys reveal superior strength properties of selected refractory HEAs. For reliable operations, the temperature of a structural component, such as a turbine blade, made from refractory alloys may need to stay below $T_{break}$. Once above $T_{break}$, phase transformations may start taking place, and the alloy may begin losing structural integrity.

**Introduction**

Metallic structural alloys with excellent mechanical properties, especially at elevated temperatures, remain in high demand, such as in the aerospace and nuclear industries. Unlike traditional alloys, which only contain one or two principal elements, high-entropy alloys (HEAs), also referred to as multi-principal-element alloys (MPEAs), multi-component alloys, or compositionally-complex alloys (CCAs), have been investigated extensively, since their inception in 2004 [1] - [8]. The carefully-designed HEAs, with either single or multiple phases, have given rise to excellent mechanical properties and vast compositional space, compared to the conventional alloys [9] - [26]. Studies of the Cantor alloy, CrMnFeCoNi, reveal that it can exhibit outstanding mechanical properties, particularly at cryogenic temperatures, with the ultimate tensile strength (UTS) approaching 1 GPa and the ductility of almost 100% at - 196 ºC [27], [28]. Moreover, the fracture toughness is well over 200 MPa√m, and the fatigue strength is nearly 300 MPa after $10^7$ cycles at room temperature [27], [29], [30].

Due to the near-infinite chemical compositional space of HEAs, finding the right chemical compositions with outstanding properties is challenging. Data analytics and machine learning (ML) can help with a rapid search of the vast compositional space, with effective tuning and optimization, and hence, expedite the development of HEAs exhibiting superior properties. It is important to keep in mind that the work on the optimization of compositions and thermomechanical treatments for HEAs has started relatively recently, and there is an enormous range of different alloy and heat-treatment options yet to be explored [27]. For background materials on ML and the motivation for the research direction proposed, on predictions of compositions yielding favorable strength properties, and on factors impacting model selection (including Occam's razor), refer to [31] - [33].

Further towards such an end, Bhandari et al. employed a ML method based on a regression technique of random forest (RF) to predict the yield strengths (YS) of HEAs at a desired temperature [34]. The yield strengths of MoNbTaTiW and HfMoNbTaTiZr were predicted at 800 ºC and 1,200 ºC using a RF model. The authors determined that the results were consistent with experiments and concluded that the RF regression model predicted the YS of HEAs at the desired temperatures with high accuracy [34]. Wen et al. presented a relationship characterizing solid-solution strengthening (SSS) for HEAs in terms of electronegativity difference of the constituent elements [35]. The authors introduced a ML model, which exhibits superior performance in predicting SSS / hardness of HEAs, compared to existing physics-based models. The ML model involves feature construction and selection, which is configured such as to capture the salient descriptors [35]. Note that the RF-prediction model of [34] and the feature construction and selection of [35] can be considered pure (or black-box) ML models, at least for the most parts, in the sense that the underlying physics are not built directly into the models. The authors of [35] proposed the following formula for SSS in HEAs:

$$\Delta \sigma_{SS} = \xi \, Z \, G \, \delta X_r, \quad (1)$$

where $Z$ represents a fitting parameter, $\xi$ a structure factor consistent with a T-model [36], $G$ the shear modulus of the HEAs, and $\delta X_r$ an electronegativity mismatch between elements. This formula, however, differs significantly from established, analytical models for SSS [see Eq. (20) in the supplementary manuscript].

Zhang et al. modeled the effect of temperature on the tensile behavior of an interstitial high-entropy alloy, with the nominal composition of $Fe_{49.5}Mn_{30}Co_{10}Cr_{10}C_{0.5}$ (atomic percent, at.%), in [37], but focused on the development of a micro-mechanism-based crystal-plasticity model (a multi-scale model), as opposed to ML. Note that while the model of [37] is physics-based, it is not a ML model. Qi et al. present in [38] a phenomenological method, where the binary-phase diagrams are analyzed to predict phase properties of HEAs. The authors introduced several phase-diagram-inspired parameters and employ ML to classify 600+ reported HEAs based on these parameters. For a further review of ML and high-throughput studies on high-entropy materials (HEMs), refer to [39] and the references listed therein.

In terms of important contributions (novelty), this study presents a bilinear log model for predicting the yield strengths of MEAs and HEAs across temperature. This model consists of separate exponentials, for a low-temperature and a high-temperature regime, with a break temperature, $T_{break}$, in between. The bilinear log model accounts for the underlying physics, in particular, diffusion processes that are required to initiate phase transformations in the high-temperature regime [40]. Furthermore, we show in Ref. [32] how a piecewise linear regression can be employed to extend the model beyond two exponentials and yield accurate fit, in case of a non-convex objective function caused by hump(s) in the data. Earlier models for the temperature dependence of yield strengths (YS) only accounted for a single exponential [41], [42]. Therefore, there was no break temperature, $T_{break}$. We consider the break point particularly important for the optimization of the high-temperature properties of alloys. For reliable operations, the temperature of turbine blades or disks made from refractory alloys may need to stay below $T_{break}$. Once the alloy temperature exceeds $T_{break}$, the alloy can lose its strength rapidly due to rapid diffusion, leading to easy dislocation motion and dissolution of strengthening phases [40]. We consider $T_{break}$ a fundamental parameter for the design of alloys with attractive high-temperature properties, one warranting the inclusion in the alloy specifications. Hence, it is important to be able to accurately estimate $T_{break}$, e.g., using the global optimization approach presented in [32].

In terms of impact, this study addresses a physics-based, i.e., not a black box, approach to ML and data analytics. Such approaches may be preferred in materials science, first because they involve fewer parameters than the black-box models, and hence, require fewer input data for accurate parameter estimation. With experiments often being time-consuming and expensive, in the case of materials science, high-quality input data tend to be a scarce resource. Secondly, the physics-based approaches to ML help establish causal links between output observations and the behavior of the underlying material system.

**Results**

*Database and Categorization of MEA and HEA Compositions*

The database described in [31] - [33] comprises the base for this study. New additions include data from [43] - [51]. The authors made a concerted effort to conduct a systematic study of the HEA literature and make the analysis as comprehensive and up to date as possible. Extending the work of [52], we define MEA or HEA compositions with 3d-transition metals only as of Type 1, compositions comprising of transition metals with large atomic-radius elements (such as Al/Ti/V/Mo) as of Type 2, compositions comprising of refractory elements of Type 3 and other compositions as of Type 4 (such as precious-element-containing HEAs and hexagonal-close-packed-structured HEAs). As **Fig. 1** illustrates, we have generalized the categorization from [52] such as to accommodate MEA compositions, defined as compositions comprised of three (3) principal elements, in addition to the HEA compositions [1]. Not accounting for separate configurations of the same composition, we analyzed three (3) distinct MEA or HEA compositions of Type 1, ten (10) of Type 2, ninety nine (99) of Type 3, and none (0) of Type 4.

*Bilinear Log Model Applied to Analysis of Yield Strengths for MEAs and HEAs*

To identify compositions with potential in retaining strengths at elevated temperatures, we summarize the reported MEAs and HEAs, according to the classifications in **Fig. 1,** in the

---

[1] Senkov et al. refer to MEAs and HEAs collectively as complex concentrated alloys (CCAs) [40].

Supplementary **Tabs. S1 – S3**, **Fig. 2**, as well as **Figs. S2 – S125** of the Supplementary Manuscript. **Fig. 2** and **Figs. S2 – S125** of the Supplementary Manuscript provide graphical insight into the superiority of the bilinear model, whereas the Supplementary **Tabs. S1 – S2** quantify the comparison with the single-exponential model. Consistent with [32], the results were derived, using the same global optimization approach, applied separately to individual alloys, to obtain a tighter fit and more accurate estimation of $T_{break}$, than for separate optimization over the low and high-temperature regimes. Note that for most of the MEA and HEA compositions, the solvus temperatures were estimated, using the rule of mixing.

*Comparison of Break Temperatures for YS vs. Ultimate Strength (US) for MEAs and HEAs*

**Fig. 3 – Fig. 5** and Supplementary **Figs. S126 - S135** summarize the comparison of the break temperatures for YS vs. US for the MEAs and HEAs. Aside from the composition of MoNbTaVW (the single outlier in **Fig. 5**), the break temperatures for the YS and the US seem quite similar. In the case of MoNbTaVW, and as the Supplementary **Fig. S132(a)** illustrates, the disparity seems to be caused by a single data point for the US at $T = 1{,}200$ °C and may highlight the need for data curation [33].

*Comparison of Yield Strengths for MEAs and HEAs vs. Superalloys*

**Fig. 6** summarizes the comparison of the yield strength for the MEAs and HEAs vs. the superalloys. The HEA compositions of $C_{0.25}Hf_{0.25}NbTaW_{0.5}$, $MoNbSi_{0.75}TaW_{0.5}$, and MoNbTaVW seem to exhibit the most graceful degradation of YS (the smallest slope) at high temperatures. But the compositions of $MoNbSi_{0.75}TaW$, CrTaVW, and $AlMo_{0.5}NbTa_{0.5}TiZr$ tend to yield the highest strength for the lower-temperature regime.

**Discussion**

The FCC + BCC composition of $C_{0.25}Hf_{0.25}NbTaW_{0.5}$ [49], the BCC + silicide composition $MoNbSi_{0.75}TaW$ [53], the BCC + Laves composition CrTaVW [54], and the BCC composition MoNbTaVW [40], [55] seem to offer favorable strength at elevated temperature (above 800 ºC), according to **Fig. 6**, whereas the BCC + silicide composition of $MoNbSi_{0.75}TaW$ [53], the BCC + Laves composition of CrTaVW [54], and the BCC + B2 composition of $AlMo_{0.5}NbTa_{0.5}TiZr$ [40] exhibit attractive strength properties at lower temperatures (25 ºC – 800 ºC). While CrTaVW and $MoNbSi_{0.75}TaW$ appear to exhibit the best performance for the low-temperature and high-temperature regimes. Note that we are basing this observation on only 3 data points, in case of CrTaVW, and 4 data points, in case of $MoNbSi_{0.75}TaW$.

As explained in [32], we expect the bilinear log model (two exponentials) to suffice for most refractory HEAs (HEA-3). As noted in [56], the mechanisms of microscopic slips (dislocation slips) can lead to the pronounced temperature dependence of mechanical strengths. It is unlikely that cross slip from {111} to (100), such as the one producing an anomalous yield stress phenomenon in CMSX-4, single-crystal, nickel-based superalloys (a hump between the low-temperature and high-temperature regimes, necessitating a tri-linear log model) will happen in refractory HEAs [32].

The loss of strengths in single-phase refractory BCC HEAs is related to the activation of the diffusion-controlled deformation mechanism, which typically occurs at $T \geq 0.5 - 0.6\ T_m$ [40]. The rapid decrease in strengths of multi-phase refractory HEAs, with increasing temperatures above 1,000 ºC, is, on the other hand, partially caused by the dissolution of secondary phases [40]. Further, Feng et al. highlight three alloy properties leading to superior high-temperature strengths, namely (1) large atomic-size and elastic-modulus mismatches, (2) insensitive temperature-dependence of elastic properties, and (3) the dominance of non-screw character dislocations [57].

*More on the Comparison between Break Temperatures for YS vs. US*

*The consistency of the break temperature across YS and US and across the HEA compositions looks very comforting. The consistency lends credence to the bilinear log model and suggests that the break temperature is indeed a universal parameter.* The coefficient of determination, $R^2$, in **Fig. 5** consolidates the comparison into a single parameter. In case of **Fig. 4**, 'Yield Strength (1)' and 'Yield Strength (2)' represent YS values corresponding to two separate fabrication runs, post-processing conditions, or possibly microstructures. For example, in the case of AlCrMoNbTi, Refs. [40] and [58] reported multi-phase BCC + Laves (almost single-phase) microstructures for samples annealed for 20 hr. at 1,300 °C, whereas Ref. [56] found similar YS for almost single-phase BCC microstructures for samples annealed for 20 hr. at 1,300 °C. In the case of MoNbTaVW and the disparity between the break temperature for the YS and US in **Fig. 5** and the Supplementary **Fig. S132**, it is worth noting that the YS of 735 MPa and the US of 943 MPa at $T = 1,200$ °C were both reported by [55]. It appears that measurement error may need to be accounted for here, first in the estimation of the YS and US and then in the assessment of the associated break temperatures. The authors noted: "Further increases in temperature to 1,400 °C and 1,600 °C led to a rapid decline in the yield stress and apparent softening" [55].

*More on the Need for a Tri-Linear Log Model*

We noticed in **Fig. 6** as well as in Supplementary **Figs. S4**, **S27**, **S69**, **S70**, and **S71** that the compositions of $Al_{0.5}CoCrCuFeNi$, AlMoNbTi, CrMoNbTi, CrMoNbV, and CrMoTaTi exhibited a relatively flat behavior in an intermediate temperature regime, in contrast to the two linear regimes. Although these cases differ from the hump reported in [32], corresponding to the anomalous yield-strength phenomenon, these cases may still necessitate a tri-linear log model. Reference [56] offers an explanation for this type of tri-linear behavior for intrinsic temperature-dependent yield strengths in BCC metals and substitutional solid solutions:

1. At the low temperatures, plastic deformation of the BCC metals and alloys is believed to be mediated by the thermally-activated formation and movement of kink pairs [56].

2. Above certain "critical temperature," a strength plateau is reached, where the strength becomes virtually strain-rate independent [56].

3. When exceeding the temperature of approx. 0.4 $T_m$, diffusional processes can lead to a rapid decrease of the yield stress, and the YS comes strain-rate dependent again [56].

In the case of the tri-linear model, we are looking at two break temperatures, $T_{break1}$ and $T_{break2}$ [32]. But [56] only references one such temperature and refers to it as "critical temperature" or "knee temperature". This seemingly corresponds to $T_{break1}$ in our model.

*Further High-Level Comparison of Yield Strengths for MEAs and HEAs vs. Superalloys*

The motivation for this study of the strengths of MEAs and HEAs is to accelerate the search for high-strength structural materials at high temperatures, for applications in thermal-protection systems and/or gas-turbine engines. The compositions of MoNbSi$_{0.75}$TaW and CrTaVW seem to exhibit the best performance (highest YS), according to **Fig. 6**, for the low-temperature and high-temperature regimes. The appearance of **Fig. 6** is affected in part by the fact that we are interpolating straight line segments between a limited number of data points for any given composition.

Out of the 124 MEA and HEA configurations, whose yield strengths that we analyzed, only 10 were conducted under tension but 114 under compression. Specifically, the yield strength measurements for C$_{0.25}$Hf$_{0.25}$NbTaW$_{0.5}$, MoNbSi$_{0.75}$TaW, CrTaVW, MoNbTaVW, and AlMo$_{0.5}$NbTa$_{0.5}$TiZr were all conducted under compression. Similarly, out of the 36 superalloy compositions whose yield strengths that we analyzed, the vast majority was conducted under tension in accordance with standards, such as the American Society for Testing and Materials (ASTM) E-8 [59]. Specifically, the yield-strength measurements for Inconel 718, Haynes 230, and Rene 95 were all conducted under tension. It is well known that the tension test is much more difficult than the compression test, since the test samples may need to be machined into dog bone shapes and polished to comply with standards, such as ASTM E8 [59]. One runs the risk of introducing defects, during polishing, and deteriorating the mechanical properties. However, according to Supplementary **Tab. 2** of [32], the ultimate strength under tension usually appears to be a little bit smaller than under compression (up to ~ 100 MPa). The difference between the strengths of the refractory HEAs and the superalloys in **Fig. 6**, on the other hand, ranges from 100 – 2,000+ MPa (in other words, much exceeds what can be explained in terms of the difference between tension and compression tests). Note, furthermore, the superalloys satisfy a broad range of additional requirements for high-temperature structural applications, including tensile ductility, fracture toughness, oxidation resistance, creep strength, fatigue strength, and processability [17].

It is important to note that **Fig. 6** exhibits significant similarity and consistency with **Fig. 20** of [17], esp. as far as superior temperature-dependent strength properties of $AlMo_{0.5}NbTa_{0.5}TiZr$ and MoNbTaVW relative to the superalloys (e.g., Inconel 718 or Haynes 230) are concerned. Inconel 718 [$(Al,Nb,Ti)5Co1Cr21Fe19Mo2Ni52$] is a precipitation-strengthened alloy used widely in the gas-turbine industry for rotating disks, but Haynes 230 ($Co4Cr27Fe3Mo1Ni60W5$) is a solid-solution (SS)-strengthened alloy used for static sheet parts [17]. After the heat treatment [55], the YS of Inconel 718 exceeds 1,000 MPa at temperatures up to 650 °C. However, the YS rapidly decreases to 138 MPa, as the temperature is increased to 982 °C, with melting occurring at ~1,210 °C. Therefore, Inconel 718 is generally used in gas-turbine engines with operating temperatures not exceeding 800 °C [55].

In further consistency with Miracle et al. [17] and Senkov et al. [55], most refractory HEAs outperform Haynes 230, and some show potential to extend the use temperature of blades and disks beyond current superalloys. $AlMo_{0.5}NbTa_{0.5}TiZr$ has sufficient room-temperature (RT) compressive plasticity (ε ≥ 10%) to suggest that tensile ductility may also be achieved [17]. The noticeable drop in YS for MoNbTaVW in **Fig. 6** at $T \geq 1{,}600\ °C$ is consistent with the rapid decrease at $T \geq 0.6\ T_m$ [55]. The rule of mixture estimates the melting temperature of MoNbTaVW as 2,724 °C.

**Conclusions and Future Research**

In conclusion, we proposed a bilinear log model for predicting the YS of MEAs and HEAs across temperatures and studied its effectiveness for 112 distinct compositions (124 configurations). We considered the break temperature, $T_{break}$, an important parameter for the design of materials with favorable high-temperature properties, one warranting inclusion in alloy specifications. For reliable operation, and not accounting for coatings, the operating temperature for the corresponding alloys may need to stay below $T_{break}$. Earlier models for the temperature dependence of the yield strength only accounted for a single exponential. Thus, there was no break temperature. We showed that the break temperature was consistent across YS and US and across a number of HEA compositions. This trend suggested that the break temperature was indeed a universal parameter, and hence, lent credence to our model. We, further, noticed that the compositions AlMoNbTi, CrMoNbTi, CrMoTaTi, and CrMoNbV exhibited a relatively-flat behavior in an intermediate temperature regime, in contrast to the two linear regimes, possibly necessitating a tri-linear log model. To elucidate these observations, we drew upon explanations from [56] regarding intrinsic temperature-dependent yield strengths in BCC metals and substitutional solid solutions. Finally, our observations regarding superior strength behavior of refractory HEAs (the category HEA-3), compared to Ni-based superalloys, were consistent with those of Miracle et al. [17], Senkov et al. [40], and Diao et al. [52]. With a caveat for the limited number of data points available, it appeared the compositions of MoNbSi$_{0.75}$TaW and CrTaVW exhibited the best performance for the low-temperature (25 ºC – 800 ºC) and the high-temperature (above 800 ºC) regimes. Future research may involve (1) measurements of yield strengths of the YS of MEAs and HEAs under tension, (2) identification of additional MEA or HEA compositions exhibiting high YS both in the low and high-temperature regimes, or (3) identification of MEA or HEA compositions exhibiting favorable balance of properties, in particular between the strength and ductility (high tensile ductility).

**Methods**

While the primary emphasis here is on the yield strength, the optimization of the mechanical properties is assumed to reside within a framework for joint optimization [32].

*Methodology for Maximization of the YS*

Our approach entails accurately capturing the input sources that contribute to variations in the YS observed (to variations in the output). We presently model the input combination as [32]

$$\text{Input} = (\text{composition}, T, \text{process}, \text{defects}, \text{grain size}, \text{microstructure}). \qquad (2)$$

"Defects" are here defined broadly such as to include inhomogeneities, impurities, dislocations, or unwanted features. "$T$" denotes temperature. Similarly, the term, "microstructure," broadly represents microstructures, at nano or micro scale, as well as phase properties, and the term "process" broadly refers to manufacturing process and post-processing. Correspondingly, the term, "grain size," generally refers to the distribution in grain sizes. Section 4.4 of [33] allows for dependence between input sources, and Section 4.5 of [33] outlines the expected dependence of the alloy strength on the individual input sources listed. We summarize the prediction model as [32]

$$\text{YS} = h[\text{composition}, T, \text{process}, \text{defects}(\text{process}, T), \text{grains}(\text{process}, T), \text{microstructure}(\text{process}, T)]. \qquad (3)$$

*Methodology for Modeling YS at Elevated Temperatures*

Inspired by **Figs. 3(c)** and **3(d)** of [32], together with physics-based insights from [40], we model the temperature dependence of the YS($T$), in terms of a bilinear log model, parametrized by the melting temperature, $T_m$, as follows:

$$YS(T) = \min(\log(YS_1(T)), \log(YS_2(T))), \quad (4)$$

$$YS_1(T) = \exp(-C_1 * T / T_m + C_2), \qquad 0 < T < T_{break}, \quad (5)$$

$$YS_2(T) = \exp(-C_3 * T / T_m + C_4), \qquad T_{break} < T < T_m. \quad (6)$$

There is a separate physics (diffusion)-induced constraint on $T_{break}$ [40]:

$$0.35 \lesssim T_{break} / T_m \lesssim 0.55, \quad (7)$$

and a continuity constraint between the low-temperature and the high-temperature regimes:

$$YS_1(T_{break}) = YS_2(T_{break}); \quad (8)$$

$$T_{break} = \frac{(C_4 - C_2)}{(C_3 - C_1)} T_m. \quad (9)$$

As explained in [32], a conceptually-simple approach for fitting the model in Eqs. (4) – (9) to the YS data available consists of first deriving the constant coefficients, $C_1$ and $C_2$, by applying linear regression to data points available to the lowest temperature region ($0 < T < 0.35\,T_m$) as well as to the intermediate region ($0.35\,T_m \leq T \leq 0.55\,T_m$). One may then derive the constants, $C_3$ and $C_4$, by applying linear regression to data points available to the intermediate ($0.35\,T_m \leq T \leq 0.55\,T_m$) and high-temperature ($T > 0.55\,T_m$) regions. Note that $T_{break}$ does not need to be known beforehand. Supported by Eq. (9), $T_{break}$ is an inherent property of a given alloy that comes out of the model as the break point between the two linear regions. The model in Eqs. (4) – (6) consists of only four (4) independent parameters, $C_1$, $C_2$, $C_3$, and $C_4$, which simply can be estimated by applying linear regression separately to low-temperature and high-temperature regimes, even to a fairly small data set. Note, moreover, that for a new alloy system, $T_m$, does not need to be known experimentally in advance either; a rough estimate for $T_m$ can be obtained, using "the rule of mixing" and a more refined estimate obtained, employing Calculation of Phase Diagram (CALPHAD) simulations [33]. A superior approach for deriving the coefficients, $C_1$, $C_2$, $C_3$, and $C_4$, entails concurrent optimization over the low-temperature and high-temperature regimes using the global optimization. Here, one seeks to minimize

$$\min_{C_1,C_2,C_3,C_4} \text{norm}_2 \left( (YS(T_i) - y_i)^2 \right), \tag{10}$$

where $y_i$ represents the YS values measured,

$$YS(T_i) = \min(\log(YS_1(T_i)), \log(YS_2(T_i))), \tag{11}$$

and $US_1(T_i)$ and $US_2(T_i)$ are modeled, using Eqs. (5) and (6), respectively. Matlab offers a function, fminunc(), for solving this type of unconstrained minimization over a generic function. Depending on the grain sizes and compositions comprising the alloys, a tri-linear log model may yield a better fit for certain alloys [28], [32]:

$$YS(T) = \min(YS_1(T), YS_2(T), YS_3(T)), \tag{12}$$

$$YS_1(T) = \exp(-C_1 * T / T_m + C_2), \quad 0 < T < T_{break1}, \tag{13}$$

$$YS_2(T) = \exp(-C_3 * T / T_m + C_4), \quad T_{break1} < T < T_{break2}, \tag{14}$$

$$YS_3(T) = \exp(-C_5 * T / T_m + C_6), \quad T_{break2} < T < T_m. \tag{15}$$

**Data Availability**

The data in this paper, including those in the Supplementary Figures, can be requested by contacting the corresponding author (baldur@imagars.com).

**Code Availability**

Matlab comprises the software package primarily used for this study. Supplementary **Figs. S7** and **S8** of [32] contain a Matlab source code for the objective functions optimized in case of the bilinear or trilinear log model, respectively.

**Acknowledgements**

XF and PKL very much appreciate the support of the U.S. Army Research Office Project (W911NF-13-1-0438 and W911NF-19-2-0049) with the program managers, Drs. M. P. Bakas, S. N. Mathaudhu, and D. M. Stepp. RF and PKL thank the support from the National Science Foundation (DMR-1611180 and 1809640) with the program directors, Drs. J. Yang, G. Shiflet, and D. Farkas. XF and PKL also appreciate the support from the Bunch Fellowship. XF and PKL would like to acknowledge funding from the State of Tennessee and Tennessee Higher Education Commission (THEC) through their support of the Center for Materials Processing (CMP). BS very much appreciates the support from the National Science Foundation (IIP-1447395 and IIP-1632408), with the program directors, Drs. G. Larsen and R. Mehta, from the U.S. Air Force (FA864921P0754), with J. Evans as the program manager, and from the U.S. Navy (N6833521C0420), with Drs. D. Shifler and J. Wolk as the program managers.

The authors also want to thank Dr. Liang Jiang for introducing them to some of the background literature on analytical modeling of the yield strength.

**Author Contributions**

B.S. and P.K.L conceived the project. B.S. performed the modeling of the temperature-dependent yield strength and put together the supplementary manuscript. X.F. helped prepare the database and carefully reviewed the manuscript for comparison with previously published work [32]. R.F. contributed towards analytical modeling. All authors edited and proofread the final manuscript and participated in discussions.

**Competing Interests**

The authors declare no competing interests (financial or non-financial).

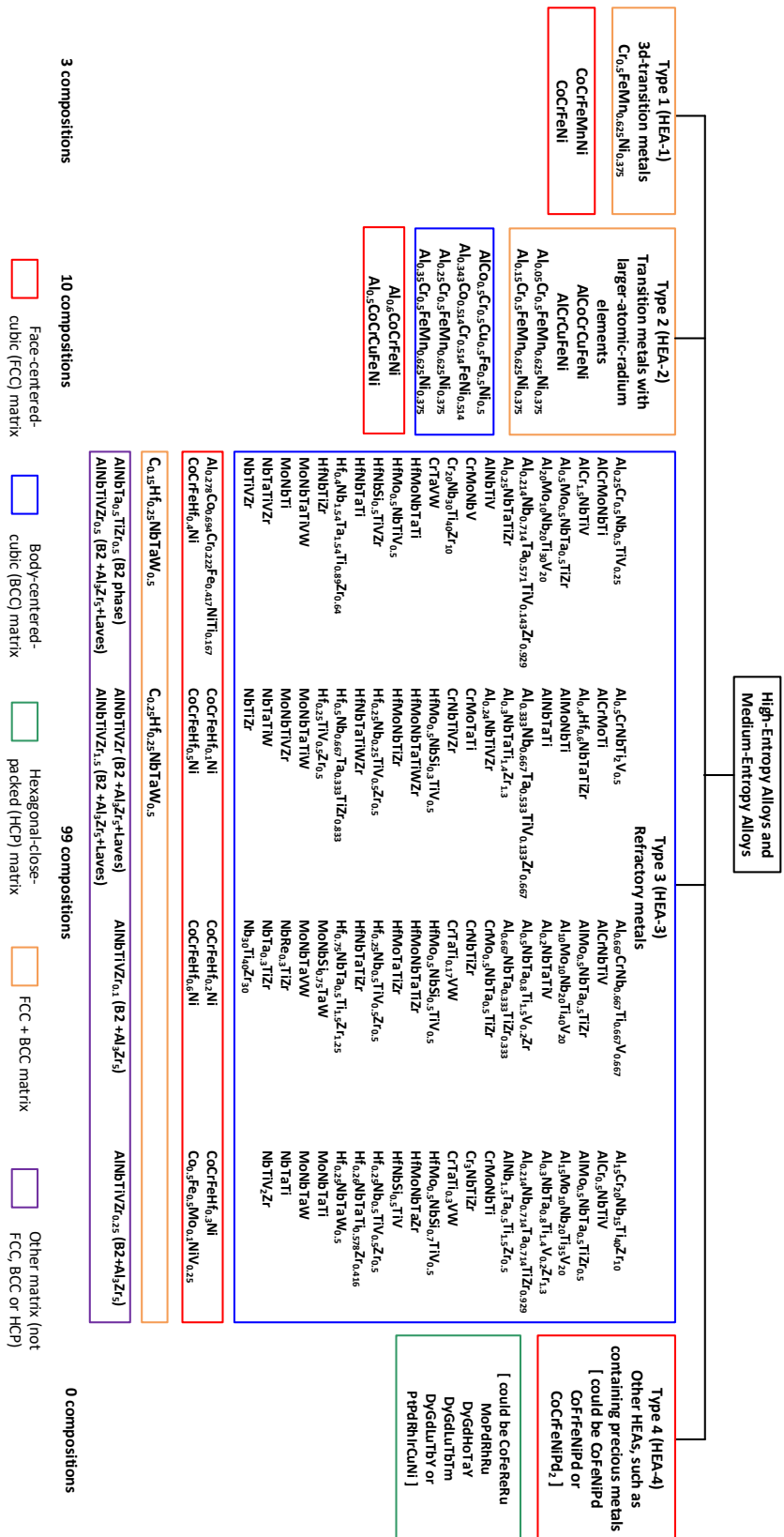

**Fig. 1**: Classification of high-entropy alloys (HEAs) and medium-entropy alloys (MEAs). The figure comprises an adapted version of **Fig. 1** from [52], now generalized such as to also include MEAs. Secondary phases, such as intermetallic phases, have been omitted for simplicity.

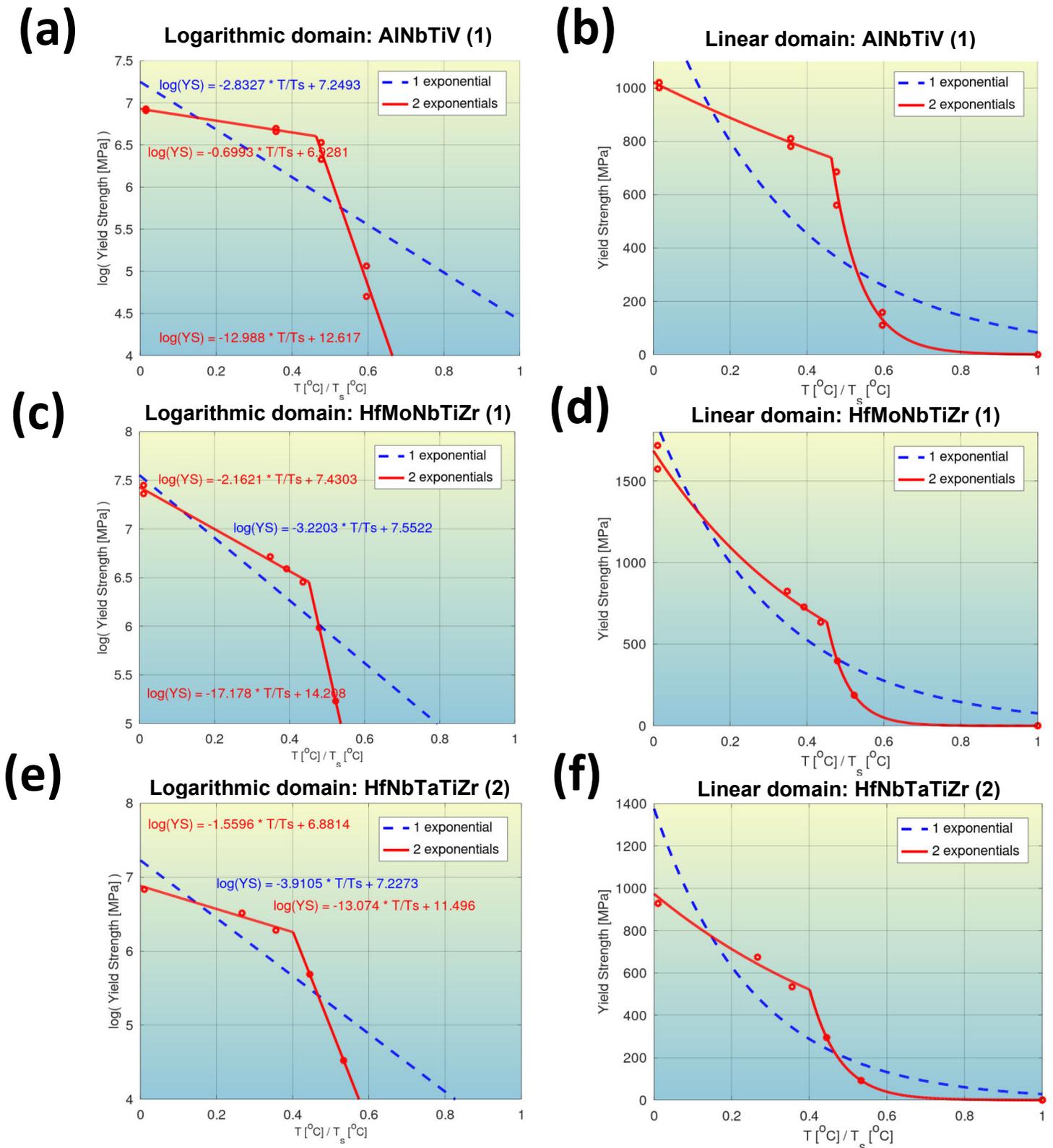

**Fig. 2**: Quantification of modeling accuracy of the bilinear log model and comparison to that of a model with a single exponential. Top row: Composition No. 42 from Supplementary **Tab. S1** (AlNbTiV, BCC phase). Middle row: Composition No. 87 from Supplementary **Tab. S2** (HfMoNbTiZr, BCC phase). Bottom row: Composition No. 98 from Supplementary **Tab. S3** [HfNbTaTiZr (2), BCC phase].

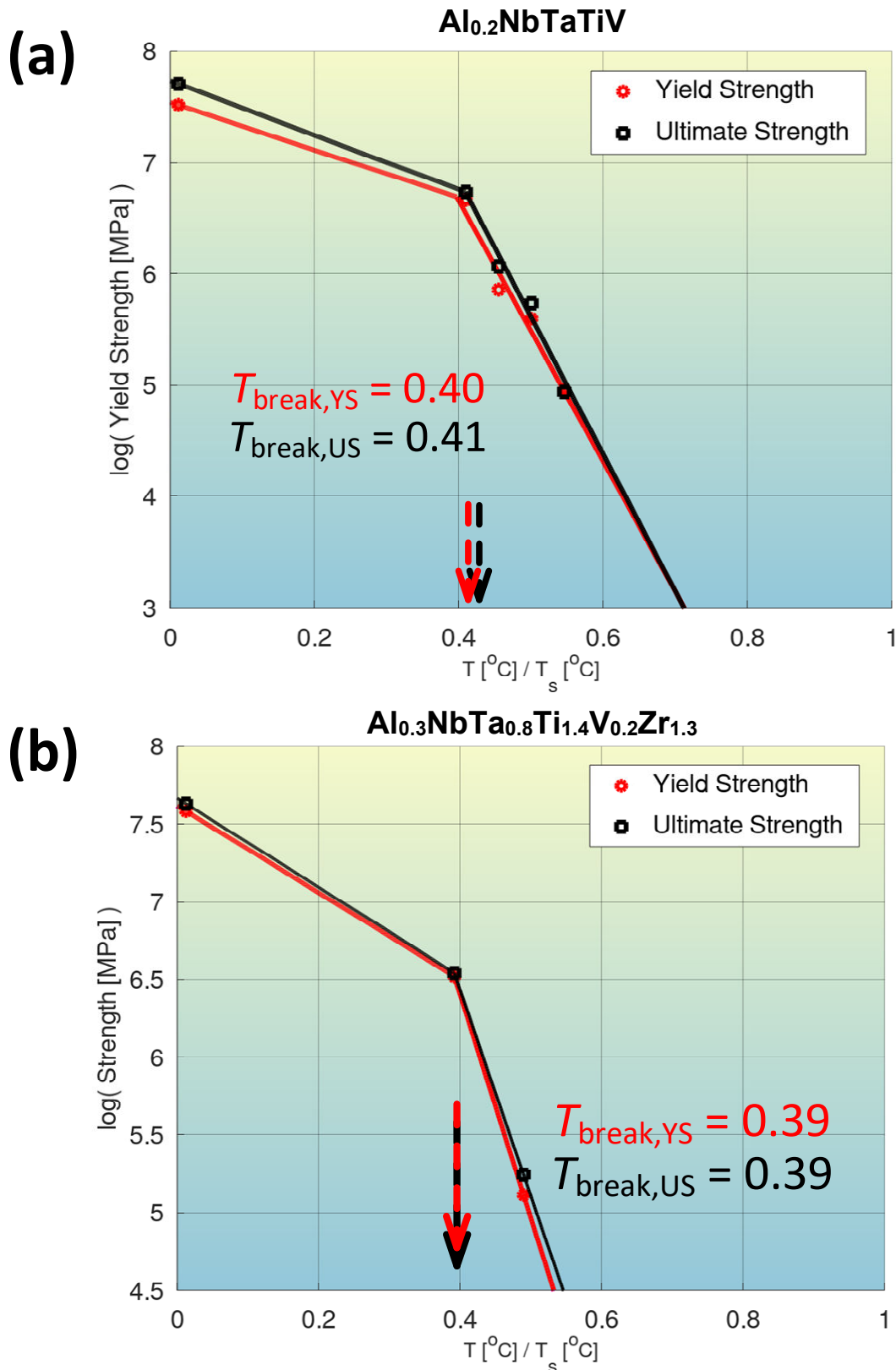

**Fig. 3**: Comparison between break temperatures for the yield strength and ultimate strength of the compositions, $Al_{0.2}NbTaTiV$ (above) and $Al_{0.3}NbTa_{0.8}Ti_{1.4}V_{0.2}Zr_{1.3}$ (below). For a similar comparison for the other compositions listed in **Fig. 4**, refer to Supplementary **Fig. S124 – S133**.

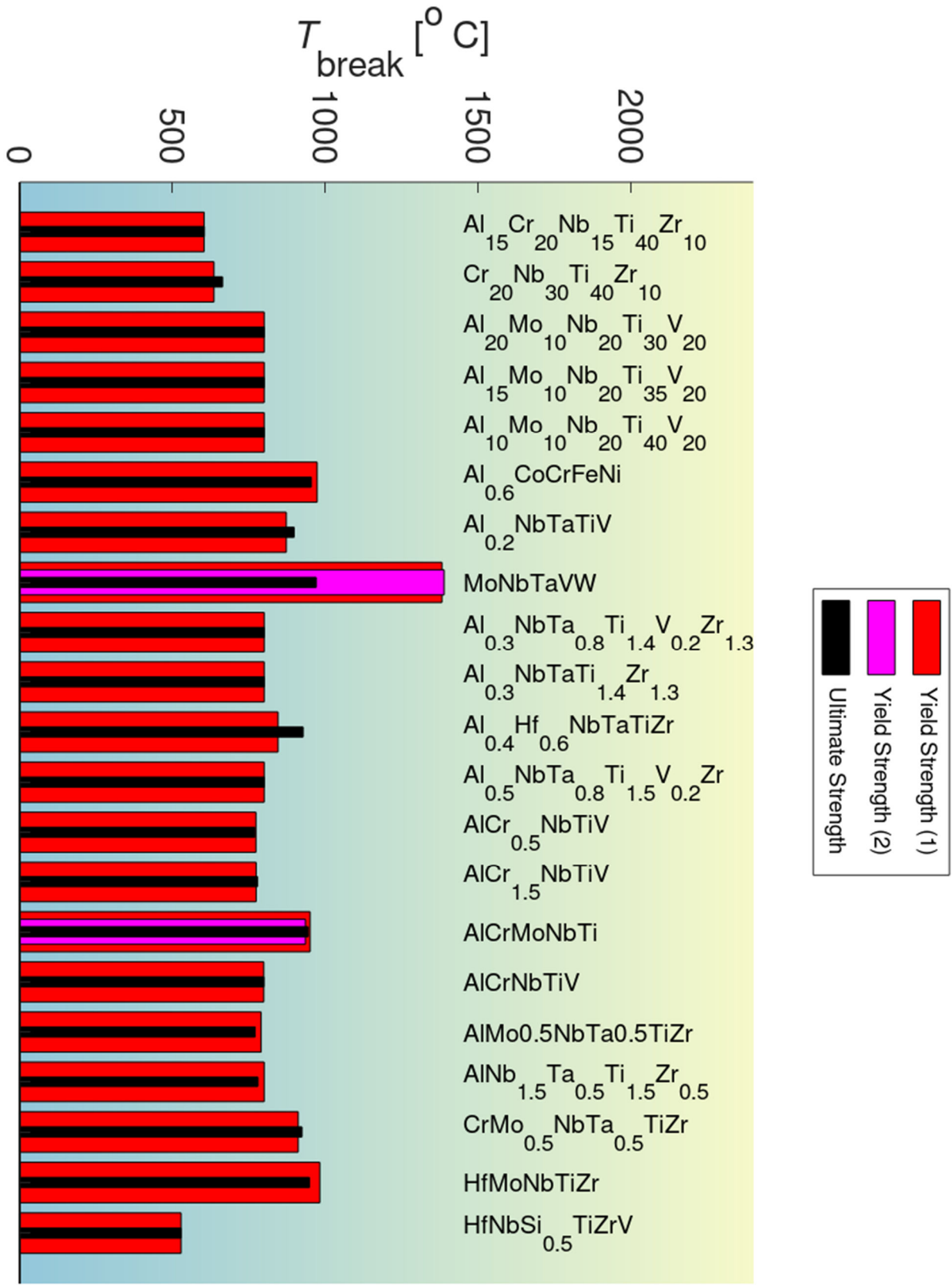

Fig. 4: Further comparison of $T_{break}$ derived from the YS to those derived from the US [32].

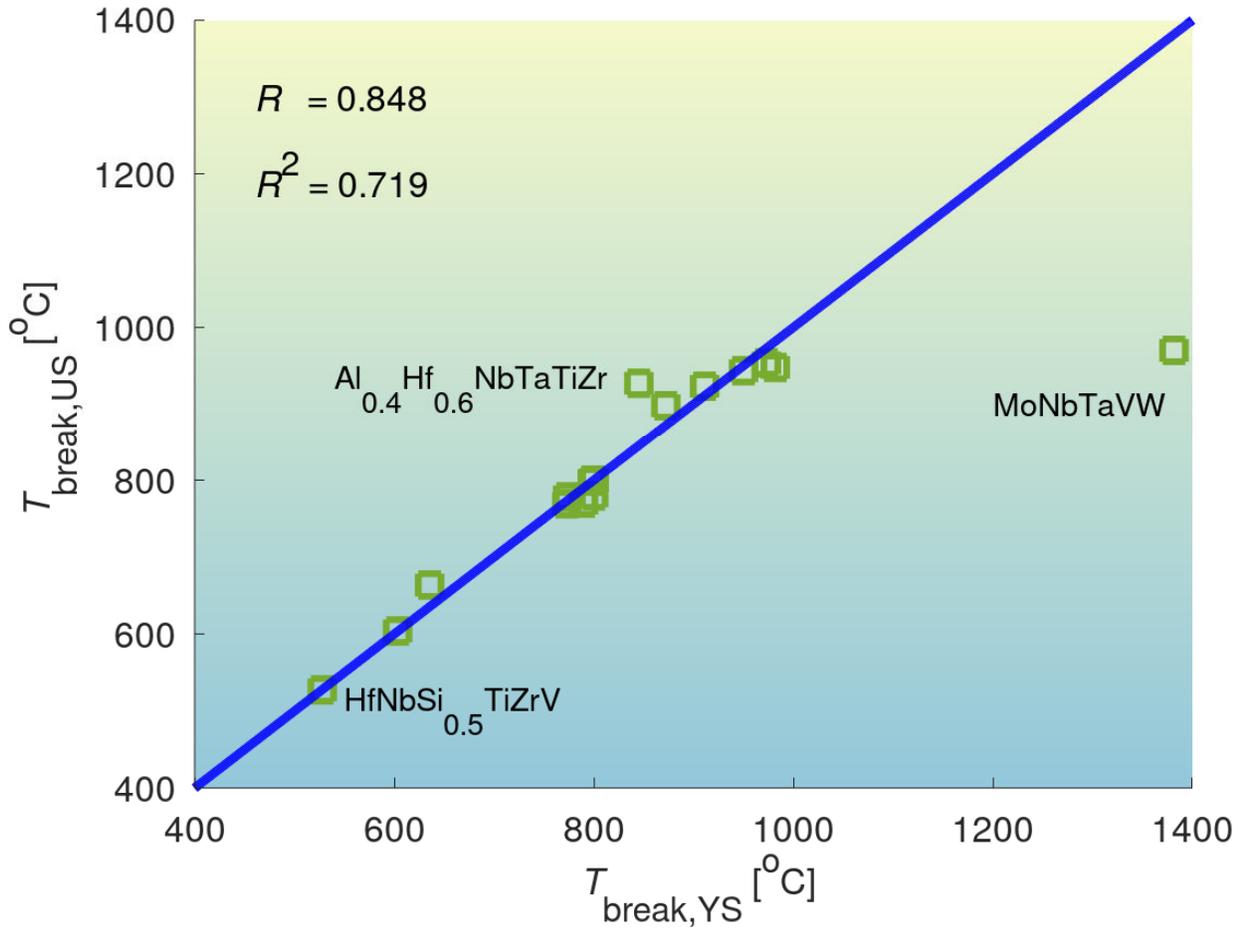

**Fig. 5**: Correlation of $T_{break}$ derived from the YS with $T_{break}$ derived from the US [32].

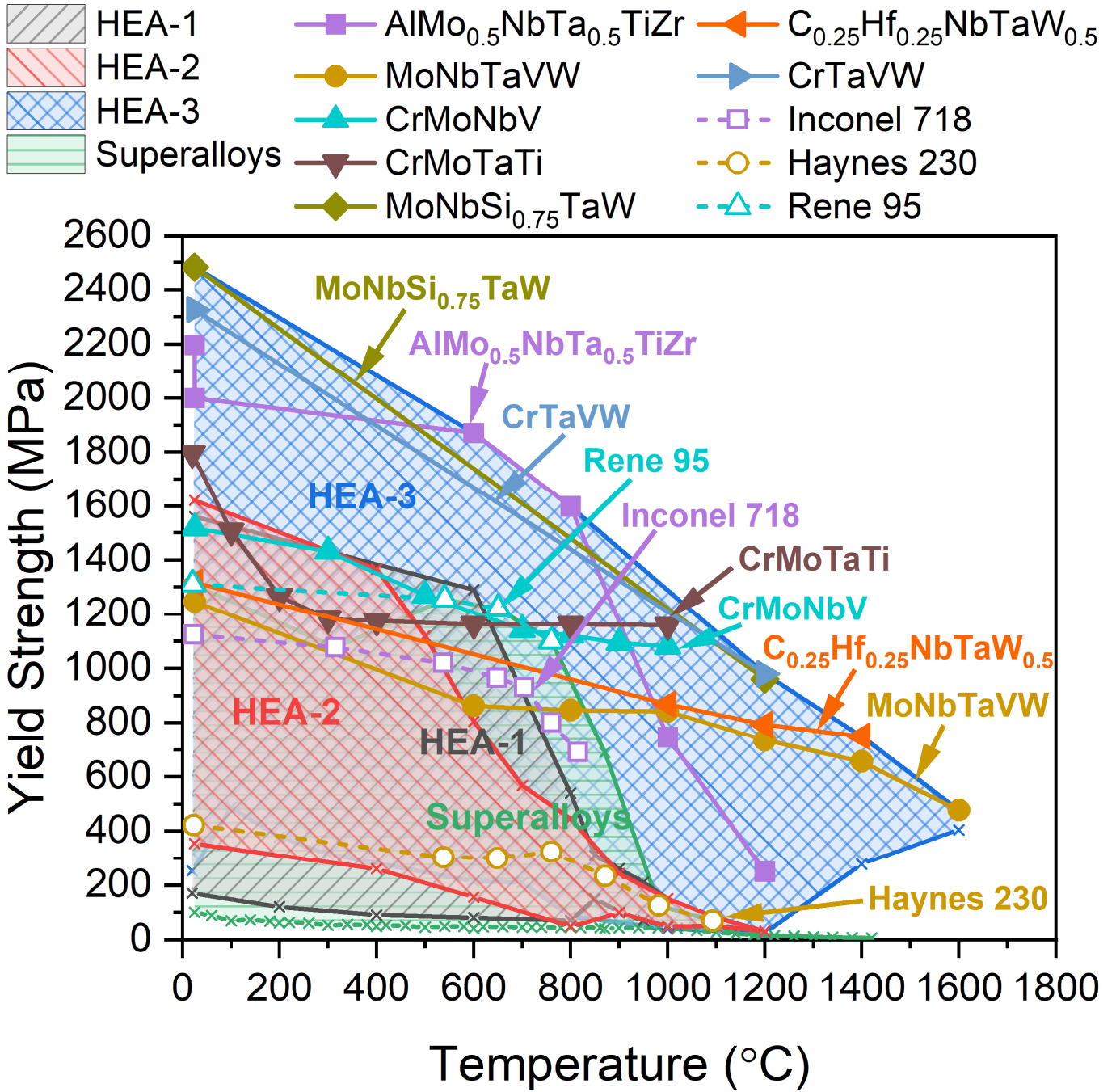

**Fig. 6**: Comparison of MEAs and HEAs yielding the best performance, in terms of temperature-dependent yield strengths, with superalloys yielding the best performance (based on the dataset from Supplementary **Tab. S1** – Supplementary **Tab. S3** as well as **Fig. 2** of [52]).

**Tab. S1:** Quantification of the ability of MEA and HEA compositions to retain ultimate strength at high temperatures. $T_{break}$ refers to the breaking temperature between the bilinear log models.

| No. | Alloy | Category | Solvus Temp. [°C] | $T_{break}$ [°C] | MSE: Two-exponentials Log | MSE: Two-exponentials Linear | MSE: Single-exponential Log | MSE: Single-exponential Linear |
|---|---|---|---|---|---|---|---|---|
| 1 | $AlCo_{0.5}Cr_{0.5}Cu_{0.5}Fe_{0.5}Ni_{0.5}$ | HEA-2 | 1,257 | 552.1 | 0.00265 | 669 | 0.14310 | 211,827 |
| 2 | $AlCoCrCuFeNi$ | HEA-2 | 1,357 | 611.5 | 0.001164 | 34 | 0.20984 | 47,809 |
| 3 | $Al_{0.5}CoCrCuFeNi$ | HEA-2 | 1,420 | 937.2 | 0.04169 | 21,655 | 0.08762 | 12,803 |
| 4 | $Al_{0.6}CoCrFeNi$ | HEA-2 | 1,476 | 973.1 | 0.0001486 | 6.5 | 0.05994 | 7,636 |
| 5 | $Al_{0.343}Co_{0.514}Cr_{0.514}FeNi_{0.514}$ | HEA-2 | 1,477 | 650.0 | 3.747e-30 | 0.0 | 0.04408 | 4,636 |
| 6 | $Al_{0.278}Co_{0.694}Cr_{0.222}Fe_{0.417}NiTi_{0.167}$ | HEA-3 | 1,447 | 782.4 | 0.006579 | 1,762 | 0.05960 | 13,342 |
| 7 | $AlCrCuFeNi$ | HEA-2 | 1,329 | 490.0 | 1.724e-29 | 0.0 | 0.50993 | 4,932 |
| 8 | $Al_{0.05}Cr_{0.5}FeMn_{0.625}Ni_{0.375}$ | HEA-2 | 1,448 | 400.0 | 4.339e-30 | 75 | 0.01140 | 1,204 |
| 9 | $Al_{0.15}Cr_{0.5}FeMn_{0.625}Ni_{0.375}$ | HEA-2 | 1,448 | 400.0 | 4.339e-30 | 182 | 0.01939 | 5,963 |
| 10 | $Al_{0.25}Cr_{0.5}FeMn_{0.625}Ni_{0.375}$ | HEA-2 | 1,448 | 400.0 | 1.282e-29 | 9 | 0.08061 | 35,700 |
| 11 | $Al_{0.35}Cr_{0.5}FeMn_{0.625}Ni_{0.375}$ | HEA-2 | 1,420 | 400.0 | 1.006e-29 | 12 | 0.06981 | 25,401 |
| 12 | $Al_{0.25}Cr_{0.5}Nb_{0.5}TiV_{0.25}$ | HEA-3 | 1,801 | 717.1 | 2.718e-11 | 0.0 | 0.30888 | 111,891 |
| 13 | $Al_{0.5}CrNbTi_2V_{0.5}$ | HEA-3 | 1,801 | 717.1 | 1.684e-12 | 0.0 | 0.30888 | 111,891 |
| 14 | $Al_{0.667}CrNb_{0.667}Ti_{0.667}V_{0.667}$ | HEA-3 | 1,741 | 778.3 | 4.380e-11 | 0.0 | 0.73434 | 405,868 |
| 15 | $Al_{15}Cr_{20}Nb_{15}Ti_{40}Zr_{10}$ | HEA-3 | 1,705 | 604.5 | 0.000494 | 40 | 0.45047 | 371,952 |
| 16 | $AlCrMoNbTi$ (1) | HEA-3 | 1,867 | 950.5 | 0.003254 | 3,194 | 0.27687 | 126,360 |
| 17 | $AlCrMoNbTi$ (2) | | | 934.9 | 6.784e-30 | 0.8 | 0.09817 | 33,249 |
| 18 | $AlCrMoTi$ | HEA-3 | 1,715 | 873.6 | 0.000326 | 325 | 0.15472 | 76,176 |
| 19 | $AlCrNbTiV$ | HEA-3 | 1,725 | 798.5 | 1.452e-29 | 0.0 | 0.68013 | 252,812 |
| 20 | $AlCr_{0.5}NbTiV$ | HEA-3 | 1,704 | 773.0 | 1.014e-29 | 0.0 | 0.86623 | 270,219 |
| 21 | $AlCr_{1.5}NbTiV$ | HEA-3 | 1,741 | 778.3 | 1.310e-29 | 0.0 | 0.73434 | 405,868 |
| 22 | $Al_{0.4}Hf_{0.6}NbTaTiZr$ | HEA-3 | 2,124 | 845.2 | 4.265e-11 | 0.0 | 0.22712 | 92,681 |
| 23 | $AlMo_{0.5}NbTa_{0.5}TiZr$ | HEA-3 | 1,896 | 789.8 | 0.003353 | 5,534 | 0.19642 | 240,956 |
| 24 | $AlMo_{0.5}NbTa_{0.5}TiZr_{0.5}$ | HEA-3 | 1,901 | Insufficient no. of points for high-temp regime | | | 7.10e-30 | 154,067 |
| 25 | $Al_{0.5}Mo_{0.5}NbTa_{0.5}TiZr$ | HEA-3 | 2,033 | Insufficient no. of points for high-temp regime | | | 7.09e-30 | 5,742 |
| 26 | $AlMoNbTi$ | HEA-3 | 1,857 | 1037.0 | 0.013444 | 6,284 | 0.061747 | 10,727 |
| 27 | $Al_{10}Mo_{10}Nb_{20}Ti_{40}V_{20}$ | HEA-3 | 1,873 | 800.0 | 9.861e-30 | 0.2 | 0.14374 | 15,617 |
| 28 | $Al_{15}Mo_{10}Nb_{20}Ti_{35}V_{20}$ | HEA-3 | 1,823 | 800.0 | 2.366e-30 | 0.2 | 0.14895 | 18,071 |
| 29 | $Al_{20}Mo_{10}Nb_{20}Ti_{30}V_{20}$ | HEA-3 | 1,772 | 800.0 | 7.010e-30 | 0.5 | 0.14601 | 23,444 |
| 30 | $AlNbTaTi$ | HEA-3 | 1,956 | Insufficient no. of points for high-temp regime | | | 1.32e-30 | 136,900 |
| 31 | $Al_{0.2}NbTaTiV$ | HEA-3 | 2,198 | 872.2 | 0.008733 | 2,031 | 0.13763 | 43,998 |
| 32 | $Al_{0.3}NbTa_{0.8}Ti_{1.4}V_{0.2}Zr_{1.3}$ | HEA-3 | 2,043 | 800.0 | 1.992e-29 | 0.0 | 0.16138 | 35,138 |
| 33 | $Al_{0.214}Nb_{0.714}Ta_{0.571}TiV_{0.143}Zr_{0.929}$ | HEA-3 | 2,043 | 800.0 | 1.124e-29 | 0.0 | 0.16138 | 35,138 |
| 34 | $Al_{0.333}Nb_{0.667}Ta_{0.533}TiV_{0.133}Zr_{0.667}$ | HEA-3 | 1,992 | 800.0 | 4.733e-30 | 0.0 | 0.13707 | 39,393 |
| 35 | $Al_{0.5}NbTa_{0.8}Ti_{1.5}V_{0.2}Zr$ | HEA-3 | 1,992 | 800..0 | 1.499e-29 | 0.0 | 0.13707 | 39,393 |
| 36 | $Al_{0.214}Nb_{0.714}Ta_{0.714}TiZr_{0.929}$ | HEA-3 | 2,088 | 800.0 | 3.352e-30 | 133 | 9.59e-06 | 123 |
| 37 | $Al_{0.25}NbTaTiZr$ | HEA-3 | 2,161 | Insufficient no. of points for high-temp regime | | | 7.10e-30 | 1,083 |
| 38 | $Al_{0.3}NbTaTi_{1.4}Zr_{1.3}$ | HEA-3 | 2,088 | 800.0 | 4.339e-30 | 133 | 9.59e-06 | 123 |
| 39 | $Al_{0.667}NbTa_{0.333}TiZr_{0.333}$ | HEA-3 | 1,863 | 800.0 | 5.719e-30 | 247 | 0.024998 | 16,165 |
| 40 | $AlNb_{1.5}Ta_{0.5}Ti_{1.5}Zr_{0.5}$ | HEA-3 | 1,863 | 800.0 | 5.719e-30 | 247 | 0.0250 | 16,157 |
| 41 | $AlNbTa_{0.5}TiZr_{0.5}$ | HEA-3 | 1,810 | Insufficient no. of points for high-temp regime | | | 7.10e-30 | 20,446 |
| 42 | $AlNbTiV$ (1) | HEA-3 | 1,679 | 777.3 | 0.00964 | 1,072 | 0.2662 | 63,550 |
| 43 | $AlNbTiV$ (2) | | | 787.3 | 4.891e-30 | 0.2 | 0.22517 | 54,137 |
| 44 | $AlNbTiVZr$ | HEA-3 | 1,714 | 600.0 | 8.086e-30 | 86 | 0.04809 | 46,044 |
| 45 | $AlNbTiVZr_{0.1}$ | HEA-3 | 1,683 | 600.0 | 6.705e-30 | 65,000 | 5.66e-05 | 77,299 |
| 46 | $AlNbTiVZr_{0.25}$ | HEA-3 | 1,683 | 600.0 | 3.353e-30 | 26,281 | 0.00206 | 75,995 |
| 47 | $AlNbTiVZr_{0.5}$ | HEA-3 | 1,698 | 600.0 | 8.086e-30 | 1,071 | 0.02061 | 44,909 |
| 48 | $AlNbTiVZr_{1.5}$ | HEA-3 | 1,727 | 600.0 | 9.861e-30 | 0.0 | 0.37004 | 153,037 |

**Tab. S2:** Continuation of quantification of ability of compositions to retain ultimate strength at high temperatures. Again, $T_{break}$ refers to the breaking temperature between the bilinear log models.

| No. | Alloy | Category | Solvus Temp. [°C] | $T_{break}$ [°C] | MSE: Two-exponentials | | MSE: Single-exponential | |
|---|---|---|---|---|---|---|---|---|
| | | | | | Log | Linear | Log | Linear |
| 49 | $Al_{0.24}NbTiVZr$ | HEA-3 | 1,903 | 1000.0 | 1.188e-29 | 0.0 | 0.0609 | 1,386 |
| 50 | $C_{0.15}Hf_{0.25}NbTaW_{0.5}$ | HEA-3 | 3,032 | 1,308.1 | 0.000249 | 8,844 | 0.000359 | 17,106 |
| 51 | $C_{0.25}Hf_{0.25}NbTaW_{0.5}$ | HEA-3 | 3,052 | Insufficient no. of points for high-temp regime | | | 0.00011 | 27,350 |
| 52 | $CoCrFeHf_{0.1}Ni$ | HEA-3 | 1,633 | 600.0 | 4.339e-30 | 843 | 0.002178 | 2,400 |
| 53 | $CoCrFeHf_{0.2}Ni$ | HEA-3 | 1,666 | 600.0 | 1.775e-30 | 1,008 | 0.006968 | 7,187 |
| 54 | $CoCrFeHf_{0.3}Ni$ | HEA-3 | 1,698 | 600.0 | 4.339e-30 | 232 | 0.031630 | 22,729 |
| 55 | $CoCrFeHf_{0.4}Ni$ | HEA-3 | 1,728 | 600.0 | 2.761e-30 | 66 | 0.06332 | 64,678 |
| 56 | $CoCrFeHf_{0.5}Ni$ | HEA-3 | 1,756 | 600.0 | 1.519e-29 | 1.9 | 0.12170 | 76,692 |
| 57 | $CoCrFeHf_{0.6}Ni$ | HEA-3 | 1,784 | 600.0 | 5.719e-30 | 2.7 | 0.10102 | 76,291 |
| 58 | CoCrFeMnNi (1) | HEA-1 | 1,528 | 606.9 | 0.000868 | 117 | 0.01796 | 1,426 |
| 59 | CoCrFeMnNi (2) | HEA-1 | 1,528 | Insufficient no. of points for high-temp regime | | | 0.002827 | 366 |
| 60 | CoCrFeMnNi (3) | HEA-1 | 1,528 | Insufficient no. of points for high-temp regime | | | 0.006329 | 214 |
| 61 | CoCrFeMnNi (4) | HEA-1 | 1,528 | 953.8 | 0.0090115 | 237.6 | 0.019301 | 562.7 |
| 62 | CoCrFeMnNi (5) | HEA-1 | 1,528 | 844.2 | 0.0052687 | 361.4 | 0.006615 | 501.3 |
| 63 | CoCrFeNi (1) | HEA-1 | 1,528 | 575.9 | 0.0077419 | 91.3 | 0.052157 | 1,106 |
| 64 | CoCrFeNi (2) | HEA-1 | 1,528 | 505.9 | 0.002166 | 315.7 | 0.002770 | 339.2 |
| 65 | $Co_{0.5}Fe_{0.5}Mo_{0.1}NiV_{0.25}$ | HEA-3 | 1,579 | 707.8 | 2.938e-05 | 991 | 0.001185 | 4,656 |
| 66 | $Cr_{0.5}FeMn_{0.625}Ni_{0.375}$ | HEA-1 | 1,520 | Insufficient no. of points for high-temp regime | | | 0.00160 | 394 |
| 67 | $CrMo_{0.5}NbTa_{0.5}TiZr$ | HEA-3 | 2,145 | 911.1 | 1.420e-29 | 0.1 | 0.18562 | 74,228 |
| 68 | CrMoNbTi | HEA-3 | 2,169 | Insufficient no. of points for high-temp regime | | | 0.00523 | 45,445 |
| 69 | CrMoNbV | HEA-3 | 2,229 | Insufficient no. of points for high-temp regime | | | 5.126e-4 | 54,193 |
| 70 | CrMoTaTi | HEA-3 | 2,304 | Insufficient no. of points for high-temp regime | | | 0.01069 | 70,558 |
| 71 | CrNbTiZr | HEA-3 | 1,977 | 588.0 | 0.002744 | 898.3 | 0.24357 | 114,016 |
| 72 | $Cr_3NbTiZr$ | HEA-3 | 1,954 | 1000.0 | 5.966e-30 | 0.0 | 0.12623 | 5,025 |
| 73 | $Cr_{20}Nb_{30}Ti_{40}Zr_{10}$ | HEA-3 | 1,977 | 635.5 | 0.006012 | 657 | 0.36123 | 100,882 |
| 74 | CrNbTiVZr | HEA-3 | 1,963 | 640.6 | 1.888e-13 | 3.2 | 0.13988 | 87,055 |
| 75 | $CrTaTi_{0.17}VW$ | HEA-3 | 2,527 | Insufficient no. of points for high-temp regime | | | 9.86e-30 | 19,678 |
| 76 | $CrTaTi_{0.3}VW$ | HEA-3 | 2,501 | Insufficient no. of points for high-temp regime | | | 7.10e-30 | 7,150 |
| 77 | CrTaVW | HEA-3 | 2,564 | Insufficient no. of points for high-temp regime | | | 9.86e-30 | 42,801 |
| 78 | $HfMo_{0.5}NbSi_{0.3}TiV_{0.5}$ | HEA-3 | 2,109 | 1000.0 | 5.719e-30 | 2.4 | 0.04603 | 4,932 |
| 79 | $HfMo_{0.5}NbSi_{0.5}TiV_{0.5}$ | HEA-3 | 2,078 | 1,000.0 | 7.482e-13 | 0.2 | 0.12429 | 21,529 |
| 80 | $HfMo_{0.5}NbSi_{0.7}TiV_{0.5}$ | HEA-3 | 2,050 | 1,000.0 | 2.099e-12 | 1.8 | 0.08886 | 21,136 |
| 81 | HfMoNbTaTi | HEA-3 | 2,404 | 1,200.0 | 5.909e-04 | 521 | 0.029685 | 17,145 |
| 82 | HfMoNbTaTiWZr | HEA-3 | 2,471 | 652.0 | 0.003678 | 8,639 | 0.006313 | 19,368 |
| 83 | HfMoNbTaTiZr (1) | HEA-3 | 2,312 | 971.6 | 5.99e-30 | 10.0 | 0.05551 | 34,619 |
| 84 | HfMoNbTaTiZr (2) | HEA-3 | | 921.9 | 8.223e-13 | 892 | 0.01247 | 22,189 |
| 85 | HfMoNbTaZr | HEA-3 | 2,441 | 1,156.4 | 4.797e-05 | 56 | 0.090513 | 44,718 |
| 86 | $HfMo_{0.5}NbTiV_{0.5}$ | HEA-3 | 2,161 | 1,000.0 | 7.807e-11 | 0.0 | 0.32574 | 17,677 |
| 87 | HfMoNbTiZr (1) | HEA-3 | | 1,036.8 | 8.247e-04 | 1,484 | 0.10005 | 27,901 |
| 88 | HfMoNbTiZr (2) | HEA-3 | 2,297 | 982.0 | 1.75e-29 | 0.0 | 0.14672 | 43,288 |
| 89 | HfMoNbTiZr (3) | HEA-3 | | 999.0 | 0.011473 | 1,932 | 0.096466 | 33,710 |
| 90 | HfMoTaTiZr | HEA-3 | 2,279 | 971.6 | 3.626e-12 | 10 | 0.055509 | 34,616 |
| 91 | $HfNbSi_{0.5}TiV$ | HEA-3 | 1,999 | 800.0 | 1.28e-29 | 0.0 | 0.17296 | 48,359 |
| 92 | $HfNbSi_{0.5}TiVZr$ | HEA-3 | 1,973 | 527.9 | 0.11598 | 11,762 | 0.28814 | 101,753 |
| 93 | $Hf_{0.25}Nb_{0.25}TiV_{0.5}Zr_{0.5}$ | HEA-3 | 1,891 | 600.0 | -1.30e-29 | 0.0 | 0.26704 | 46,402 |
| 94 | $Hf_{0.25}Nb_{0.5}TiV_{0.5}Zr_{0.5}$ | HEA-3 | 1,944 | 600.0 | 1.381e-29 | 0.0 | 0.23429 | 55,590 |
| 95 | HfNbTaTi | HEA-3 | 2,349 | Insufficient no. of points for high-temp regime | | | 0.004187 | 2,811 |
| 96 | HfNbTaTiWZr | HEA-3 | 2,445 | 800.0 | 0.006052 | 838 | 0.001269 | 1,052 |

**Tab. S3:** Continuation of quantification of ability of compositions to retain ultimate strength at high temperatures. Again, $T_{break}$ refers to the breaking point between the bilinear log models.

| No. | Alloy | Category | Solvus Temp. [ºC] | $T_{break}$ [ºC] | MSE: Two-exponentials Log | MSE: Two-exponentials Linear | MSE: Single-exponential Log | MSE: Single-exponential Linear |
|---|---|---|---|---|---|---|---|---|
| 97 | HfNbTaTiZr (1) | HEA-3 | 2,250 | 923.7 | 0.05267? | 55,965? | 0.08006 | 57,689 |
| 98 | HfNbTaTiZr (2) | HEA-3 | | 901.8 | 8.995e-04 | 420 | 0.17110 | 39,622 |
| 99 | $Hf_{0.26}NbTaTi_{0.578}Zr_{0.416}$ | HEA-3 | 2,400 | Insufficient no. of points for high-temp regime | | | 0.000169 | 282 |
| 100 | $Hf_{0.4}Nb_{1.54}Ta_{1.54}Ti_{0.89}Zr_{0.64}$ | HEA-3 | 2,402 | Insufficient no. of points for high-temp regime | | | 1.64e-04 | 349 |
| 101 | $Hf_{0.5}Nb_{0.667}Ta_{0.333}TiZr_{0.833}$ | HEA-3 | 2,095 | Insufficient no. of points for high-temp regime | | | 0.01306 | 14,952 |
| 102 | $Hf_{0.75}NbTa_{0.5}Ti_{1.5}Zr_{1.25}$ | HEA-3 | 2,096 | Insufficient no. of points for high-temp regime | | | 0.003079 | 6,877 |
| 103 | $Hf_{0.25}NbTaW_{0.5}$ | HEA-3 | 2,999 | 1,260.3 | 0.001394 | 1,739 | 0.002777 | 8,333 |
| 104 | HfNbTiZr | HEA-3 | 2058 | 800.0 | 8.28e-30 | 5 | 0.017099 | 1,227 |
| 105 | $Hf_{0.25}TiV_{0.5}Zr_{0.5}$ | HEA-3 | 1,826 | 600.0 | 5.719e-30 | 0.0 | 0.16210 | 15,545 |
| 106 | $MoNbSi_{0.75}TaW$ | HEA-3 | 2,653 | Insufficient no. of points for high-temp regime | | | 3.16e-30 | 29,312 |
| 107 | MoNbTaTi | HEA-3 | 2,404 | Insufficient no. of points for high-temp regime | | | 0.009497 | 8,162 |
| 108 | MoNbTaTiVW | HEA-3 | 2,520 | Insufficient no. of points for high-temp regime | | | 0.001441 | 11,184 |
| 109 | MoNbTaTiW | HEA-3 | 2,641 | Insufficient no. of points for high-temp regime | | | 0.01091 | 10,642 |
| 110 | MoNbTaVW (1) | HEA-3 | 2,690 | 1,381.6 | 0.0016761 | 2,419 | 0.007094 | 15,052 |
| 111 | MoNbTaVW (2) | HEA-3 | | 1388.7 | 0.0016290 | 2,225 | 0.007099 | 15,074 |
| 112 | MoNbTaW (1) | HEA-3 | ,2,885 | Insufficient no. of points for high-temp regime | | | 0.008323 | 6,991 |
| 113 | MoNbTaW (2) | HEA-3 | | Insufficient no. of points for high-temp regime | | | 0.009019 | 8,367 |
| 114 | MoNbTi | HEA-3 | 2,256 | 1000.0 | 8.875e-30 | 247 | 0.010637 | 16,165 |
| 115 | MoNbTiVZr | HEA-3 | 2,107 | 1000.0 | 5.325e-30 | 0.0 | 0.16997 | 17,795 |
| 116 | $NbRe_{0.3}TiZr$ | HEA-3 | 2,108 | 1000.0 | 1.302e-29 | 0.0 | 0.13719 | 7,441 |
| 117 | NbTaTi | HEA-3 | 2,387 | 1000.0 | 0.018186 | 1,691 | 0.01829 | 2,029 |
| 118 | NbTaTiVZr | HEA-3 | 2,185 | 1000.0 | 0.001323 | 4,697 | 0.02112 | 7,439 |
| 119 | NbTaTiW | HEA-3 | 2,646 | 1209 | 0.0045432 | 3,383 | 0.004626 | 3,712 |
| 120 | $NbTa_{0.3}TiZr$ | HEA-3 | 2,092 | 1000.0 | 6,705e-30 | 0.4 | 0.07497 | 3,090 |
| 121 | $NbTiV_2Zr$ | HEA-3 | 1,964 | 664.9 | 2.858e-11 | 0.0 | 0.18977 | 33,200 |
| 122 | NbTiVZr | HEA-3 | 1,978 | 588.4 | 0.003697 | 733.4 | 0.35011 | 110,362 |
| 123 | NbTiZr | HEA-3 | 2,000 | 821.2 | 0.015832 | 507 | 0.17549 | 20,873 |
| 124 | $Nb_{30}Ti_{40}Zr_{30}$ | HEA-3 | 1,967 | 481.8 | 0.029842 | 1,288.4 | 0.052614 | 3,805 |
| **Average** (cases of insufficient no. of data points for high-temp regime omitted): | | | | | 0.003496 | 2,144 | 0.145413 | 56,509 |

**Tab. S4:** Quantification of ability of compositions to retain ultimate strengths at high temperatures. $T_{break}$ refers to the breaking point between bilinear log models. The dashed line refers to the information not available. The compositions are numbered in a fashion consistent with **Tab. S1**. The slope in the high-temperature regime refers to **Fig. 2(a)**, **2(c)** and **2(e)** as well as corresponding Figures in the Supplementary Manuscript (Supplementary **Fig. S2** – Supplementary **Fig. S123**).

| No. | Alloy | Solvus Temperature [ºC] | $T_{break}$ [ºC] | Yield Strength $C_3$: Slope for high-temp. regime in Figs. 2(a) –4(a) | $T_{break}$ [ºC] | Ultimate Strength $C_3$: Slope for high-temp. regime in Figs. 2(a) –4(a) |
|---|---|---|---|---|---|---|
| 4 | $Al_{0.6}CoCrFeNi$ | 1,476 | 973.1 | -8.11 | 953.9 | -8.11 |
| 15 | $Al_{15}Cr_{20}Nb_{15}Ti_{40}Zr_{10}$ | 1,705 | 604.5 | -13.10 | 604.0 | -13.20 |
| 16 | AlCrMoNbTi (1) | 1,867 | 950.1 | -16.14 | 943.7 | -15.80 |
| 17 | AlCrMoNbTi (2) | 1,867 | 934.9 | -12.13 | | |
| 19 | AlCrNbTiV | 1,725 | 798.5 | -22.27 | 798.9 | -19.35 |
| 20 | $AlCr_{0.5}NbTiV$ | 1,704 | 773.0 | -23.62 | 769.3 | -18.30 |
| 21 | $AlCr_{1.5}NbTiV$ | 1,741 | 773.8 | -22.28 | 778.3 | -20.00 |
| 22 | $Al_{0.4}Hf_{0.6}NbTaTiZr$ | 2,124 | 845.2 | -12.83 | 927.3 | -12.90 |
| 23 | $AlMo_{0.5}NbTa_{0.5}TiZr$ | 1,896 | 789.8 | -8.79 | 770.1 | -8.93 |
| 27 | $Al_{10}Mo_{10}Nb_{20}Ti_{40}V_{20}$ | 1,873 | 800.0 | -11.34 | 800.0 | -9.47 |
| 28 | $Al_{15}Mo_{10}Nb_{20}Ti_{35}V_{20}$ | 1,823 | 800.0 | -11.25 | 800.0 | -10.94 |
| 29 | $Al_{20}Mo_{10}Nb_{20}Ti_{30}V_{20}$ | 1,772 | 800.0 | -11.02 | 800.0 | -8.37 |
| 30 | $Al_{0.2}NbTaTiV$ | 2,198 | 872.2 | -11.70 | 898.0 | -12.53 |
| 32 | $Al_{0.3}NbTa_{0.8}Ti_{1.4}V_{0.2}Zr_{1.3}$ | 2,043 | 800.0 | -14.37 | 800.0 | -13.77 |
| 35 | $Al_{0.5}NbTa_{0.8}Ti_{1.5}V_{0.2}Zr$ | 1,992 | 800..0 | -12.81 | 800.0 | -13.25 |
| 38 | $Al_{0.3}NbTaTi_{1.4}Zr_{1.3}$ | 2,088 | 800.0 | -4.47 | 800.0 | -4.95 |
| 40 | $AlNb_{1.5}Ta_{0.5}Ti_{1.5}Zr_{0.5}$ | 1,863 | 800.0 | -5.51 | 779.5 | -8.19 |
| 67 | $CrMo_{0.5}NbTa_{0.5}TiZr$ | 2,145 | 911.1 | -12.51 | 923.5 | -12.86 |
| 73 | $Cr_{20}Nb_{30}Ti_{40}Zr_{10}$ | 1,977 | 635.5 | -14.27 | 663.7 | -15.62 |
| 88 | HfMoNbTiZr (2) | 2,297 | 982.0 | -14.04 | 948.0 | -13.96 |
| 92 | $HfNbSi_{0.5}TiVZr$ | 1,973 | 527.85 | -13.20 | 527.5 | -17.81 |
| 110 | MoNbTaVW (1) | 2,690 | 1,381.6 | -4.29 | 970.2 | -4.85 |
| 111 | MoNbTaVW (2) | | 1388.7 | -4.29 | | |

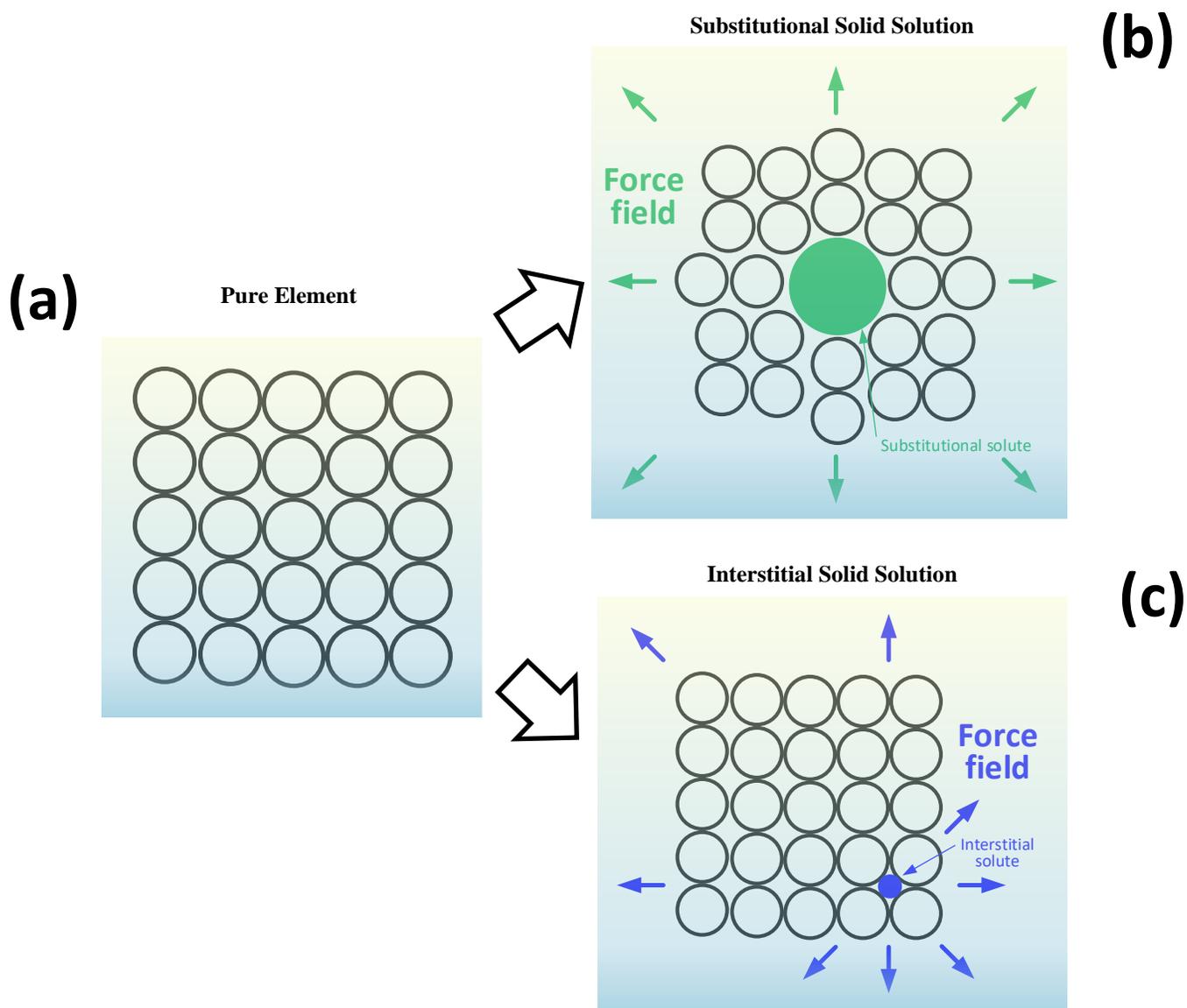

**Fig. S1**: Schematic diagram of a solid solution. The figure to the left describes a pure element, for which all the atoms are equi-sized. The figure to the top right represents a substitutional solid solution (solute), where the central atom has been substituted by a larger atom. The green element is considered a substitutional solute. The figure to the bottom right describes an interstitial solid solution (solute), where the blue atom has been inserted into the lattice creating a local force (strain) field.

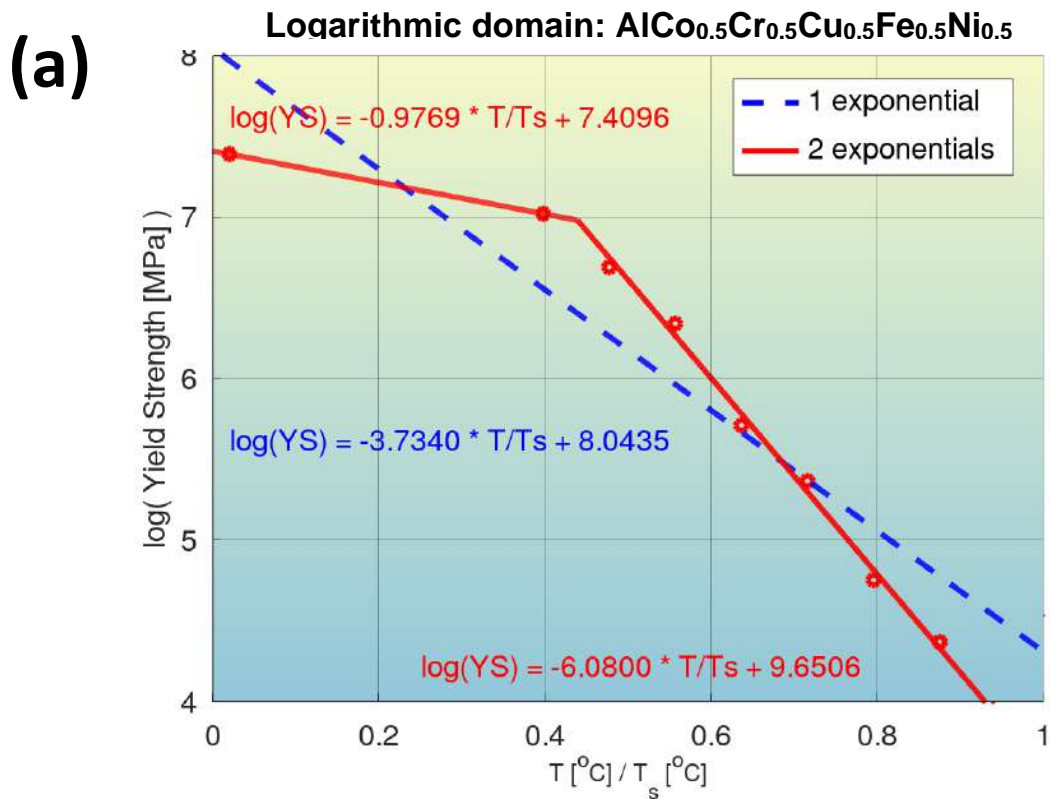

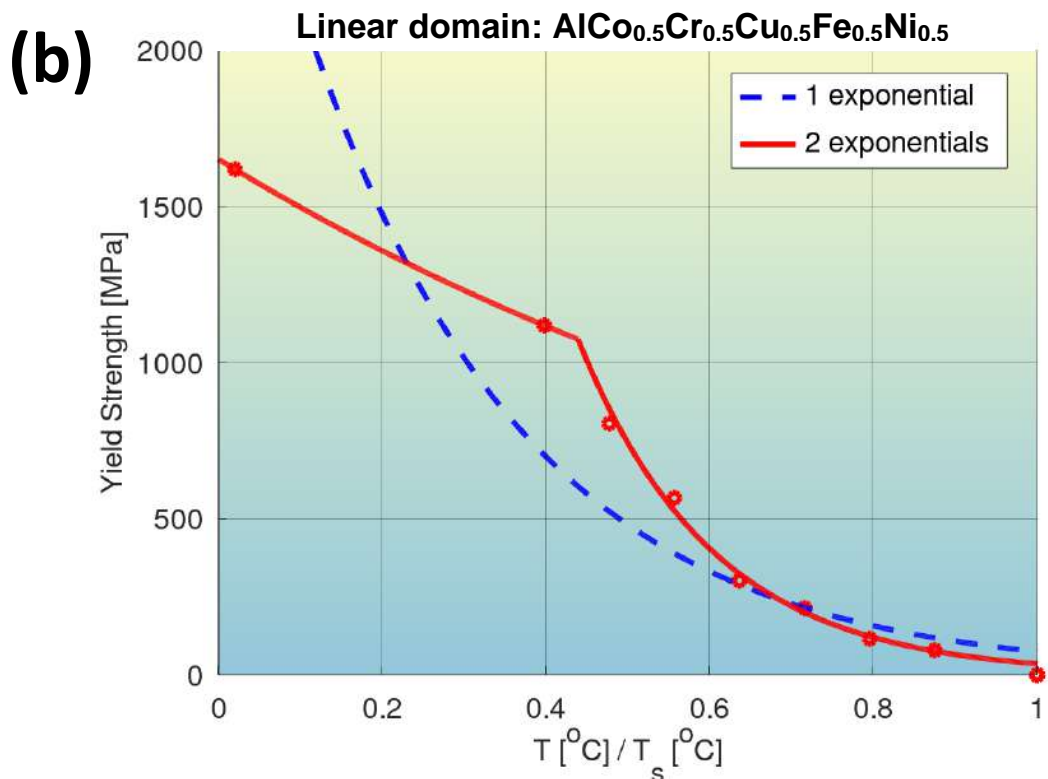

**Fig. S2**: Quantification of modeling accuracy of the bilinear log model, for composition No. 1 from **Tab. S1** (AlCo$_{0.5}$Cr$_{0.5}$Cu$_{0.5}$Fe$_{0.5}$Ni$_{0.5}$, BCC phase), and comparison to that of a model with a single exponential.

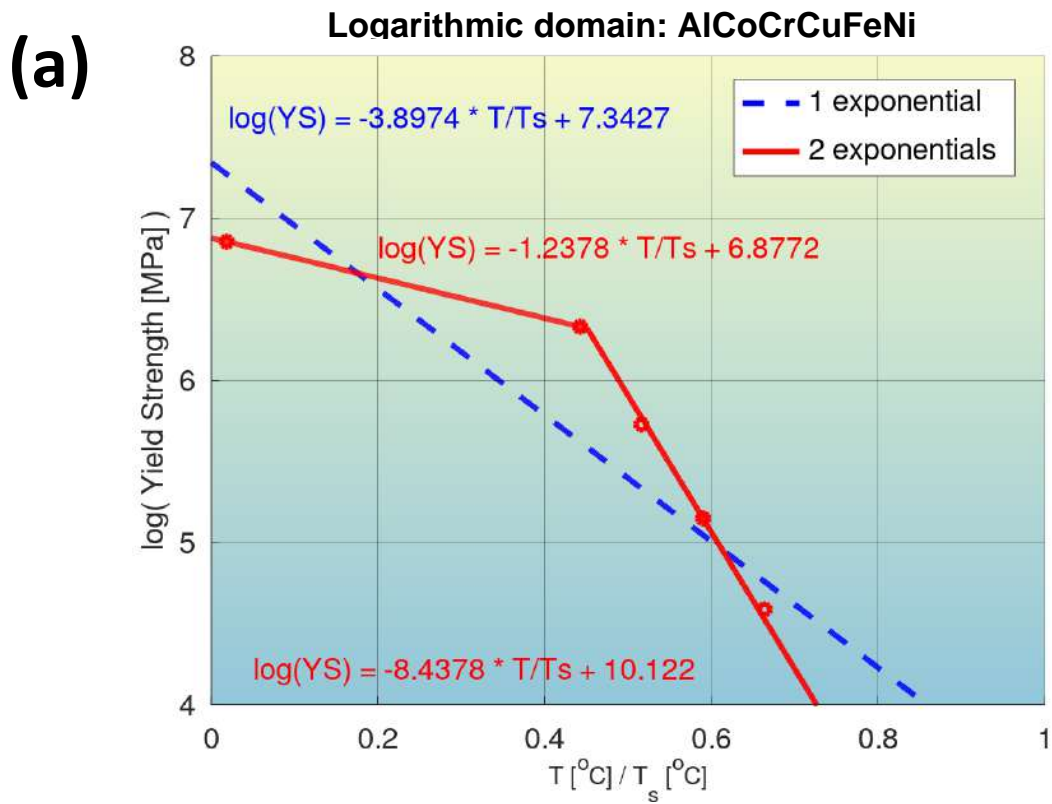

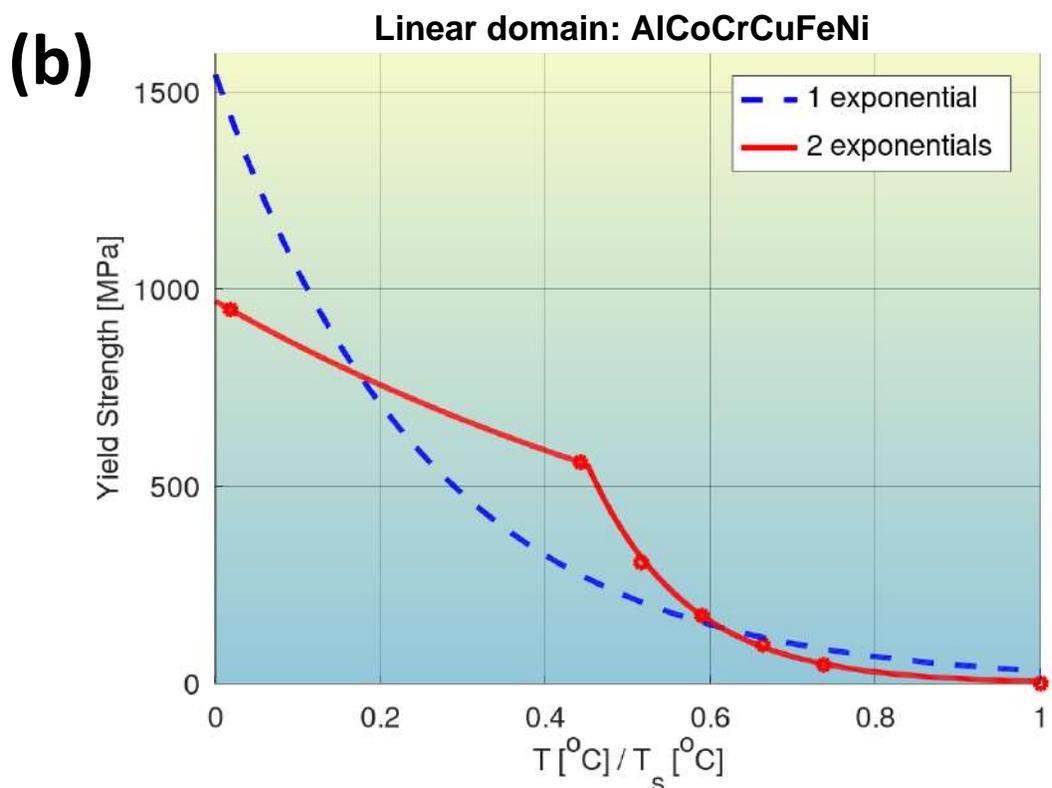

**Fig. S3**: Quantification of modeling accuracy of the bilinear log model, for composition No. 2 from **Tab. S1** (AlCoCrCuFeNi, FCC+BCC phase), and comparison to that of a model with a single exponential.

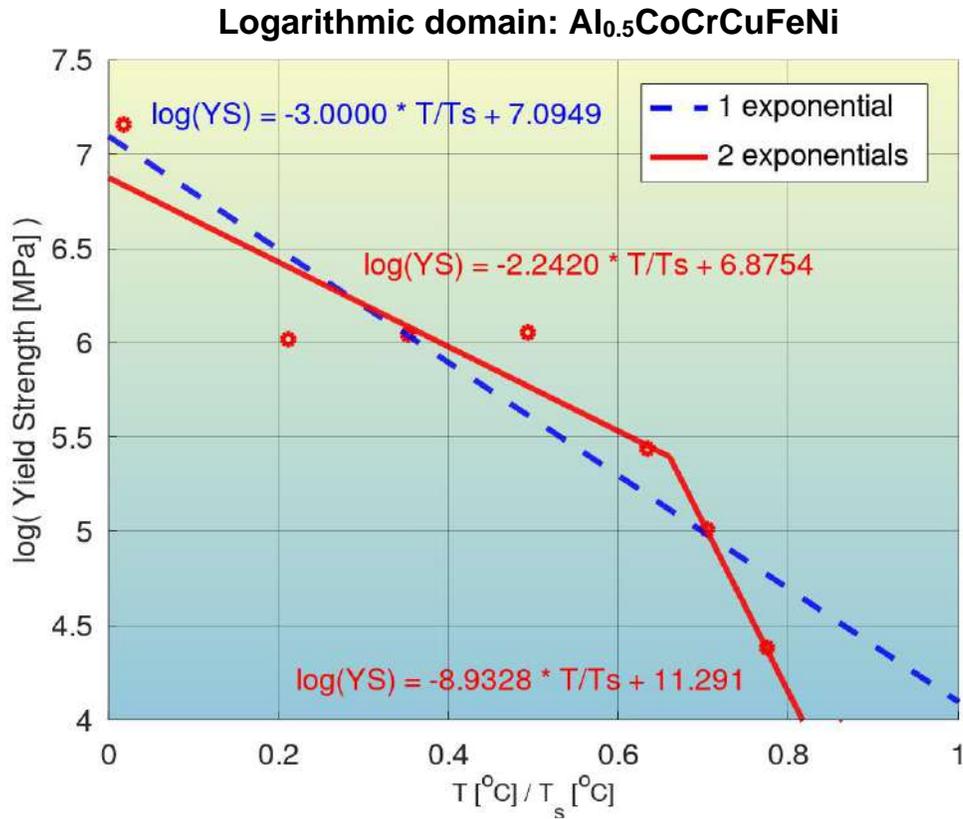

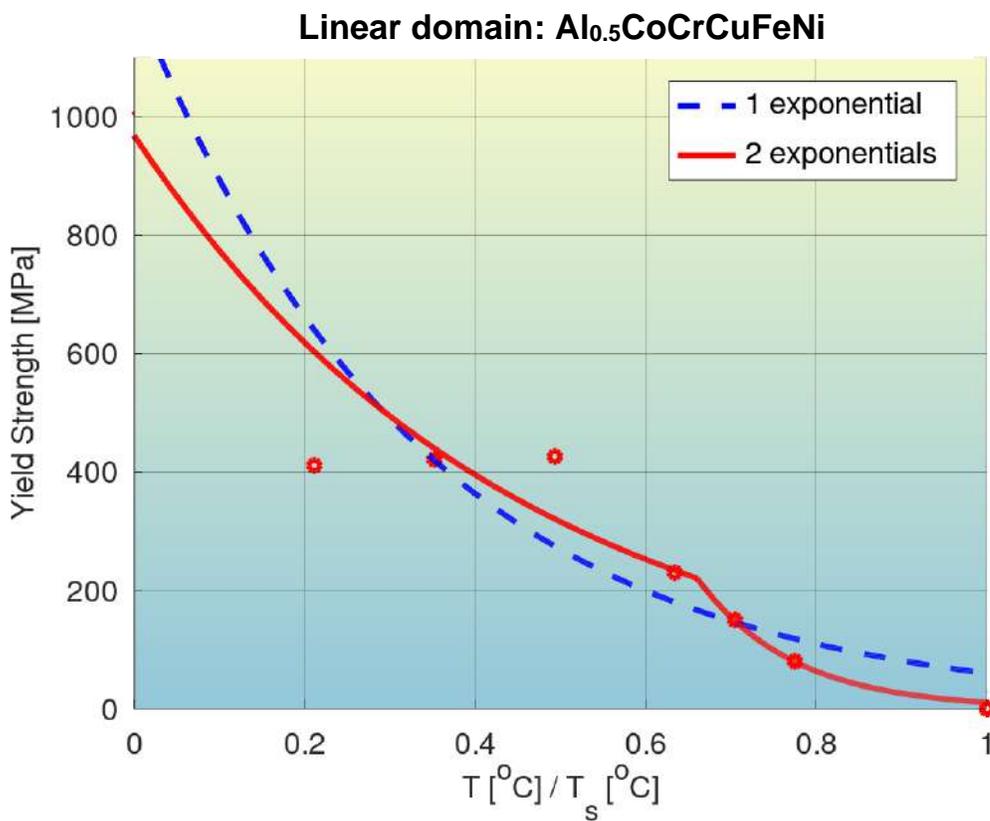

**Fig. S4**: Quantification of modeling accuracy of the bilinear log model, for composition No. 3 from **Tab. S1** (Al$_{0.5}$CoCrCuFeNi, FCC phase), and comparison to that of a model with a single exponential.

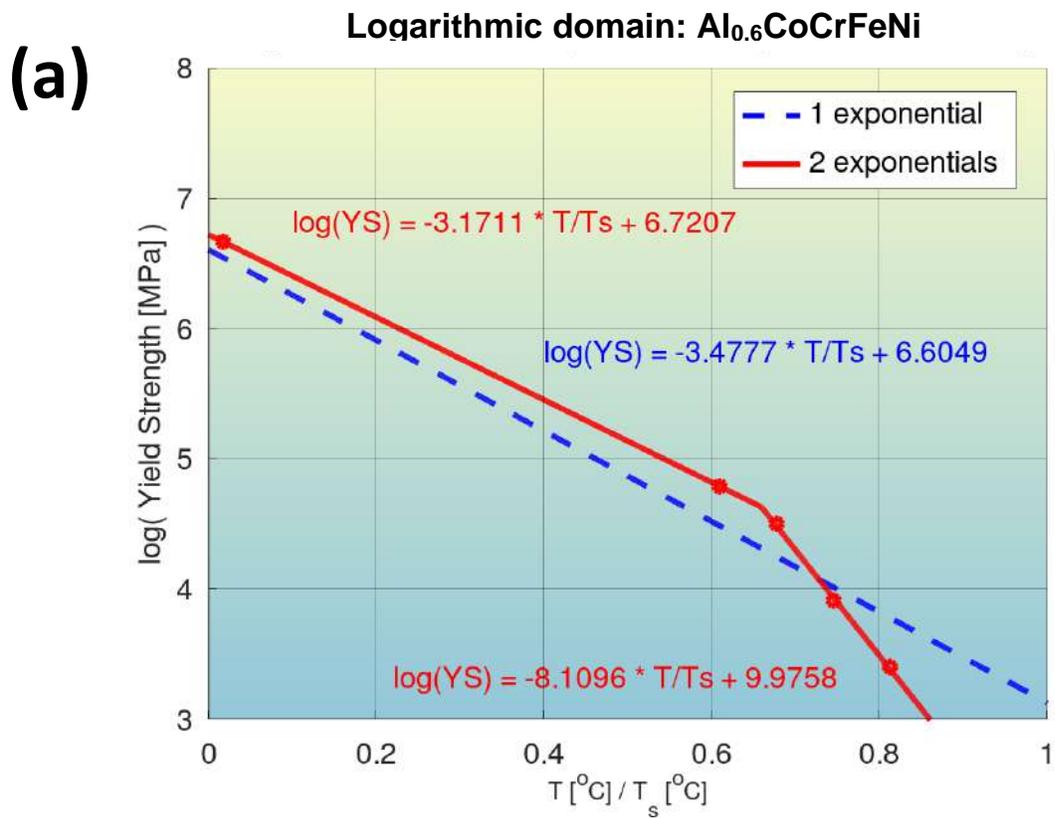

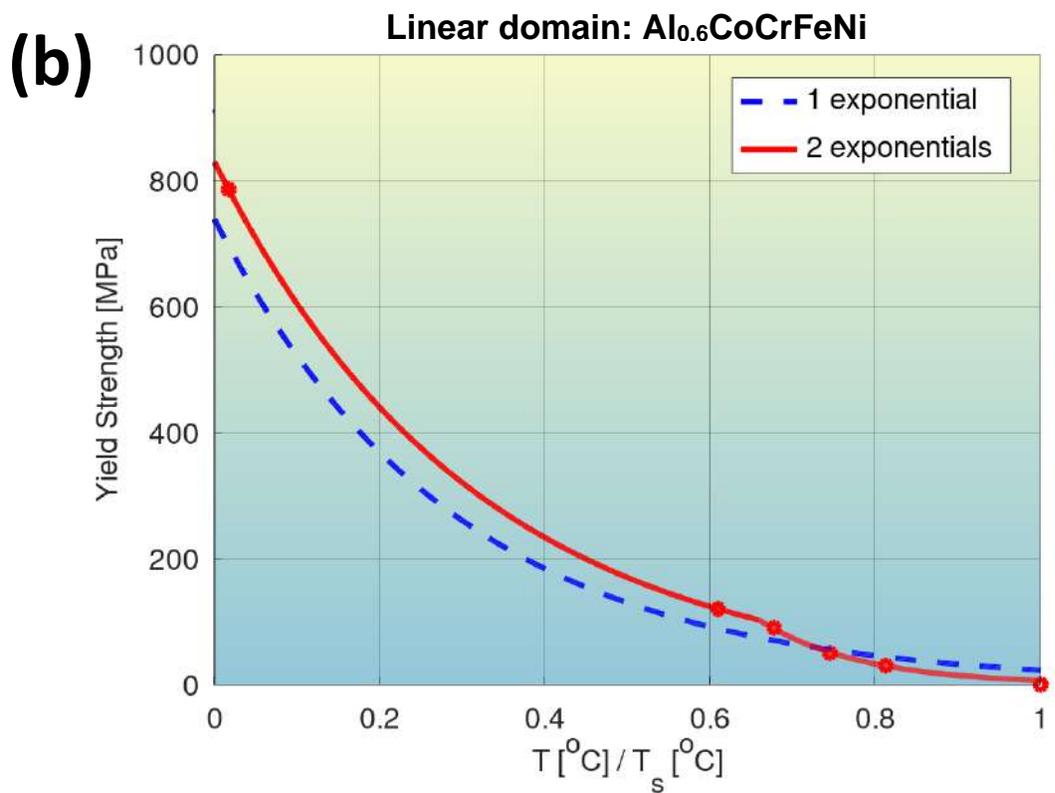

**Fig. S5**: Quantification of modeling accuracy of the bilinear log model, for composition No. 4 from **Tab. S1** (Al$_{0.6}$CoCrFeNi, FCC+BCC phase, strain rate = 0.001), and comparison to that of a model with a single exponential.

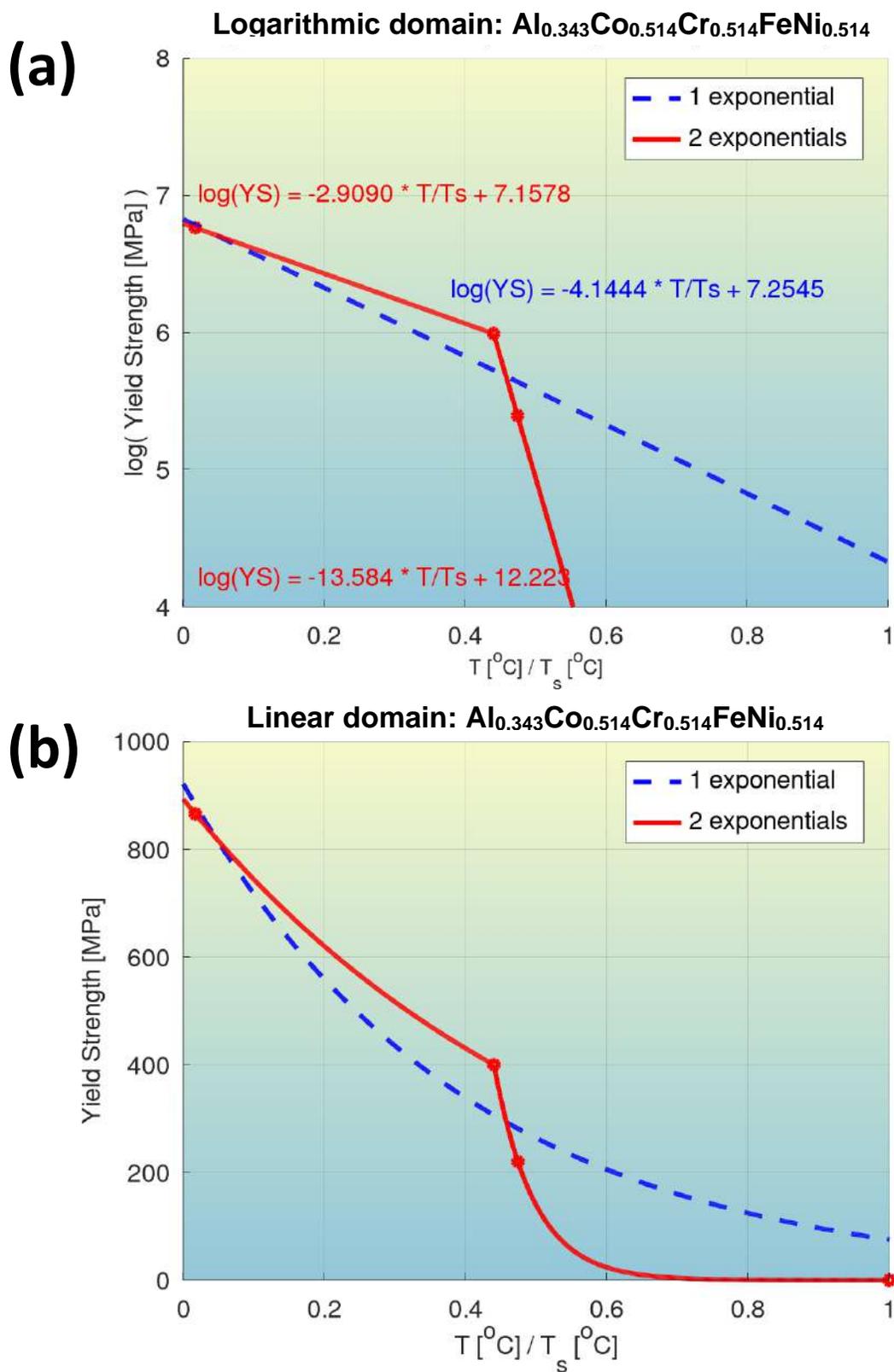

**Fig. S6**: Quantification of modeling accuracy of the bilinear log model, for composition No. 5 from **Tab. S1** ($Al_{0.343}Co_{0.514}Cr_{0.514}FeNi_{0.514}$, BCC+B2 phase), and comparison to that of a model with a single exponential.

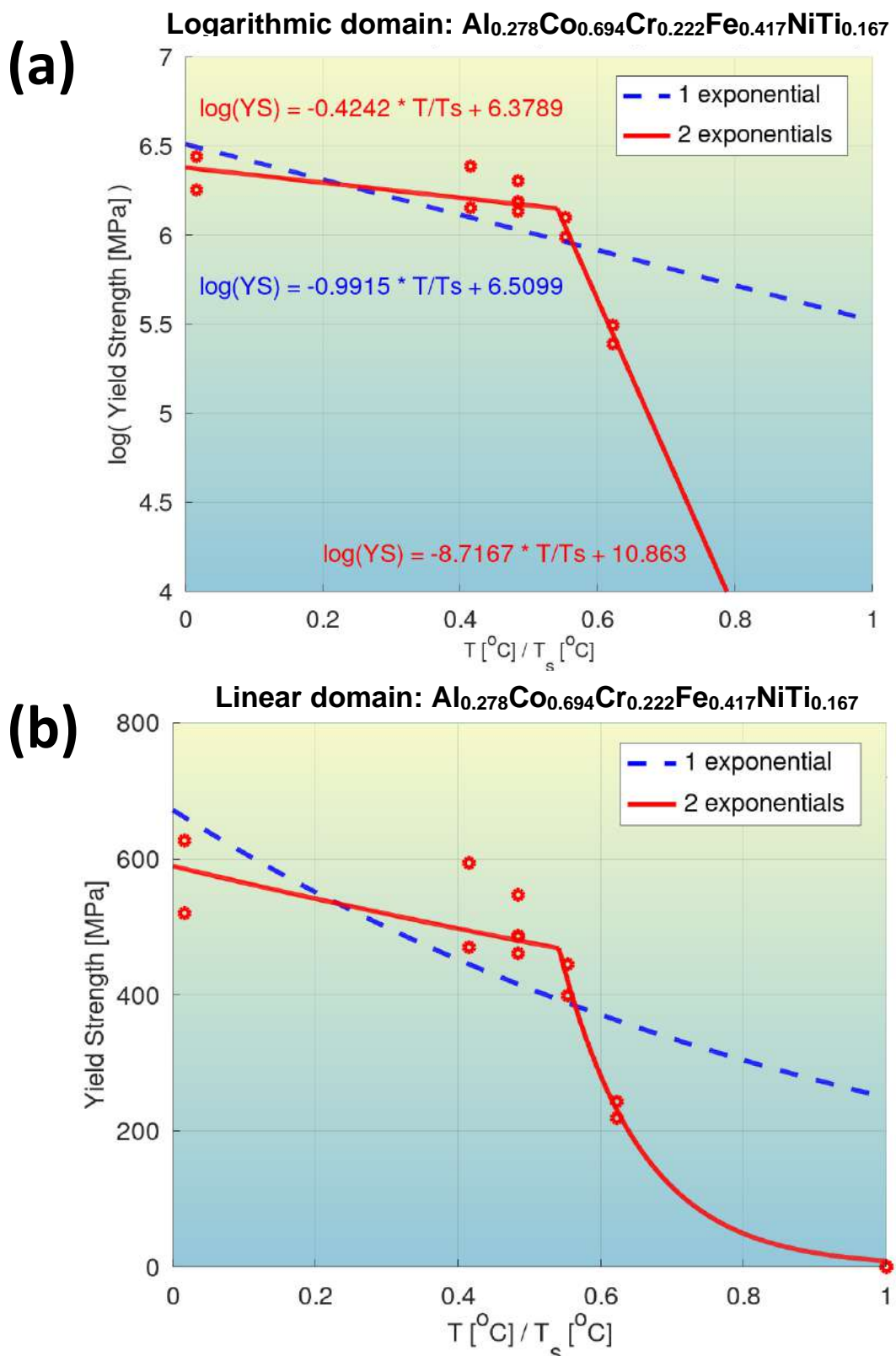

**Fig. S7**: Quantification of modeling accuracy of the bilinear log model, for composition No. 6 from **Tab. S1** (Al$_{0.278}$Co$_{0.694}$Cr$_{0.222}$Fe$_{0.417}$NiTi$_{0.167}$, FCC phase), and comparison to that of a model with a single exponential.

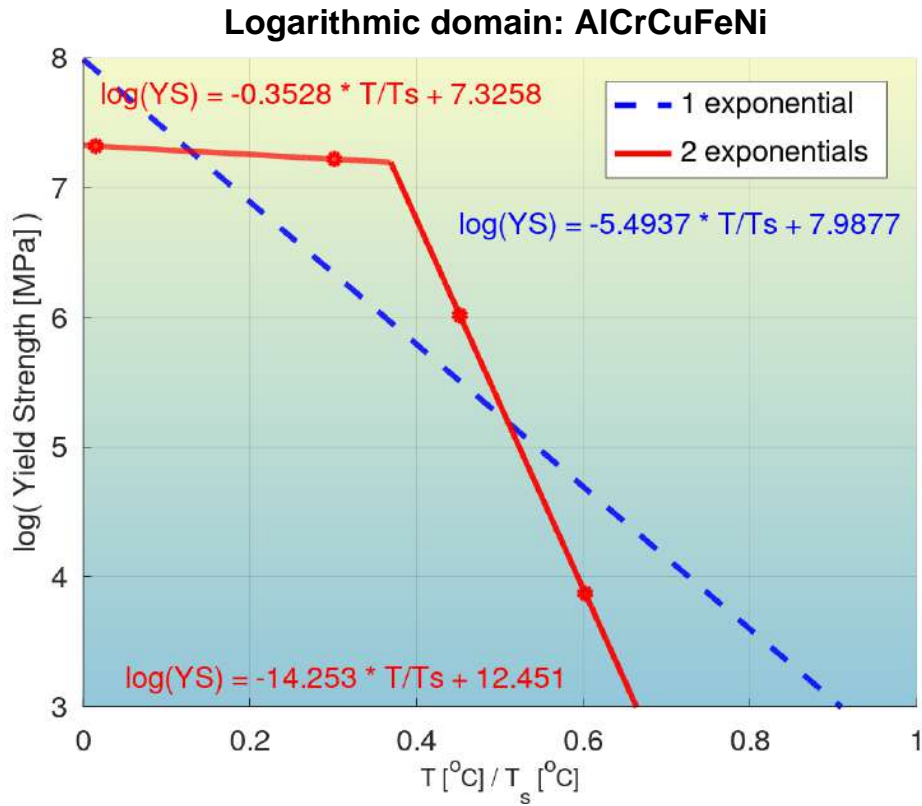

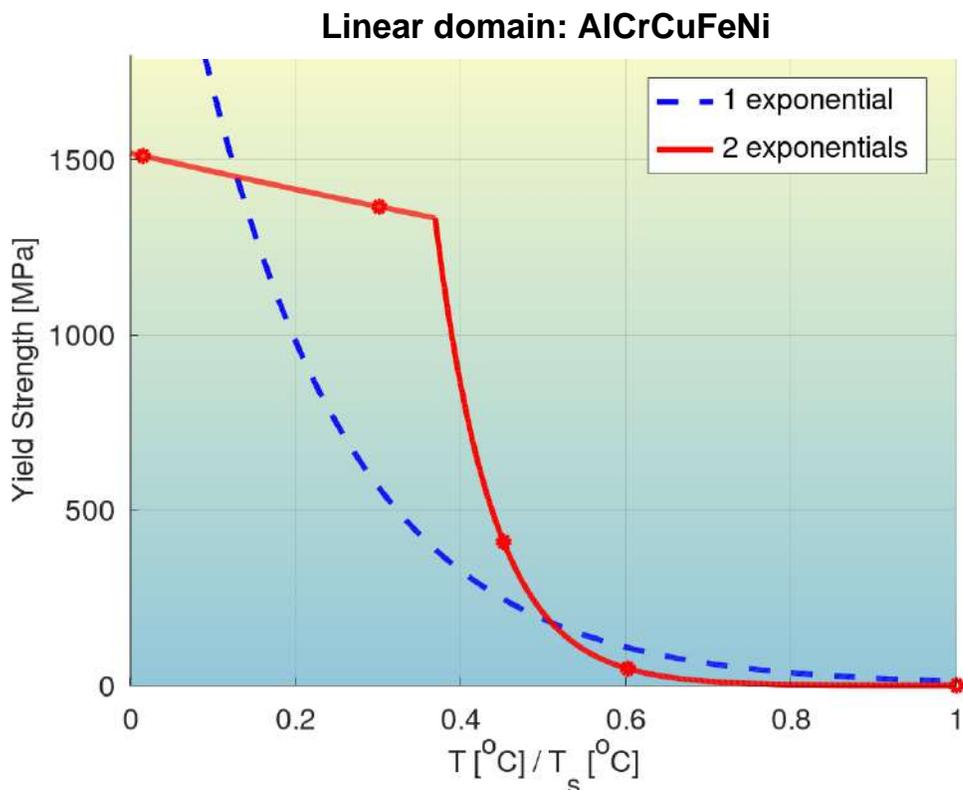

**Fig. S8**: Quantification of modeling accuracy of the bilinear log model, for composition No. 7 from **Tab. S1** (AlCrCuFeNi, FCC+BCC or FCC+BCC+IM phase), and comparison to that of a model with a single exponential.

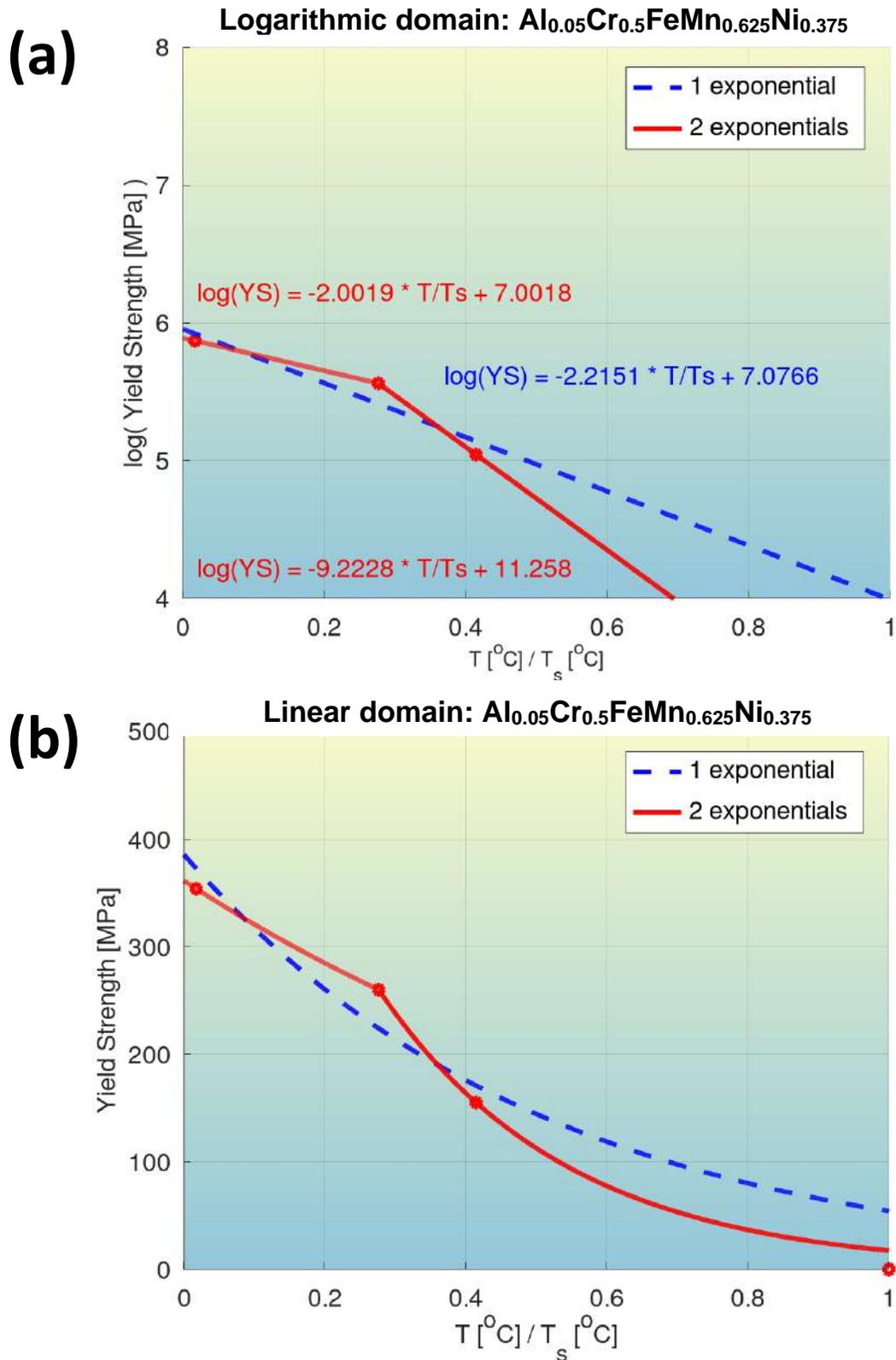

**Fig. S9**: Quantification of modeling accuracy of the bilinear log model, for composition No. 8 from **Tab. S1** (Al$_{0.05}$Cr$_{0.5}$FeMn$_{0.625}$Ni$_{0.375}$, FCC+BCC phase), and comparison to that of a model with a single exponential.

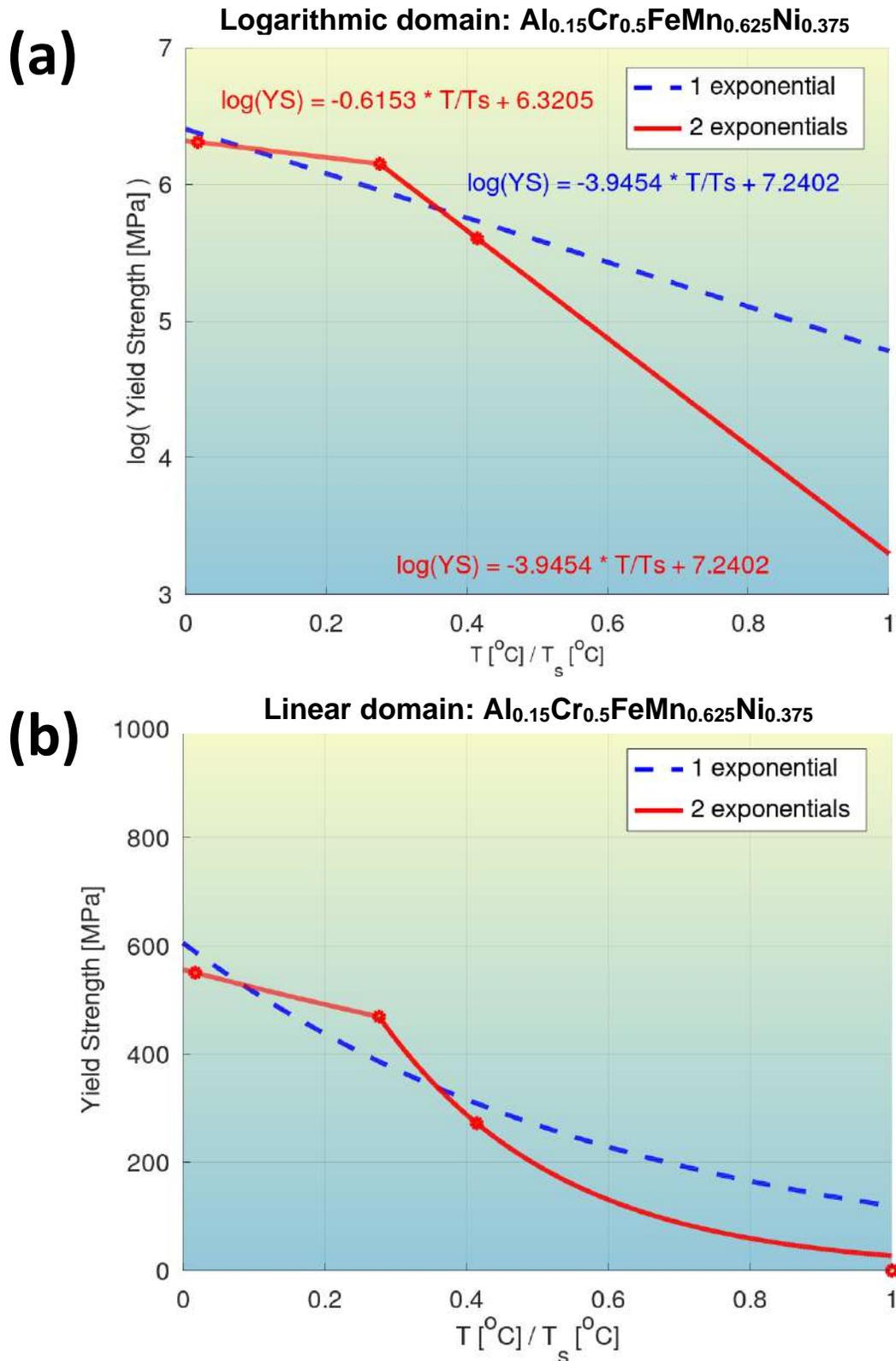

**Fig. S10**: Quantification of modeling accuracy of the bilinear log model, for composition No. 9 from **Tab. S1** (Al$_{0.15}$Cr$_{0.5}$FeMn$_{0.625}$Ni$_{0.375}$, FCC+BCC phase), and comparison to that of a model with a single exponential.

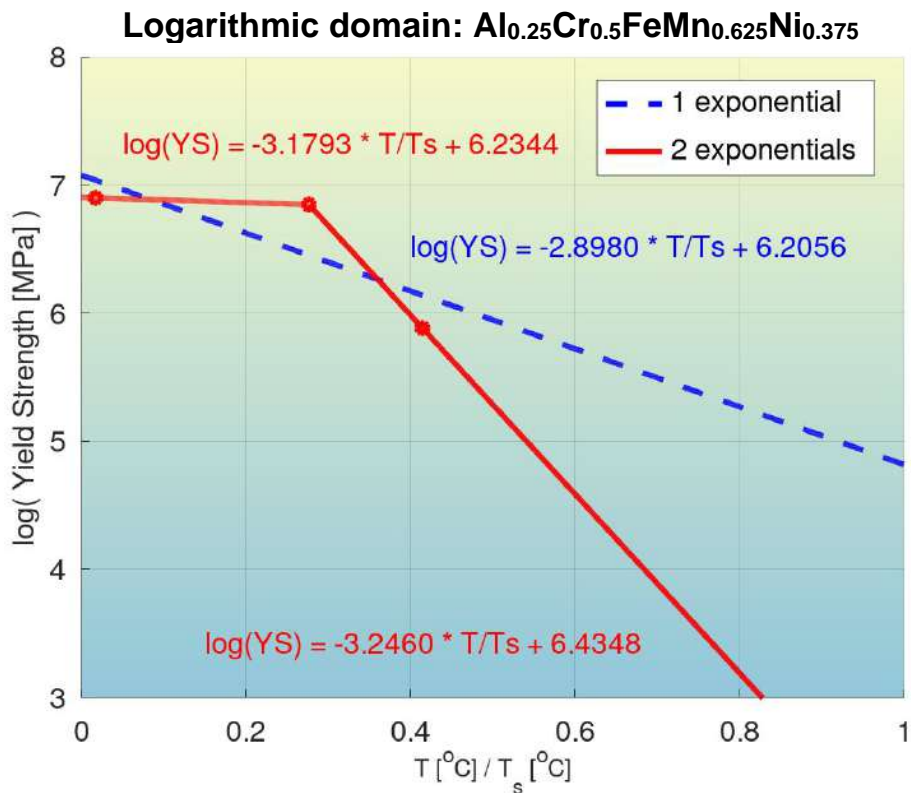
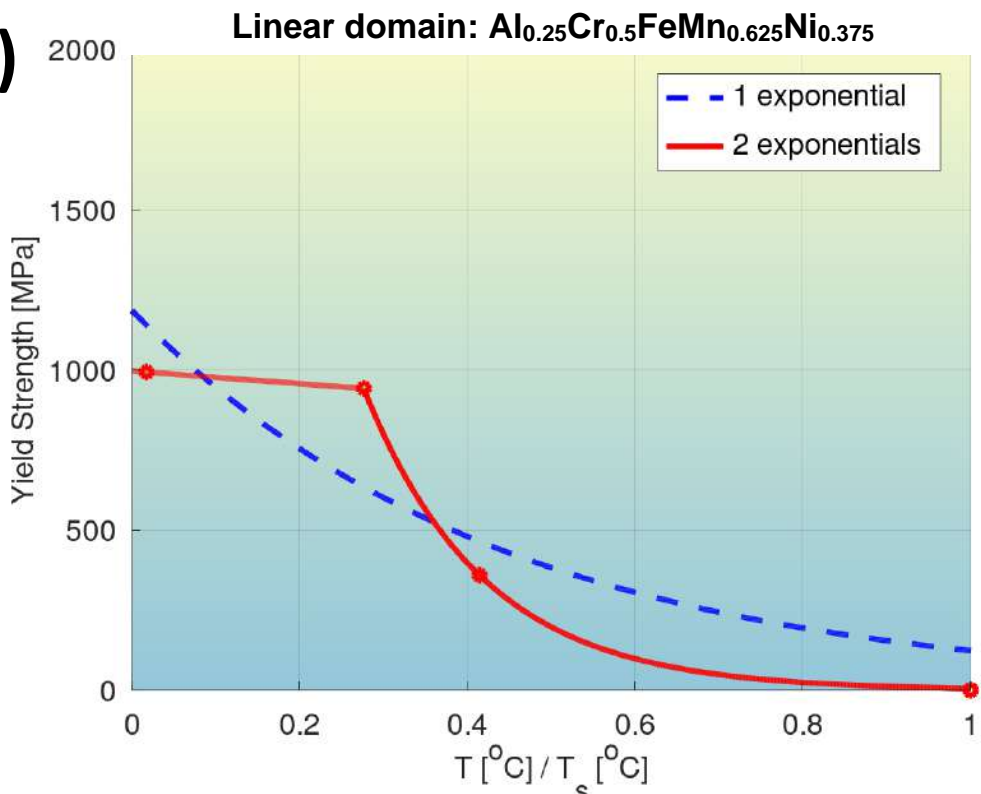

**Fig. S11**: Quantification of modeling accuracy of the bilinear log model, for composition No. 10 from **Tab. S1** ($Al_{0.25}Cr_{0.5}FeMn_{0.625}Ni_{0.375}$, BCC+B2 phase, and comparison to that of a model with a single exponential.
15

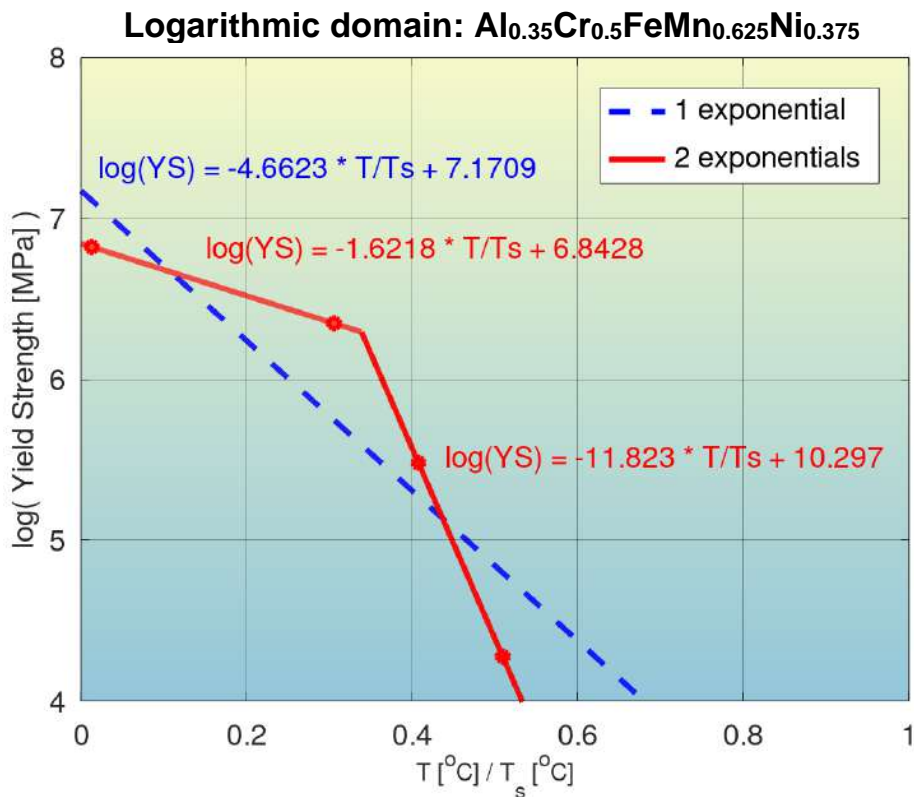

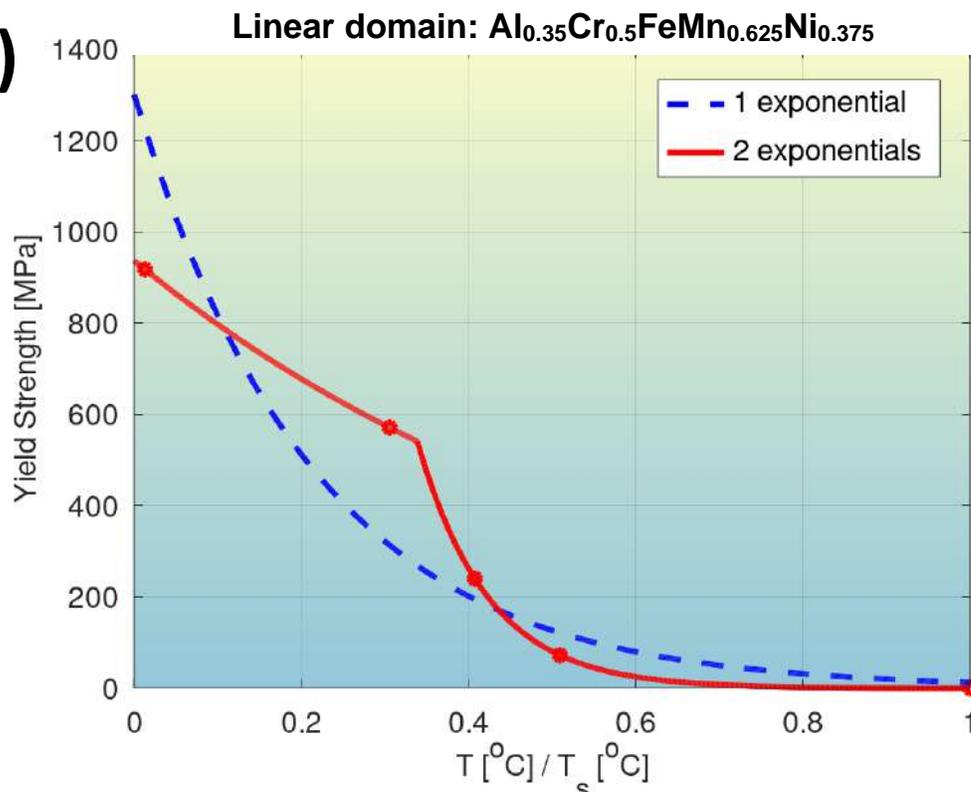

**Fig. S12**: Quantification of modeling accuracy of the bilinear log model, for composition No. 11 from **Tab. S1** (Al$_{0.35}$Cr$_{0.5}$FeMn$_{0.625}$Ni$_{0.375}$, BCC+B2 phase), and comparison to that of a model with a single exponential.

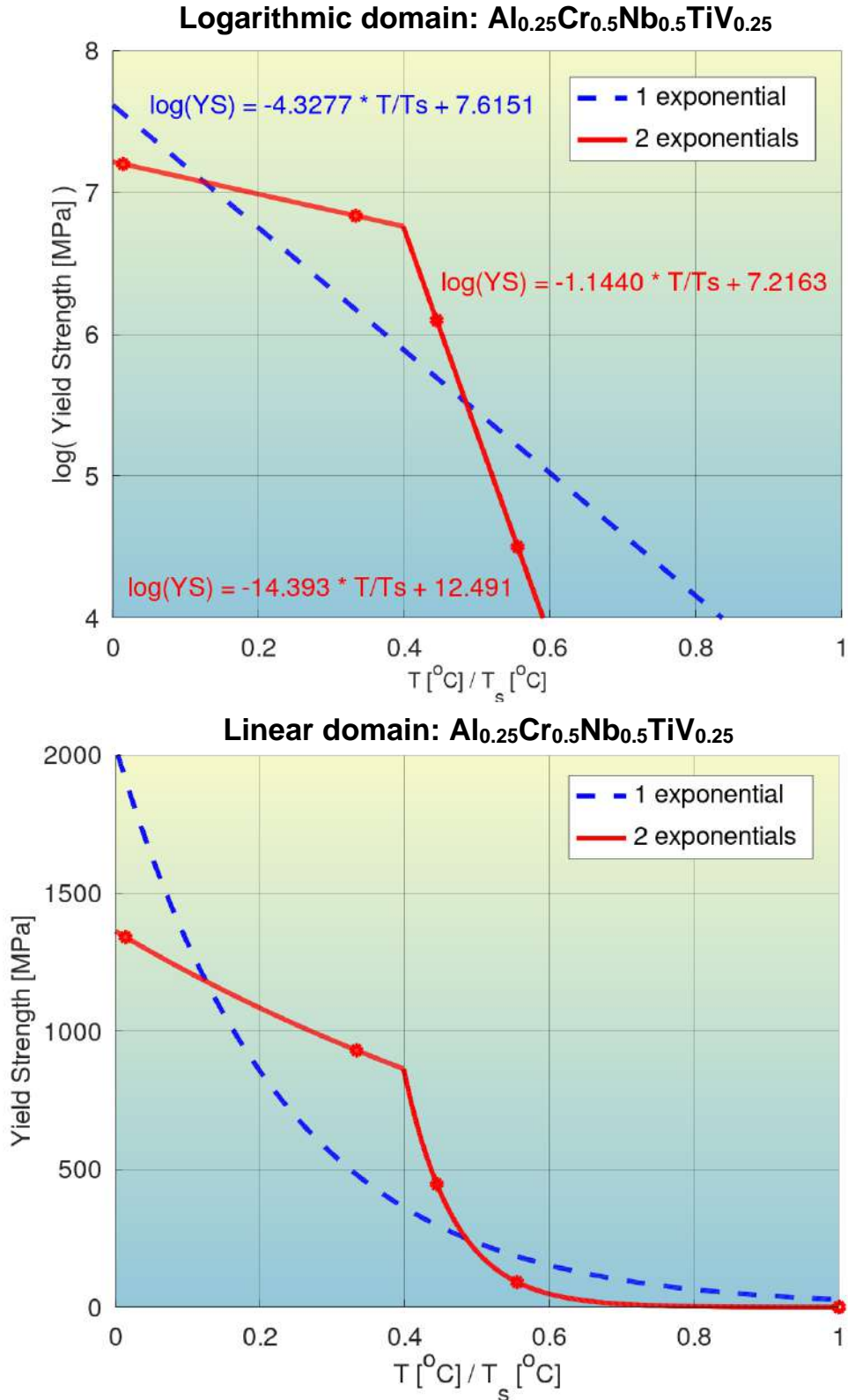

**Fig. 13**: Quantification of modeling accuracy of the bilinear log model, for composition No. 12 from **Tab. S1** ($Al_{0.25}Cr_{0.5}Nb_{0.5}TiV_{0.25}$, BCC+Laves phase), and comparison to that of a model with a single exponential.

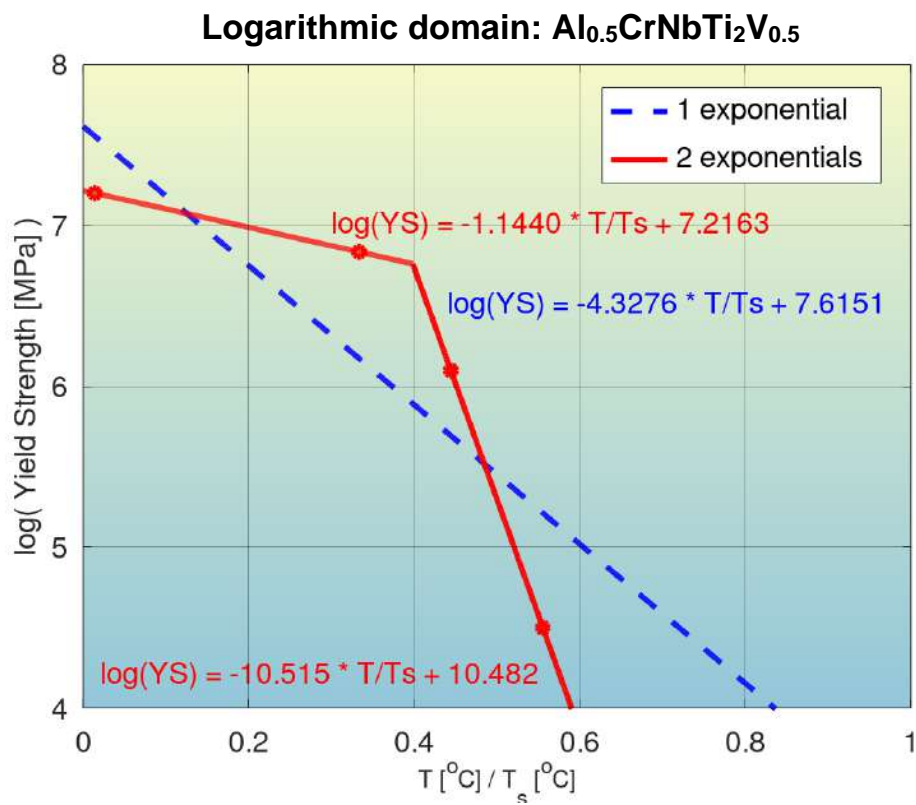

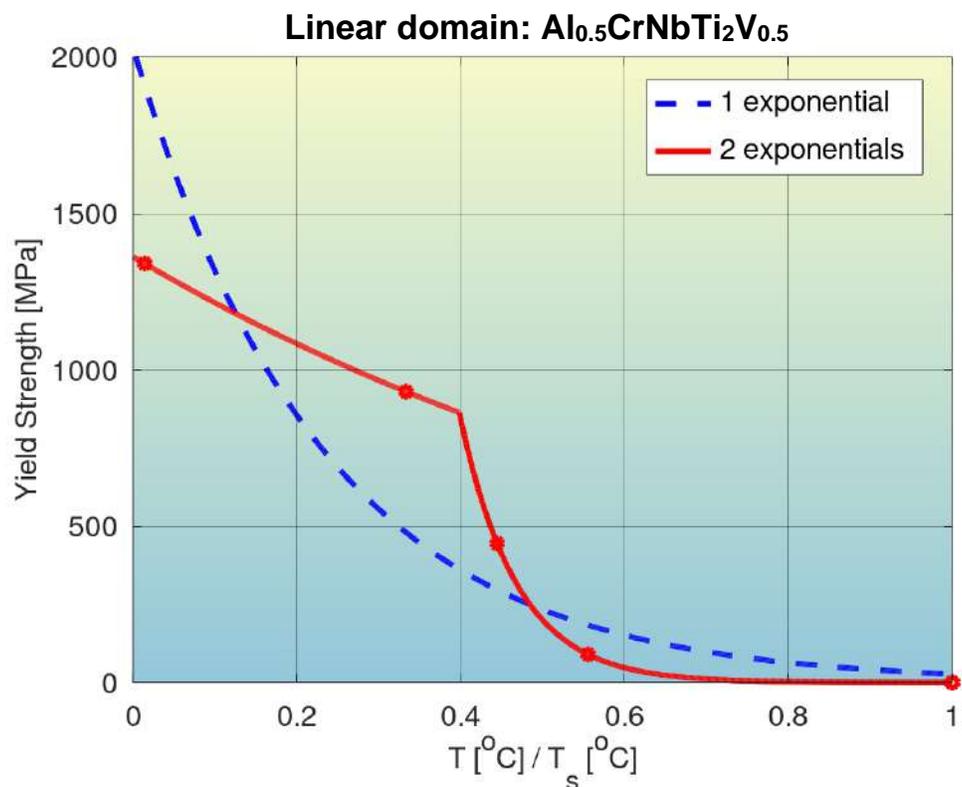

**Fig. S14**: Quantification of modeling accuracy of the bilinear log model, for composition No. 13 from **Tab. S1** (Al$_{0.5}$CrNbTi$_2$V$_{0.5}$, BCC+Laves phase), and comparison to that of a model with a single exponential.

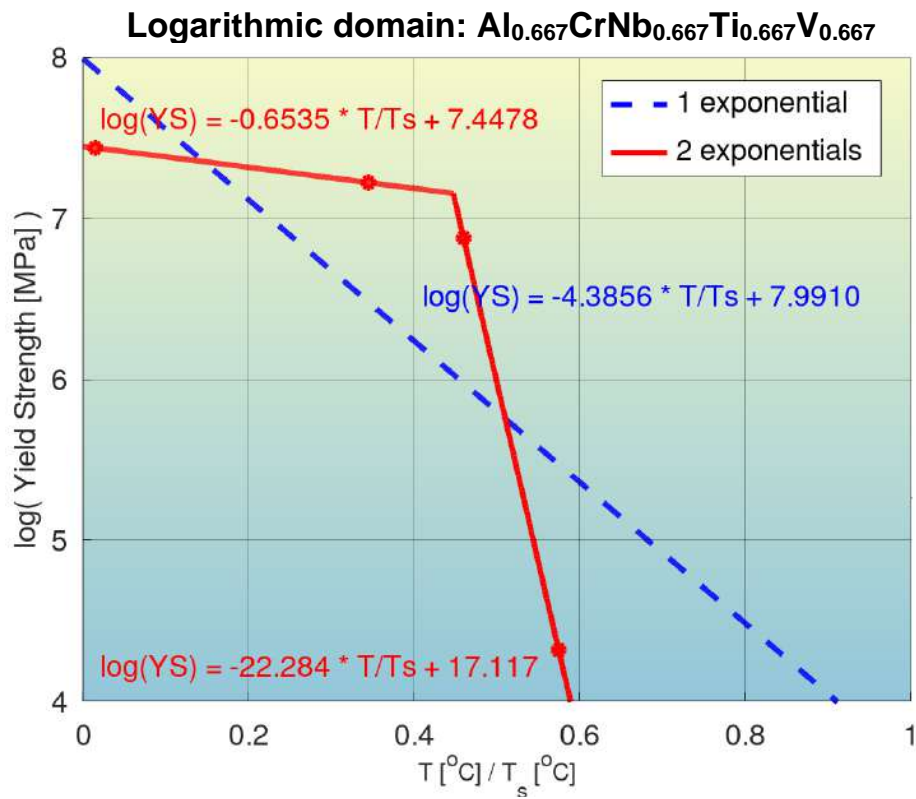

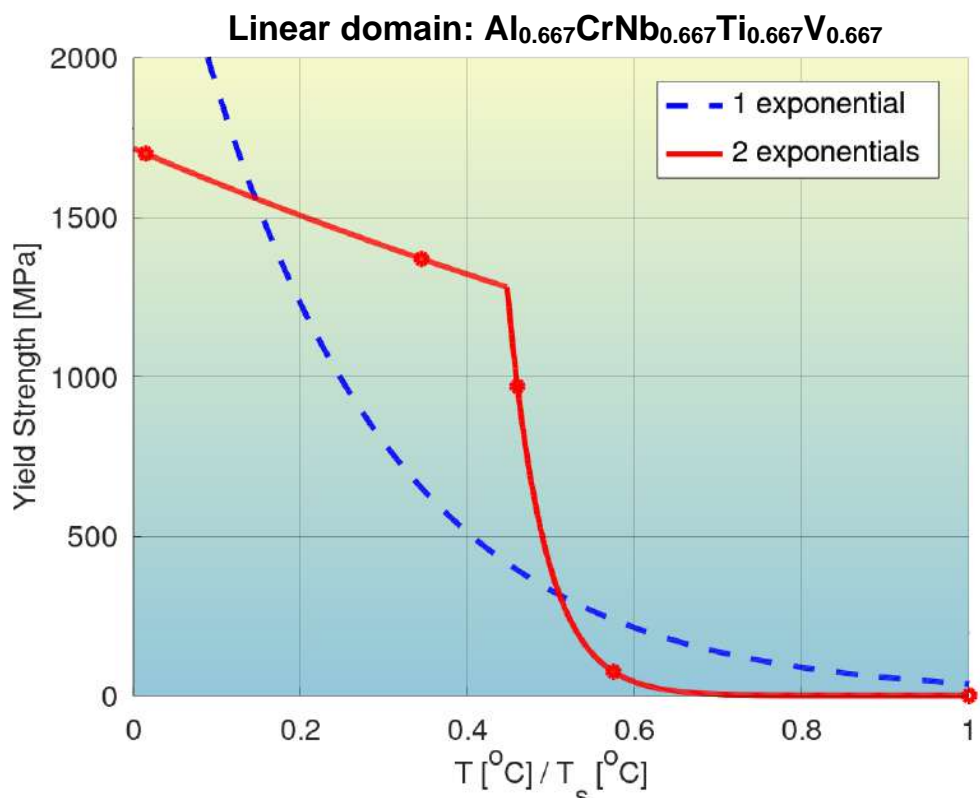

**Fig. S15**: Quantification of modeling accuracy of the bilinear log model, for composition No. 14 from **Tab. S1** ($Al_{0.667}CrNb_{0.667}Ti_{0.667}V_{0.667}$, BCC+Laves phase), and comparison to that of a model with a single exponential.

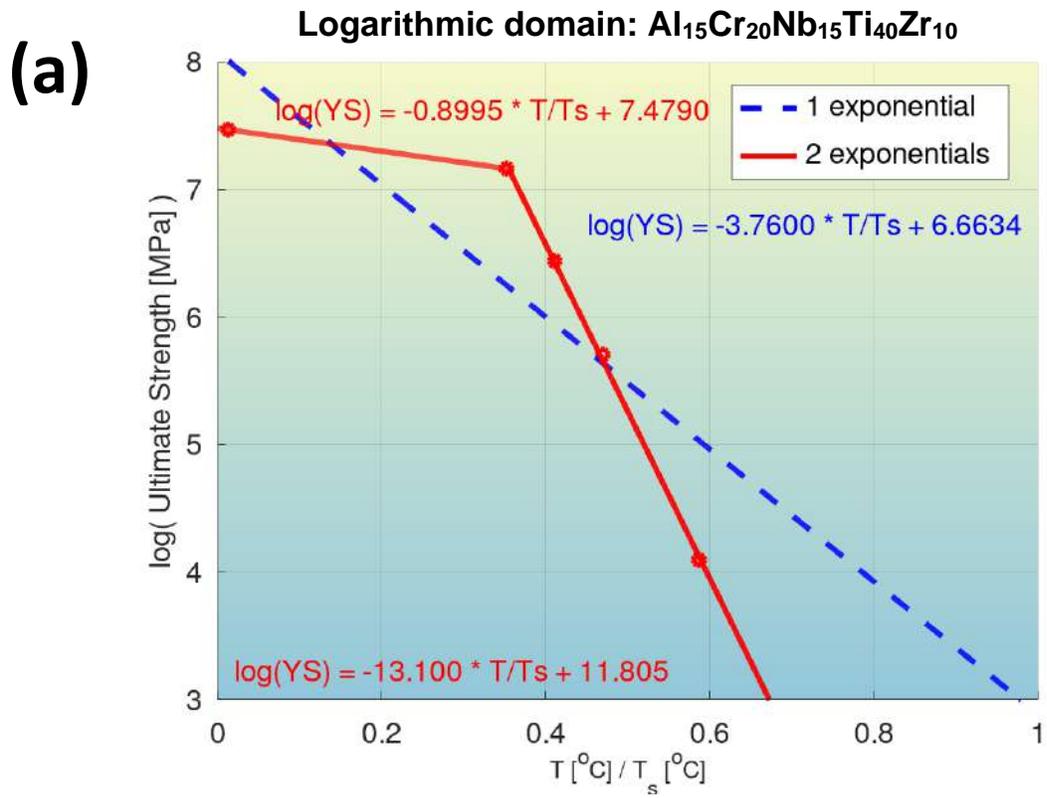

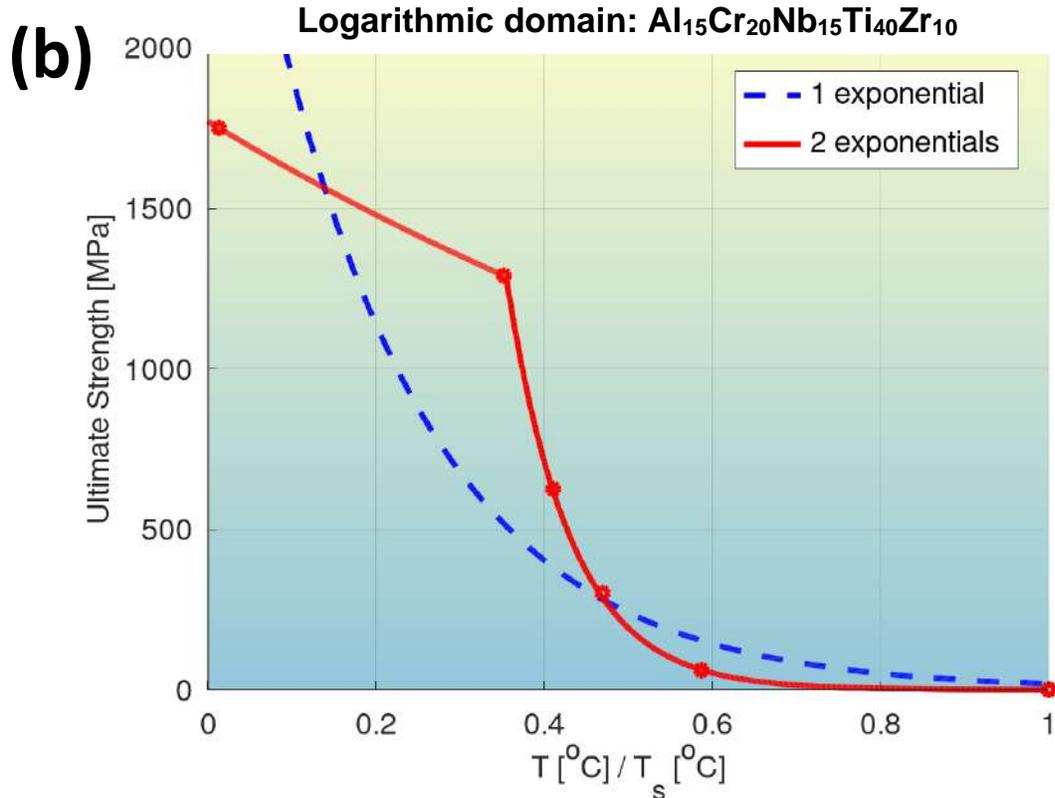

**Fig. S16**: Quantification of modeling accuracy of the bilinear log model, for composition No. 15 from **Tab. S1** ($Al_{15}Cr_{20}Nb_{15}Ti_{40}Zr_{10}$, B2+Laves phase), and comparison to that of a model with a single exponential.

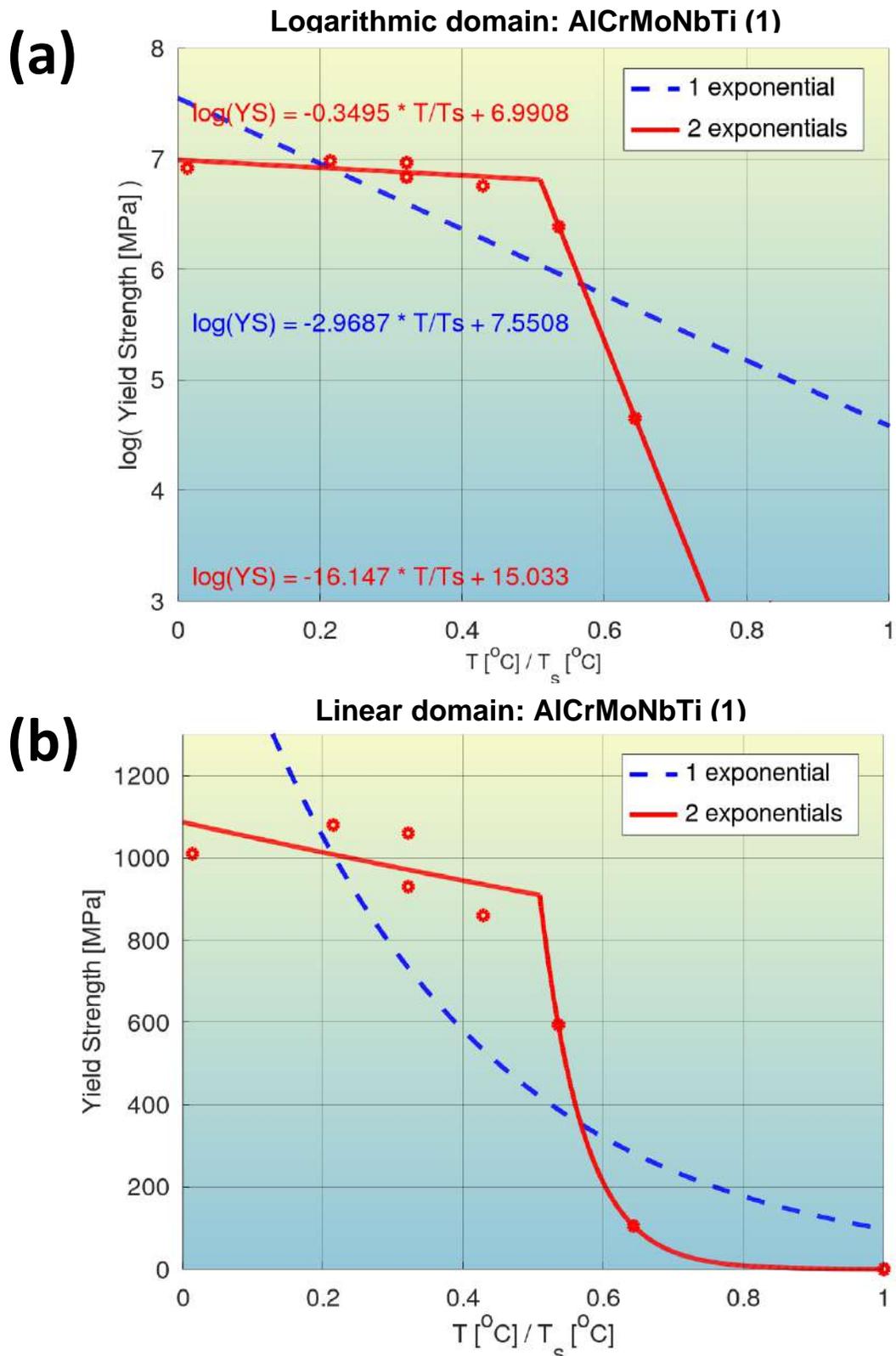

**Fig. S17**: Quantification of modeling accuracy of the bilinear log model, for composition No. 16 from **Tab. S1** (AlCrMoNbTi (1), BCC+unknown phase), and comparison to that of a model with a single exponential.

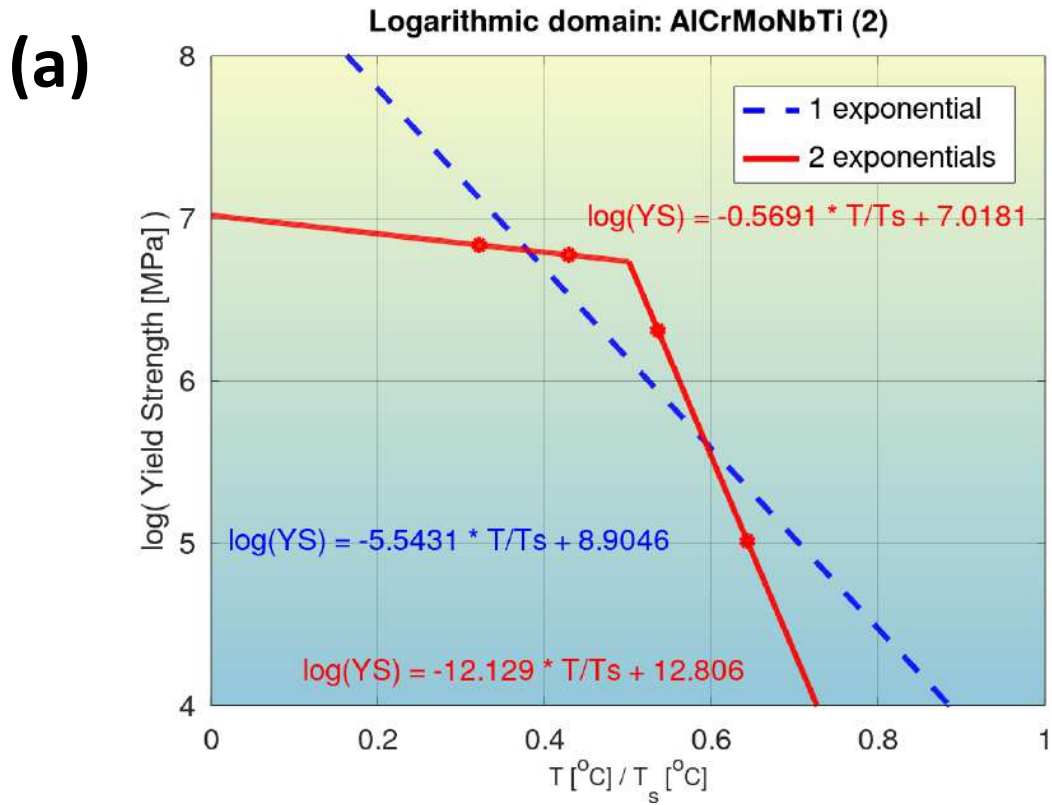

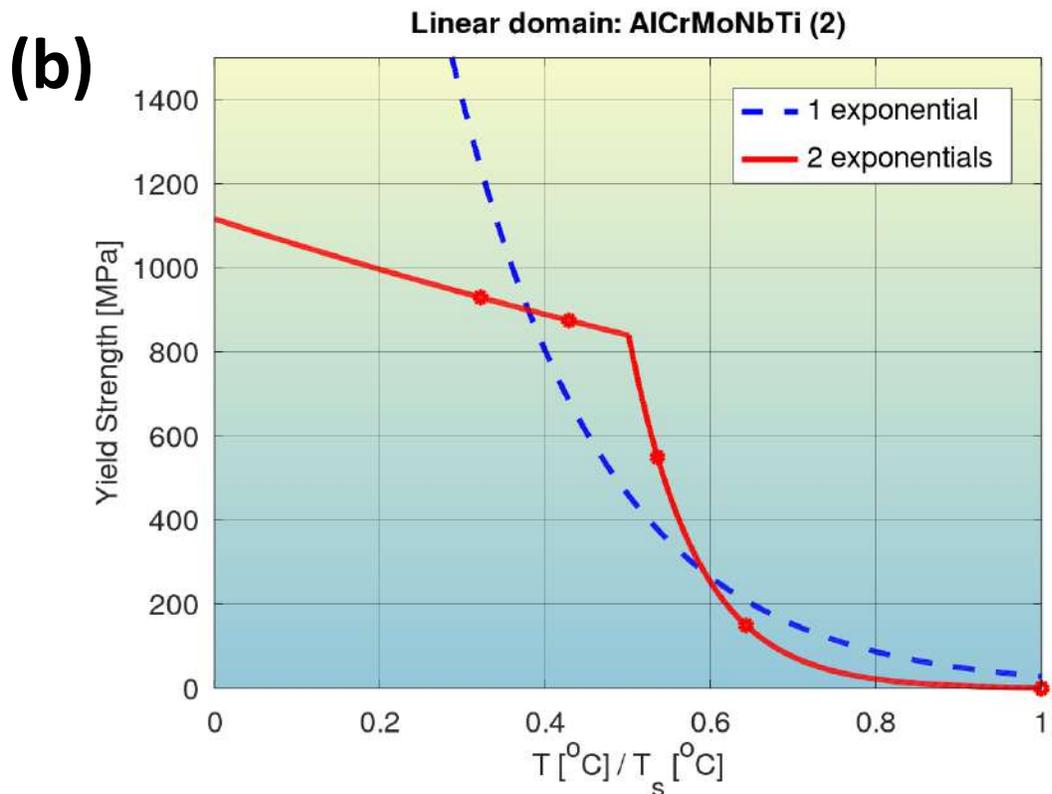

**Fig. S18**: Quantification of modeling accuracy of the bilinear log model, for composition No. 17 from **Tab. S1** (AlCrMoNbTi (2), BCC phase), and comparison to that of a model with a single exponential.

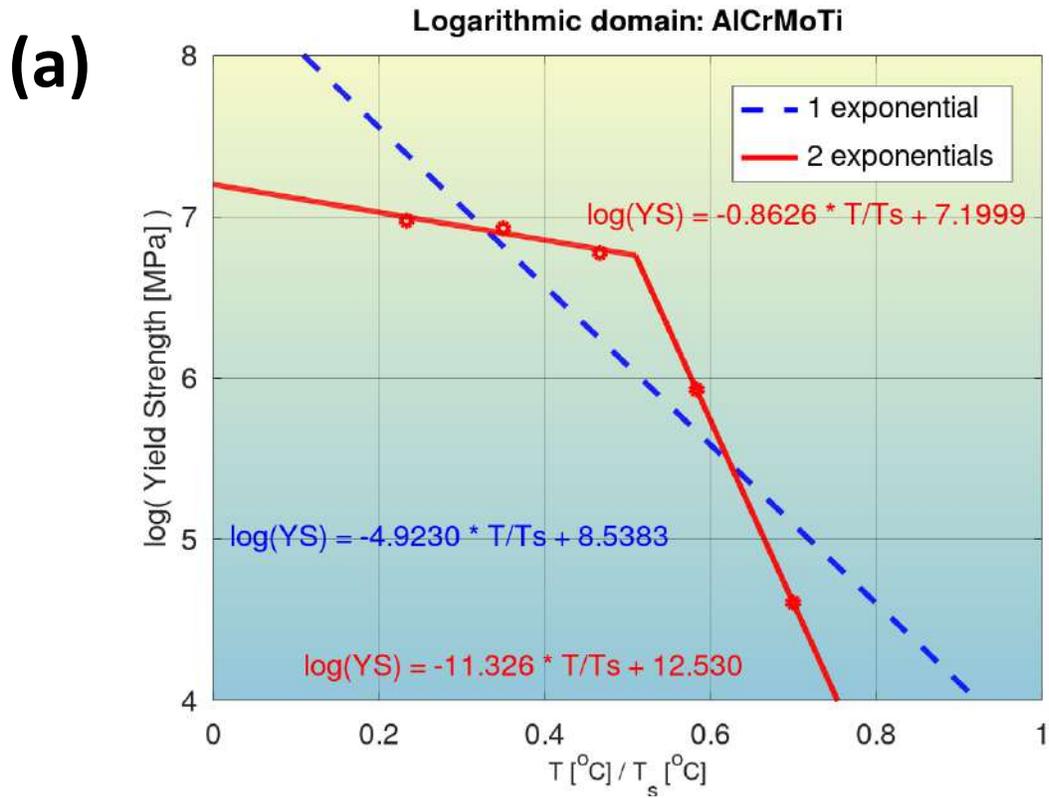

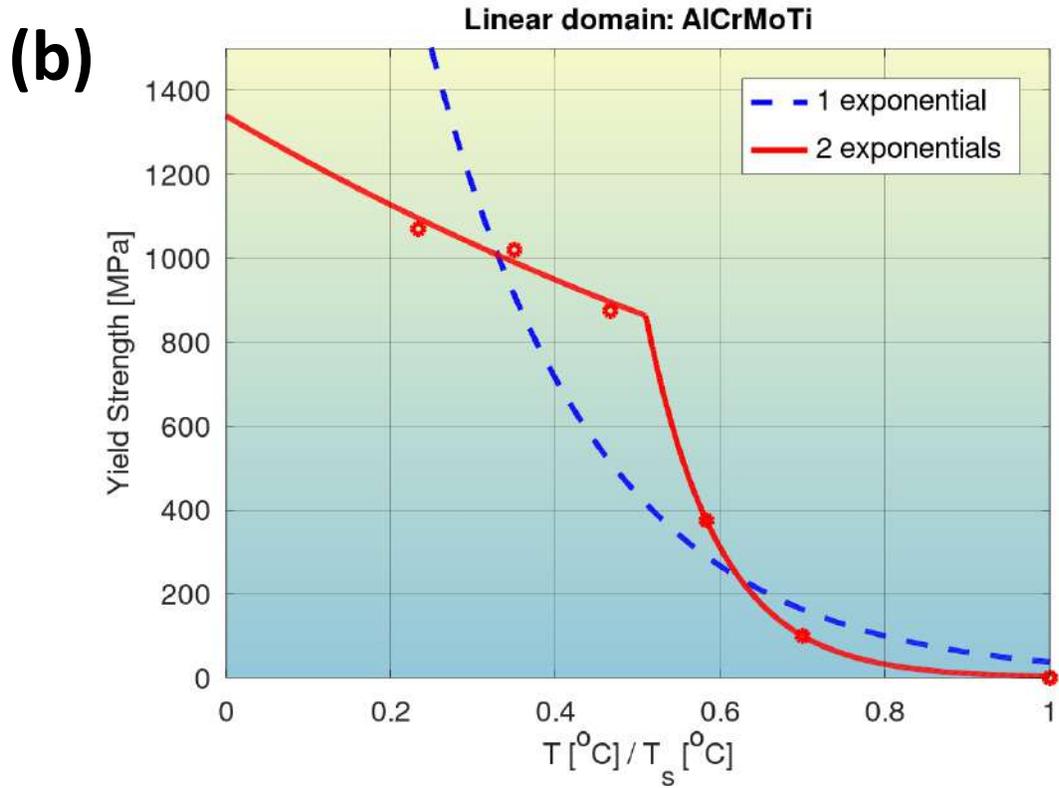

**Fig. S19**: Quantification of modeling accuracy of the bilinear log model, for composition No. 18 from **Tab. S1** (AlCrMoTi, BCC phase), and comparison to that of a model with a single exponential.

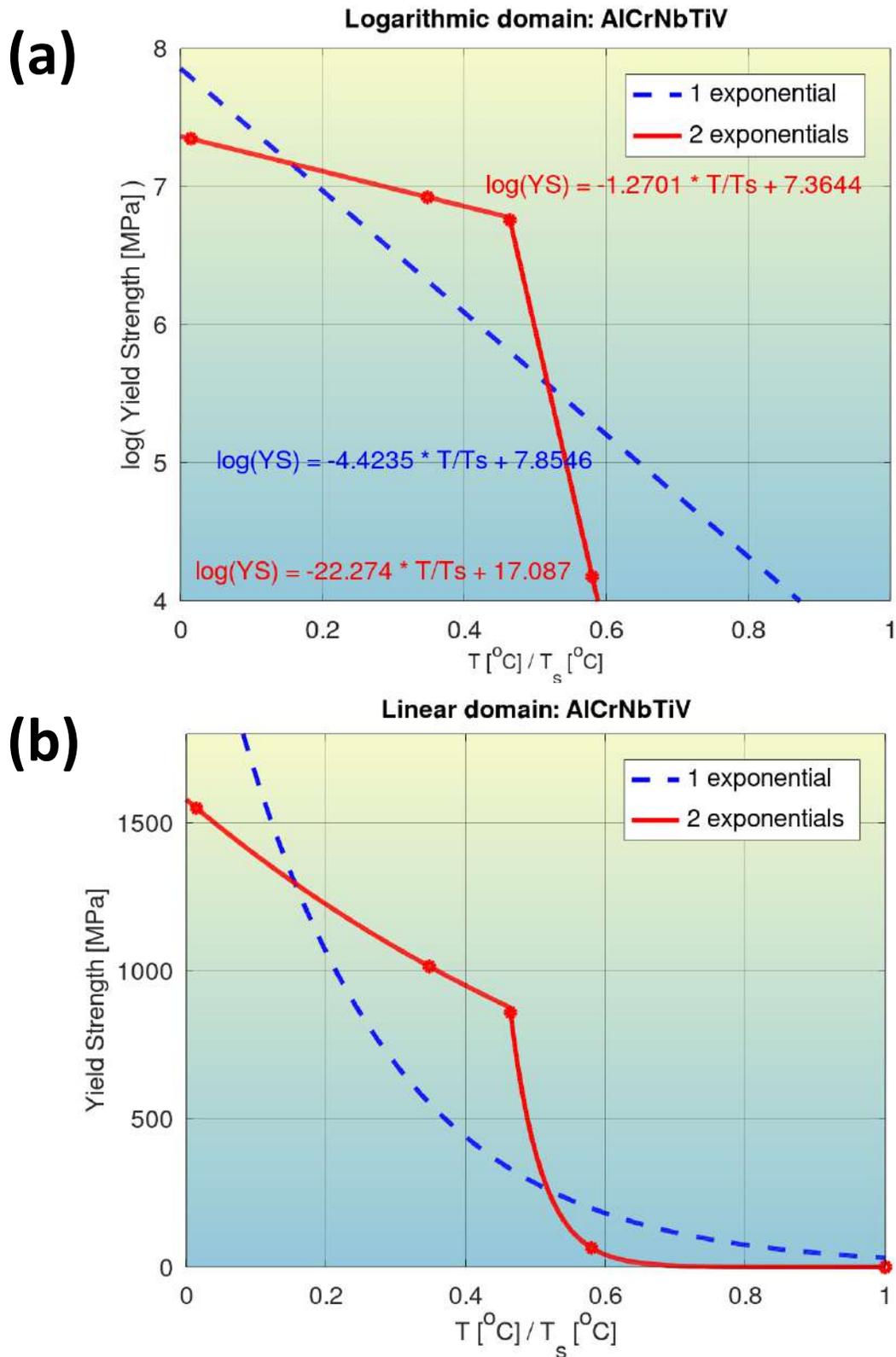

**Fig. S20**: Quantification of modeling accuracy of the bilinear log model, for composition No. 19 from **Tab. S1** (AlCrNbTiV, BCC+Laves phase), and comparison to that of a model with a single exponential.

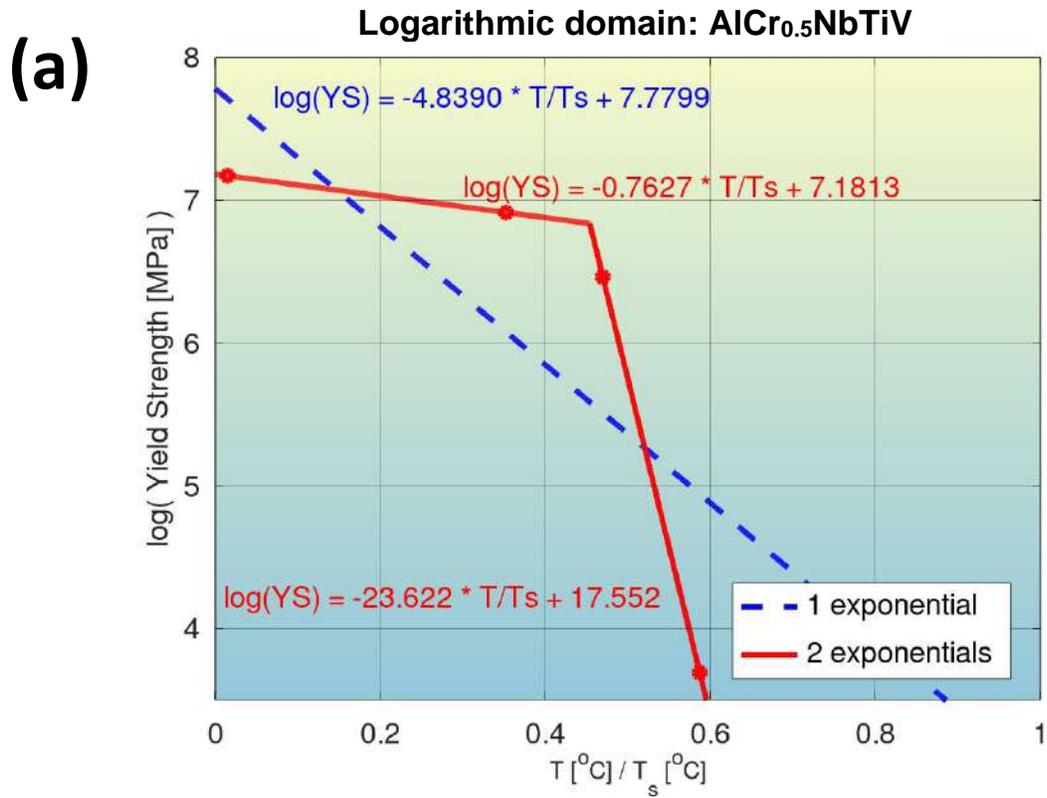

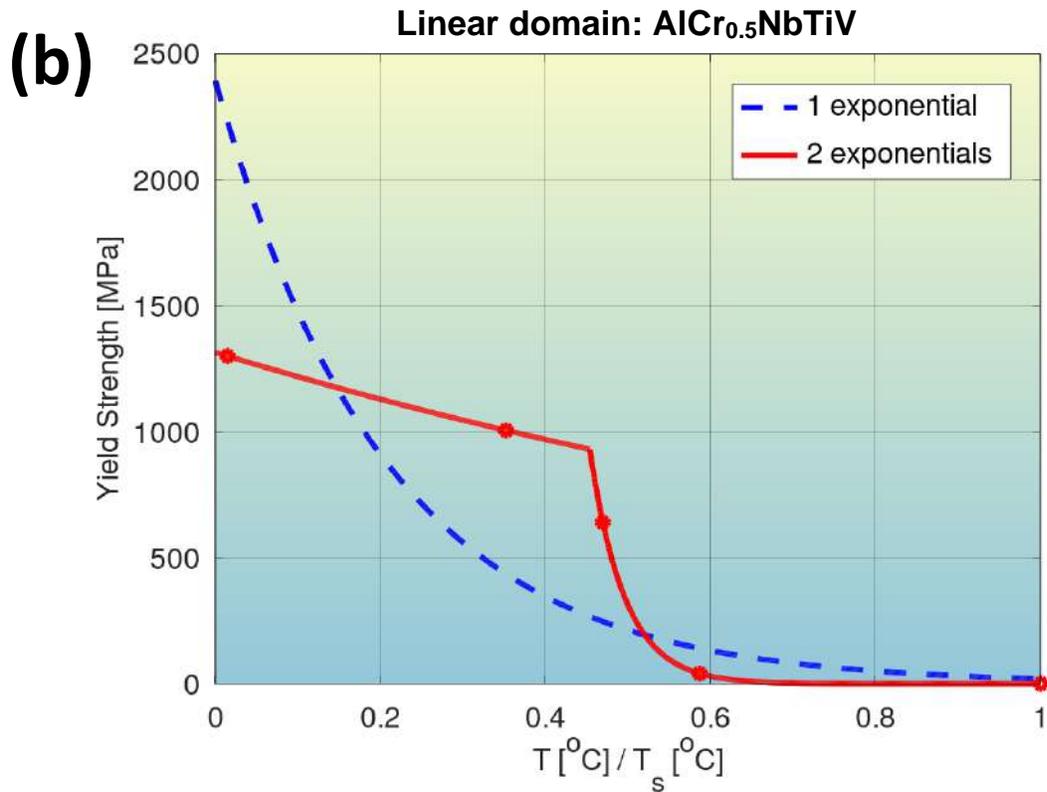

**Fig. S21**: Quantification of modeling accuracy of the bilinear log model, for composition No. 20 from **Tab. S1** (AlCr$_{0.5}$NbTiV, BCC phase), and comparison to that of a model with a single exponential.

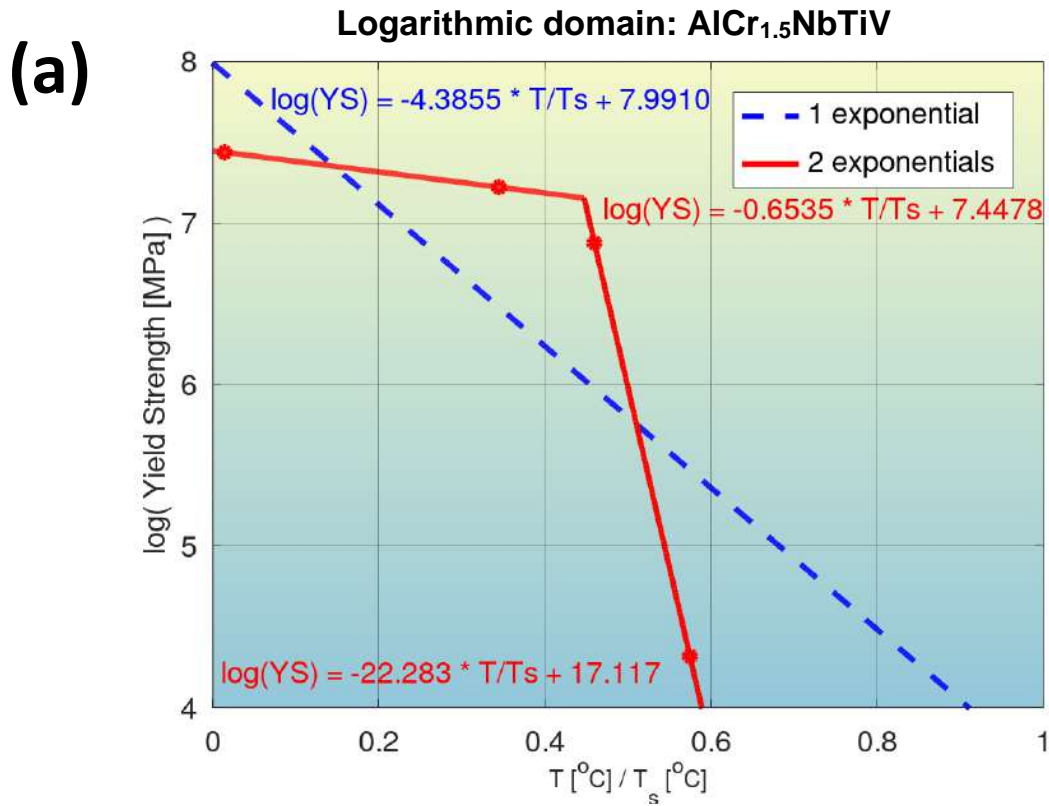

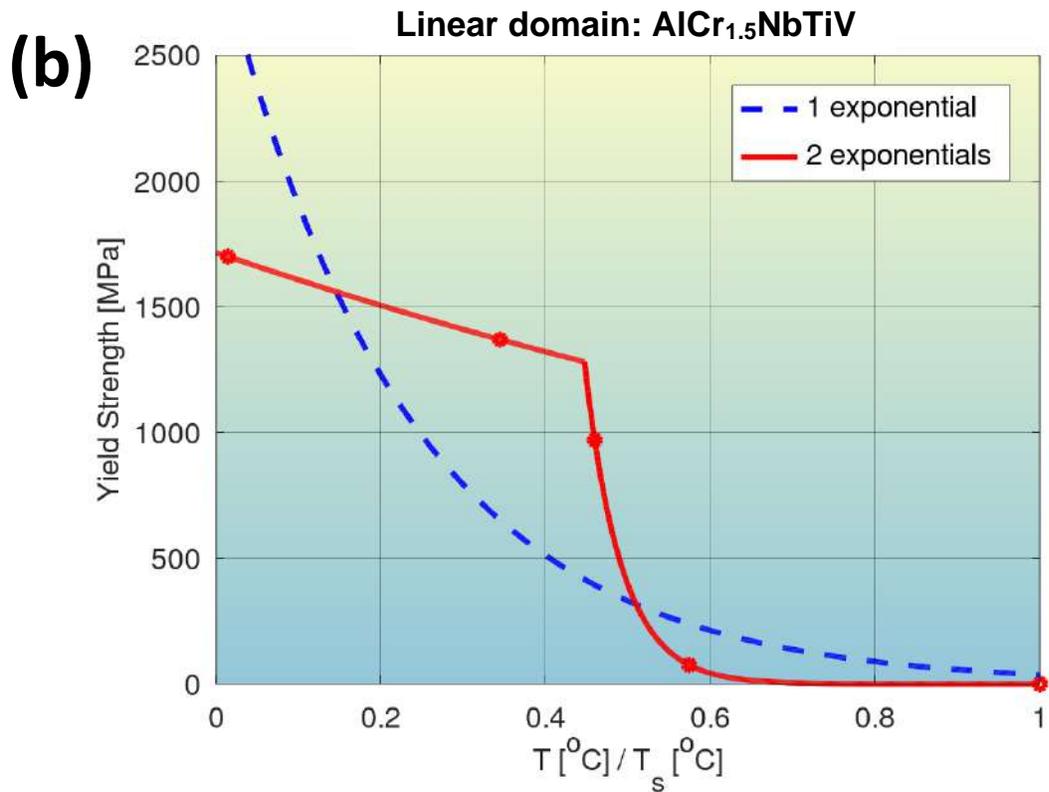

**Fig. S22**: Quantification of modeling accuracy of the bilinear log model, for composition No. 21 from **Tab. S1** (AlCr$_{1.5}$NbTiV, BCC+Laves phase), and comparison to that of a model with a single exponential.

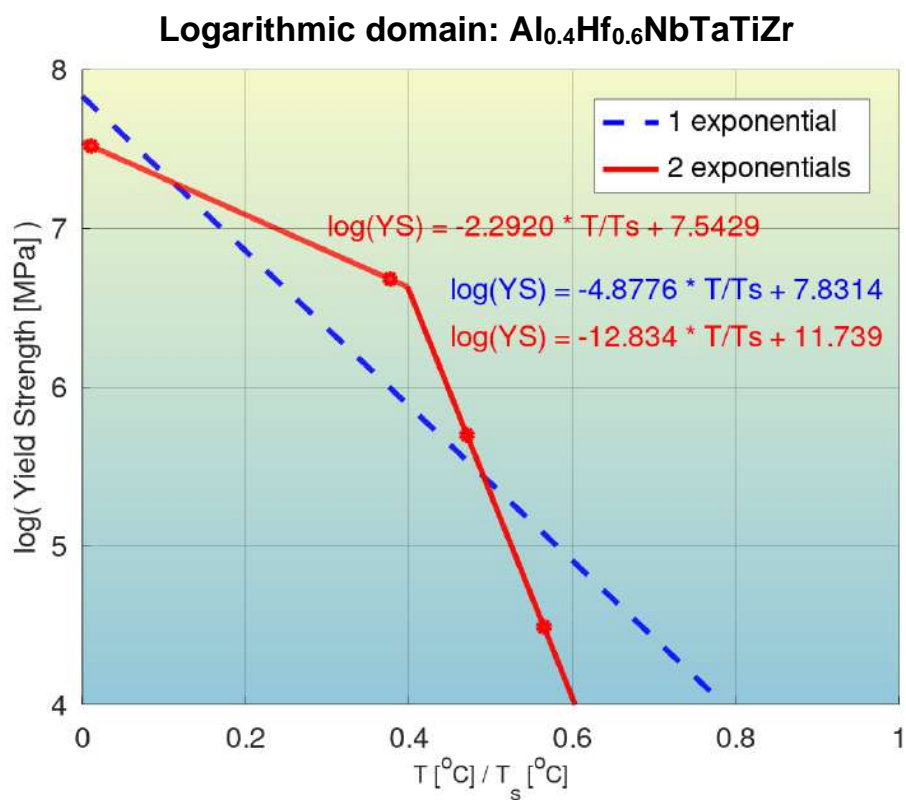

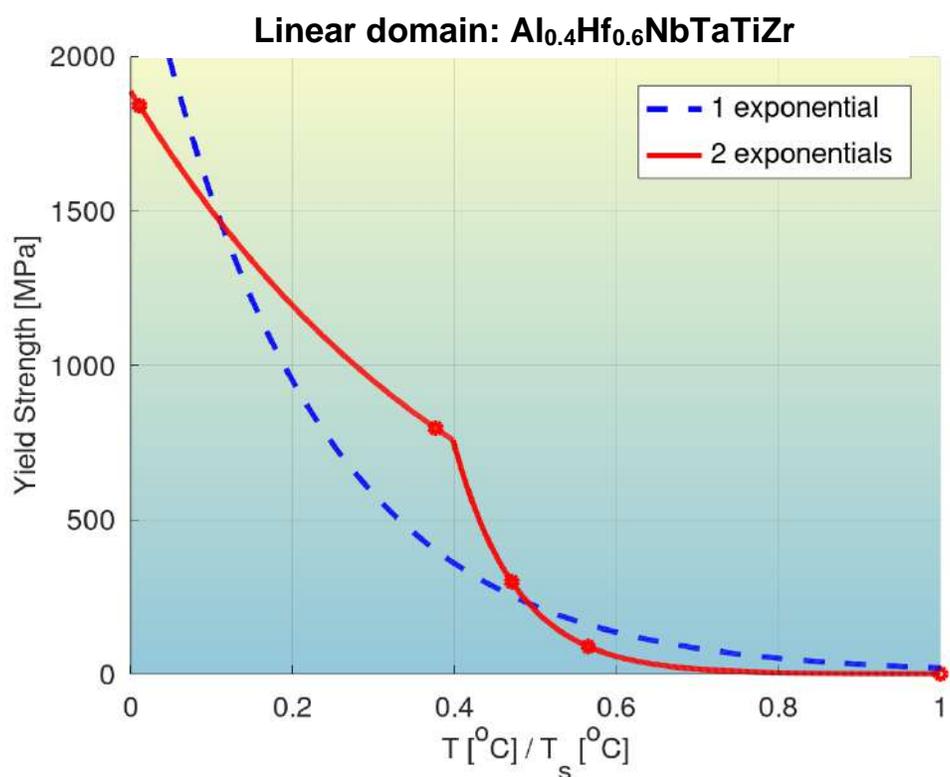

**Fig. S23**: Quantification of modeling accuracy of the bilinear log model, for composition No. 22 from **Tab. S1** ($Al_{0.4}Hf_{0.6}NbTaTiZr$, BCC phase), and comparison to that of a model with a single exponential.

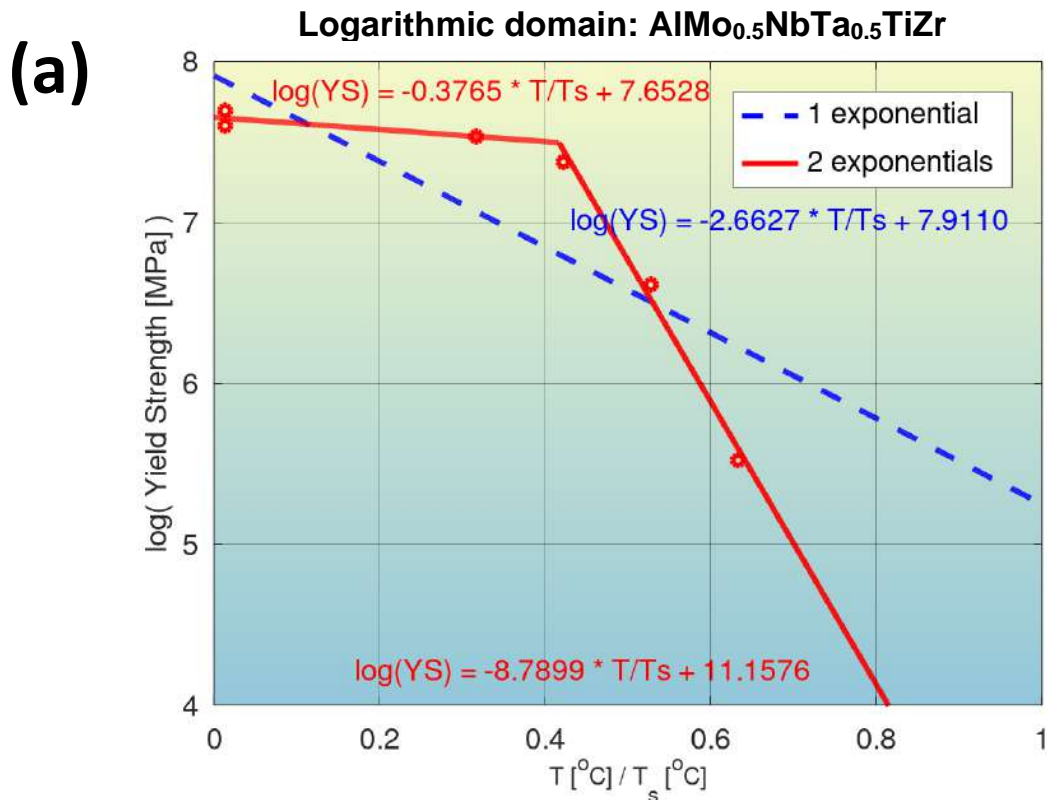

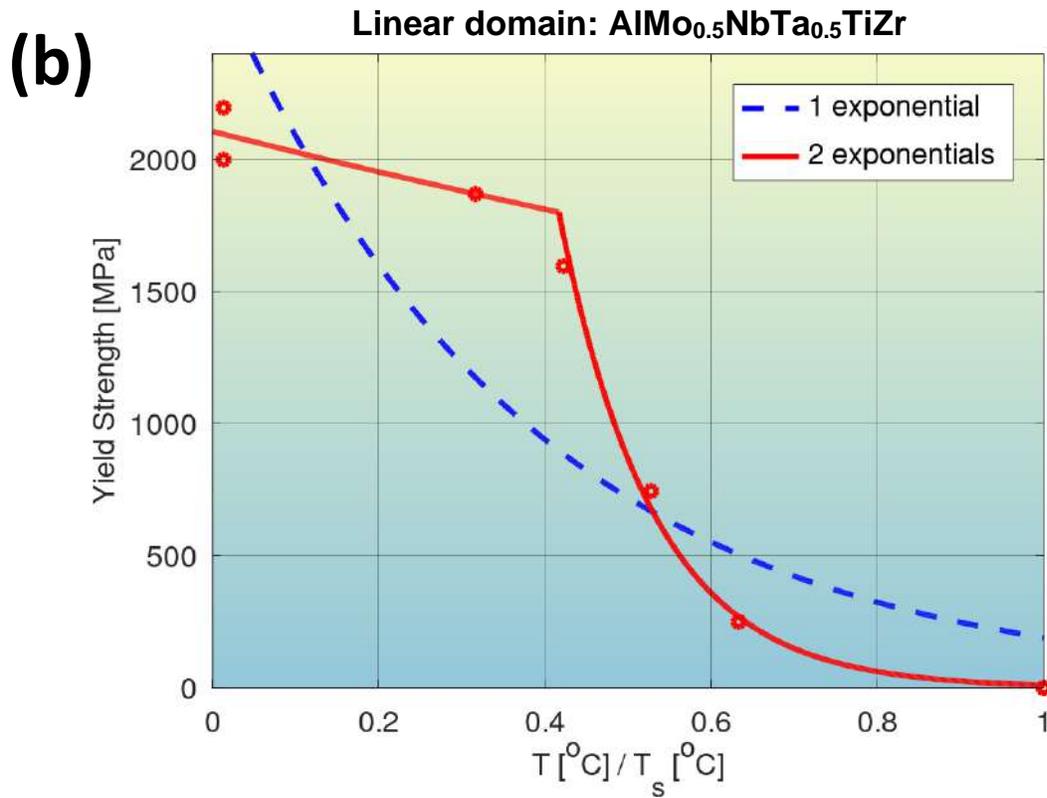

**Fig. S24**: Quantification of modeling accuracy of the bilinear log model, for composition No. 23 from **Tab. S1** (AlMo$_{0.5}$NbTa$_{0.5}$TiZr, BCC+B2 phase), and comparison to that of a model with a single exponential.

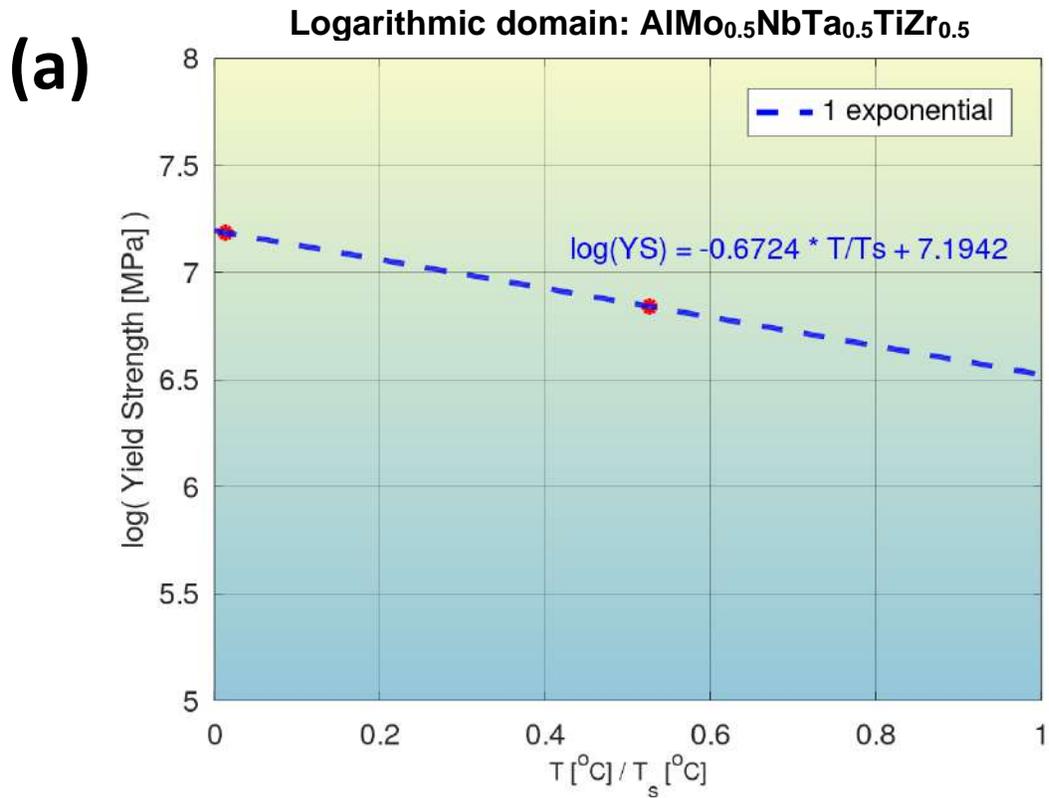
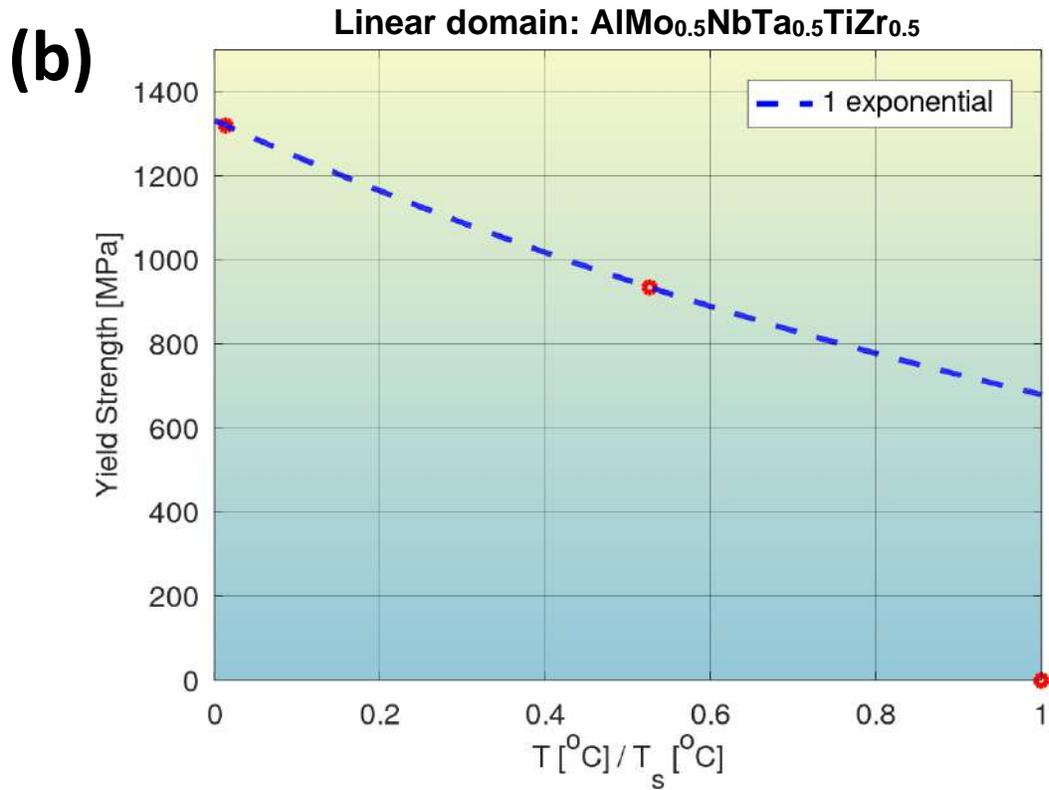

**Fig. S25**: Quantification of modeling accuracy of the bilinear log model, for composition No. 24 from **Tab. S1** (AlMo$_{0.5}$NbTa$_{0.5}$TiZr$_{0.5}$, B2 phase), and comparison to that of a model with a single exponential.

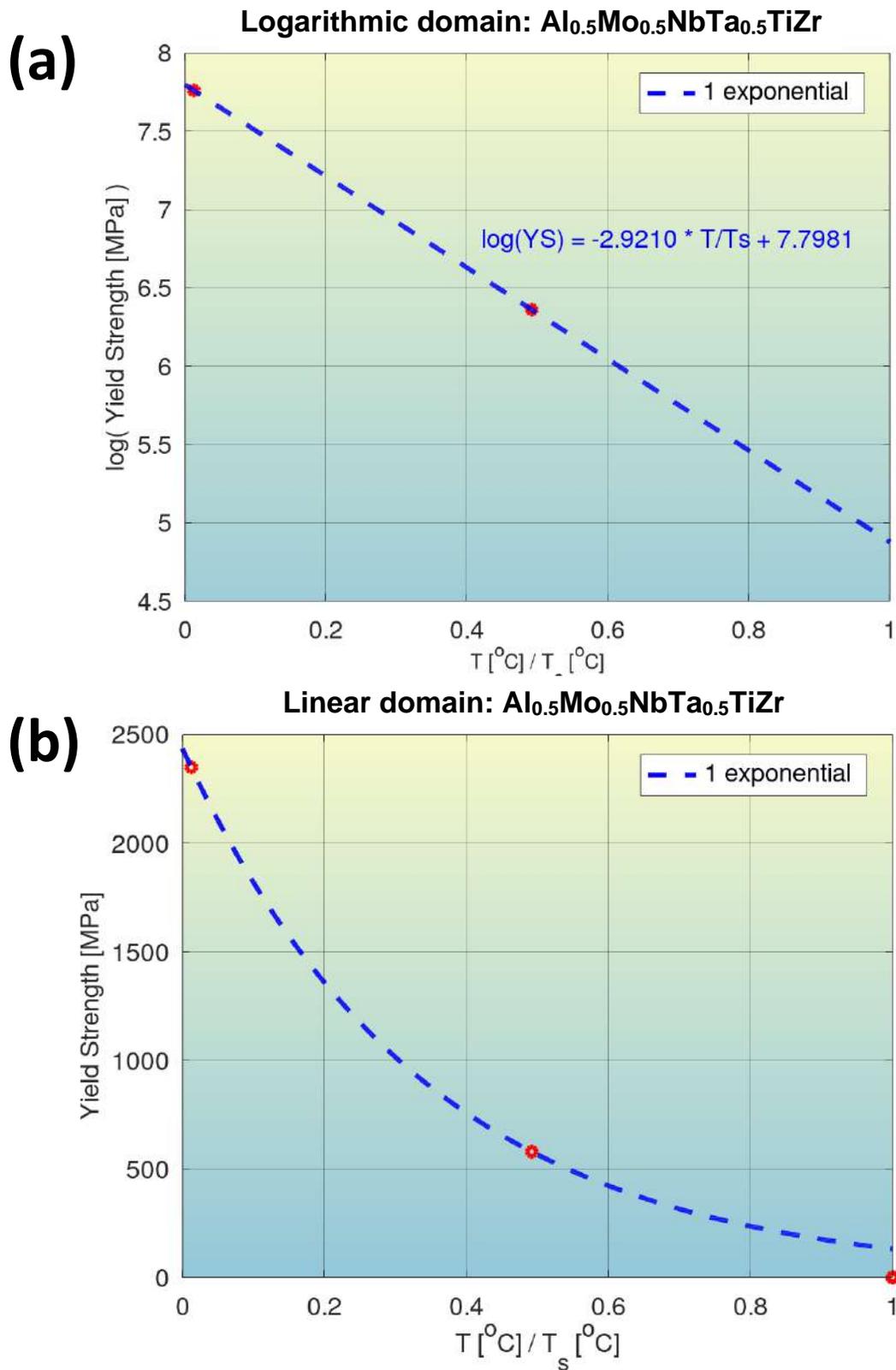

**Fig. S26**: Quantification of modeling accuracy of the bilinear log model, for composition No. 25 from **Tab. S1** (Al$_{0.5}$Mo$_{0.5}$NbTa$_{0.5}$TiZr, BCC+B2 phase), and comparison to that of a model with a single exponential.

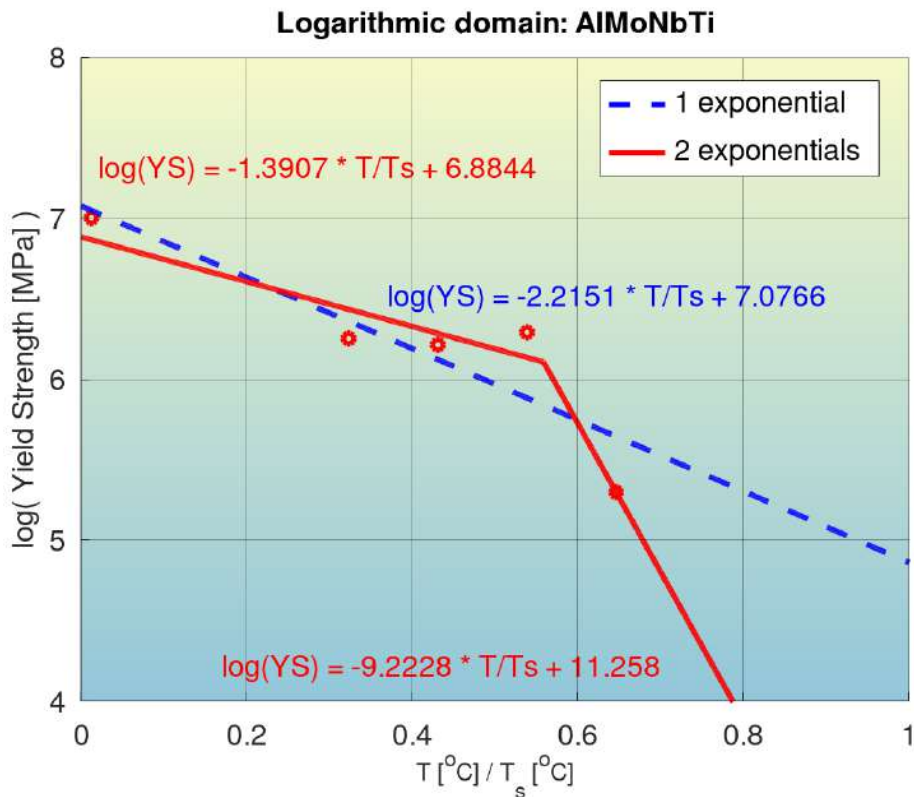

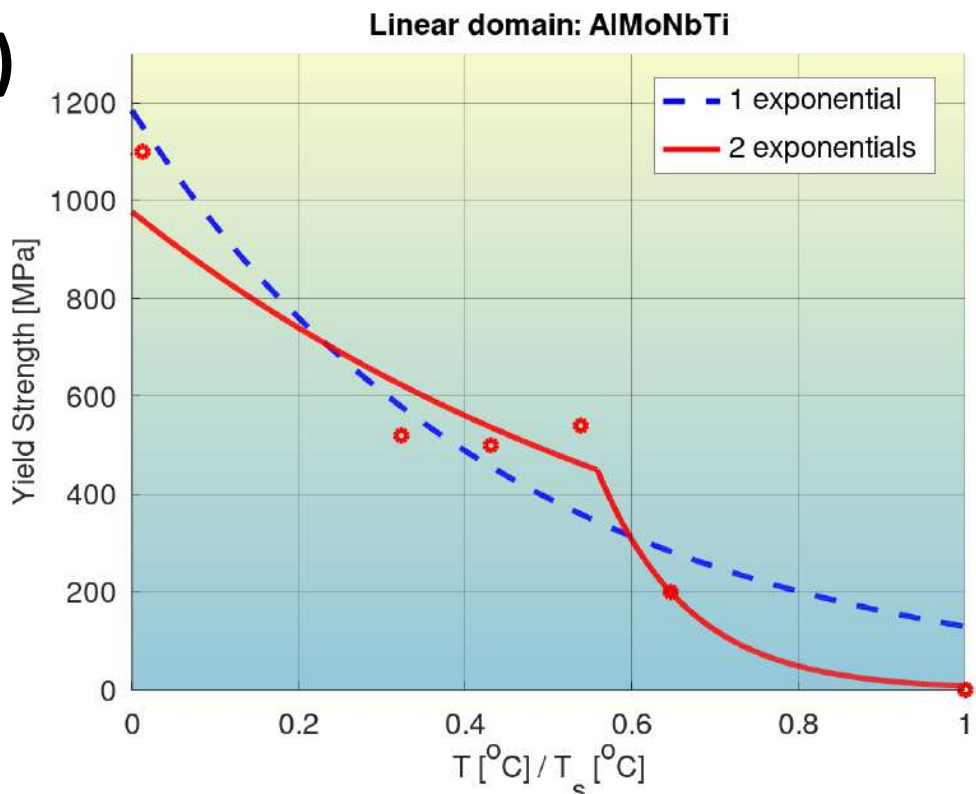

**Fig. S27**: Quantification of modeling accuracy of the bilinear log model, for composition No. 26 from **Tab. S1** (AlMoNbTi, BCC phase), and comparison to that of a model with a single exponential.

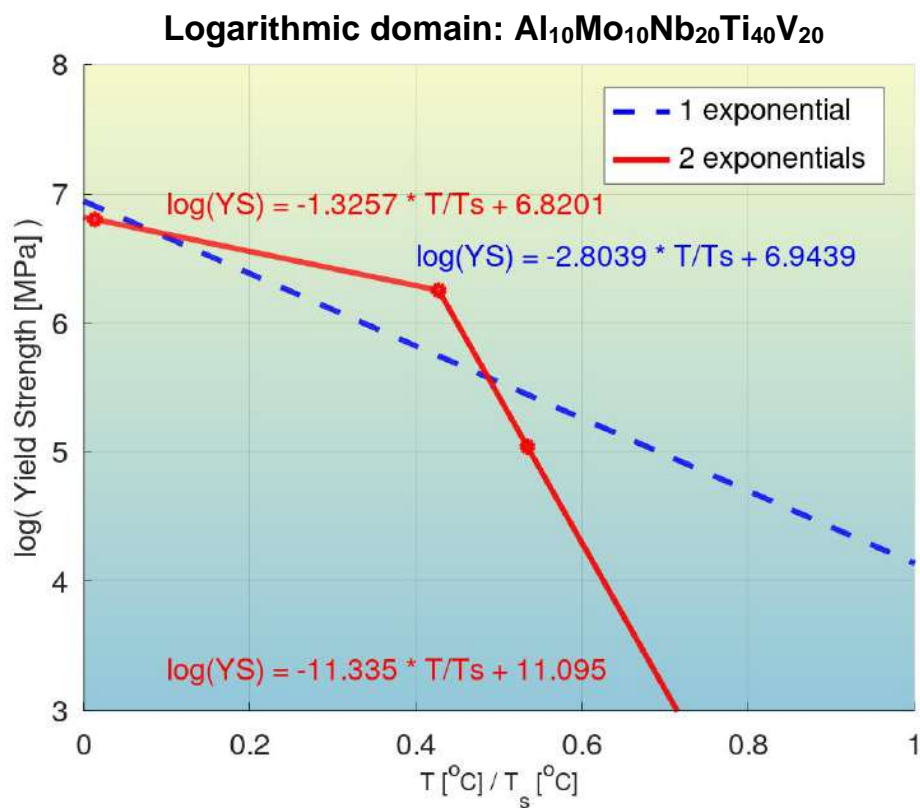
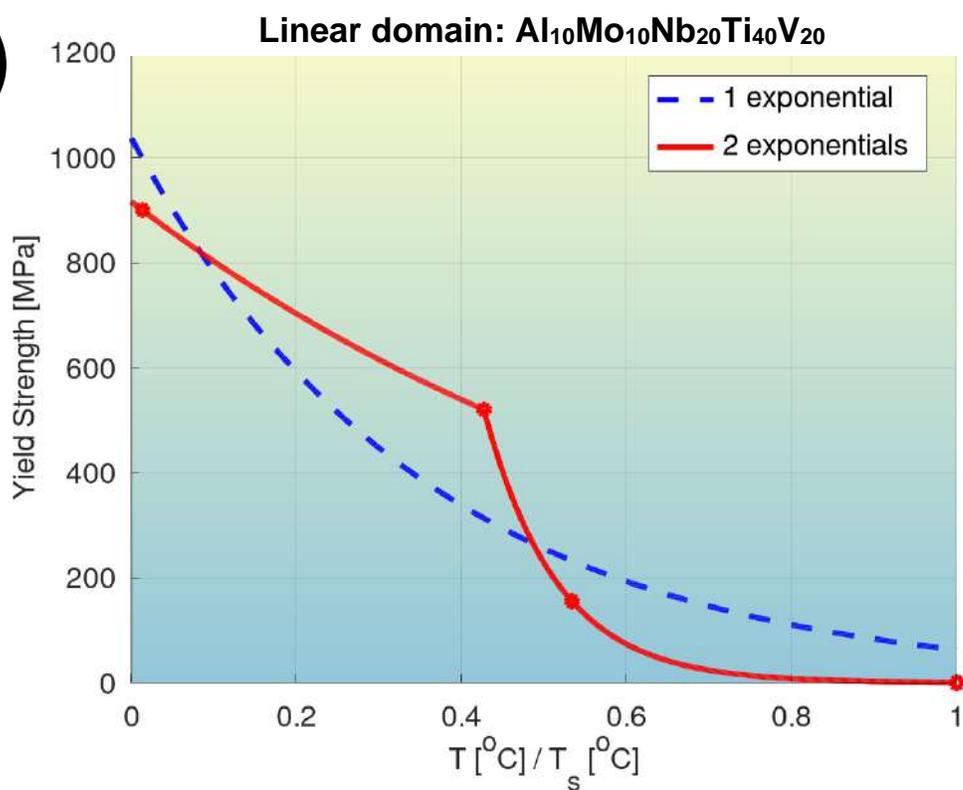

**Fig. S28**: Quantification of modeling accuracy of the bilinear log model, for composition No. 27 from **Tab. S1** (Al$_{10}$Mo$_{10}$Nb$_{20}$Ti$_{40}$V$_{20}$, BCC phase), and comparison to that of a model with a single exponential.

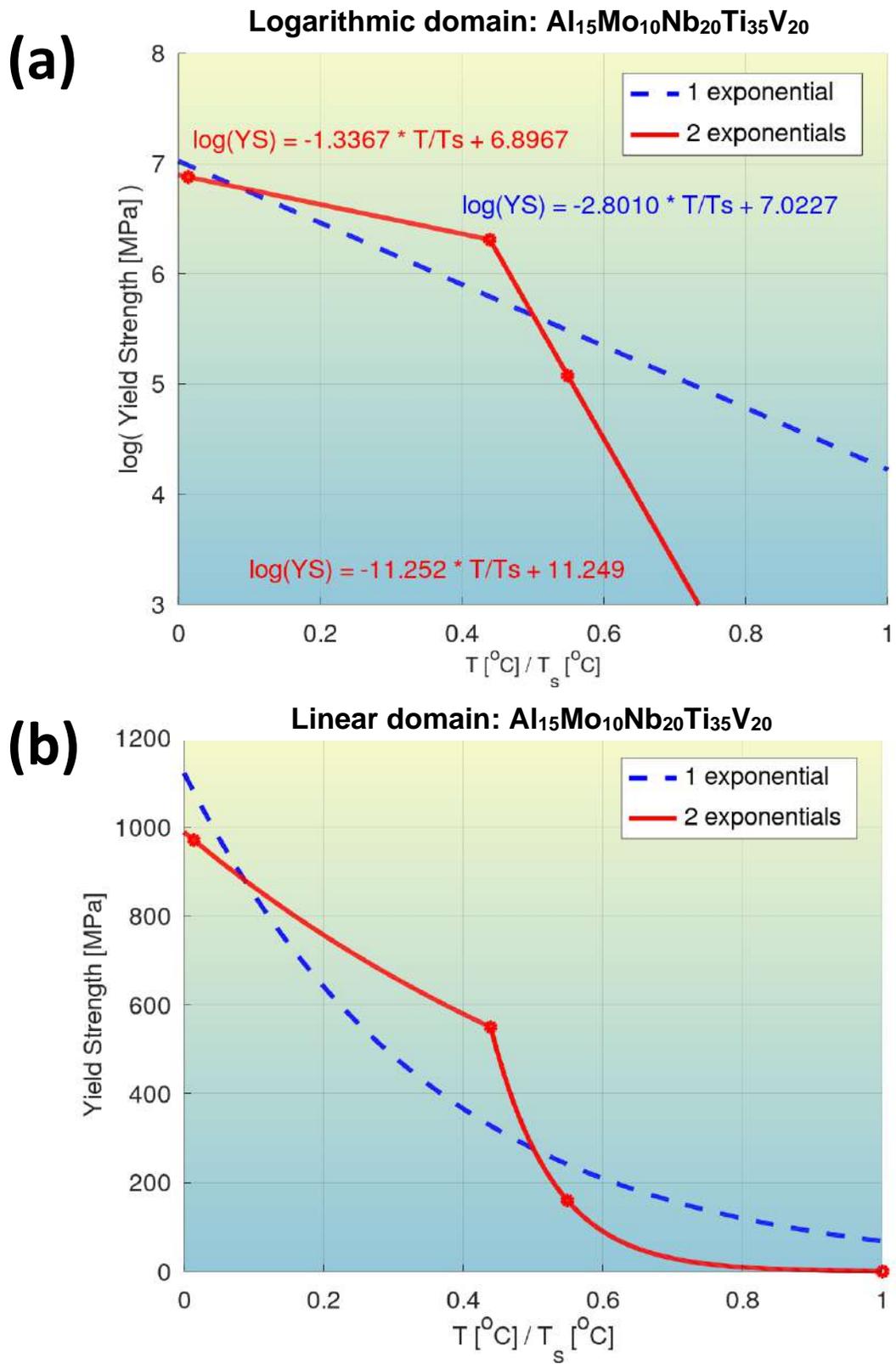

**Fig. S29**: Quantification of modeling accuracy of the bilinear log model, for composition No. 28 from **Tab. S1** ($Al_{15}Mo_{10}Nb_{20}Ti_{35}V_{20}$, BCC phase), and comparison to that of a model with a single exponential.

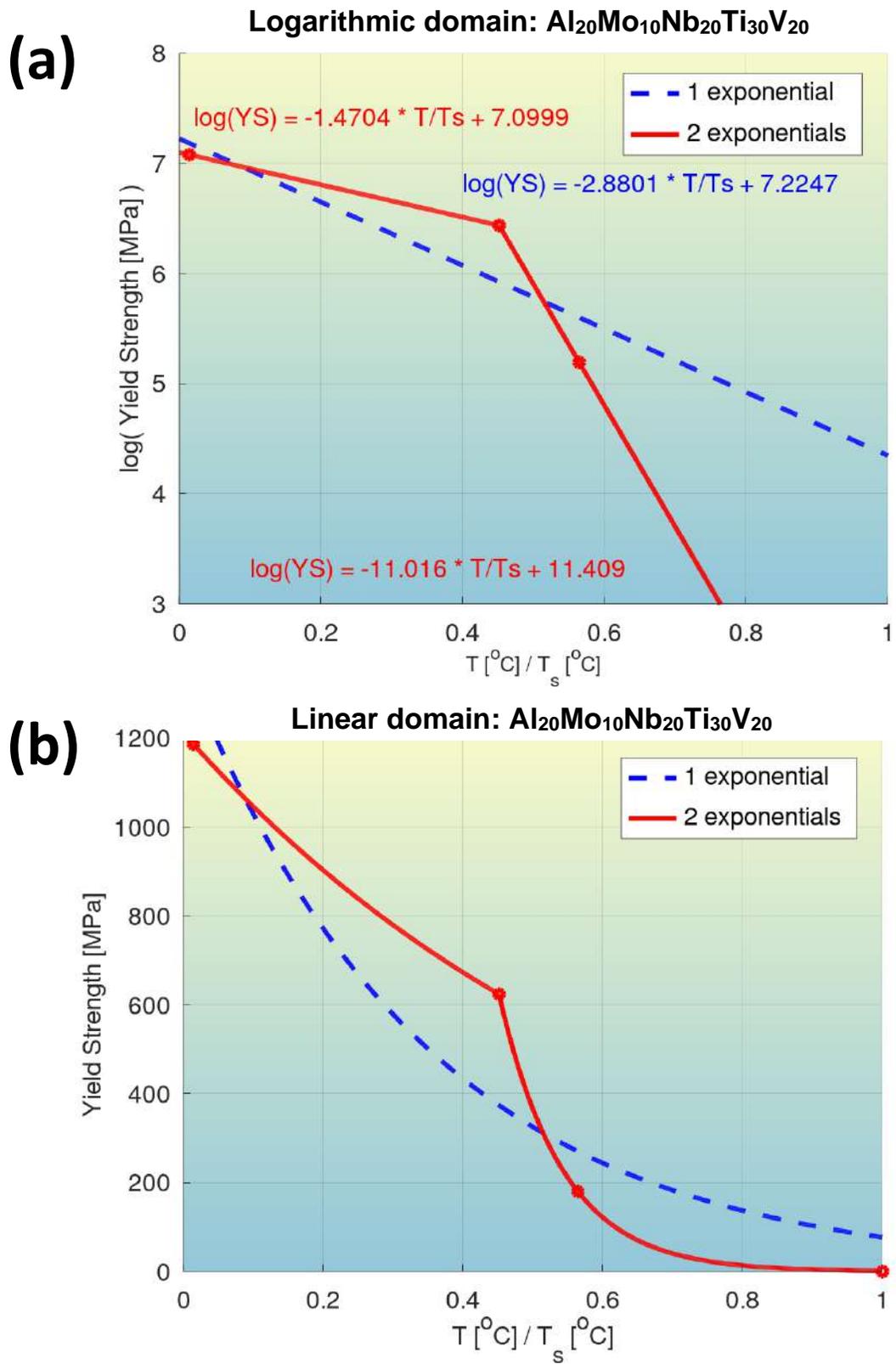

**Fig. S30**: Quantification of modeling accuracy of the bilinear log model, for composition No. 29 from **Tab. S1** (Al$_{20}$Mo$_{10}$Nb$_{20}$Ti$_{30}$V$_{20}$, BCC phase), and comparison to that of a model with a single exponential.

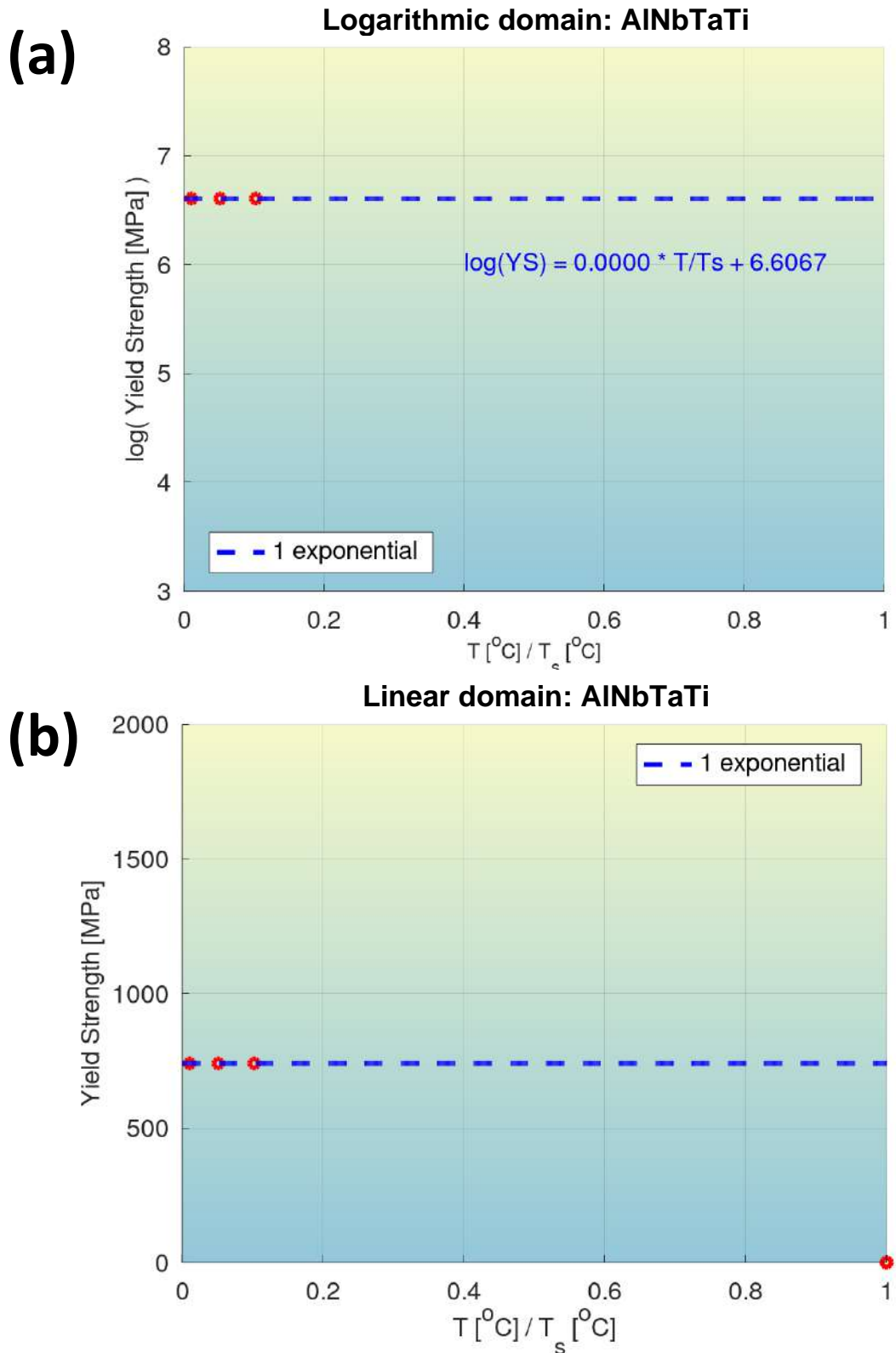

**Fig. S31**: Quantification of modeling accuracy of the bilinear log model, for composition No. 30 from **Tab. S1** (AlNbTaTi, BCC phase), and comparison to that of a model with a single exponential.

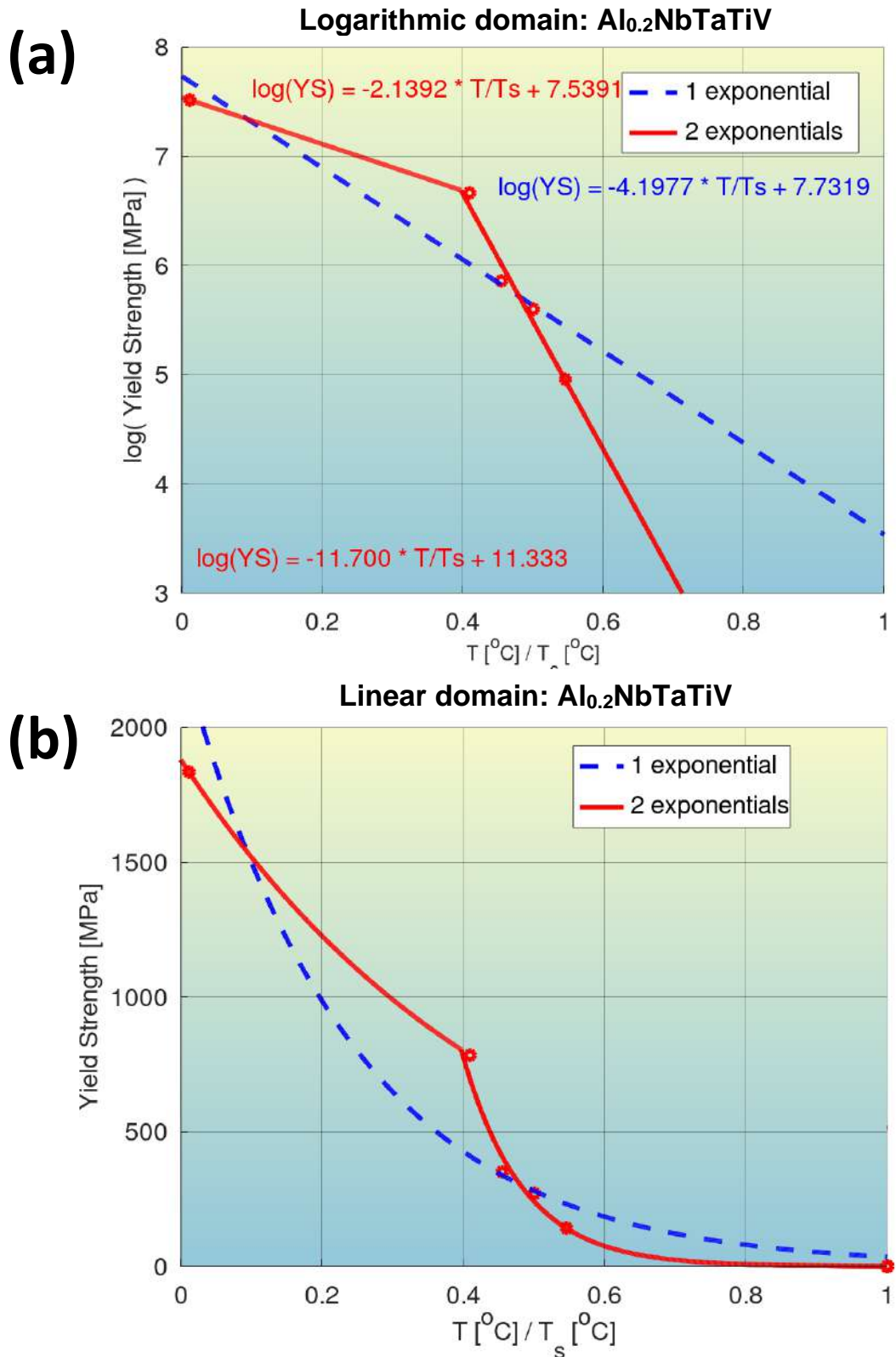

**Fig. S32**: Quantification of modeling accuracy of the bilinear log model, for composition No. 31 from **Tab. S1** (Al$_{0.2}$NbTaTiV, BCC phase), and comparison to that of a model with a single exponential.

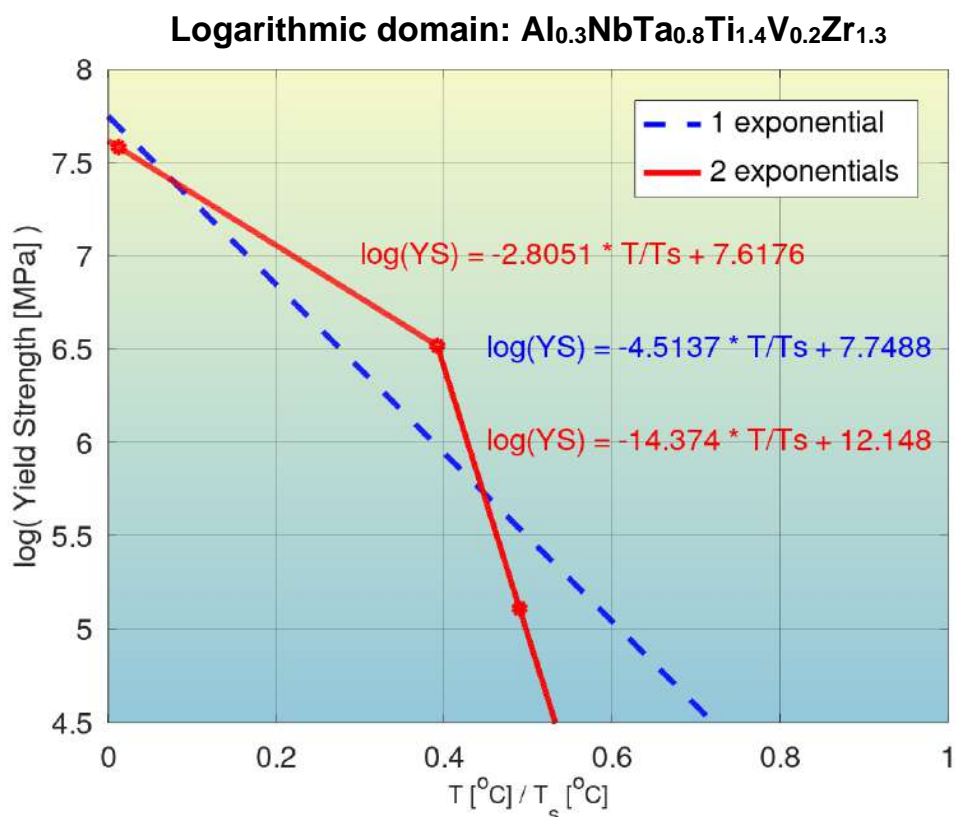

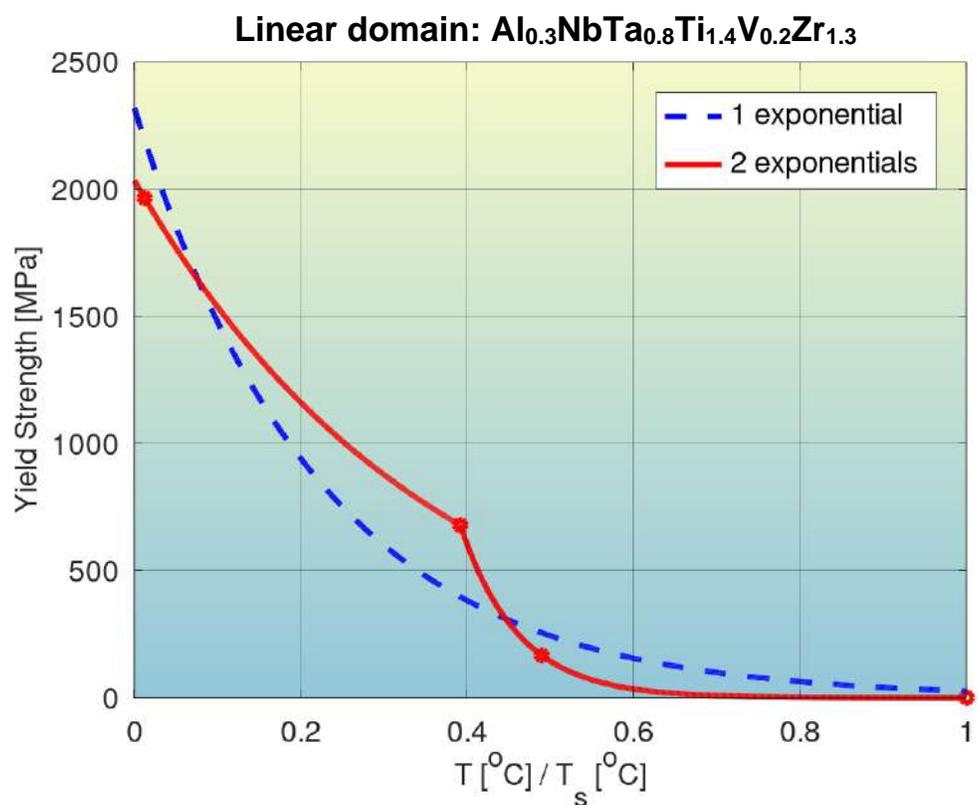

**Fig. S33**: Quantification of modeling accuracy of the bilinear log model, for composition No. 32 from **Tab. S1** ($Al_{0.3}NbTa_{0.8}Ti_{1.4}V_{0.2}Zr_{1.3}$, BCC phase), and comparison to that of a model with a single exponential.

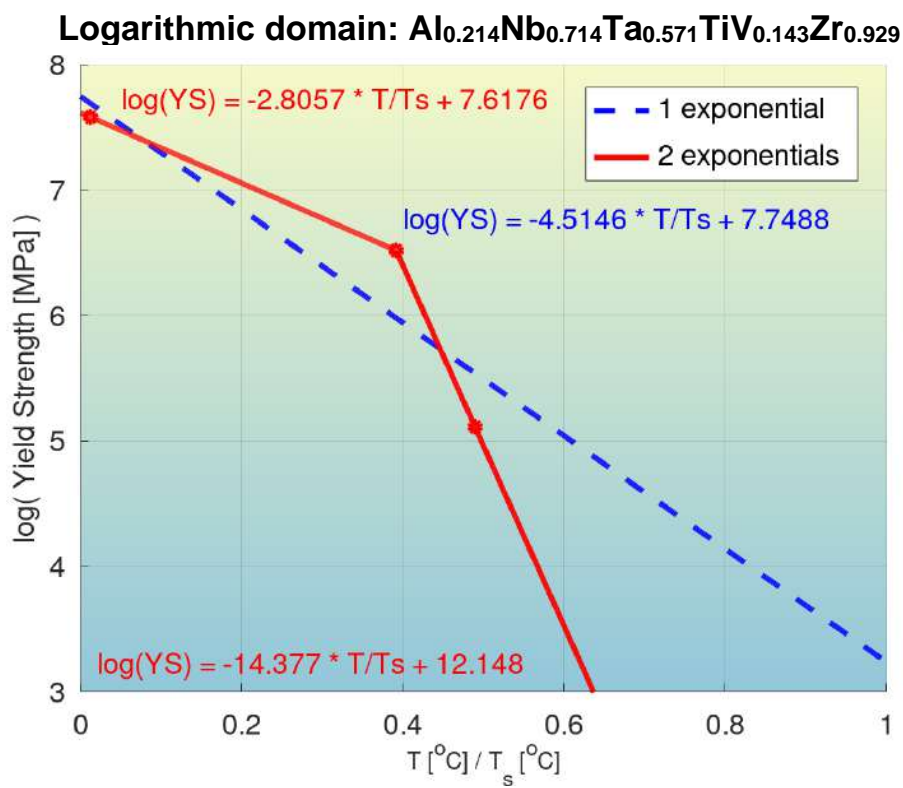
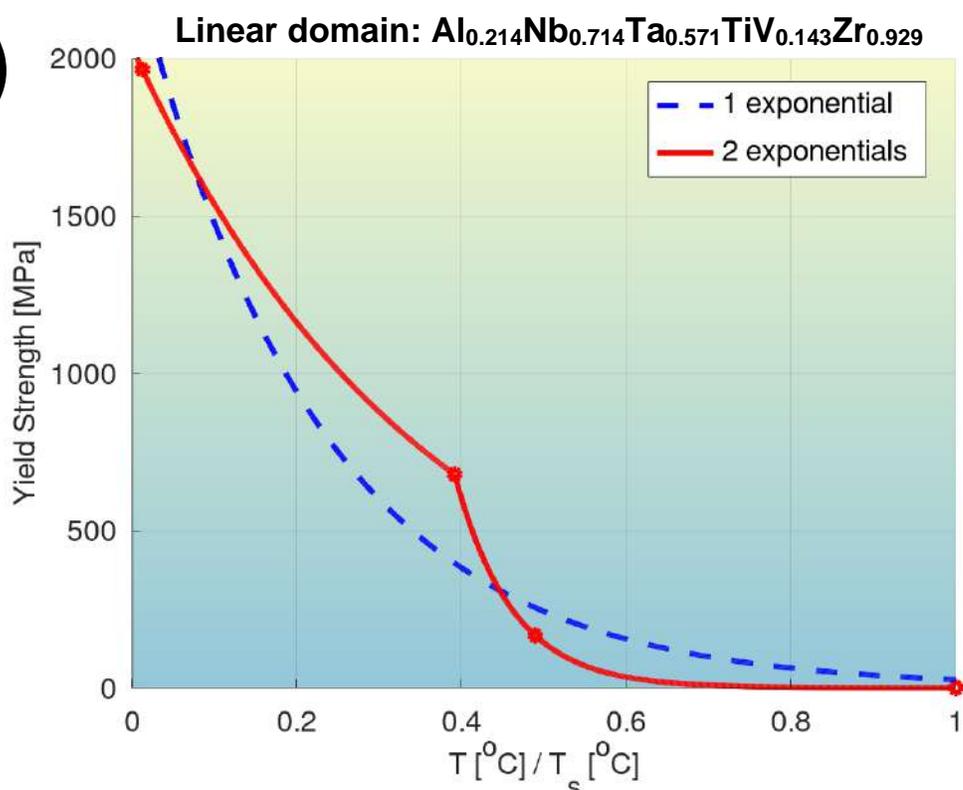

Fig. S34: Quantification of modeling accuracy of the bilinear log model for composition No. 33 from Tab. S1 (Al$_{0.214}$Nb$_{0.714}$Ta$_{0.571}$TiV$_{0.143}$Zr$_{0.929}$, BCC phase), and comparison to that of a model with a single exponential.

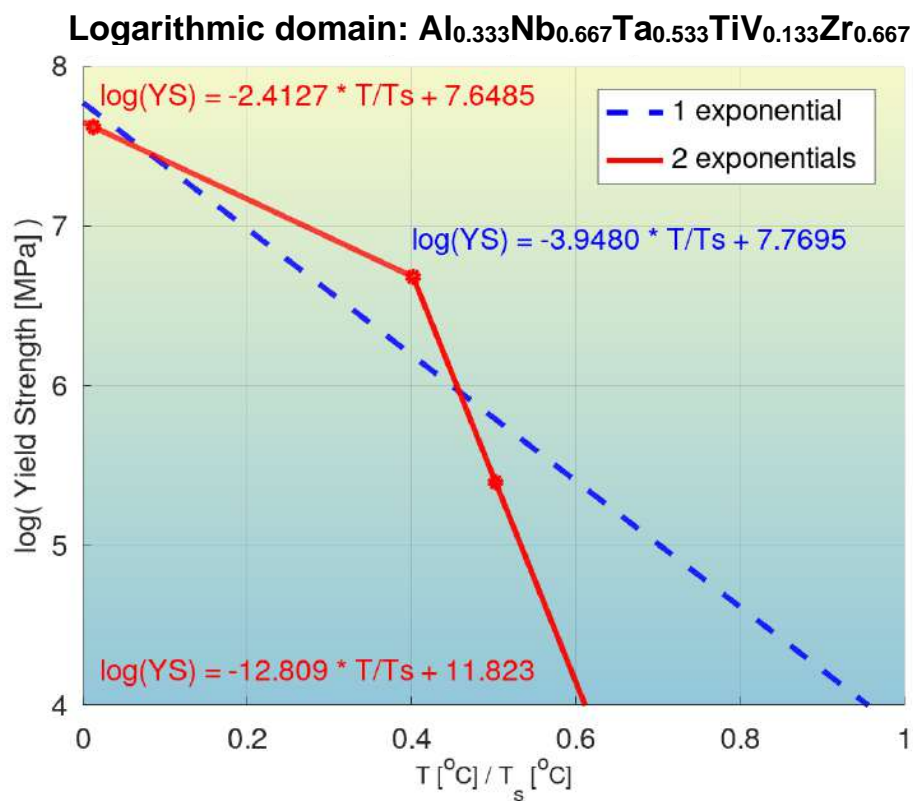
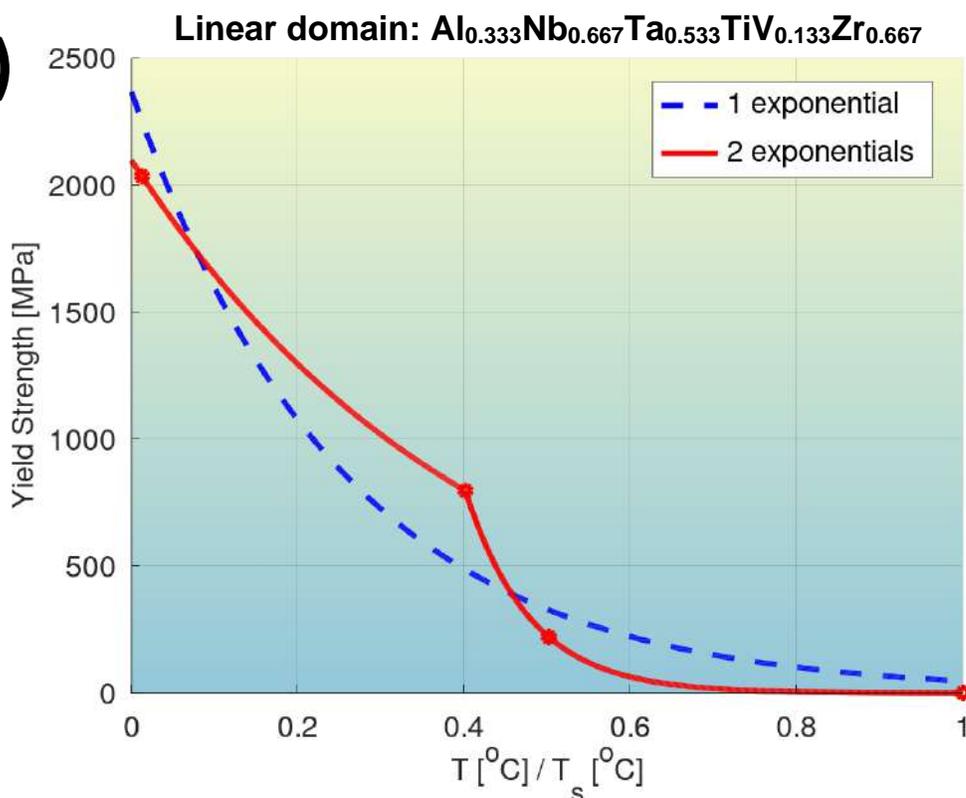

**Fig. S35**: Quantification of modeling accuracy of the bilinear log model for composition No. 34 from **Tab. S1** (Al$_{0.333}$Nb$_{0.667}$Ta$_{0.533}$TiV$_{0.133}$Zr$_{0.667}$, BCC phase), and comparison to that of a model with a single exponential.

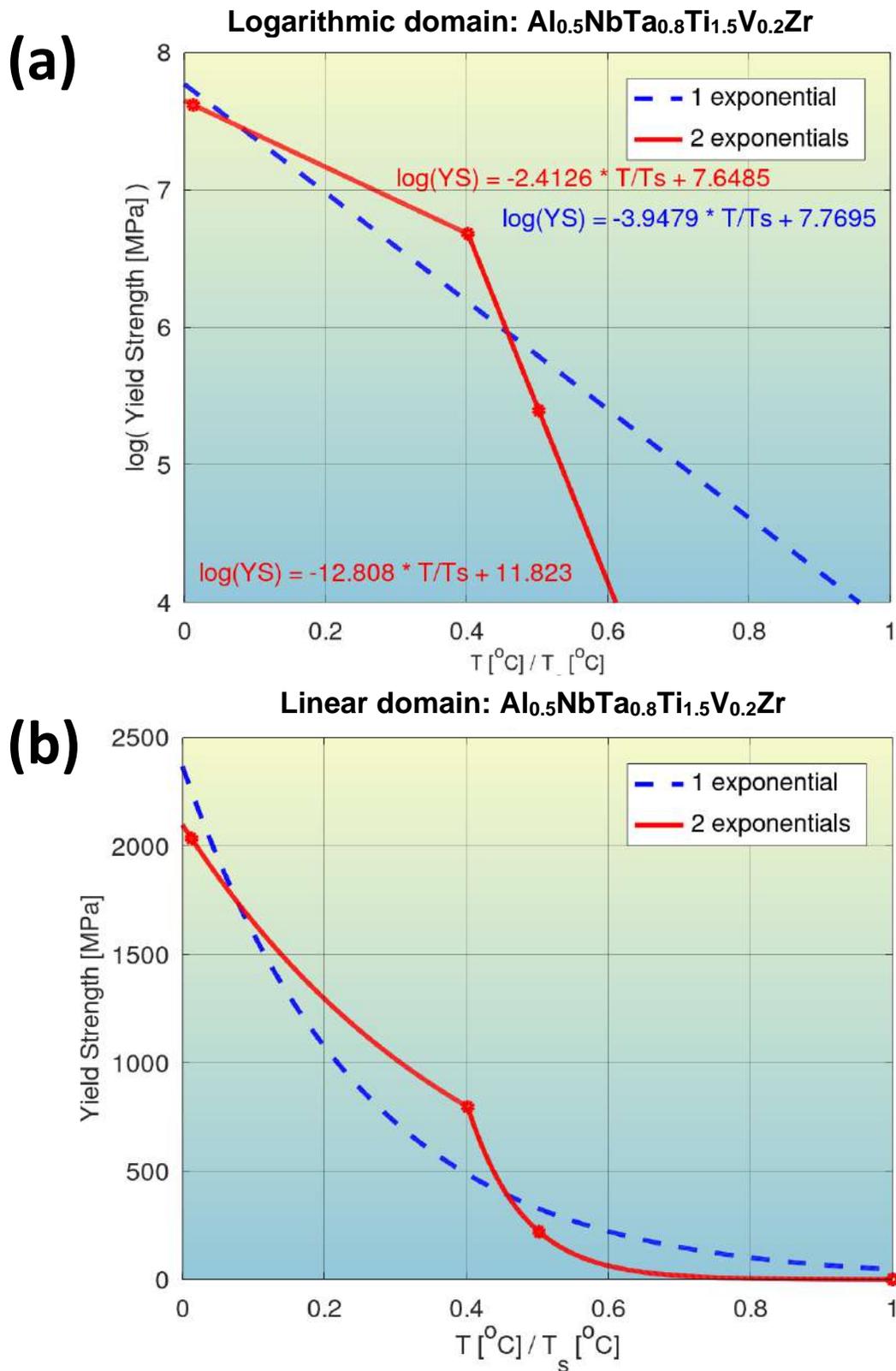

**Fig. S36**: Quantification of modeling accuracy of the bilinear log model, for composition No. 35 from **Tab. S1** (Al$_{0.5}$NbTa$_{0.8}$Ti$_{1.5}$V$_{0.2}$Zr, BCC+B2 phase), and comparison to that of a model with a single exponential.

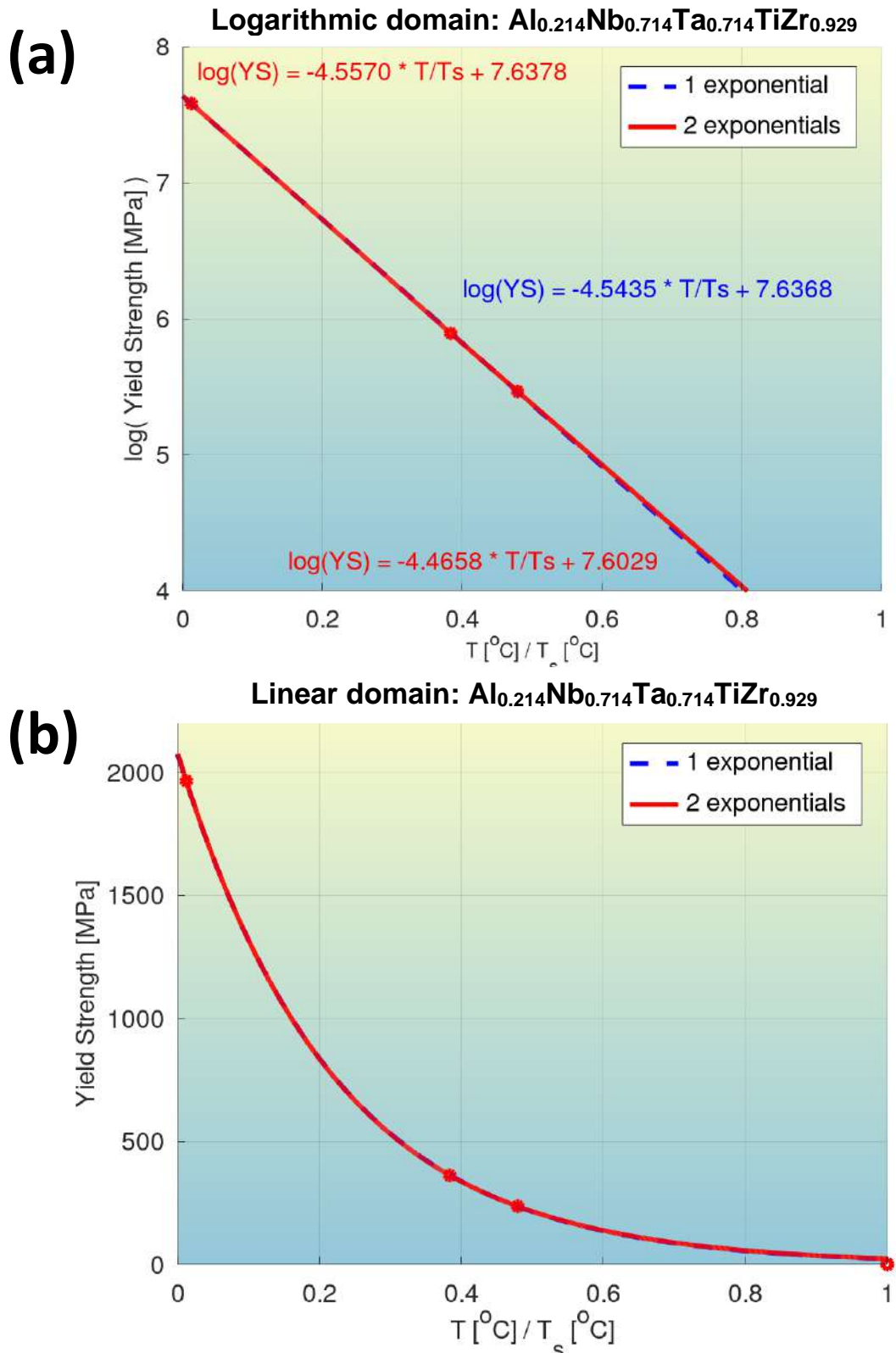

**Fig. S37**: Quantification of modeling accuracy of the bilinear log model, for composition No. 36 from **Tab. S1** (Al$_{0.214}$Nb$_{0.714}$Ta$_{0.714}$TiZr$_{0.929}$, BCC phase), and comparison to that of a model with a single exponential.

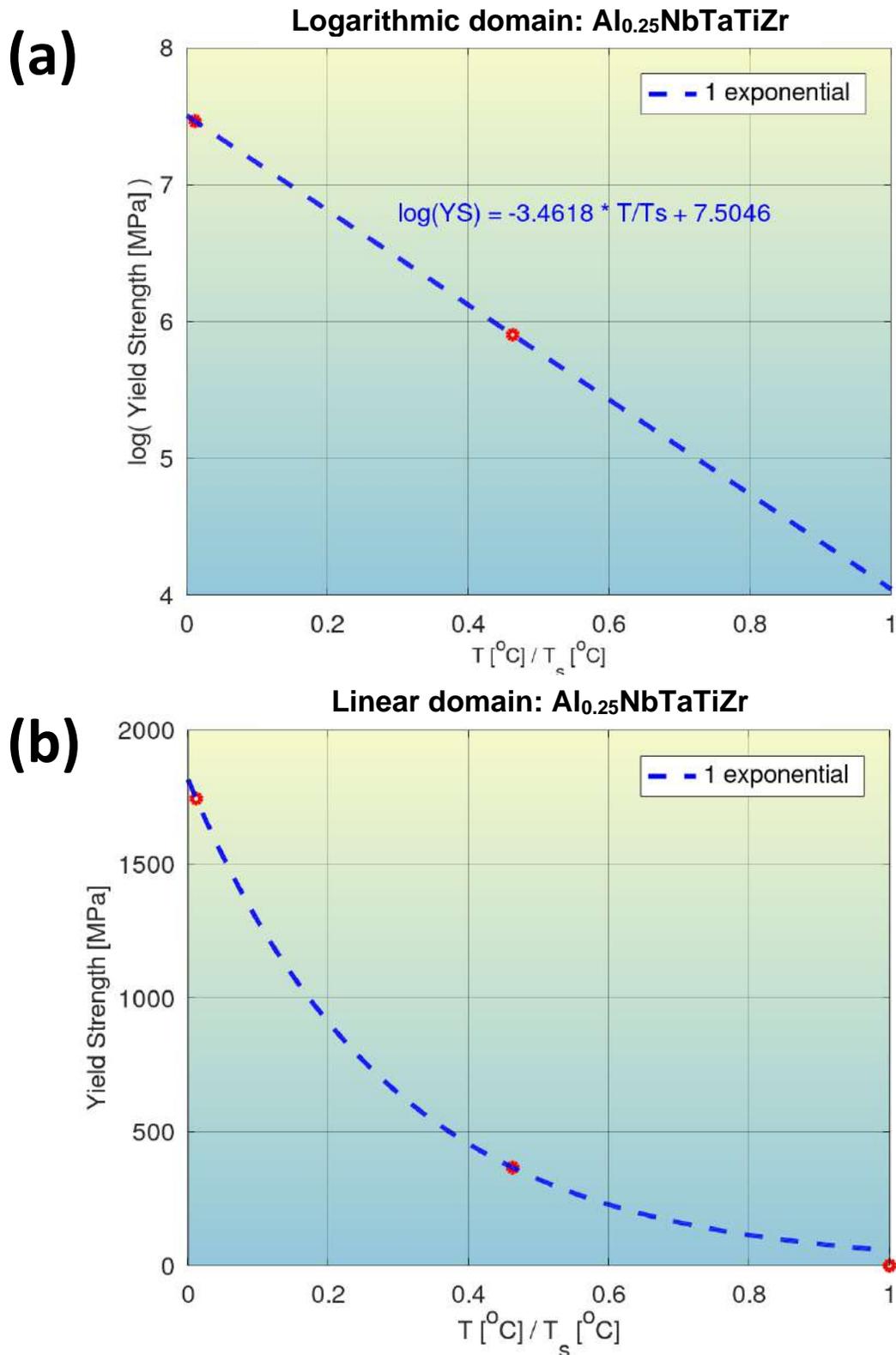

**Fig. S38**: Quantification of modeling accuracy of the bilinear log model, for composition No. 37 from **Tab. S1** (Al$_{0.25}$NbTaTiZr, BCC+B2 phase), and comparison to that of a model with a single exponential.

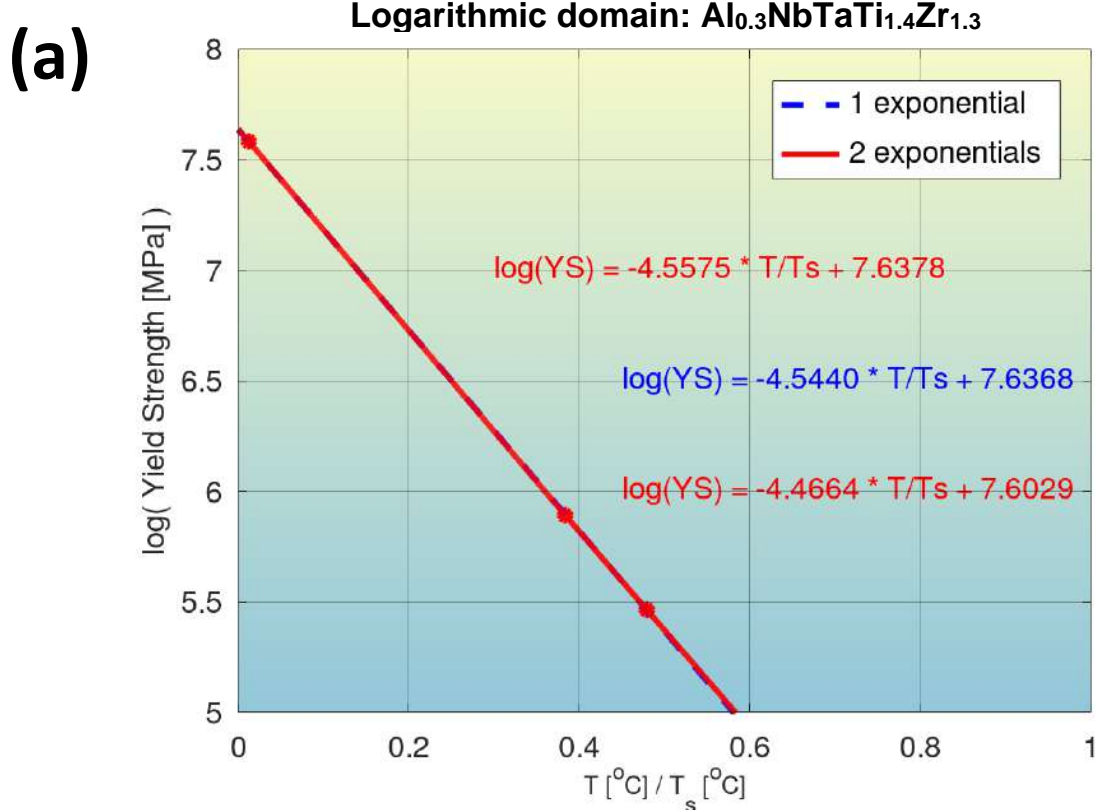

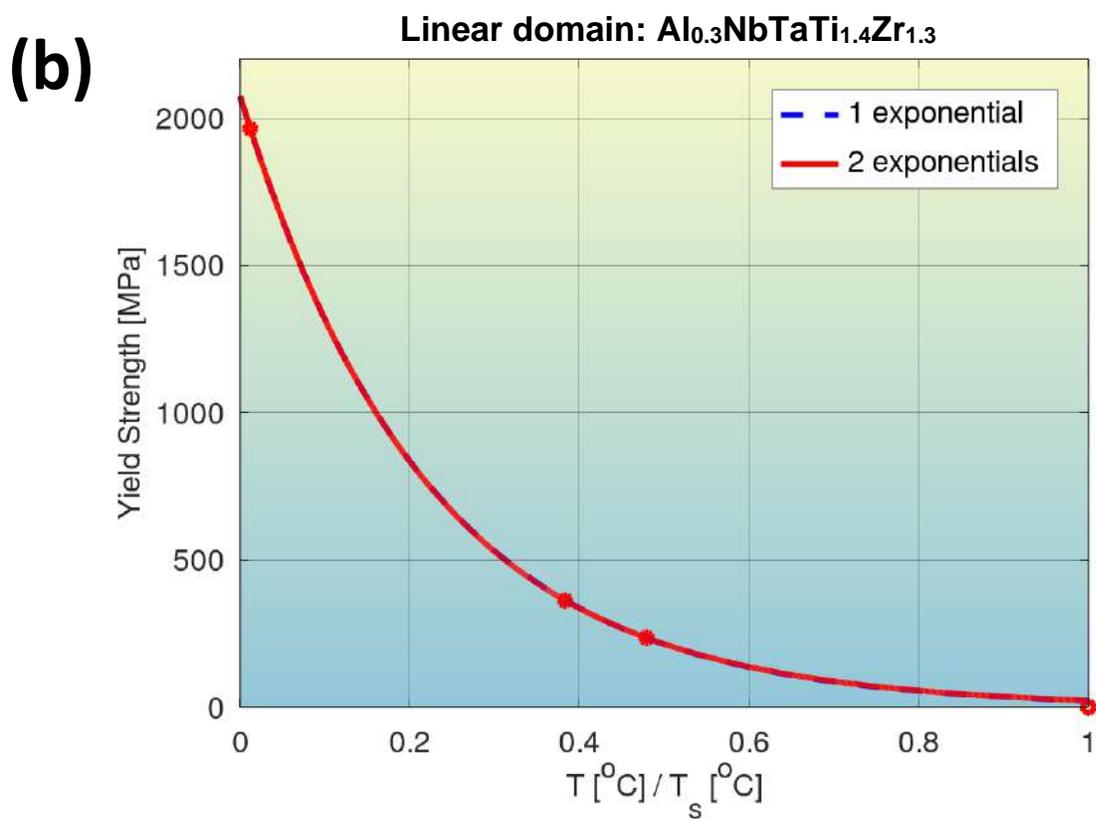

**Fig. S39**: Quantification of modeling accuracy of the bilinear log model, for composition No. 38 from **Tab. S1** (Al$_{0.3}$NbTaTi$_{1.4}$Zr$_{1.3}$, BCC+B2 phase, and comparison to that of a model with a single exponential.

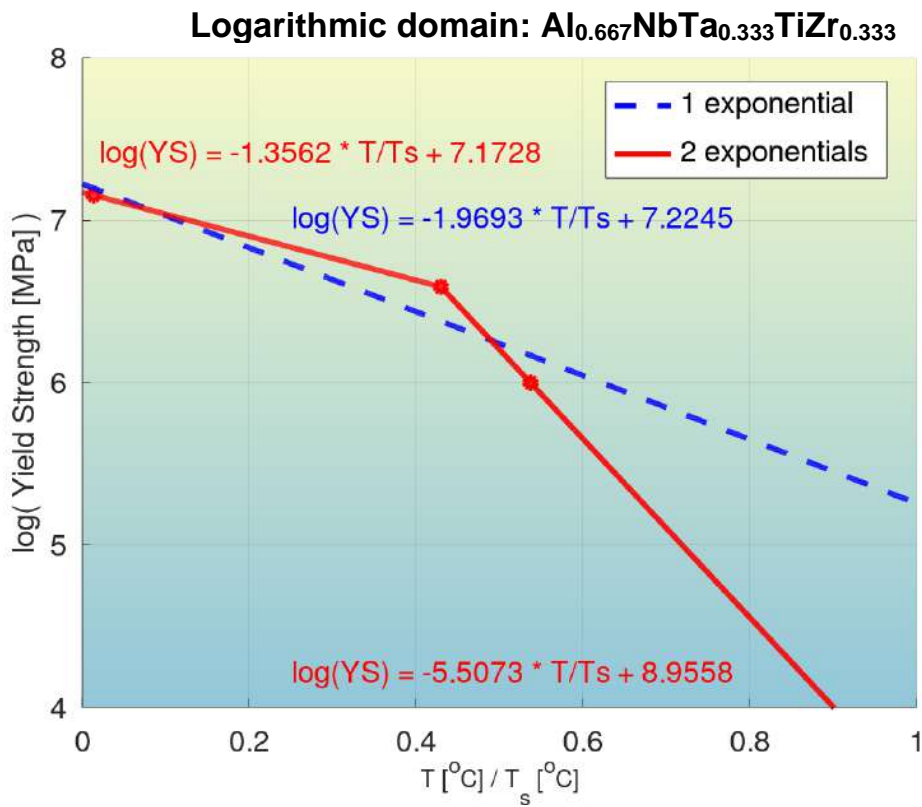

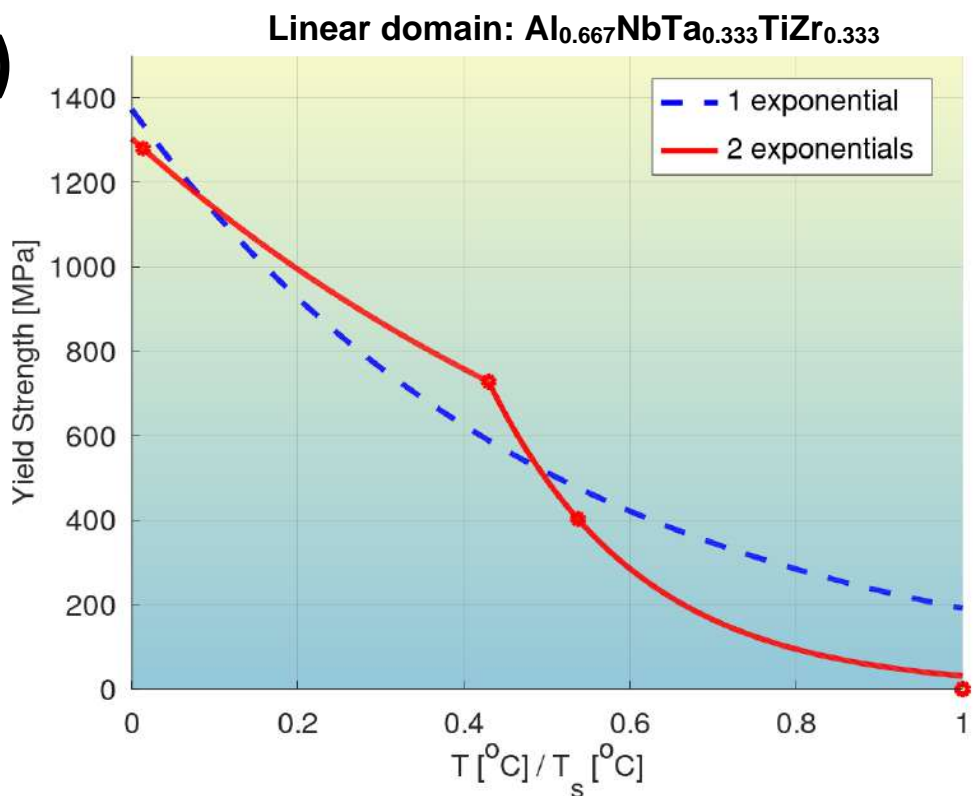

**Fig. S40**: Quantification of modeling accuracy of the bilinear log model, for composition No. 39 from **Tab. S1** (Al$_{0.667}$NbTa$_{0.333}$TiZr$_{0.333}$, BCC phase), and comparison to that of a model with a single exponential.

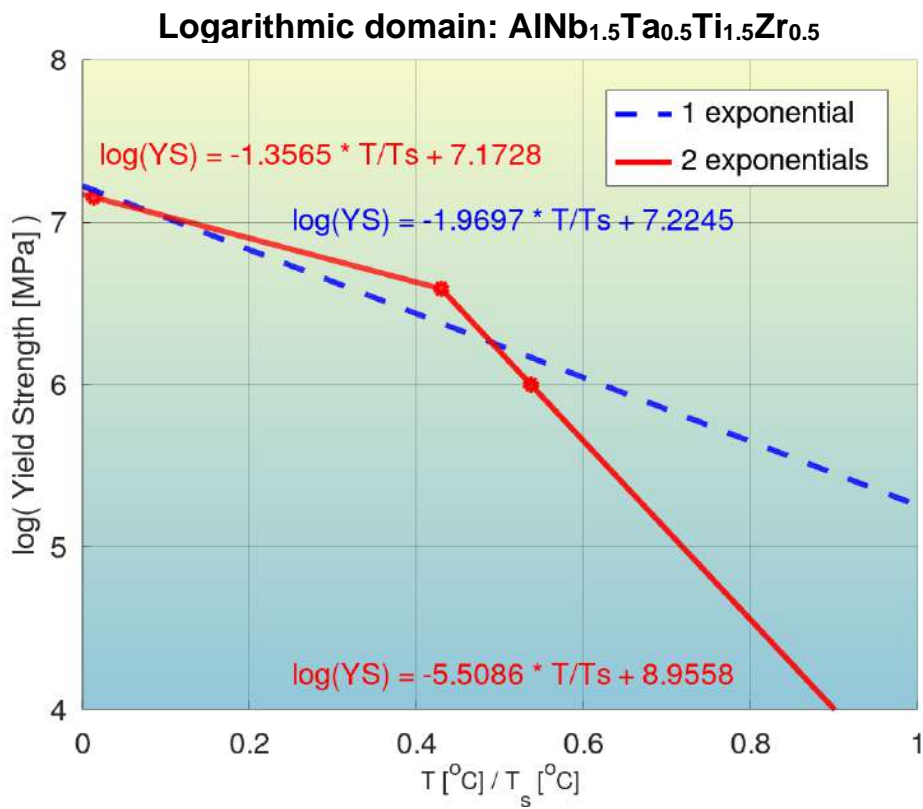

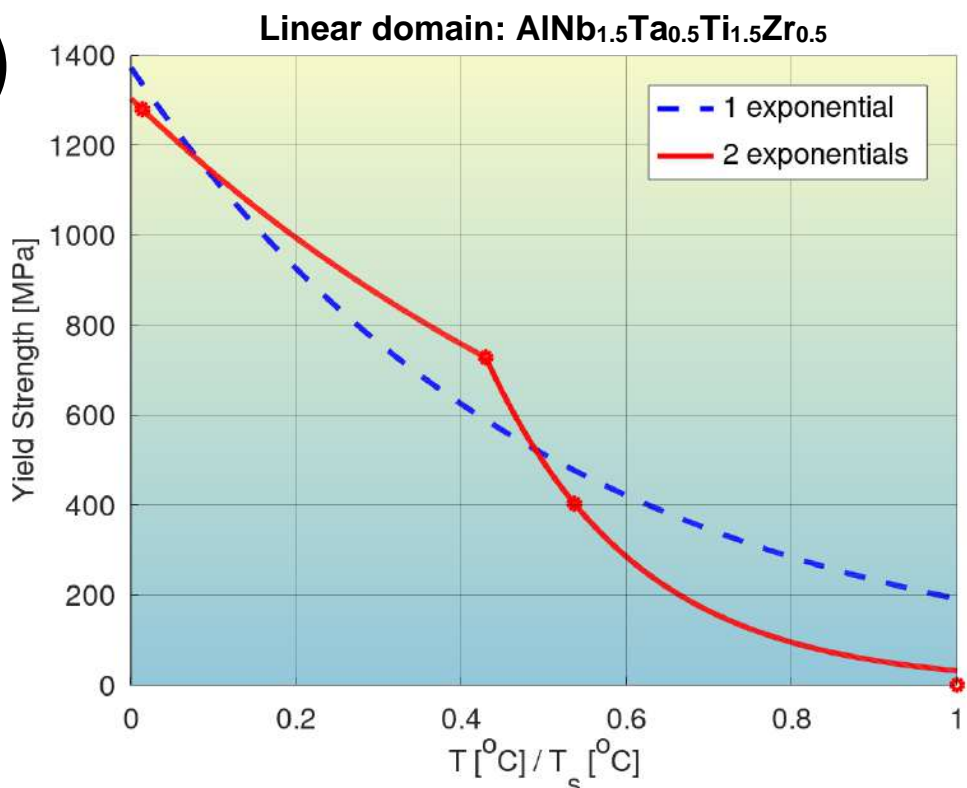

**Fig. S41**: Quantification of modeling accuracy of the bilinear log model, for composition No. 40 from **Tab. S1** (AlNb$_{1.5}$Ta$_{0.5}$Ti$_{1.5}$Zr$_{0.5}$, BCC phase), and comparison to that of a model with a single exponential.

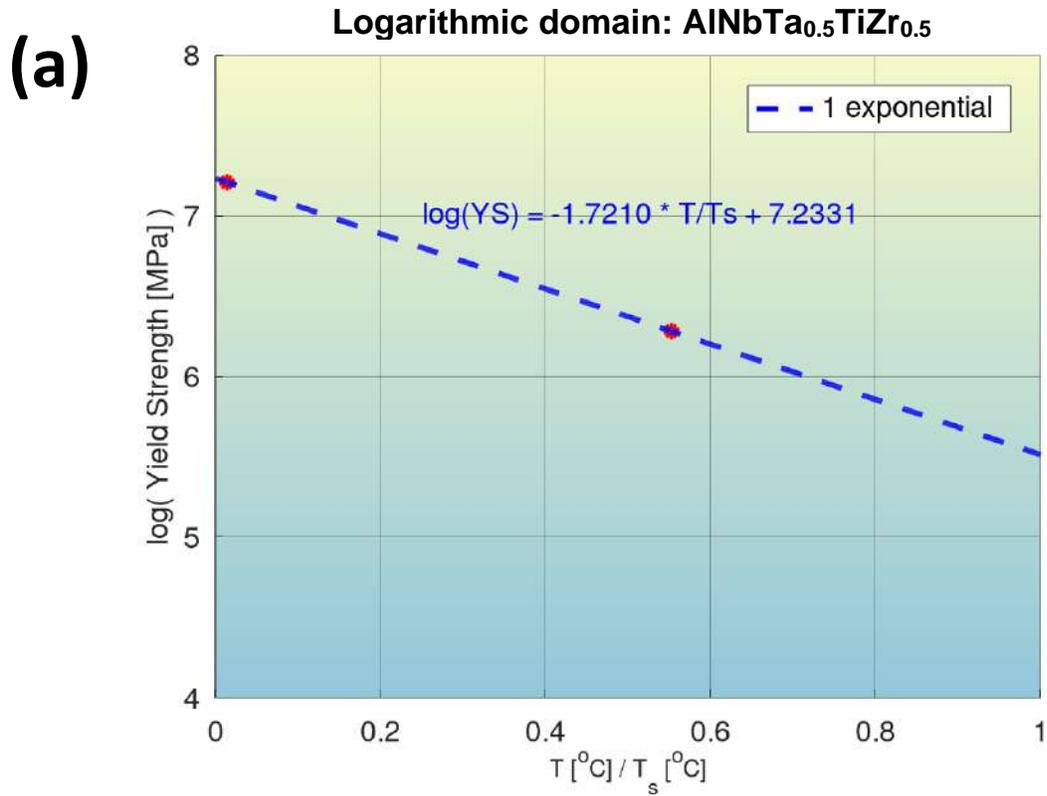

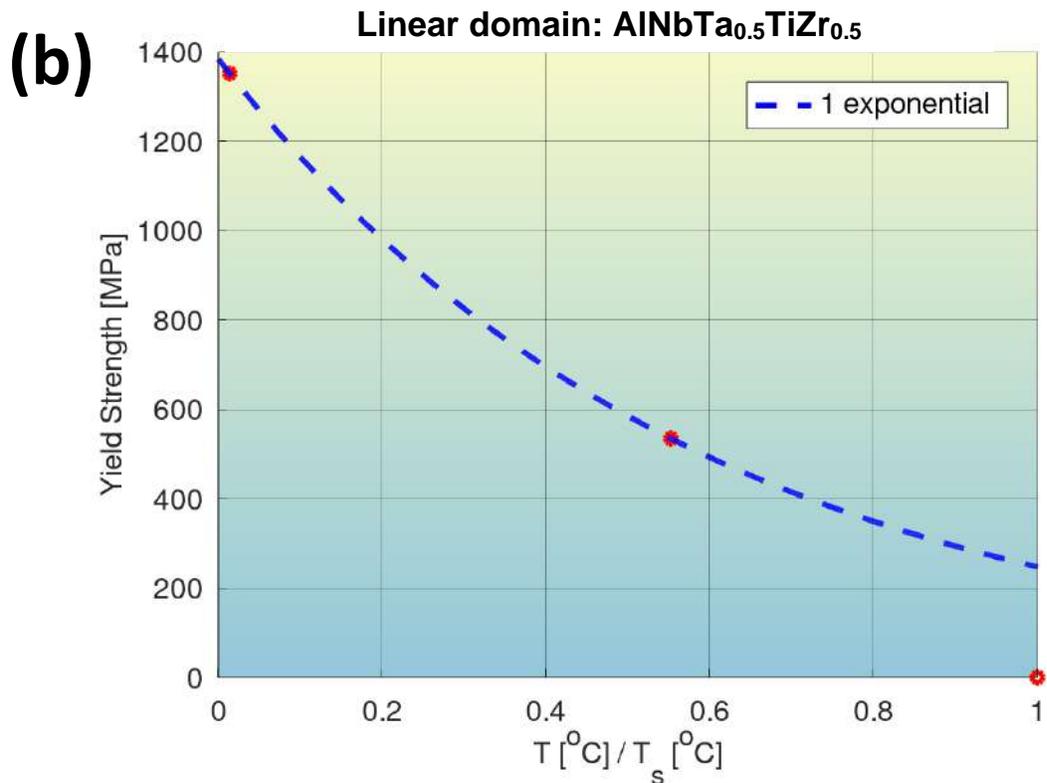

**Fig. S42**: Quantification of modeling accuracy of the bilinear log model for composition No. 41 from **Tab. S1** (AlNbTa$_{0.5}$TiZr$_{0.5}$, B2 phase), and comparison to that of a model with a single exponential.

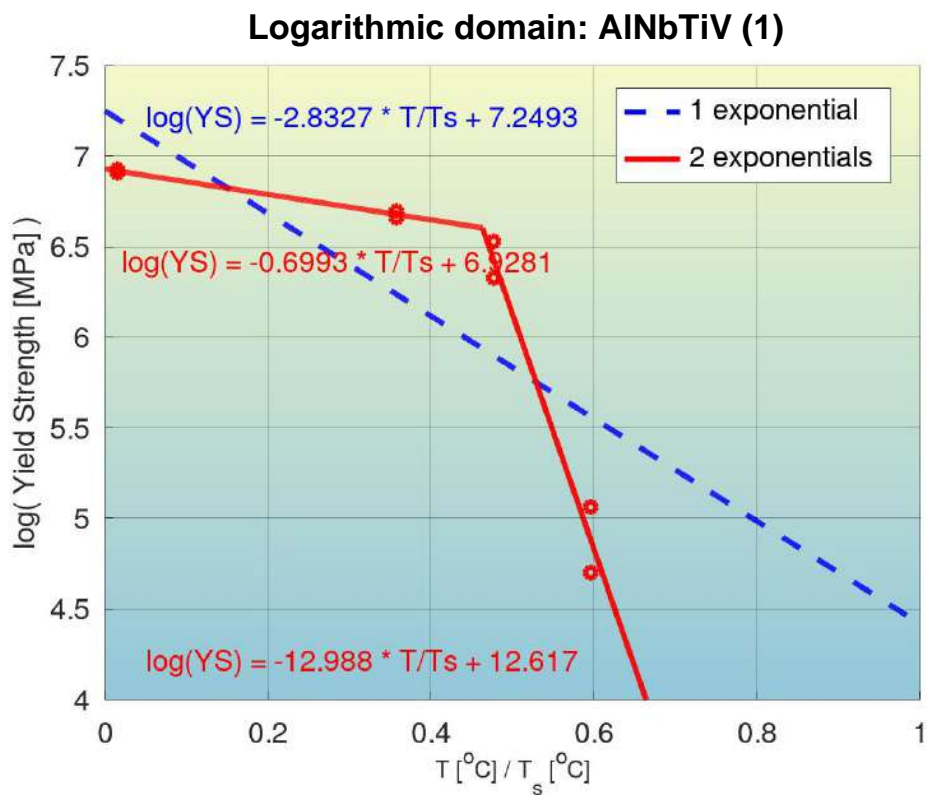

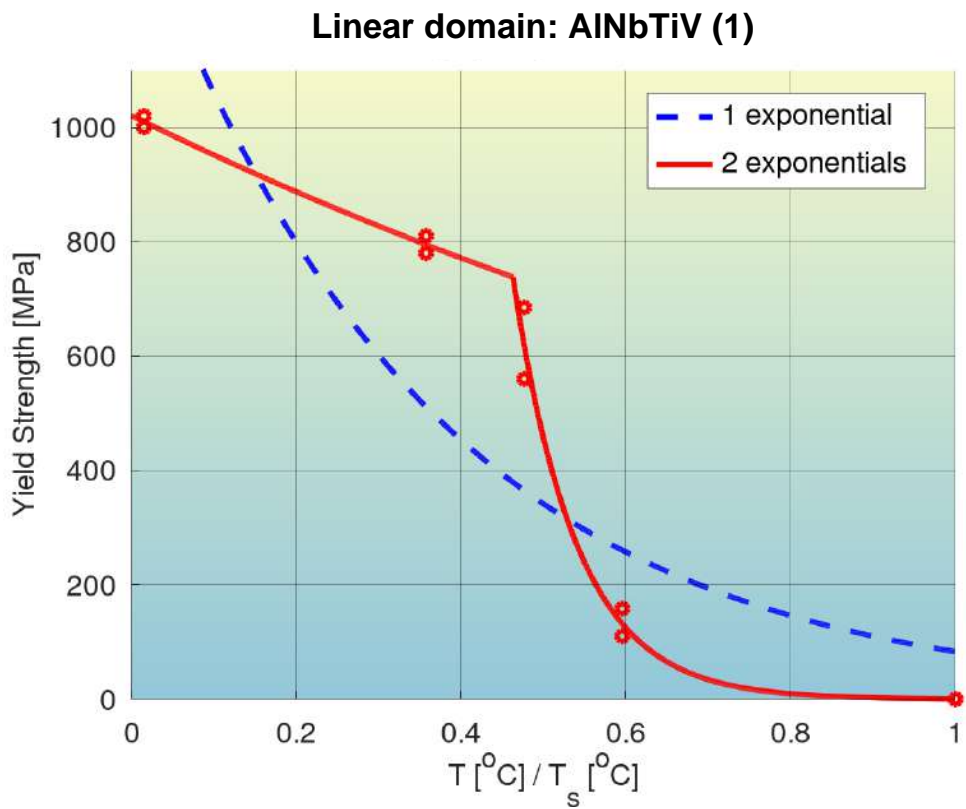

**Fig. S43**: Quantification of modeling accuracy of the bilinear log model, for composition No. 42 from **Tab. S1** (AlNbTiV (1), BCC phase), and comparison to that of a model with a single exponential.

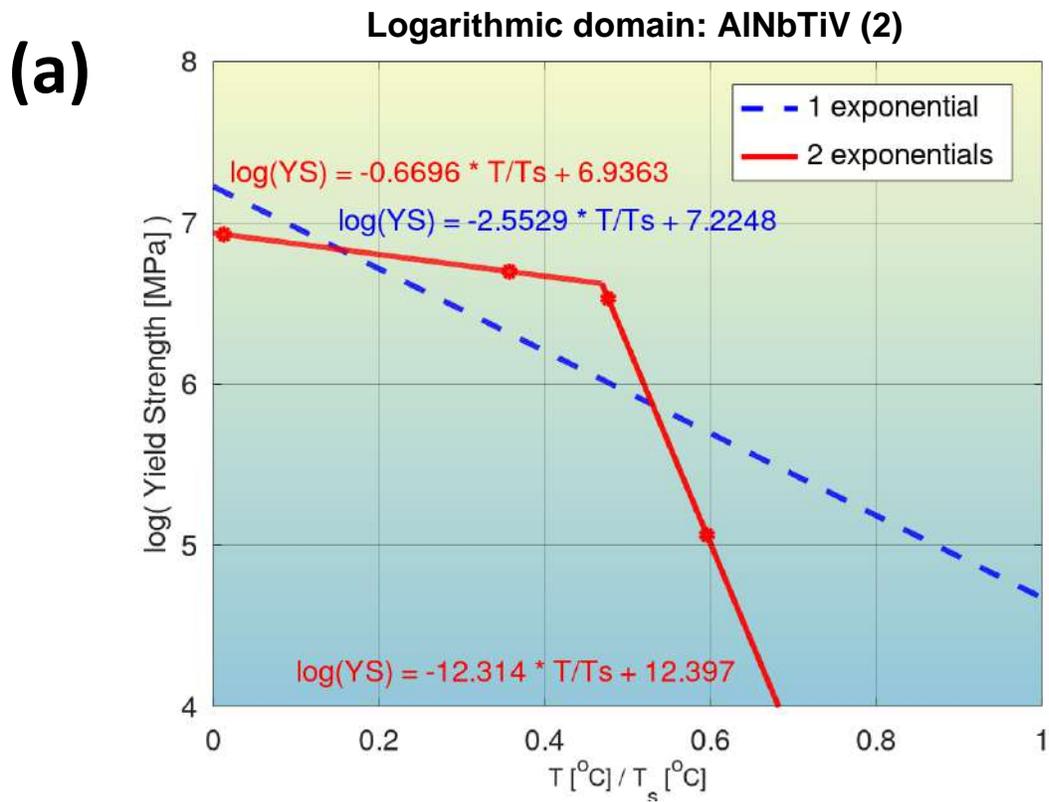

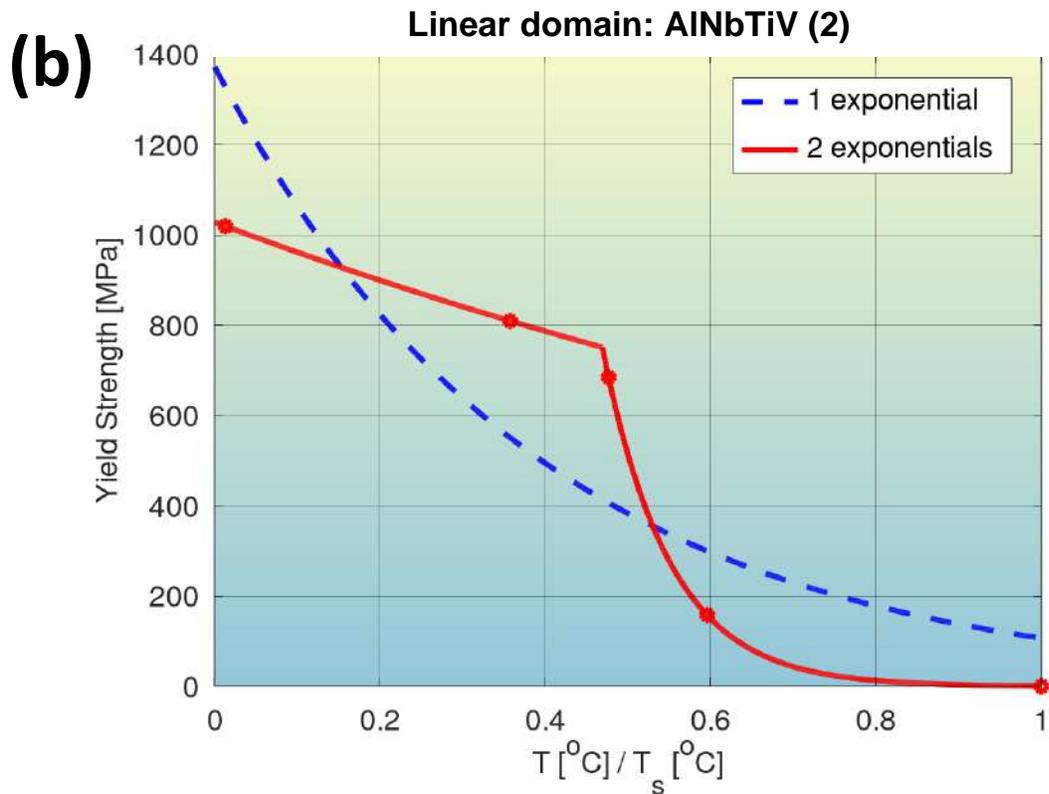

**Fig. S44**: Quantification of modeling accuracy of the bilinear log model for composition No. 43 from **Tab. S1** (AlNbTiV (2), BCC phase), and comparison to that of a model with a single exponential.

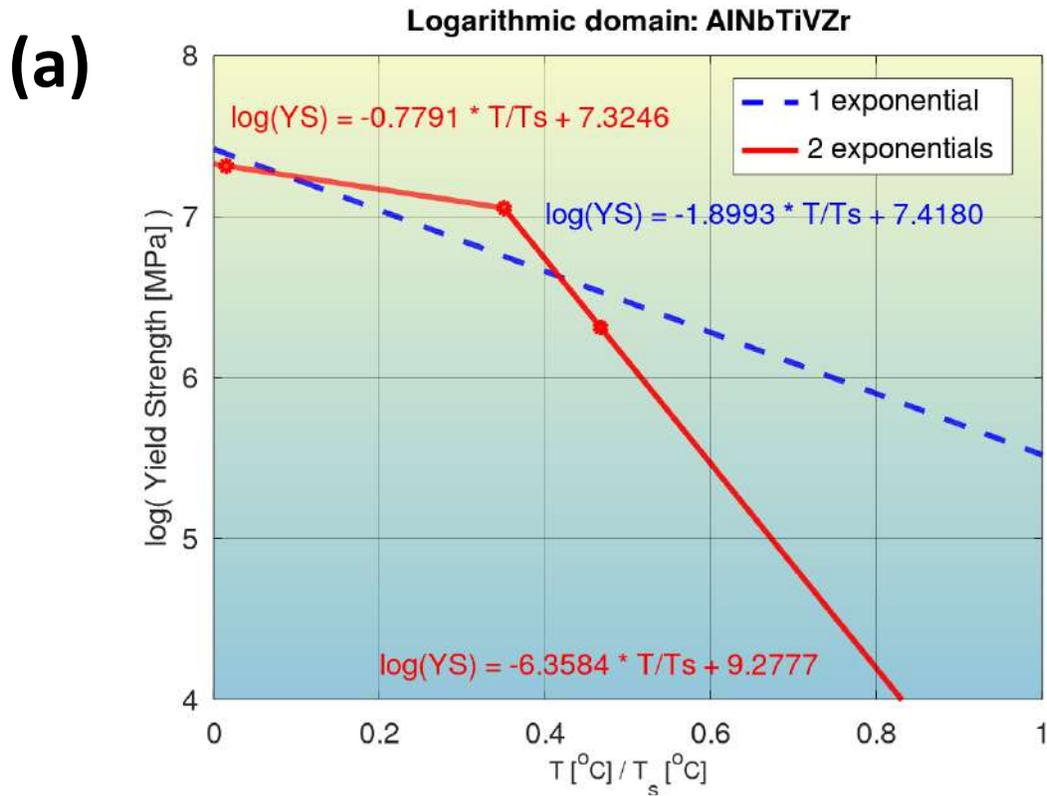

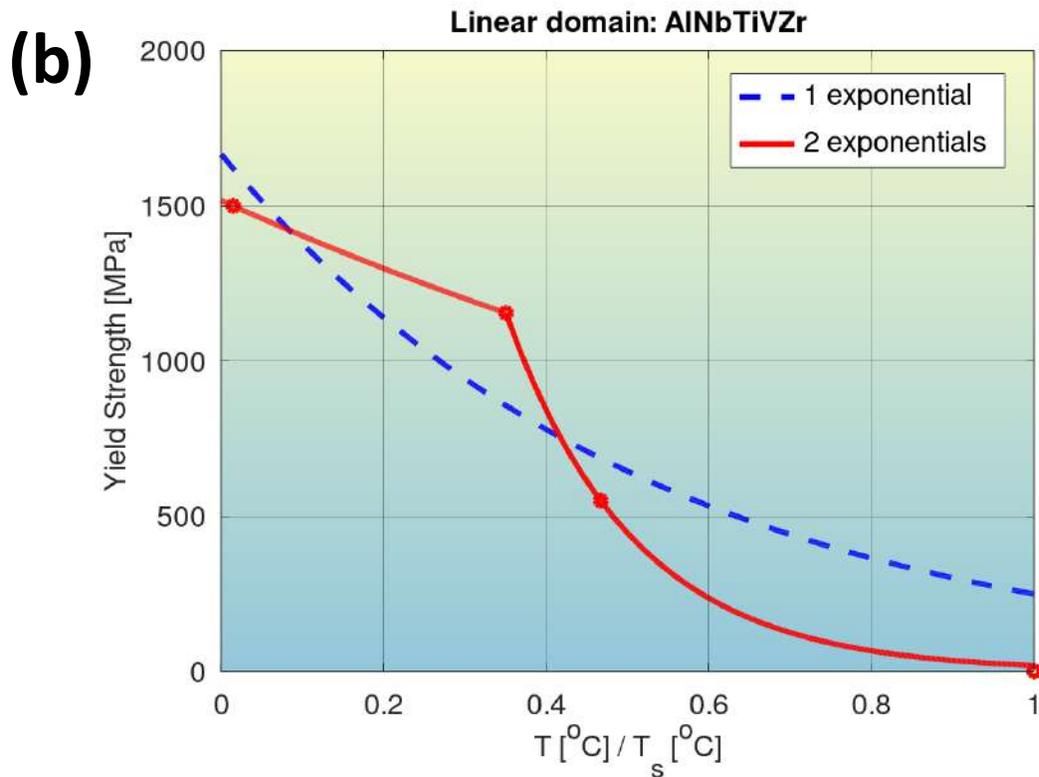

**Fig. S45**: Quantification of modeling accuracy of the bilinear log model for composition No. 44 from **Tab. S1** (AlNbTiVZr, B2+$Al_3Zr_5$+Laves phase), and comparison to that of a model with a single exponential.

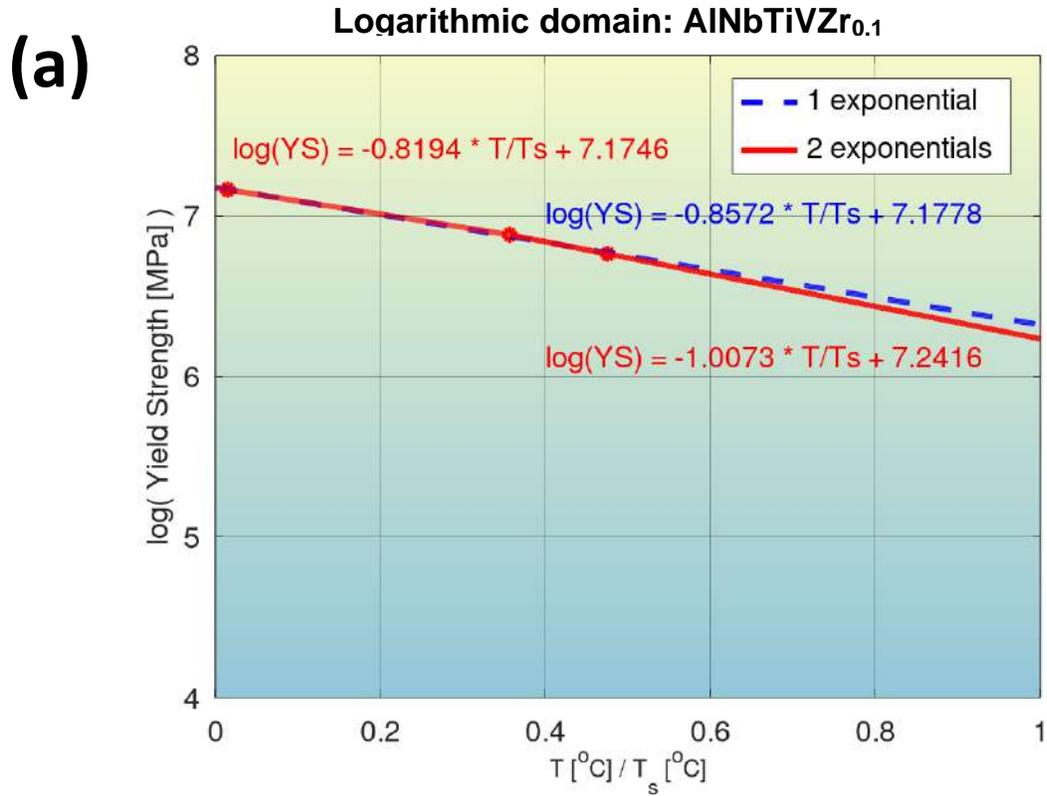

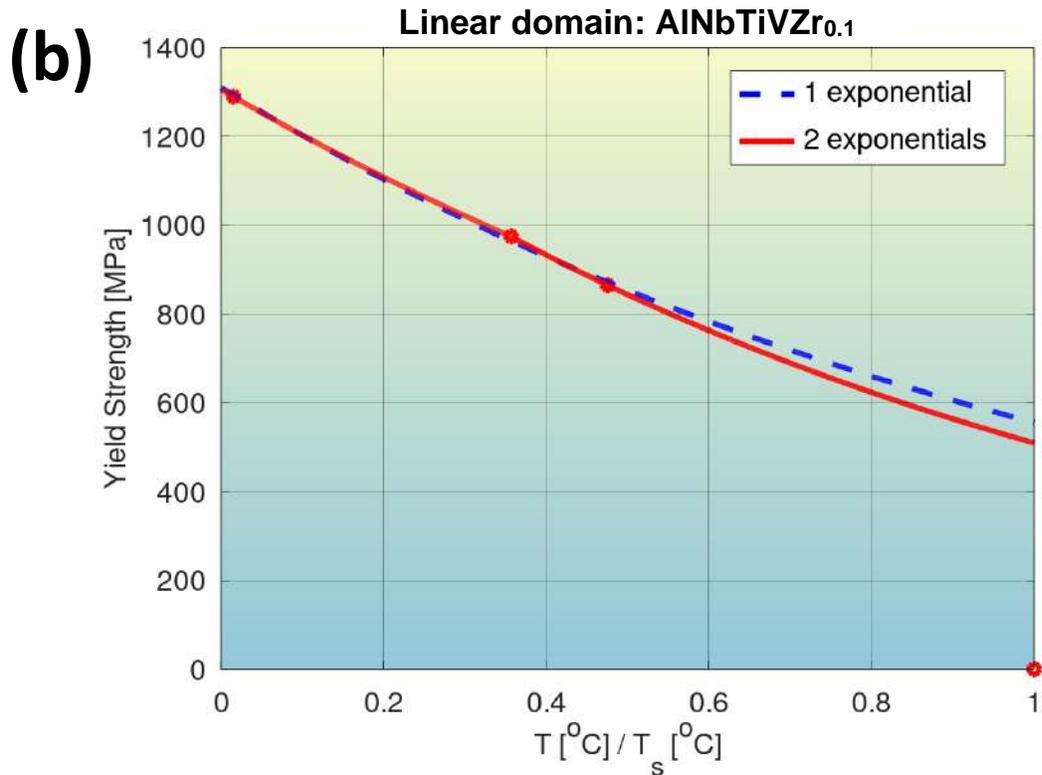

**Fig. S46**: Quantification of modeling accuracy of the bilinear log model, for composition No. 45 from **Tab. S1** (AlNbTiVZr$_{0.1}$, B2 + Al$_3$Zr$_5$ phase), and comparison to that of a model with a single exponential.

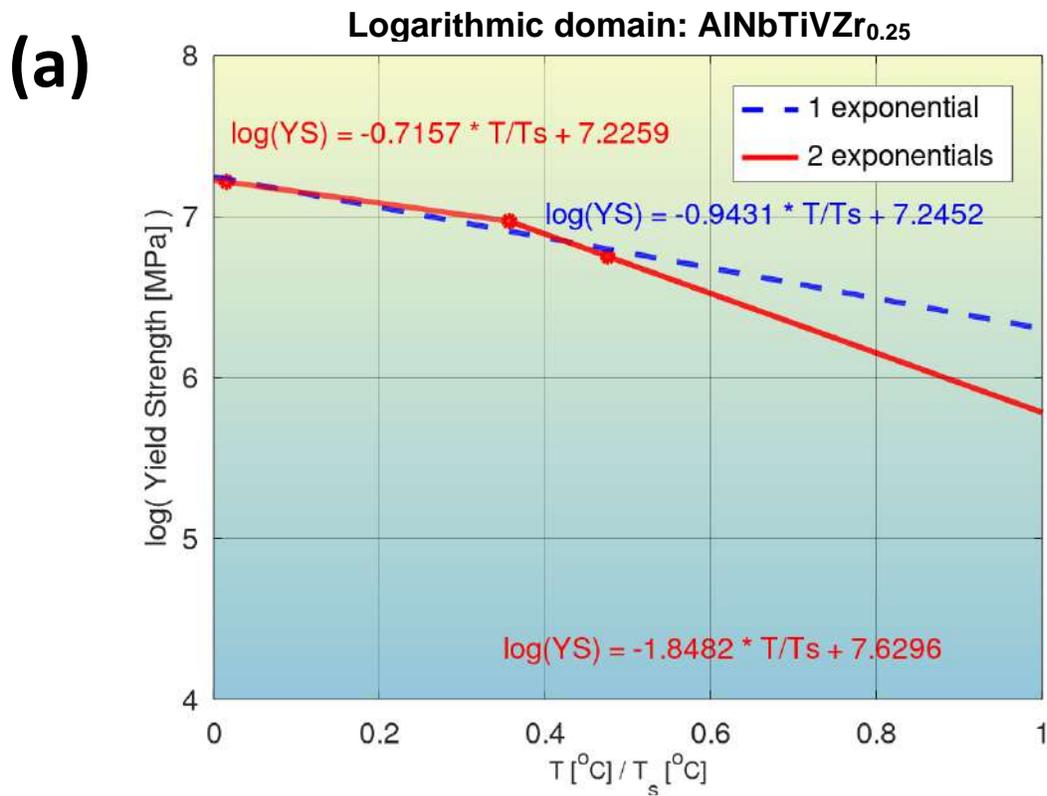
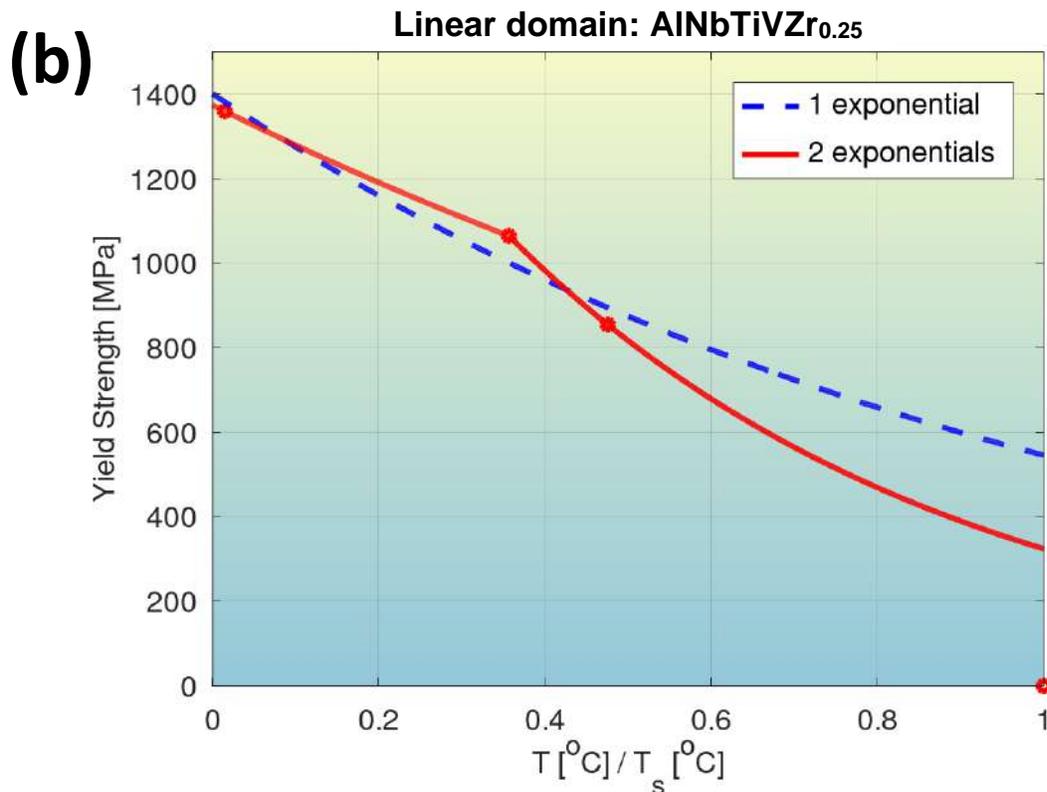

**Fig. S47**: Quantification of modeling accuracy of the bilinear log model, for composition No. 46 from **Tab. S1** (AlNbTiVZr$_{0.25}$, B2+Al$_3$Zr$_5$ phase), and comparison to that of a model with a single exponential.

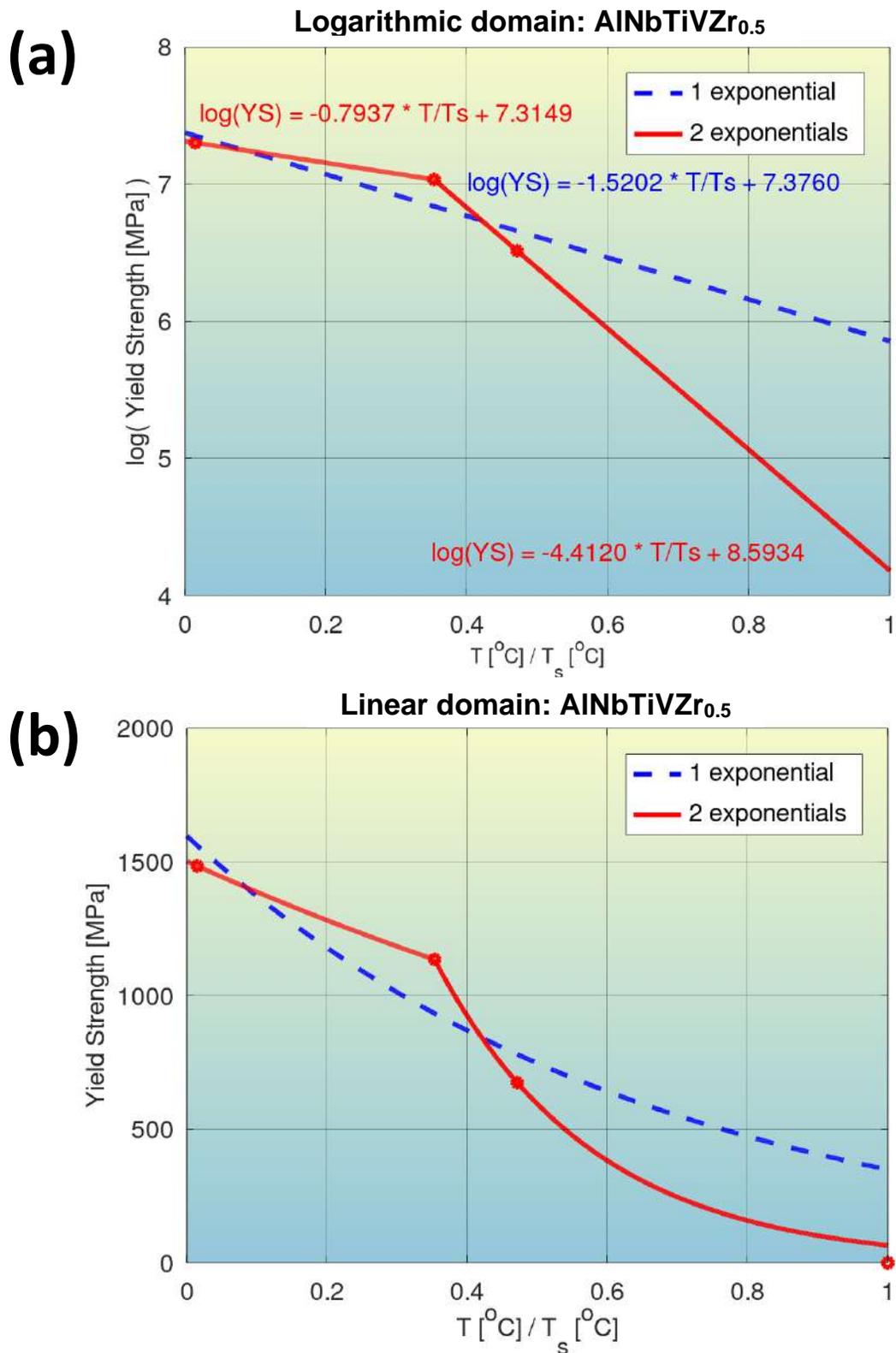

**Fig. S48**: Quantification of modeling accuracy of the bilinear log model, for composition No. 47 from **Tab. S1** (AlNbTiVZr$_{0.5}$, B2+Al$_3$Zr$_5$+Laves phase), and comparison to that of a model with a single exponential.

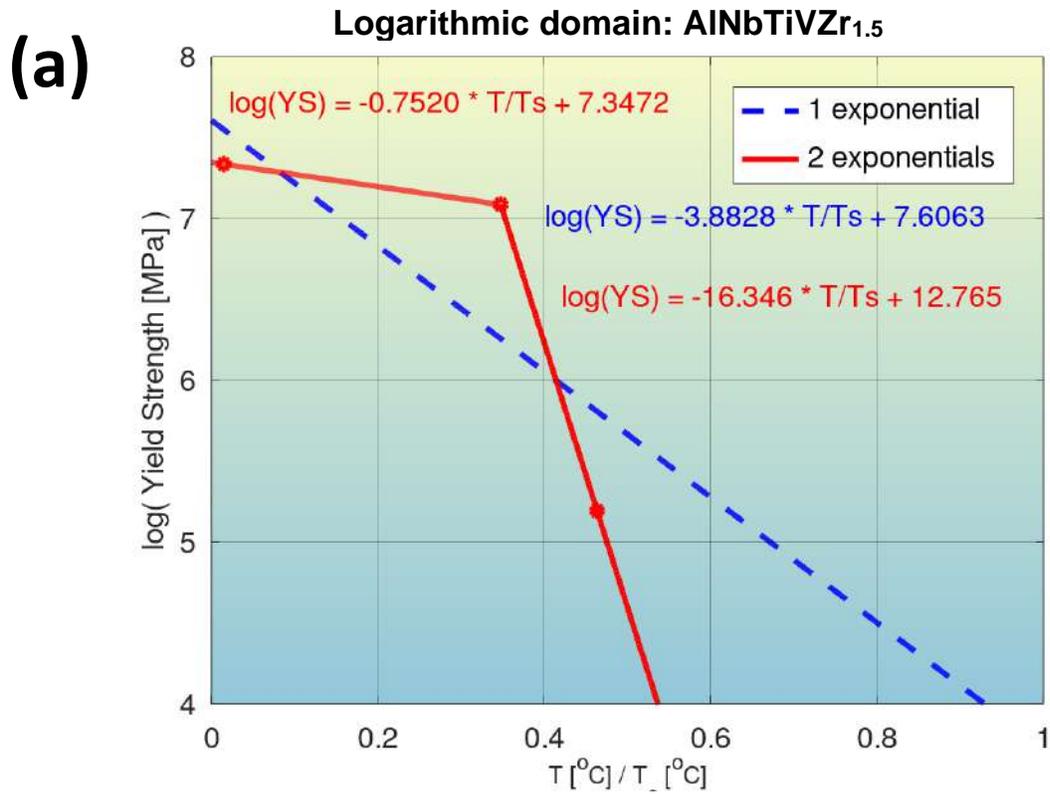

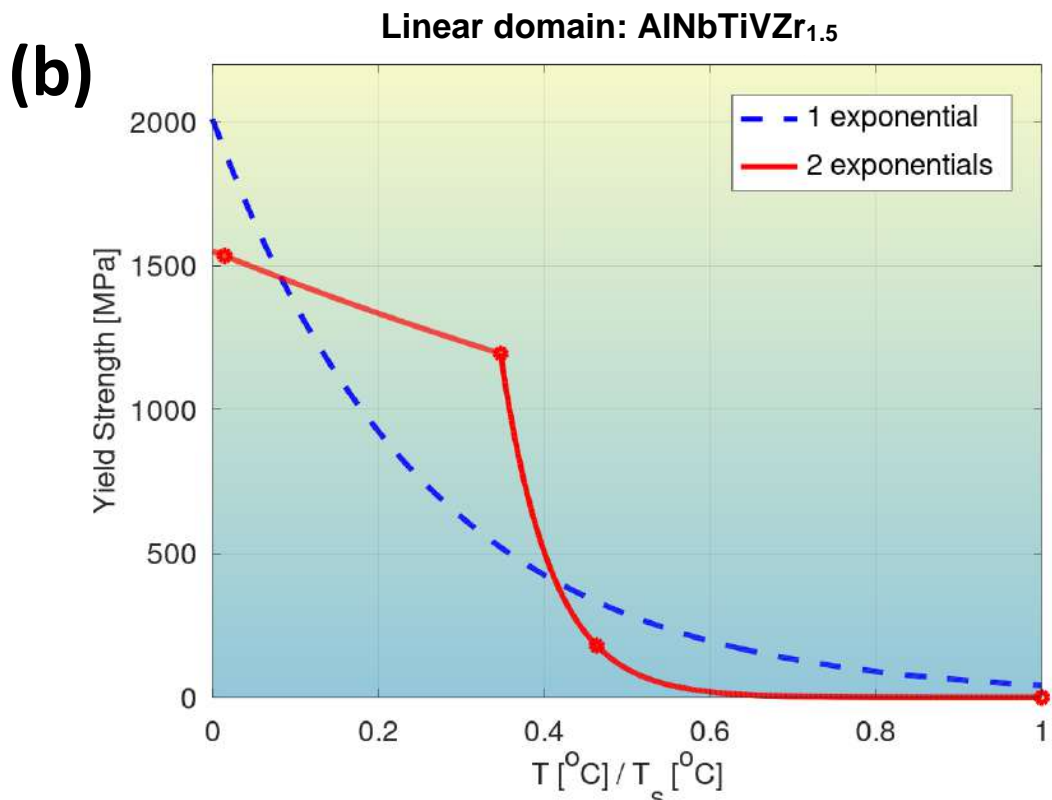

**Fig. S49**: Quantification of modeling accuracy of the bilinear log model, for composition No. 48 from **Tab. S1** (AlNbTiVZr$_{1.5}$, B2+Al$_3$Zr$_5$+Laves phase), and comparison to that of a model with a single exponential.

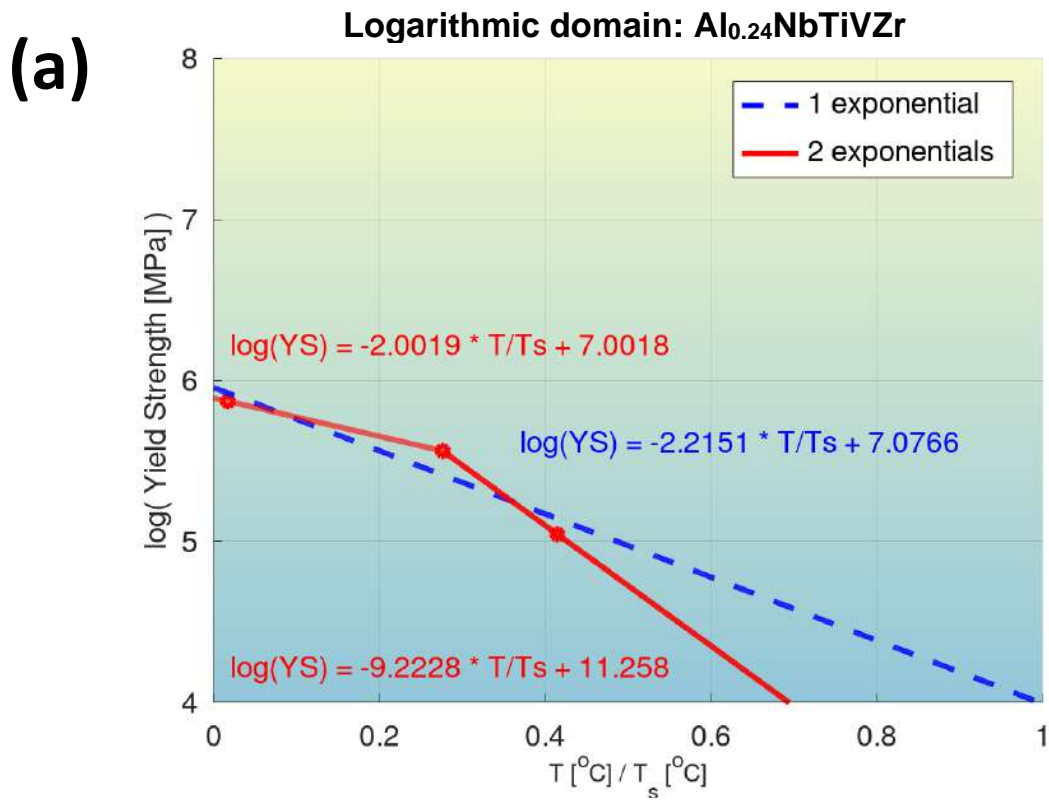

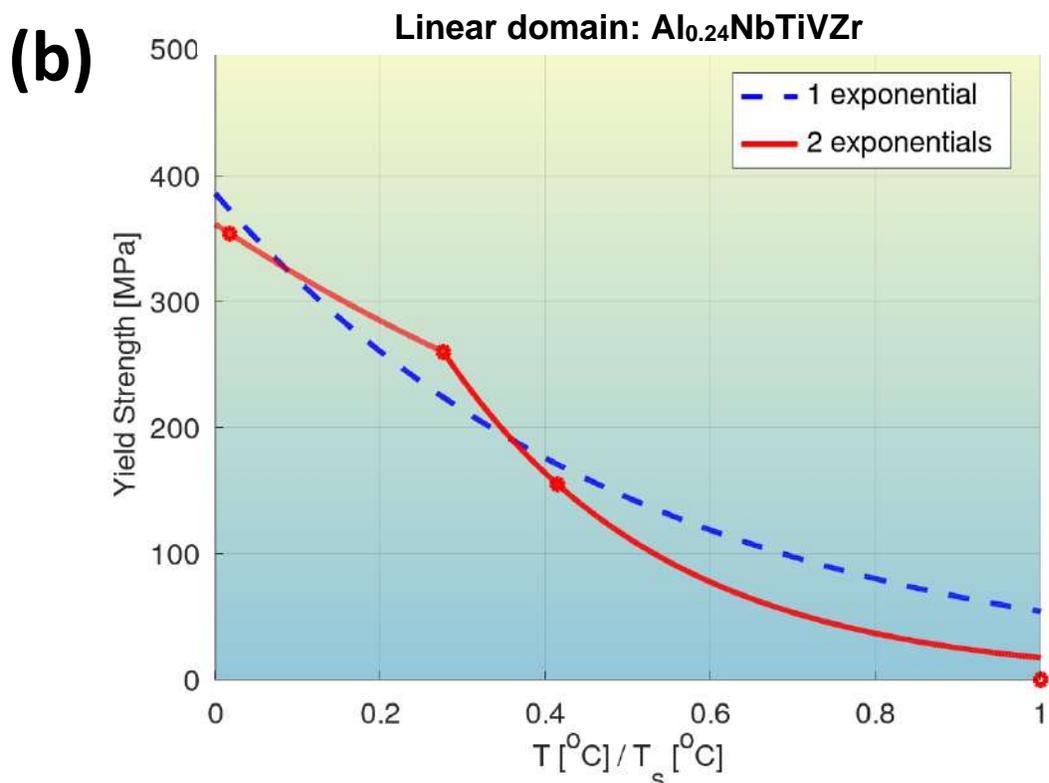

**Fig. S50**: Quantification of modeling accuracy of the bilinear log model, for composition No. 49 from **Tab. S2** (Al$_{0.24}$NbTiVZr, BCC+Laves phase), and comparison to that of a model with a single exponential.

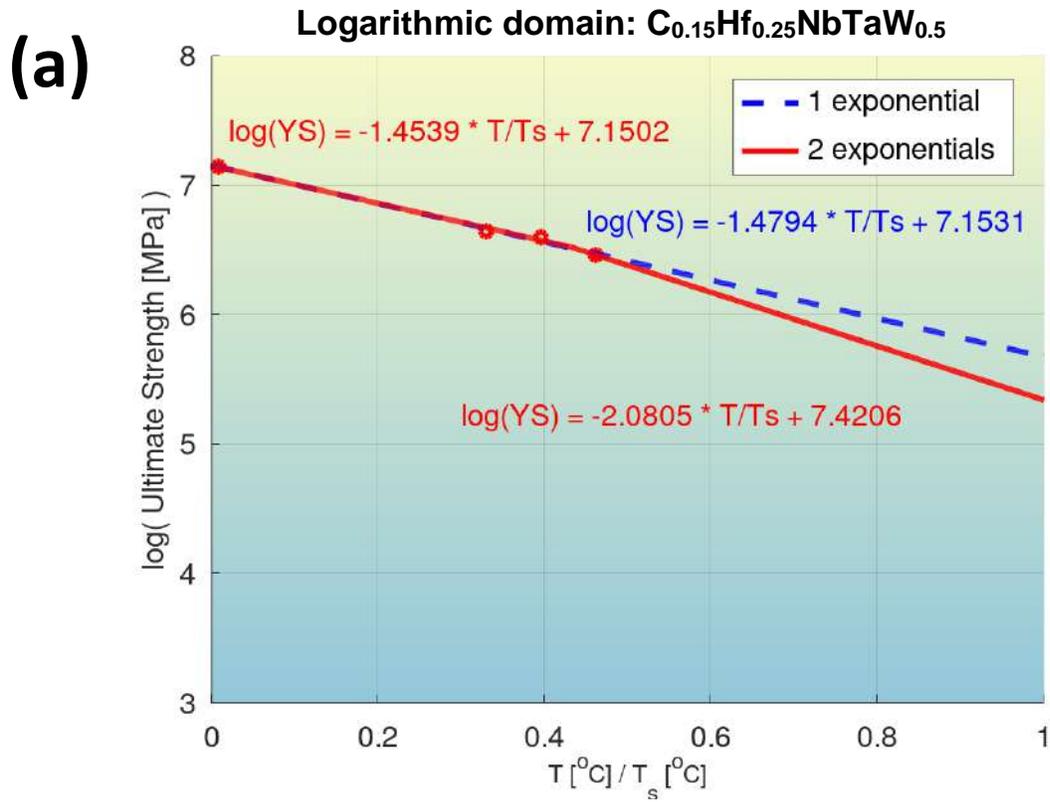

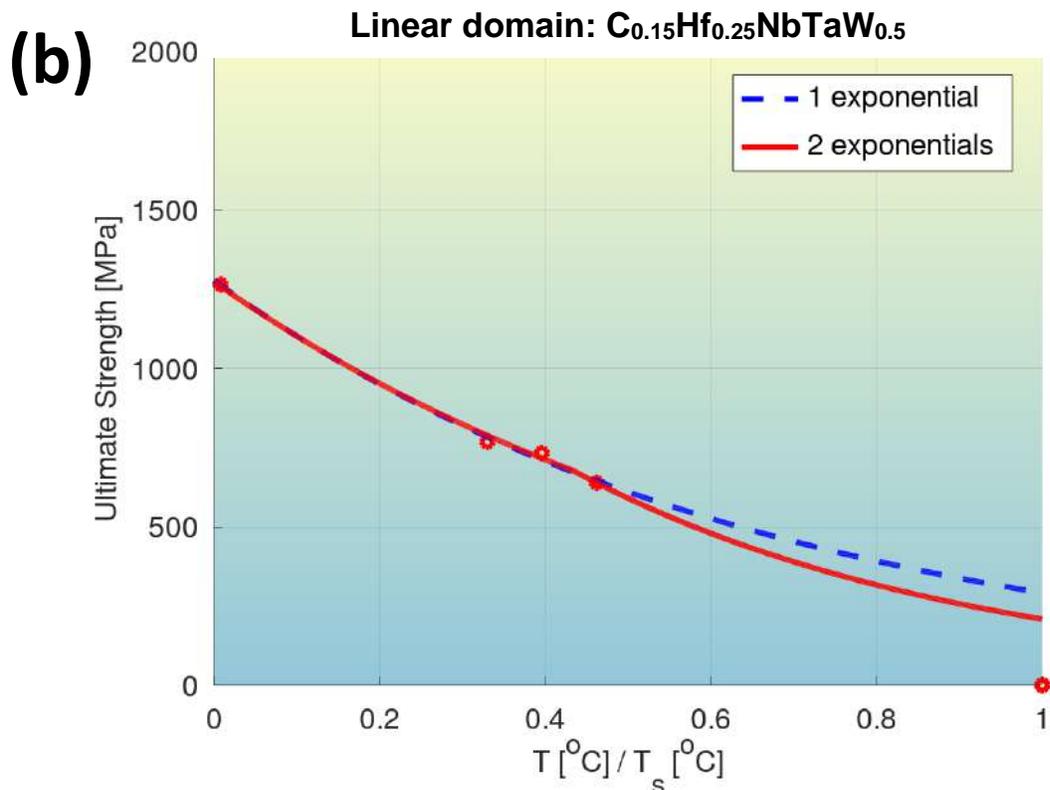

**Fig. S51**: Quantification of modeling accuracy of the bilinear log model, for composition No. 50 from **Tab. S2** ($C_{0.15}Hf_{0.25}NbTaW_{0.5}$, FCC+BCC), and comparison to that of a model with a single exponential.

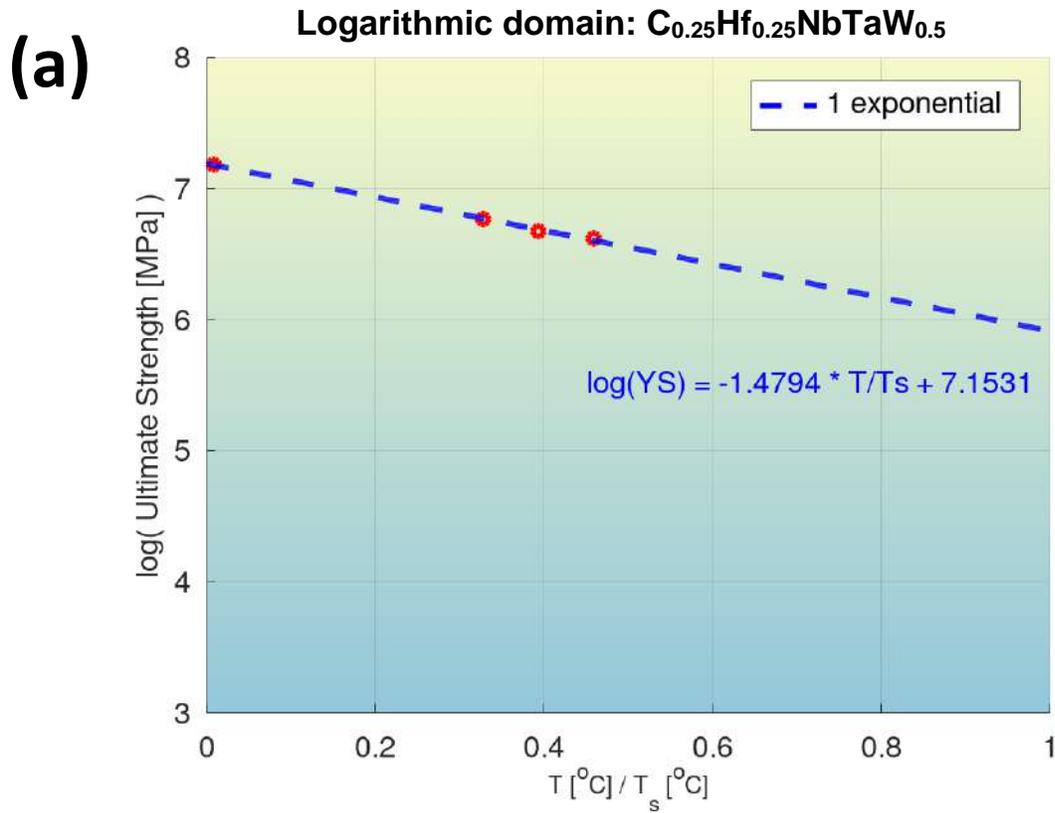

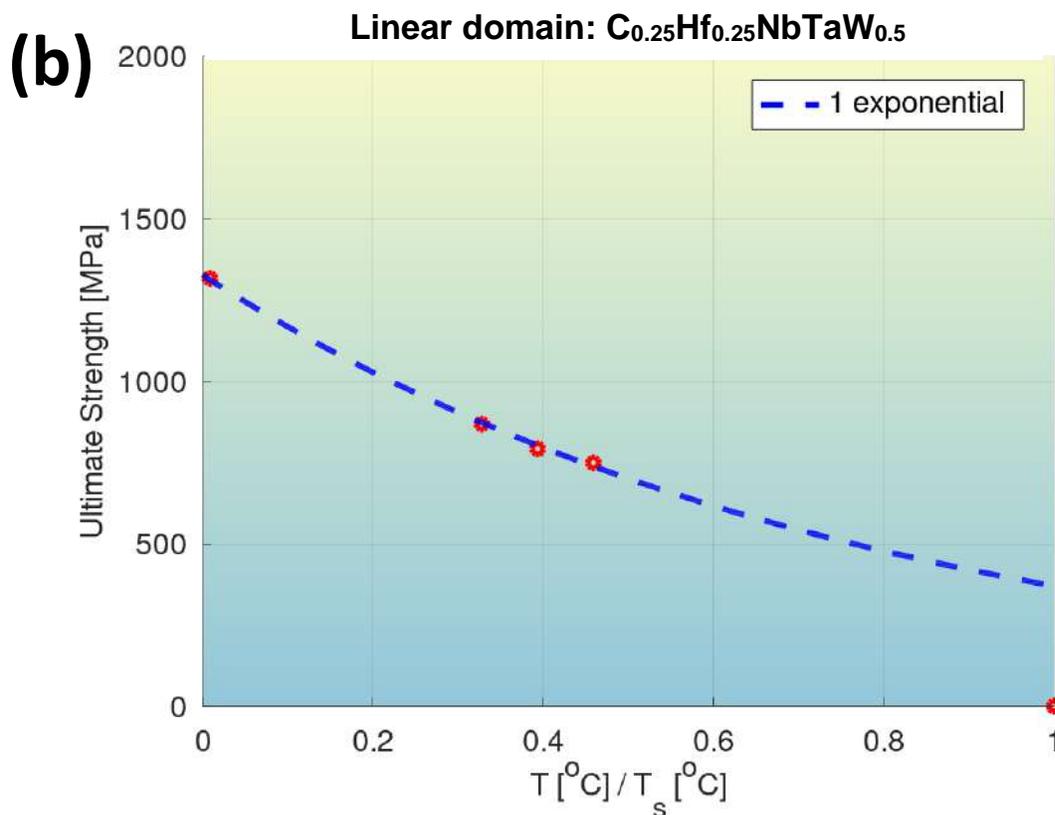

**Fig. S52**: Quantification of modeling accuracy of the bilinear log model, for composition No. 51 from **Tab. S2** ($C_{0.25}Hf_{0.25}NbTaW_{0.5}$, FCC+BCC), and comparison to that of a model with a single exponential.

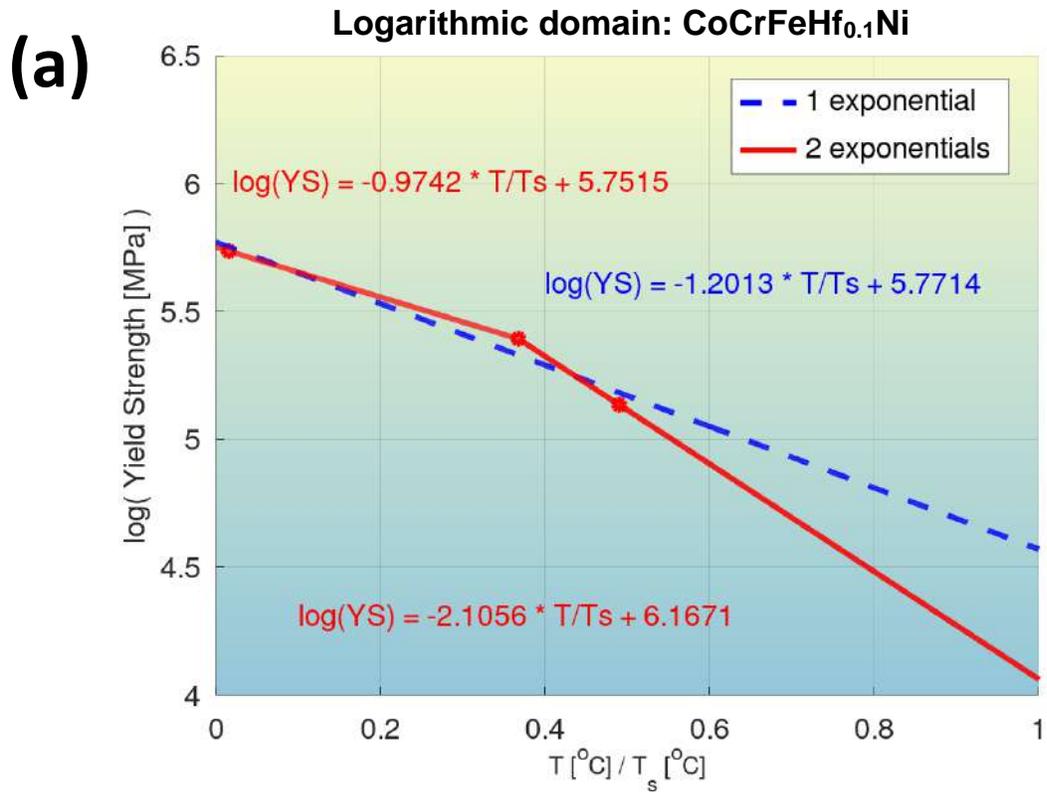
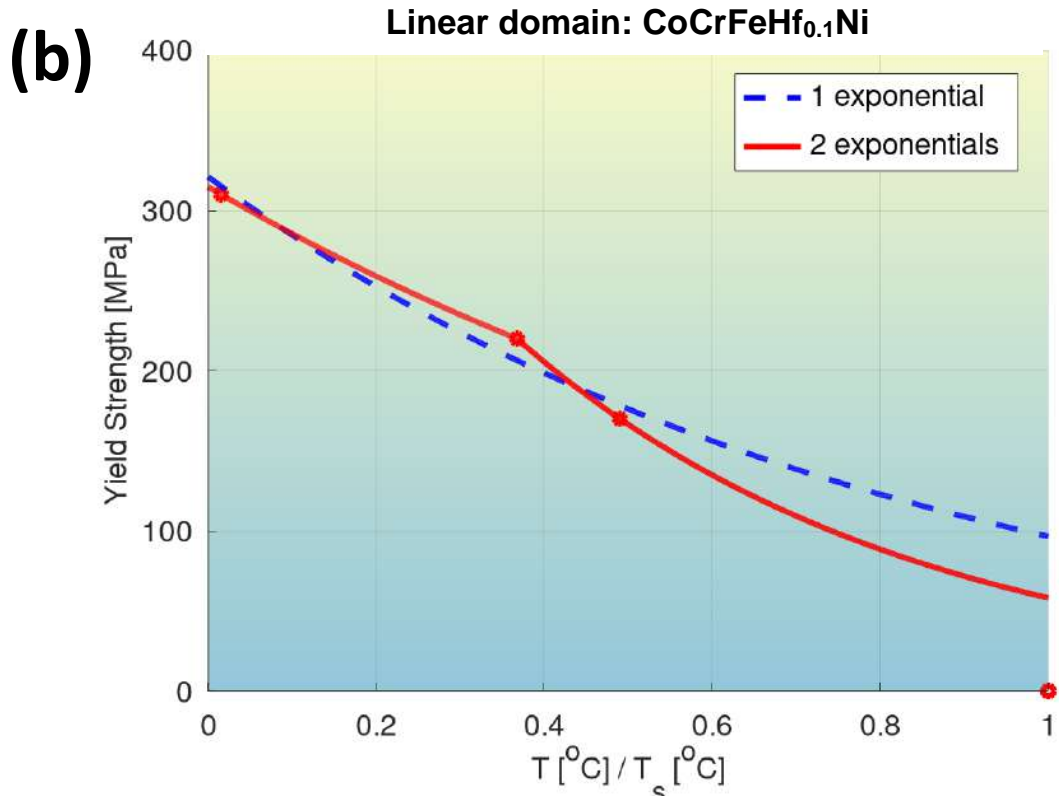

**Fig. S53**: Quantification of modeling accuracy of the bilinear log model, for composition No. 52 from **Tab. S2** (CoCrFeHf$_{0.1}$Ni, FCC+Laves phases), and comparison to that of a model with a single exponential.

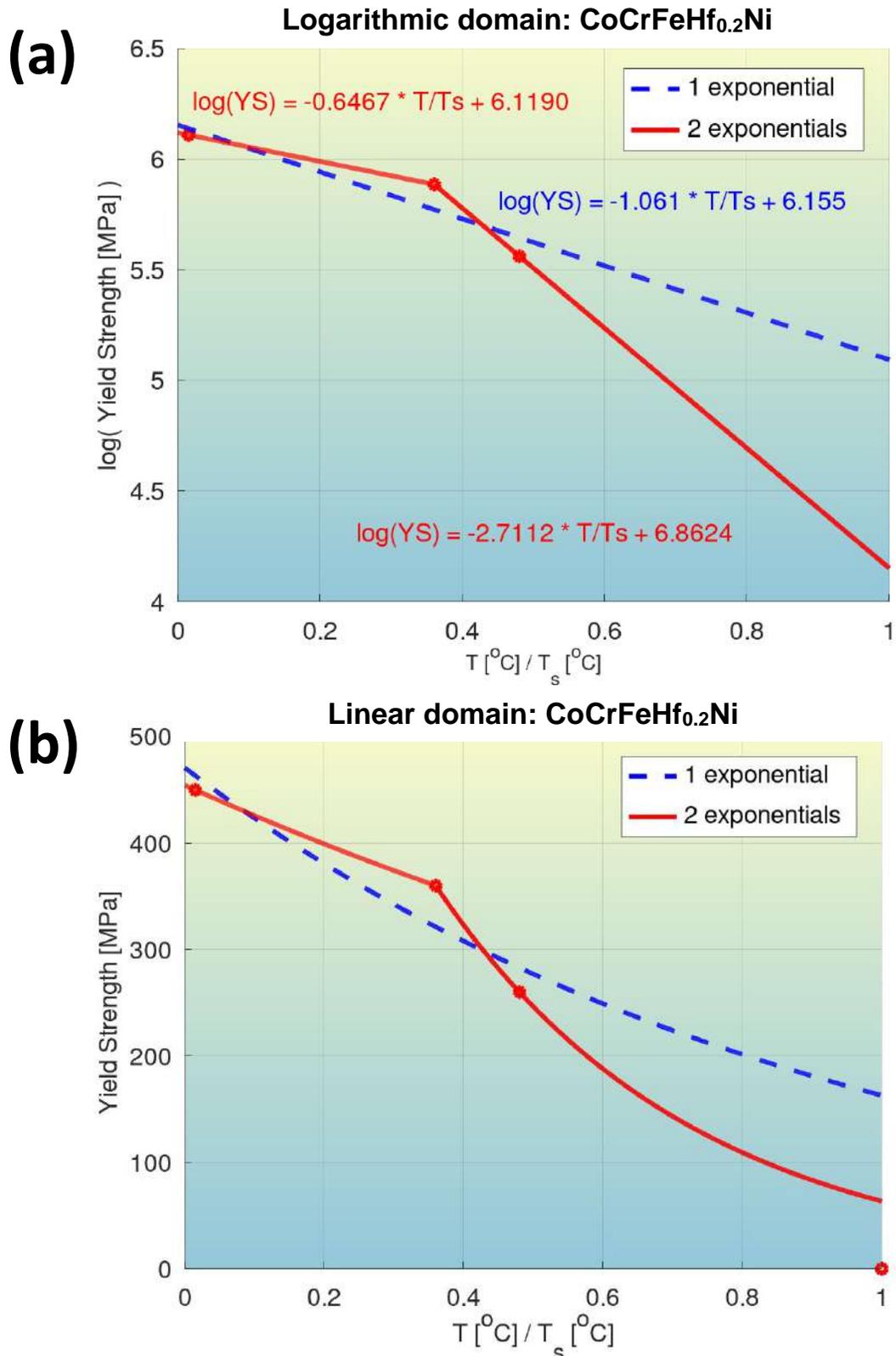

**Fig. S54**: Quantification of modeling accuracy of the bilinear log model, for composition No. 53 from **Tab. S2** (CoCrFeHf$_{0.2}$Ni, FCC+Laves phases), and comparison to that of a model with a single exponential.

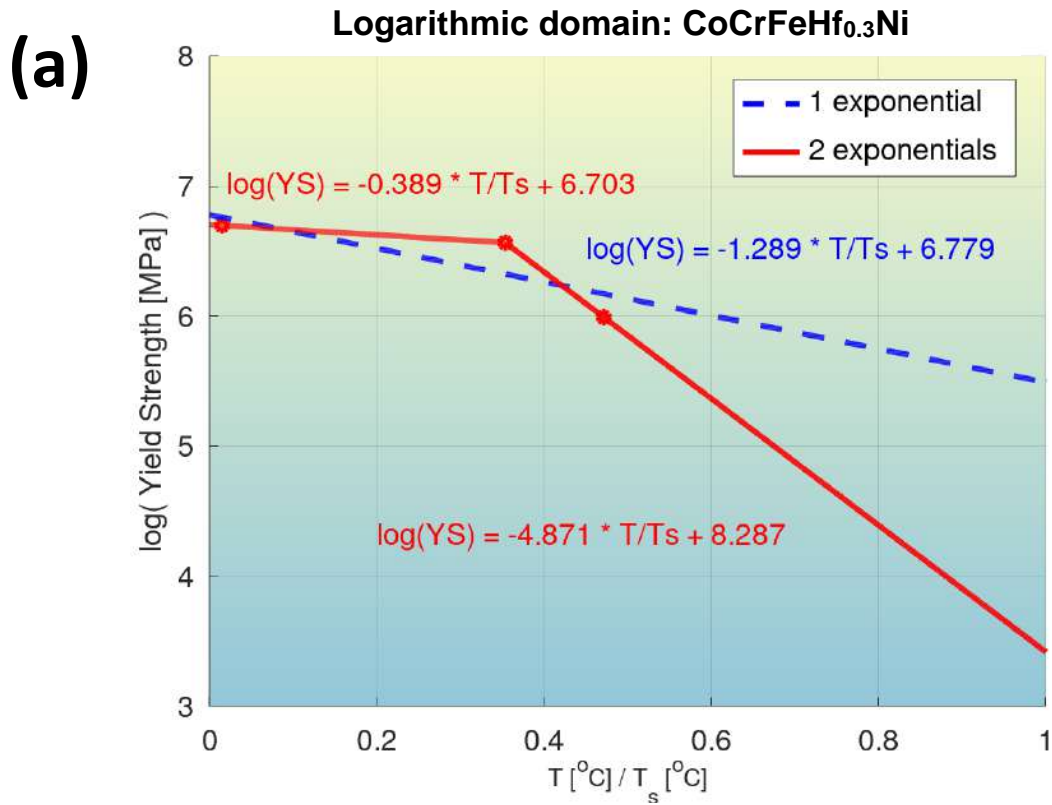

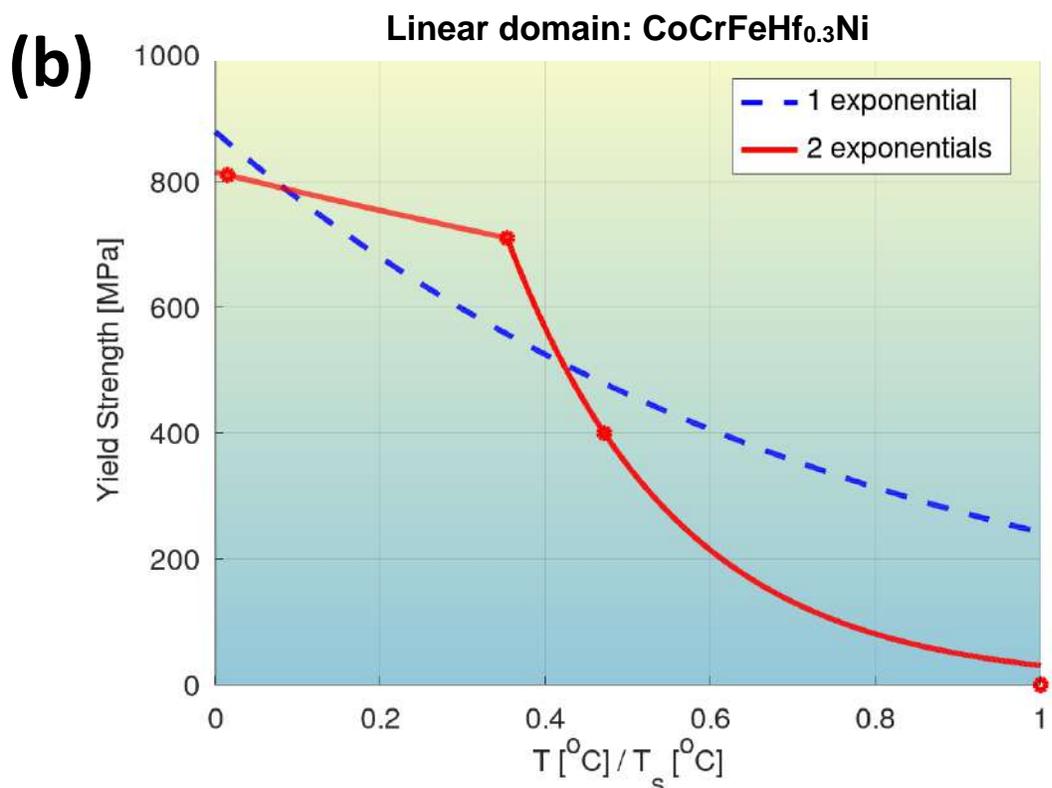

**Fig. S55**: Quantification of modeling accuracy of the bilinear log model, for composition No. 54 from **Tab. S2** (CoCrFeHf$_{0.3}$Ni, FCC+Laves phases), and comparison to that of a model with a single exponential.

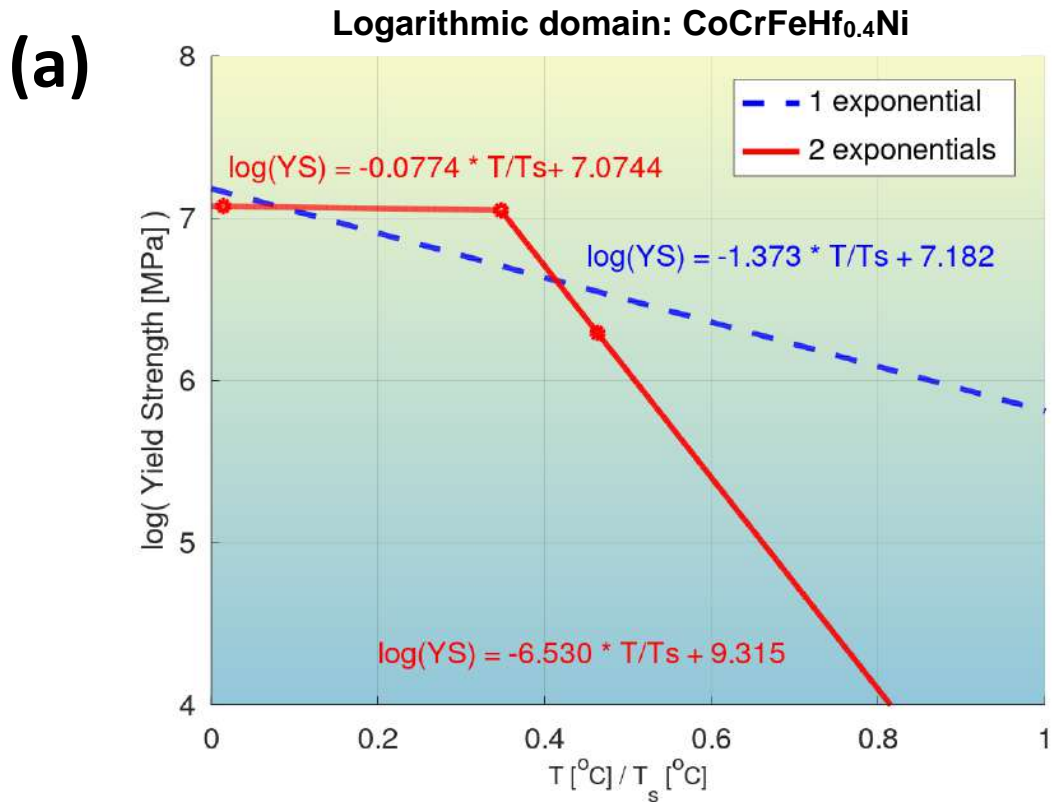

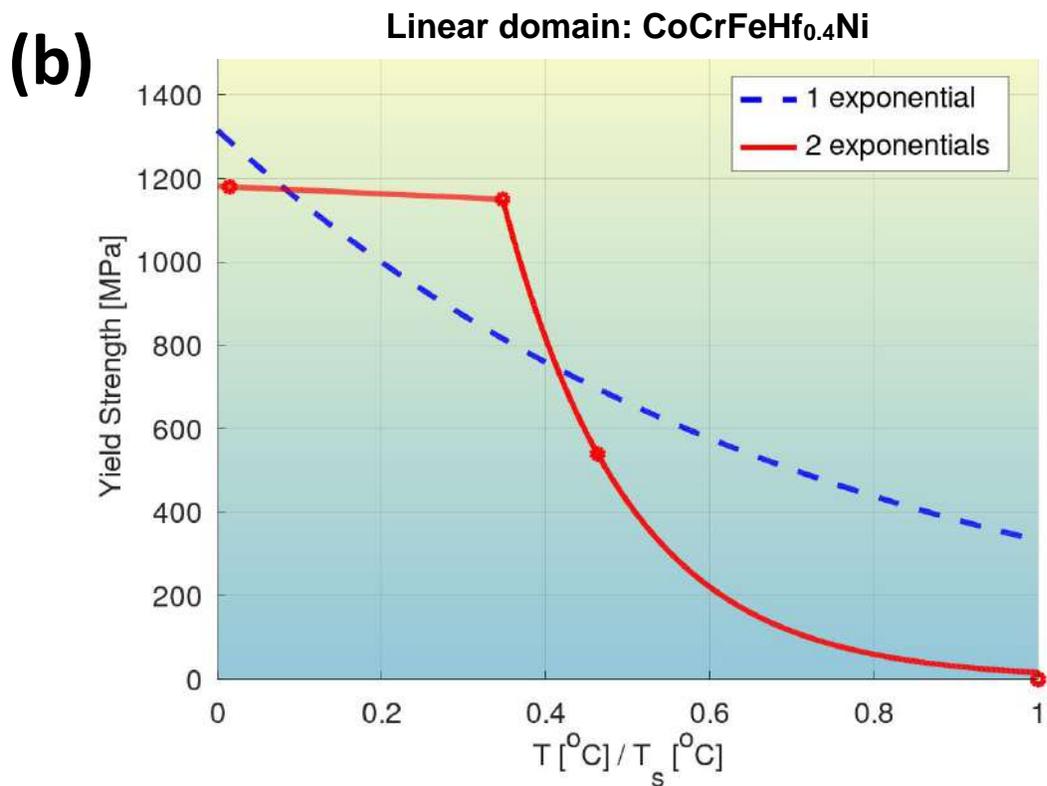

**Fig. S56**: Quantification of modeling accuracy of the bilinear log model, for composition No. 55 from **Tab. S2** (CoCrFeHf$_{0.4}$Ni, FCC+Laves phases), and comparison to that of a model with a single exponential.

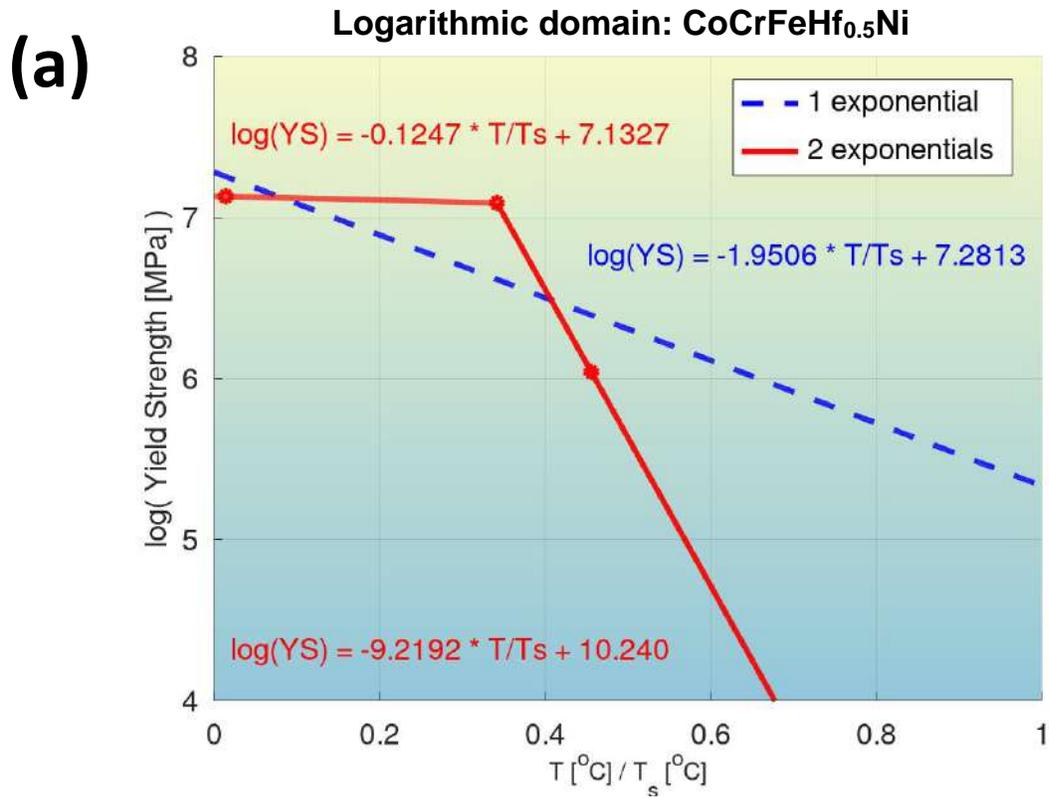
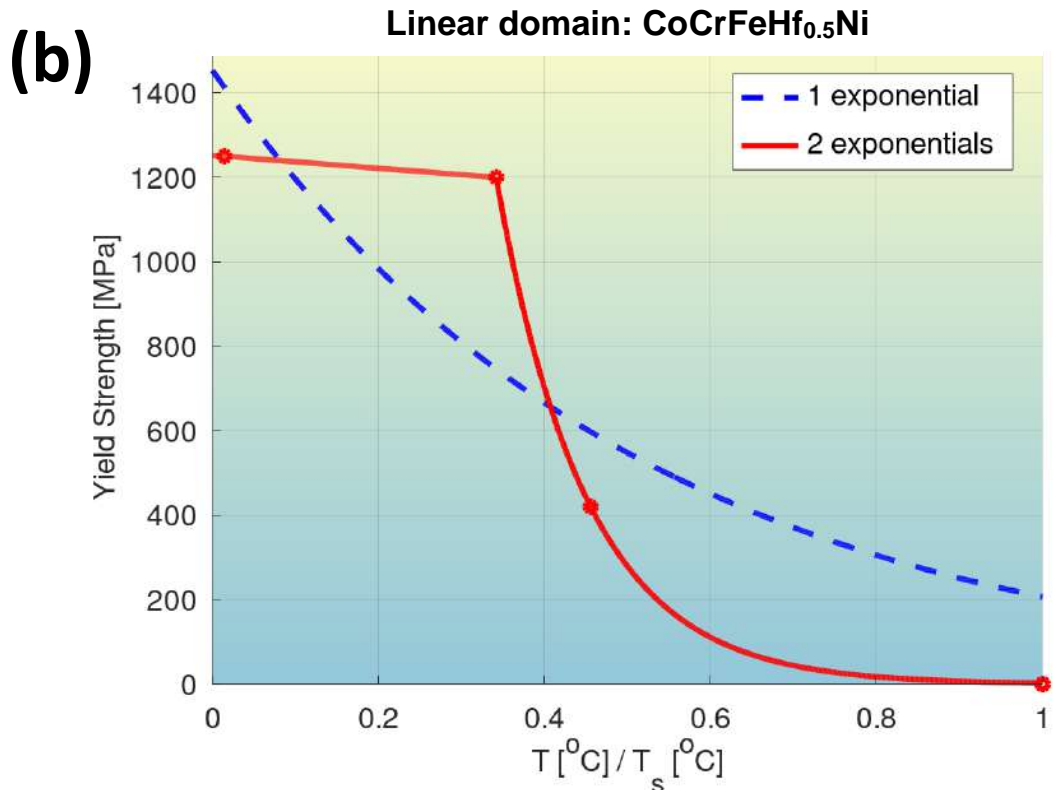

**Fig. S57**: Quantification of modeling accuracy of the bilinear log model, for composition No. 56 from **Tab. S2** (CoCrFeHf$_{0.5}$Ni, FCC+Laves phases), and comparison to that of a model with a single exponential.

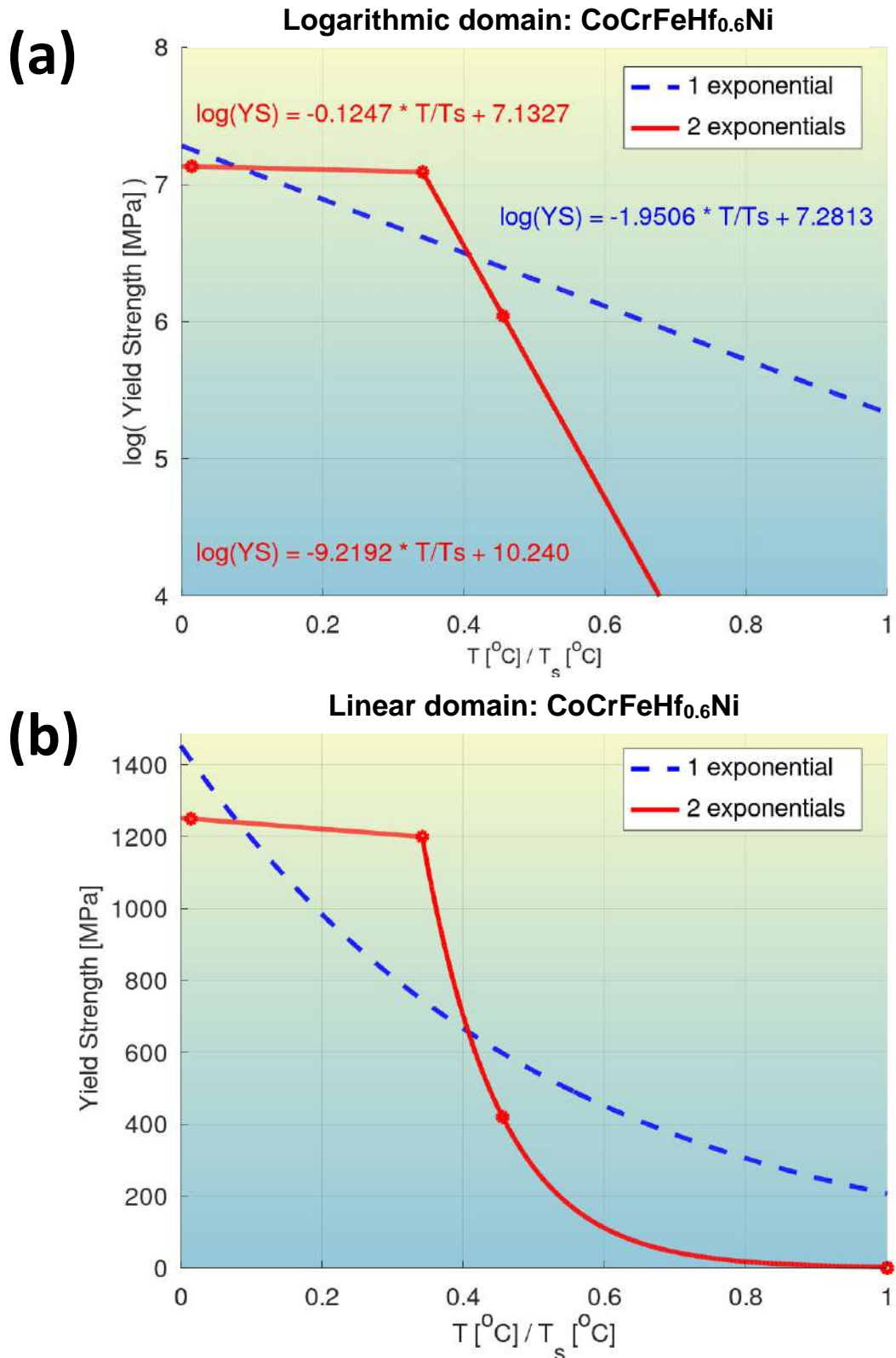

**Fig. S58**: Quantification of modeling accuracy of the bilinear log model, for composition No. 57 from **Tab. S2** (CoCrFeHf$_{0.6}$Ni, FCC+Laves phases), and comparison to that of a model with a single exponential.

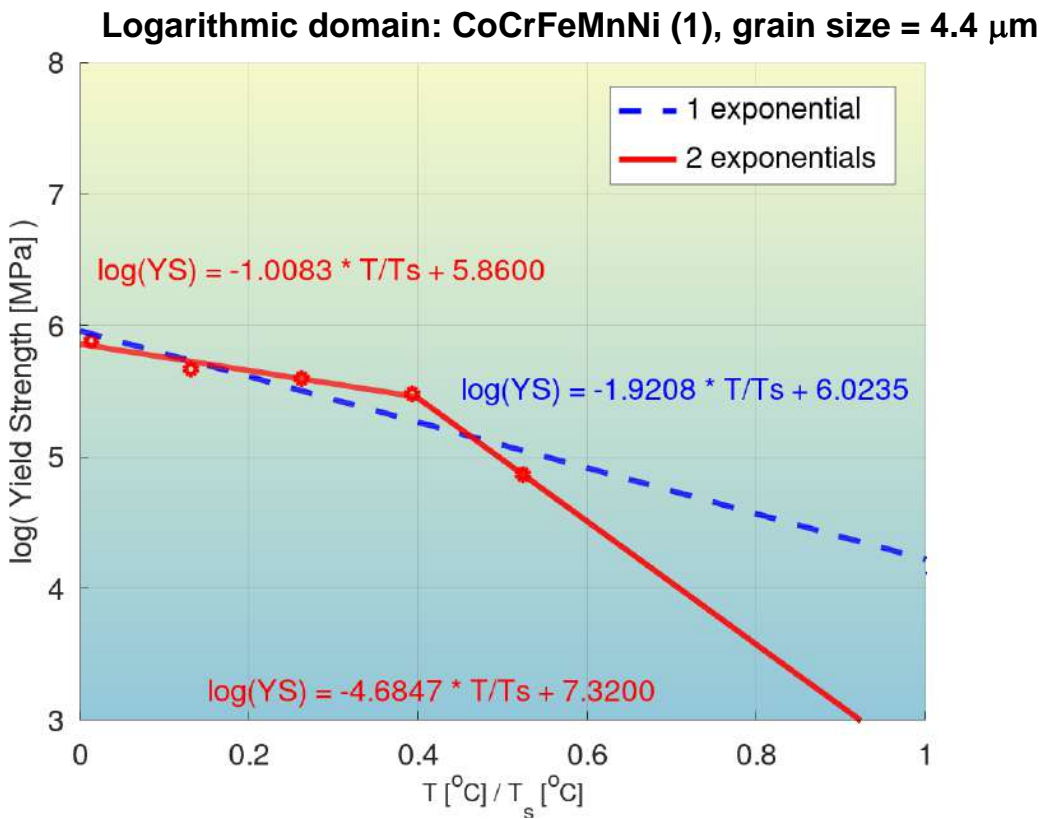

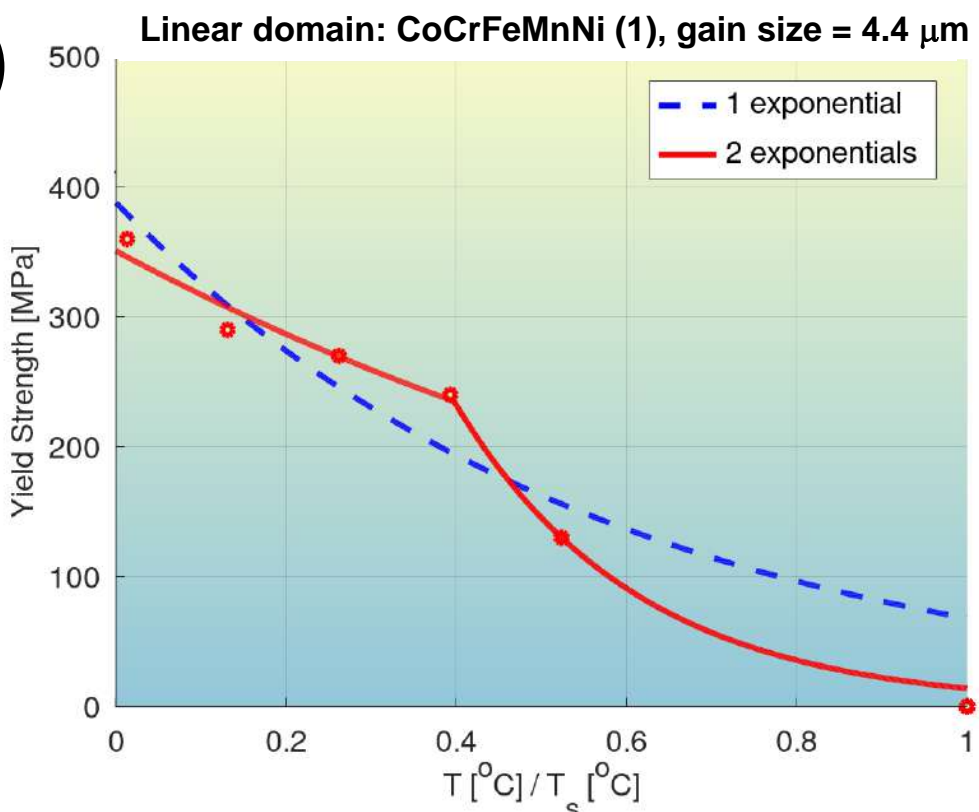

**Fig. S59**: Quantification of modeling accuracy of the bilinear log model, for composition No. 58 from **Tab. S2** (CoCrFeMnNi (1), FCC phases), and comparison to that of a model with a single exponential.

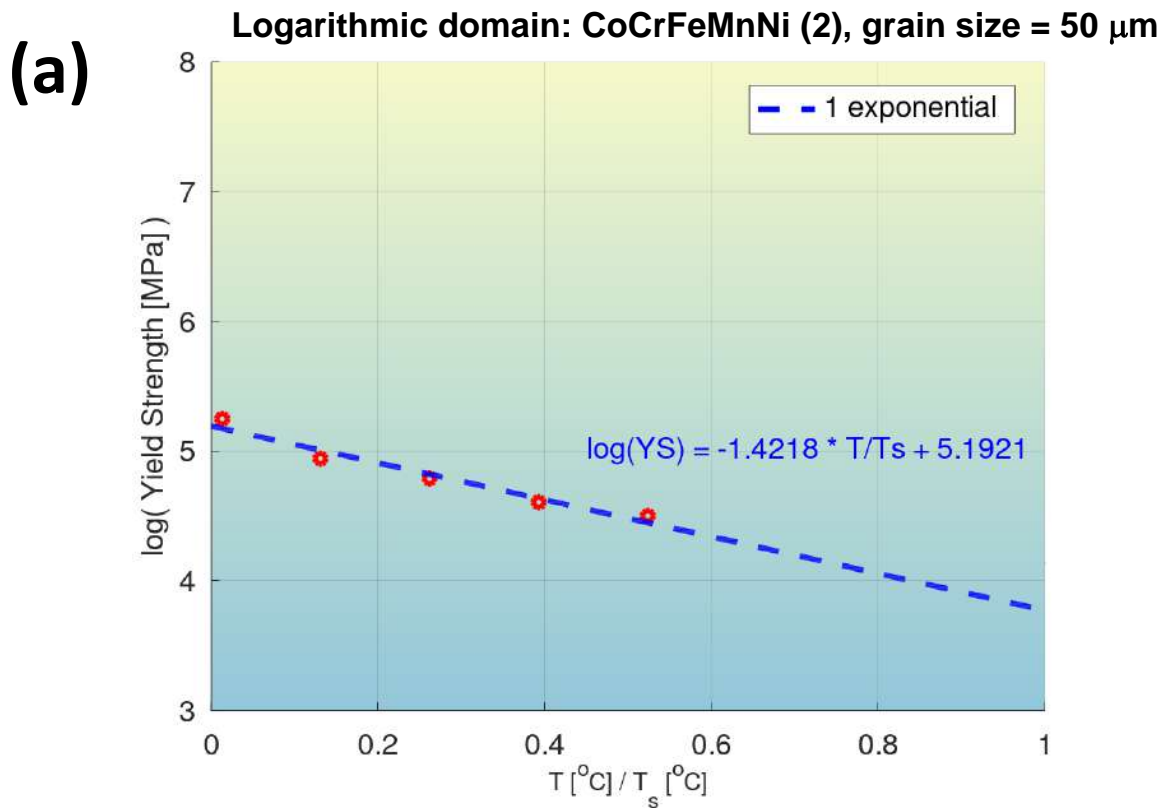

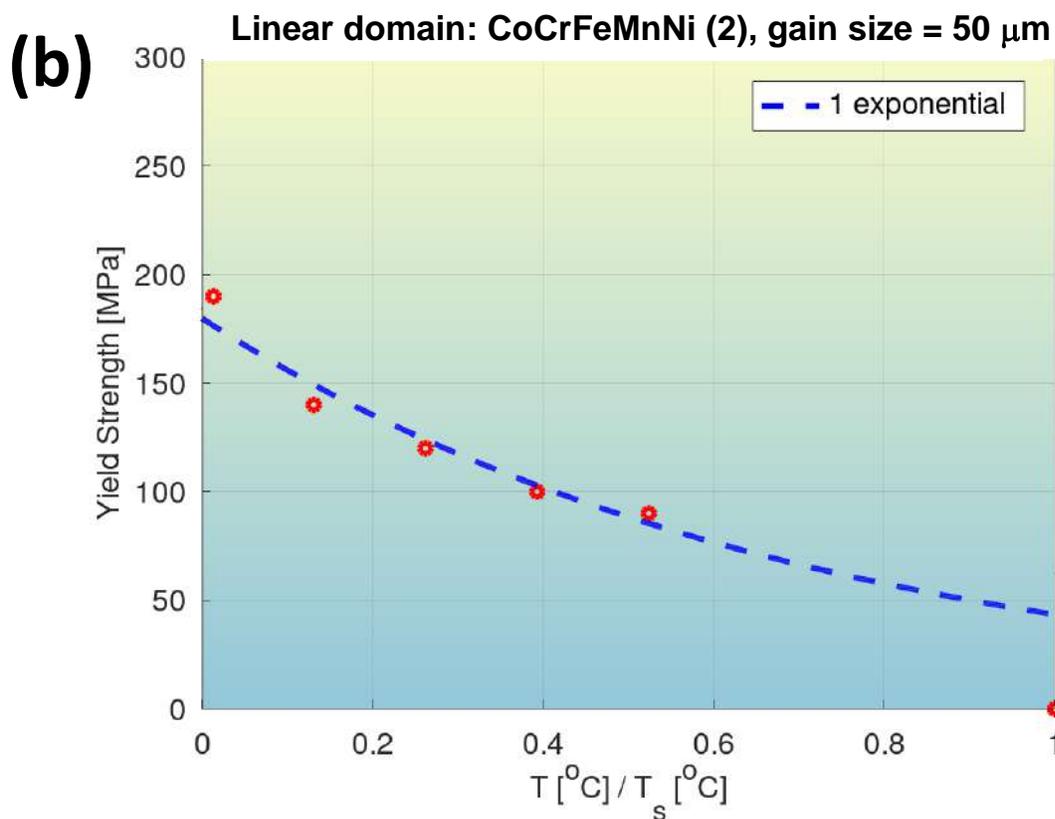

**Fig. S60**: Quantification of modeling accuracy of the bilinear log model, for composition No. 59 from **Tab. S2** (CoCrFeMnNi (2), FCC phases), and comparison to that of a model with a single exponential.

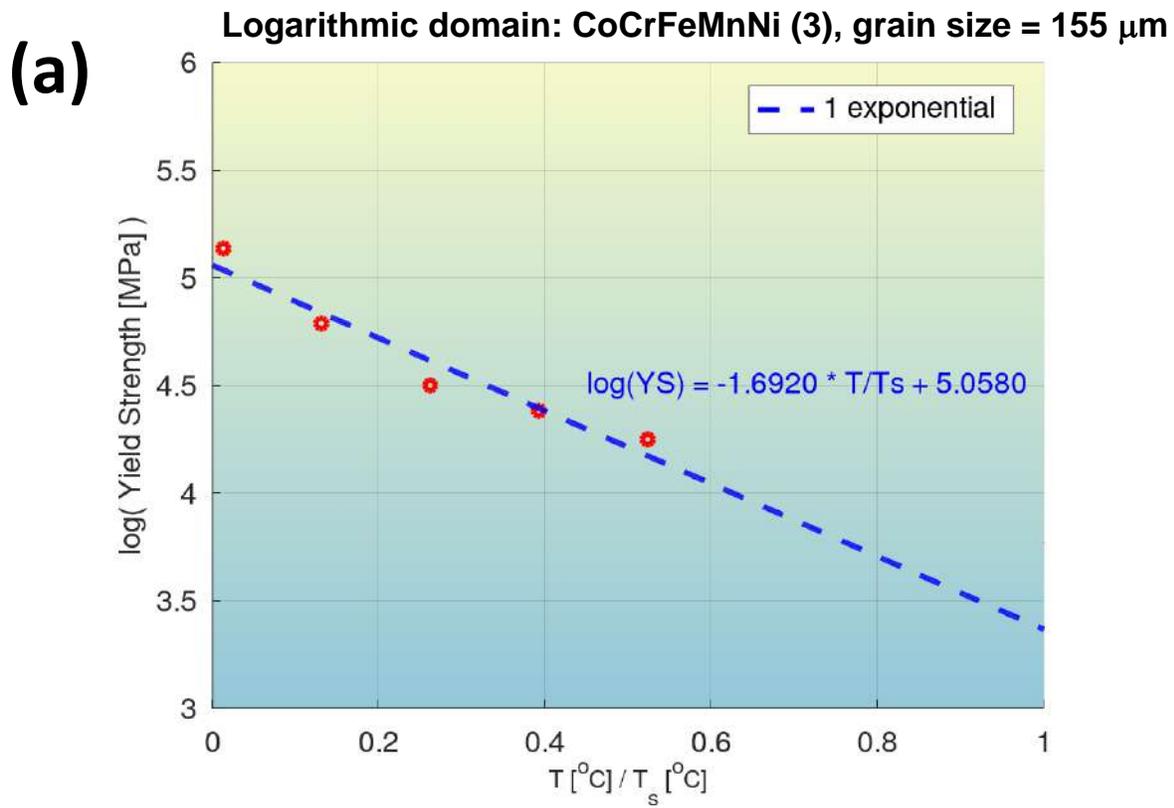

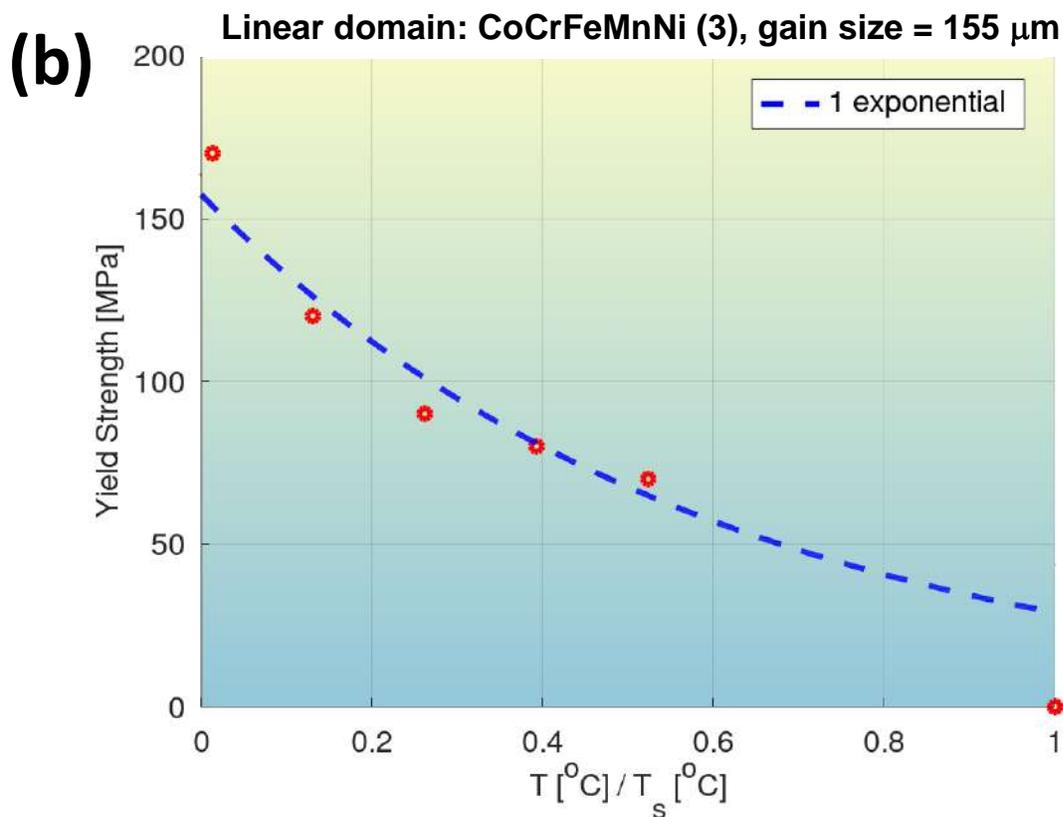

**Fig. S61**: Quantification of modeling accuracy of the bilinear log model, for composition No. 60 from **Tab. S2** (CoCrFeMnNi (3), FCC phases), and comparison to that of a model with a single exponential.

**Logarithmic domain: CoCrFeMnNi (4), grain size = 35 μm, strain rate = 1.0e-03 sec⁻¹**

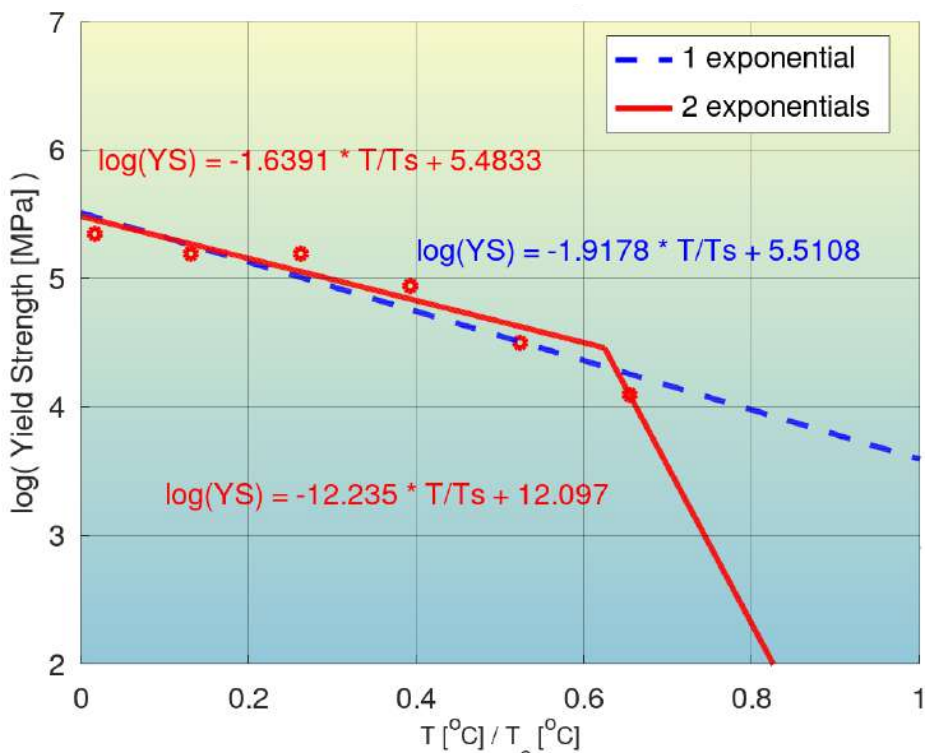

**Linear domain: CoCrFeMnNi (4), grain size = 35 μm, strain rate = 1.0e-03 sec⁻¹**

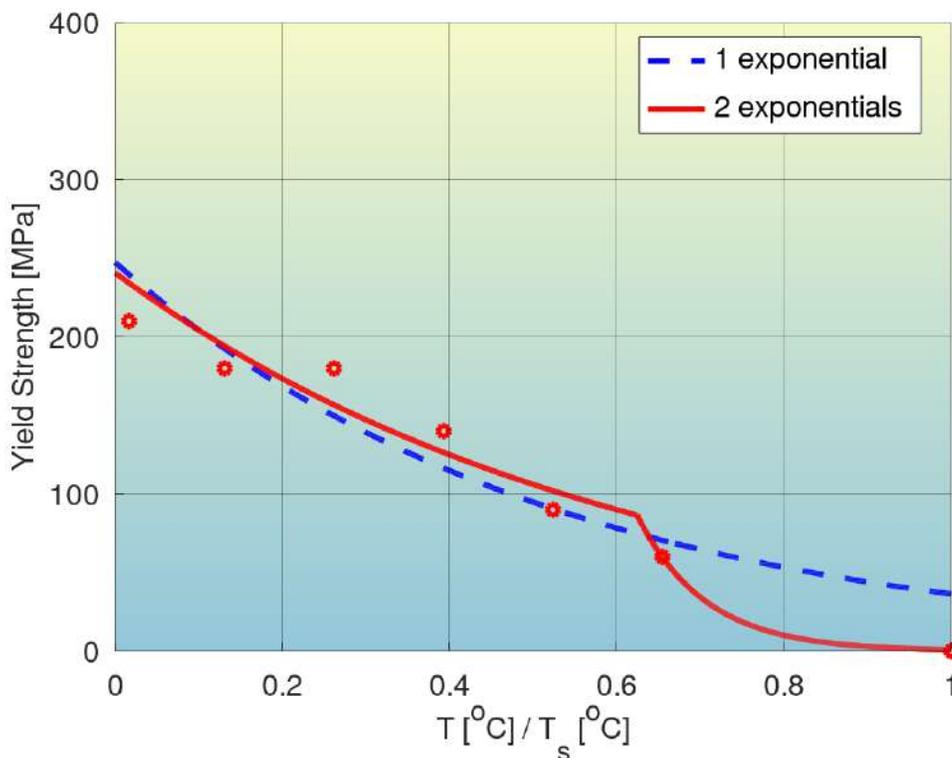

**Fig. S62**: Quantification of modeling accuracy of the bilinear log model, for composition No. 61 from **Tab. S2** (CoCrFeMnNi (4), FCC phases), and comparison to that of a model with a single exponential.

**Logarithmic domain: CoCrFeMnNi (5), grain size = 35 μm, strain rate = 1.0e-01 sec⁻¹**

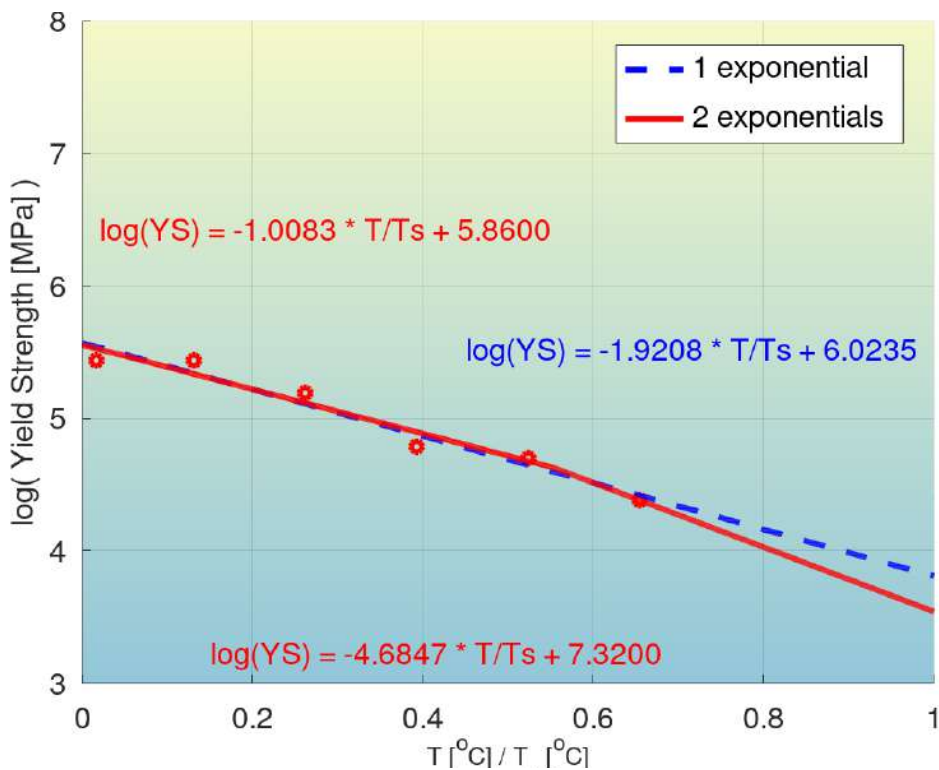

**Linear domain: CoCrFeMnNi (5), grain size = 35 μm, strain rate = 1.0e-01 sec⁻¹**

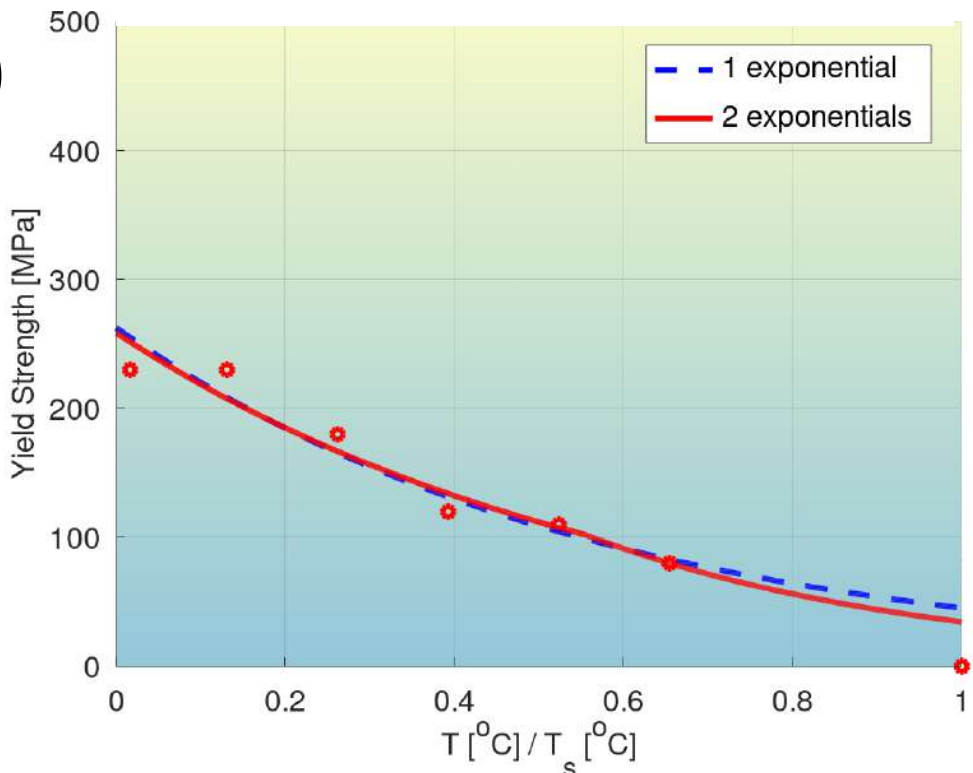

**Fig. S63**: Quantification of modeling accuracy of the bilinear log model, for composition No. 62 from **Tab. S2** (CoCrFeMnNi (4), FCC phases), and comparison to that of a model with a single exponential.

**Logarithmic domain: CoCrFeNi (1), grain size = 14 μm, strain rate = 1.0e-03 sec⁻¹**

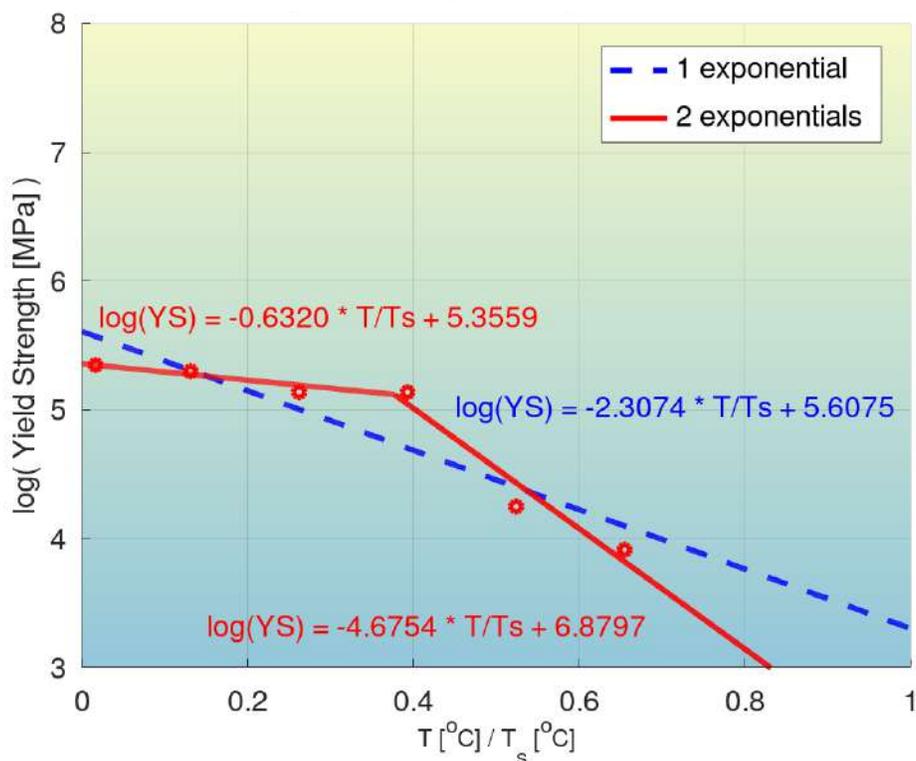

**Linear domain: CoCrFeNi (1), grain size = 14 μm, strain rate = 1.0e-03 sec⁻¹**

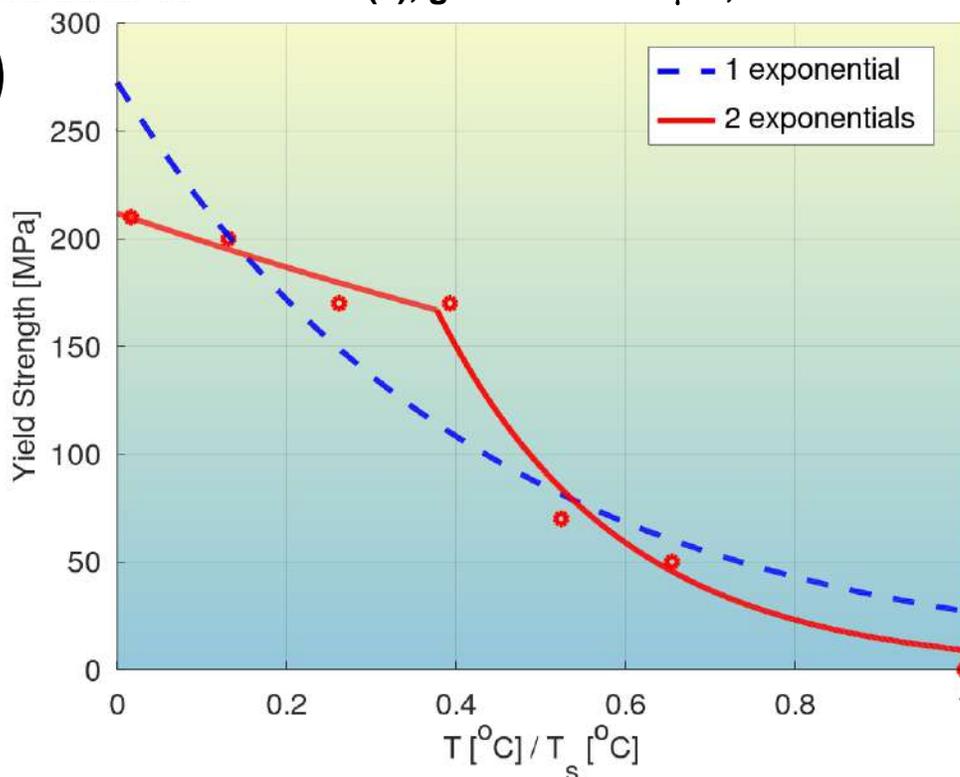

**Fig. S64**: Quantification of modeling accuracy of the bilinear log model, for composition No. 63 from **Tab. S2** (CoCrFeNi (1), FCC phases), and comparison to that of a model with a single exponential.

**Logarithmic domain: CoCrFeNi (2), grain size = 14 μm, strain rate = 1.0e-01 sec$^{-1}$**

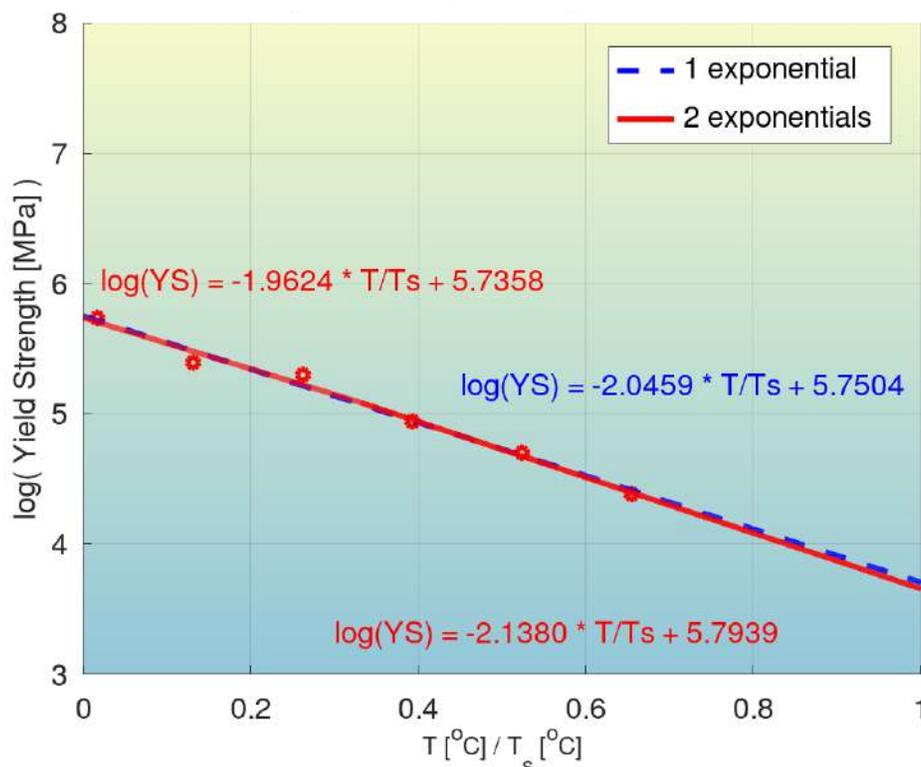

**Linear domain: CoCrFeNi (2), grain size = 14 μm, strain rate = 1.0e-01 sec$^{-1}$**

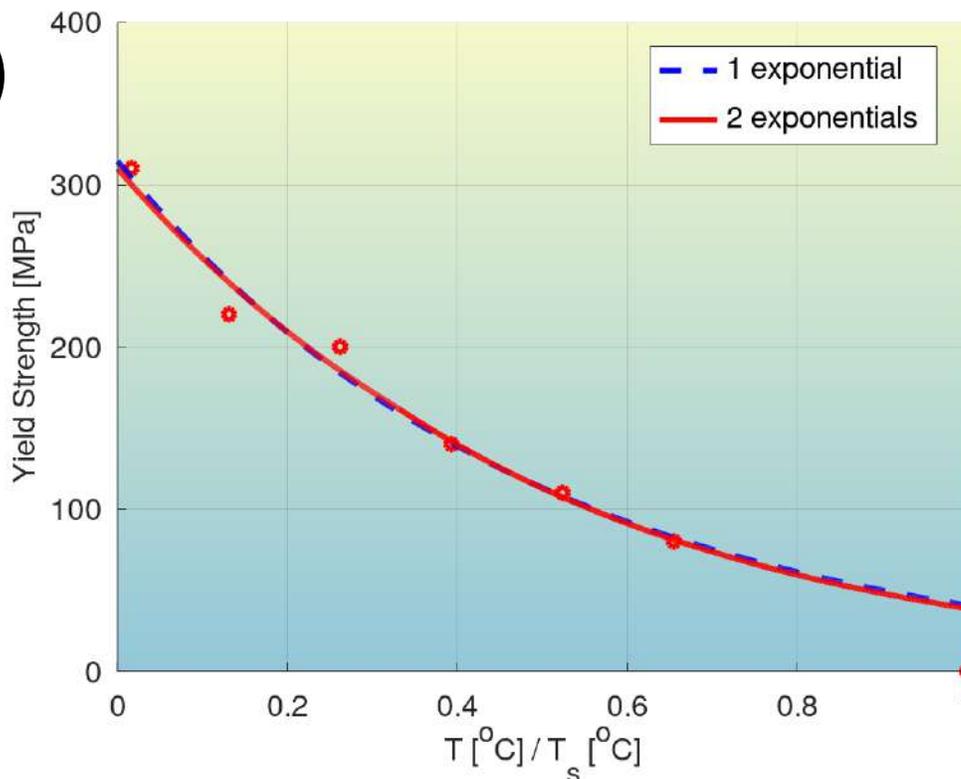

**Fig. S65**: Quantification of modeling accuracy of the bilinear log model, for composition No. 64 from **Tab. S2** (CoCrFeNi (2), FCC phases), and comparison to that of a model with a single exponential.

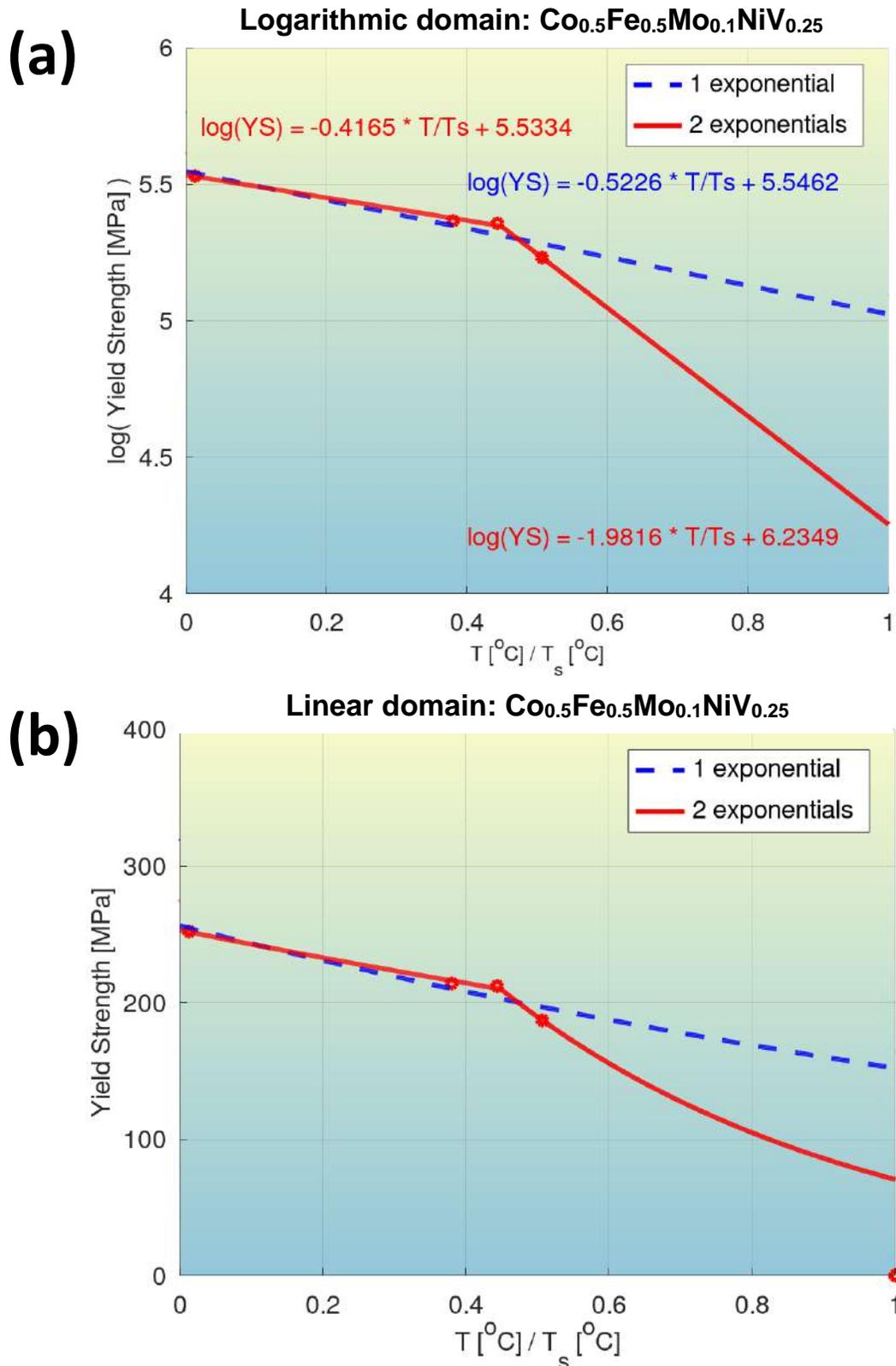

**Fig. S66**: Quantification of modeling accuracy of the bilinear log model, for composition No. 65 from **Tab. S2** ($Co_{0.5}Fe_{0.5}Mo_{0.1}NiV_{0.25}$, FCC phase), and comparison to that of a model with a single exponential.

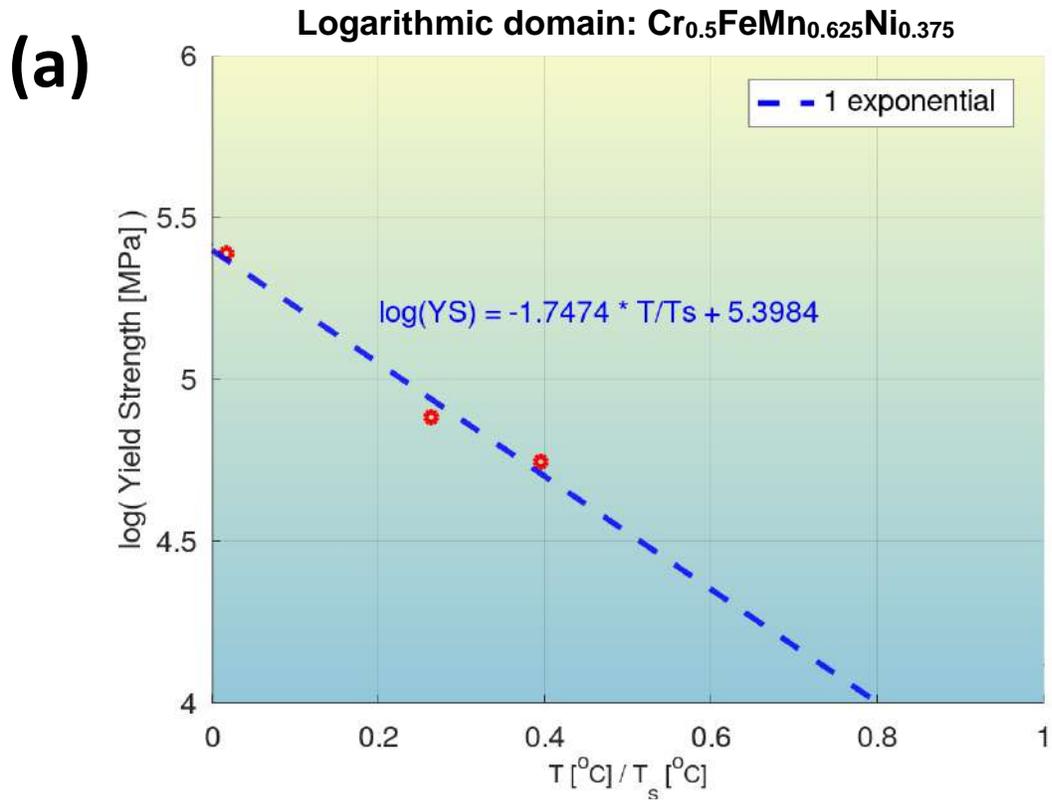

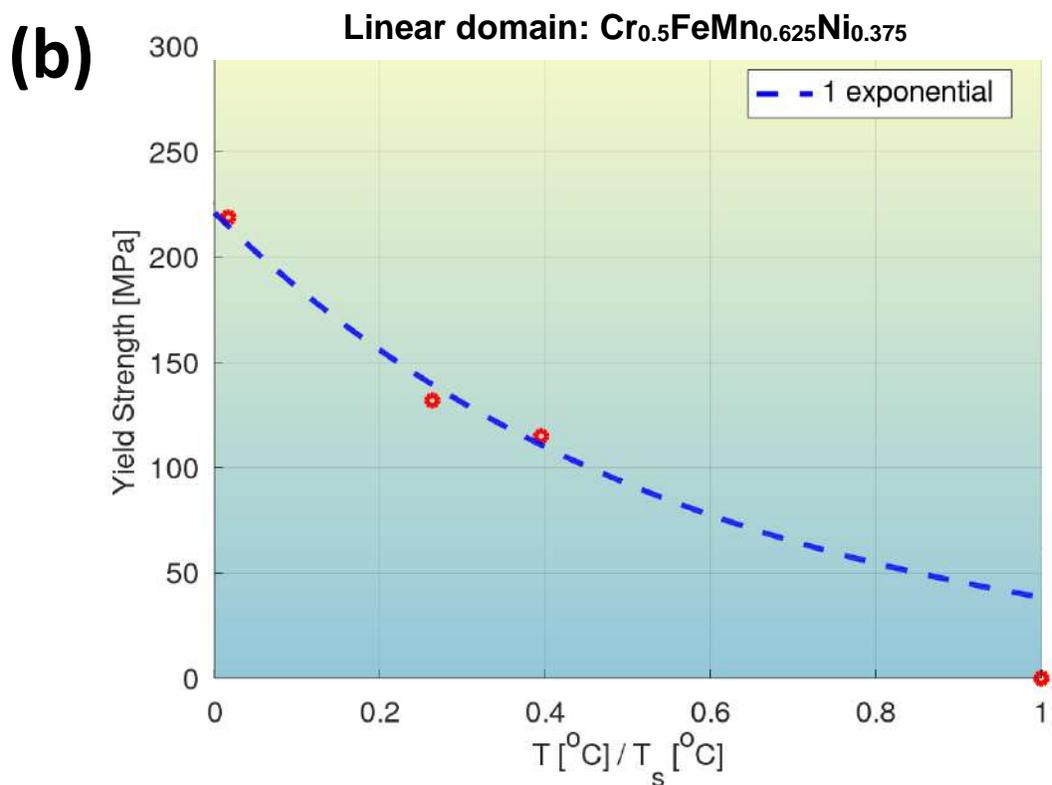

**Fig. S67**: Quantification of modeling accuracy of the bilinear log model, for composition No. 66 from **Tab. S2** (Cr$_{0.5}$FeMn$_{0.625}$Ni$_{0.375}$, FCC+BCC phases, and comparison to that of a model with a single exponential.

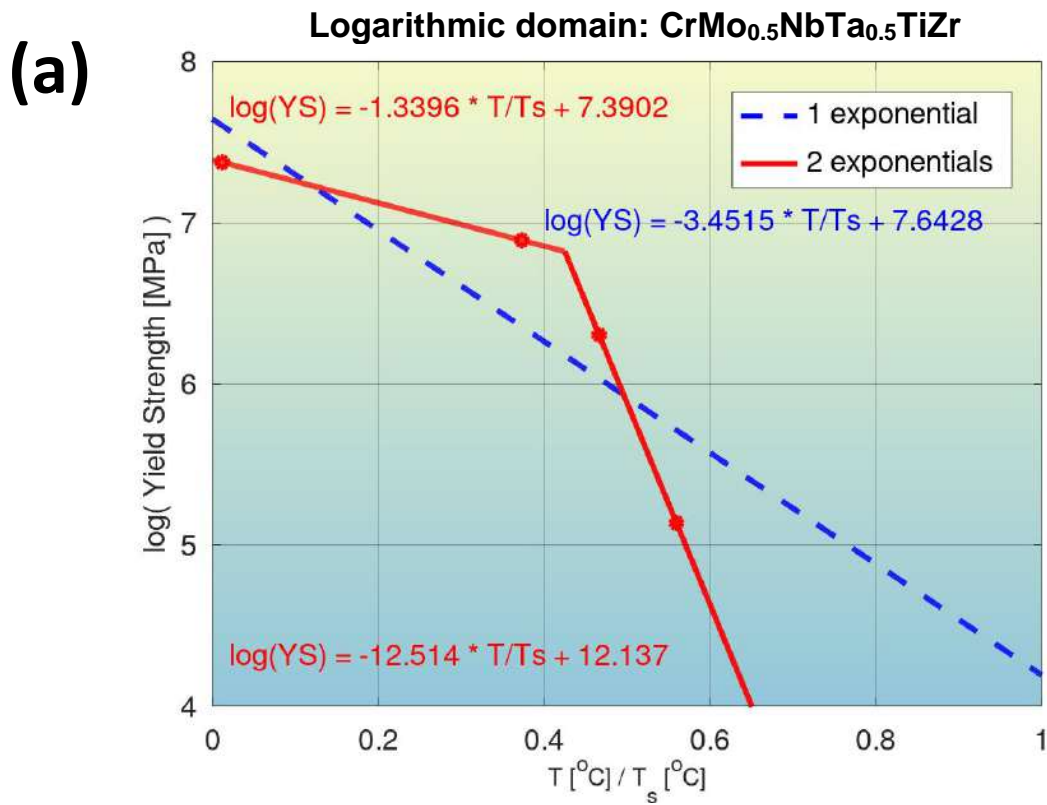

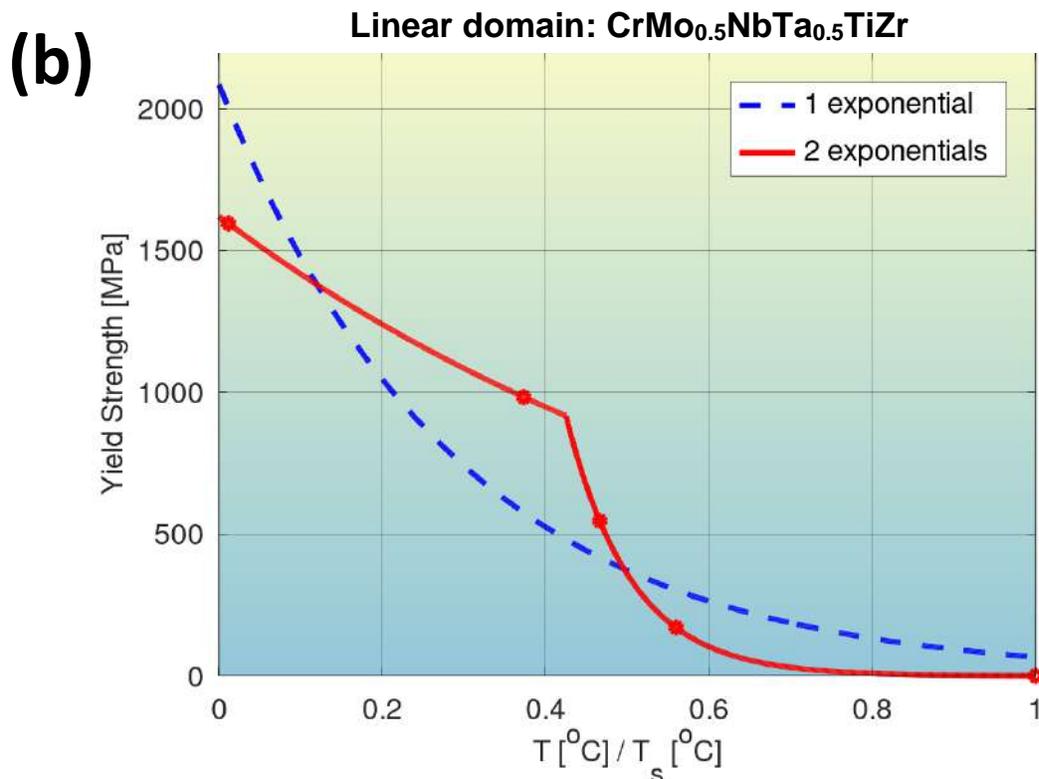

**Fig. S68**: Quantification of modeling accuracy of the bilinear log model, for composition No. 67 from **Tab. S2** (CrMo$_{0.5}$NbTa$_{0.5}$TiZr, 2BCC+Laves phases), and comparison to that of a model with a single exponential.

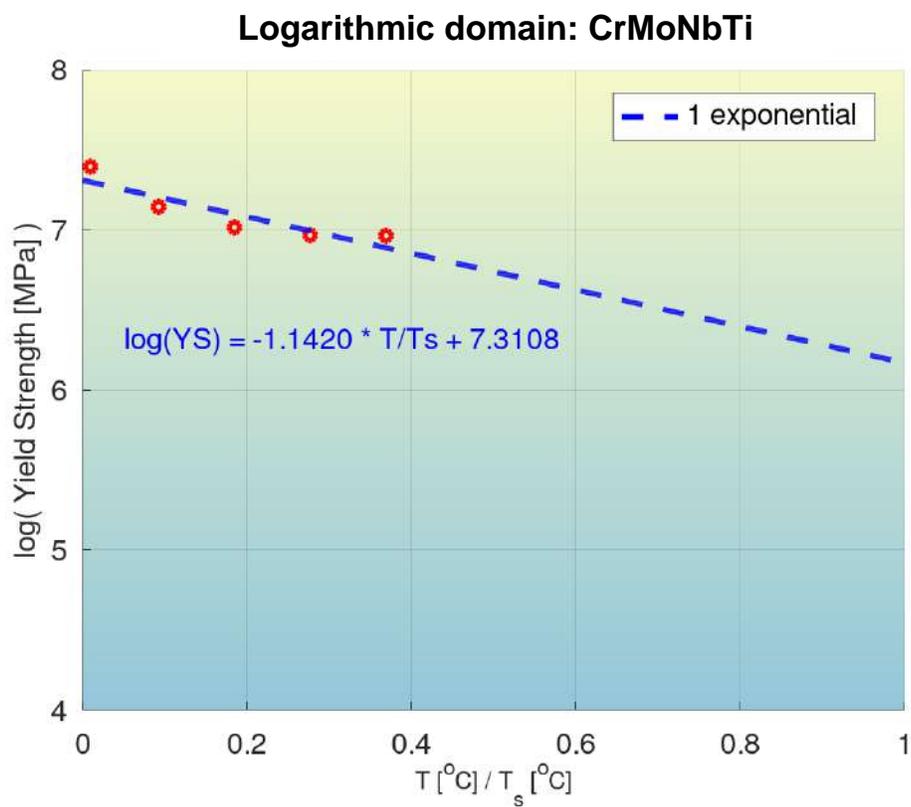

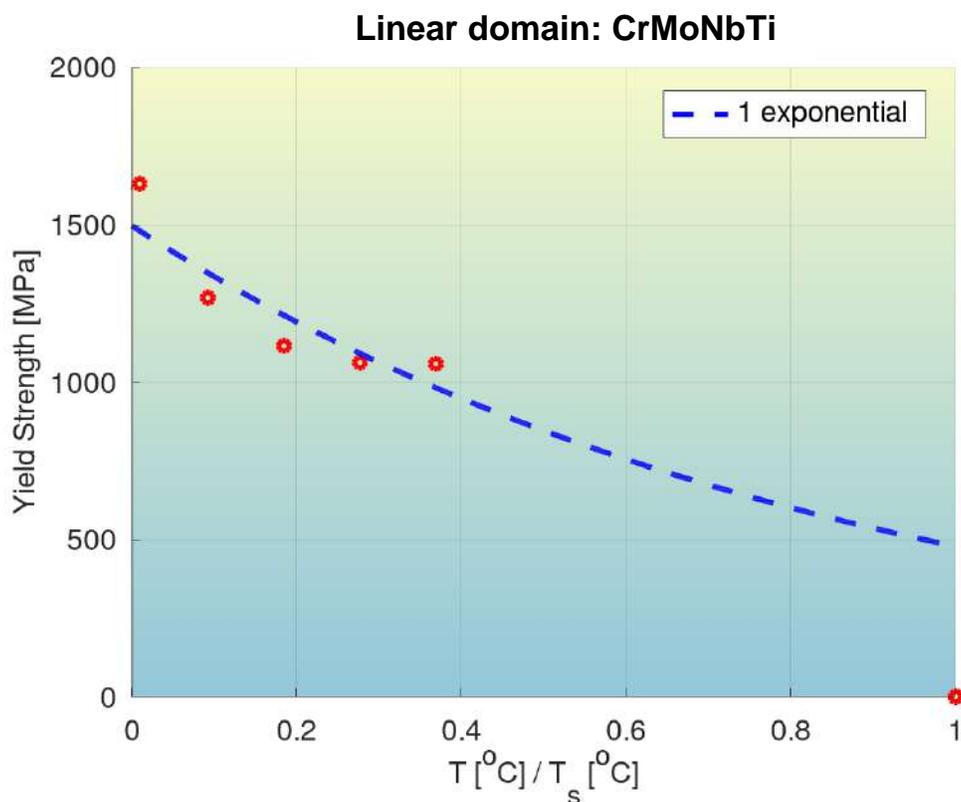

**Fig. S69**: Quantification of modeling accuracy of the bilinear log model, for composition No. 68 from **Tab. S2** (CrMoNbTi, BCC phase), and comparison to that of a model with a single exponential.

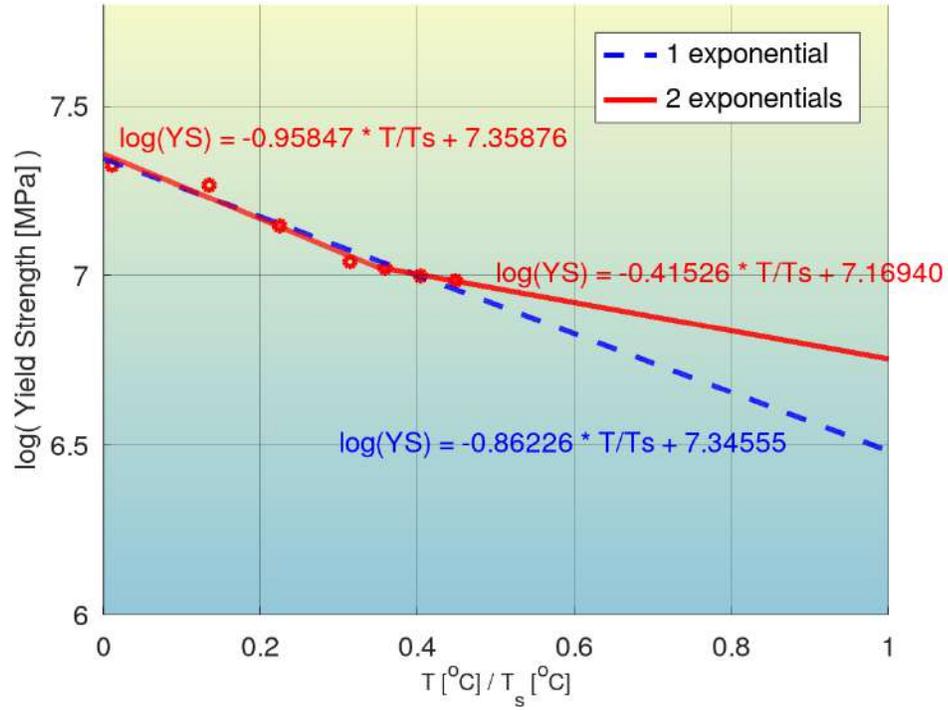

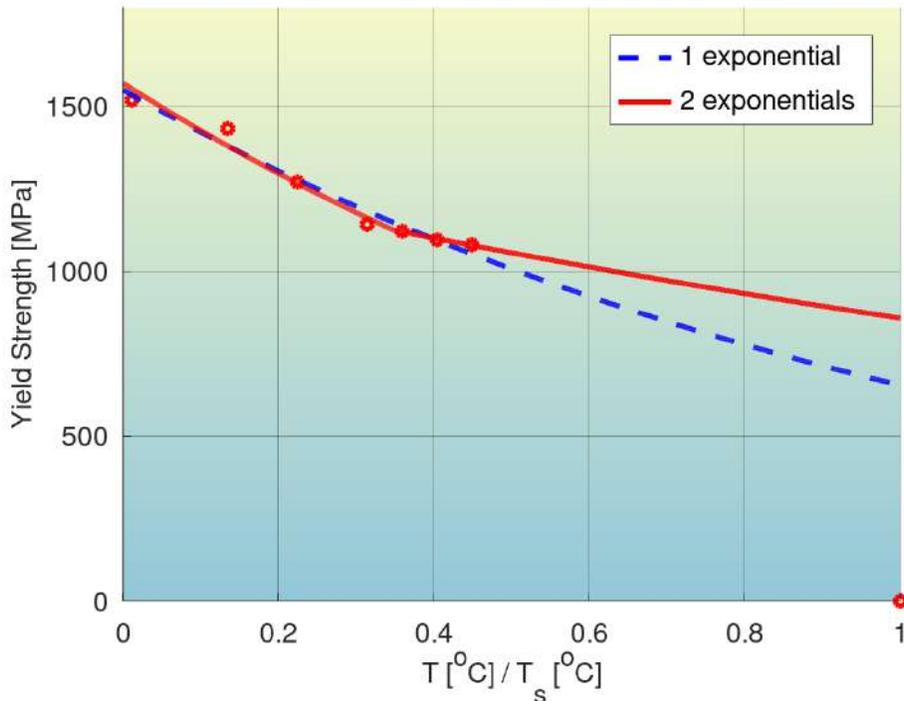

**Fig. S70**: Quantification of modeling accuracy of the bilinear log model, for composition No. 69 from **Tab. S2** (CrMoNbV, BCC phase), and comparison to that of a model with a single exponential.

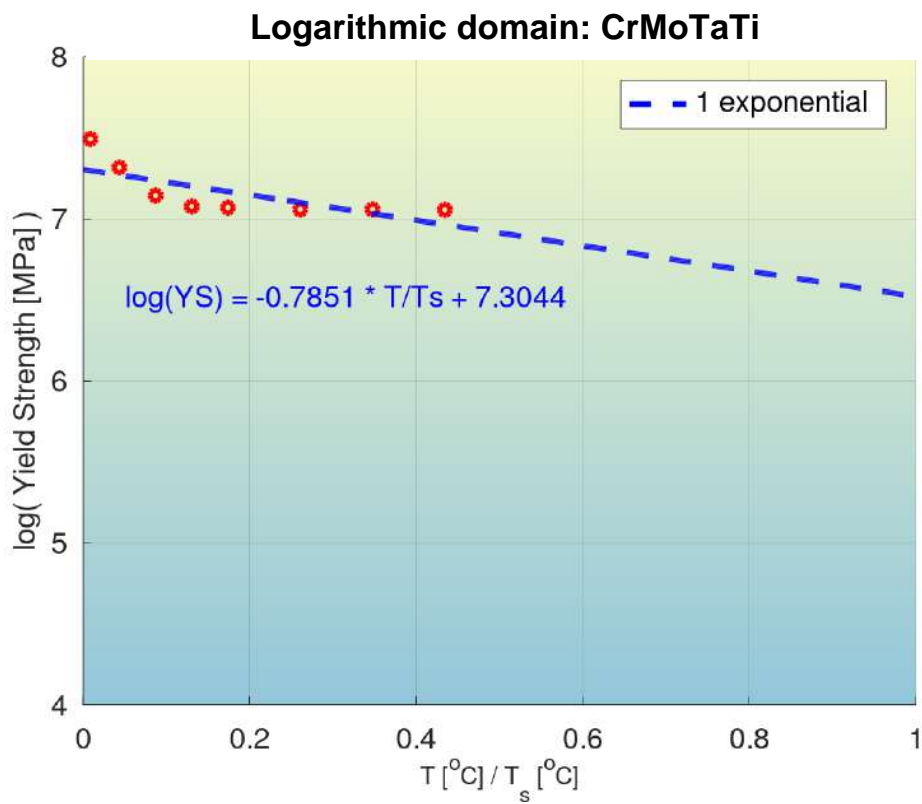

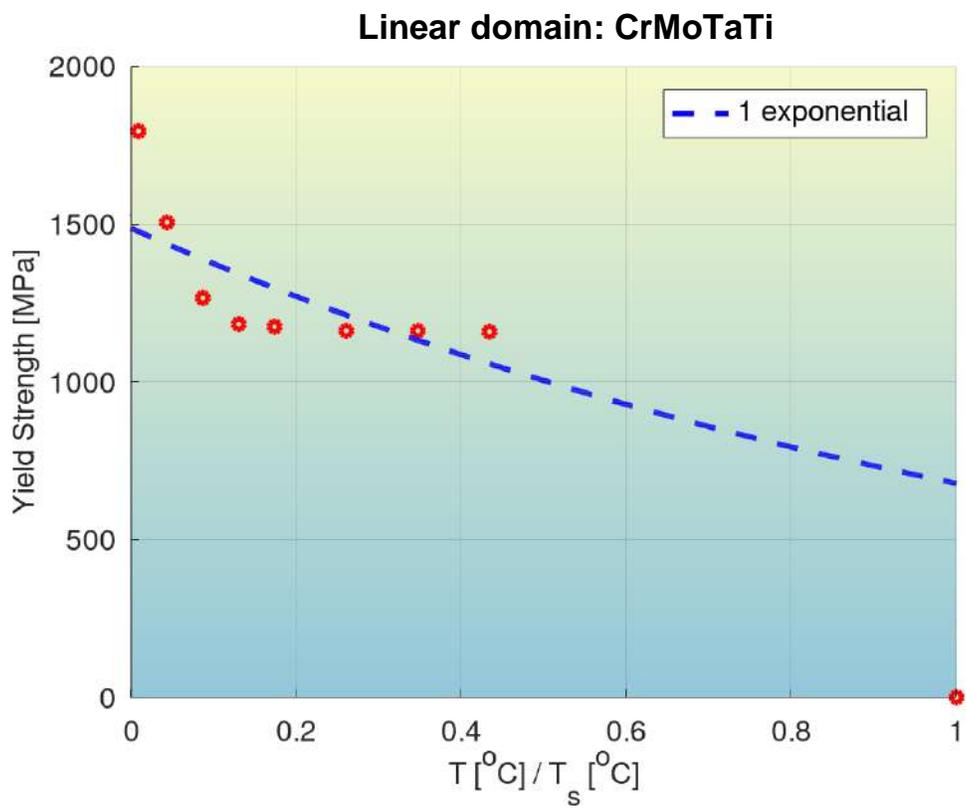

**Fig. S71**: Quantification of modeling accuracy of the bilinear log model, for composition No. 70 from **Tab. S2** (CrMoTaTi, BCC+Laves phases), and comparison to that of a model with a single exponential.

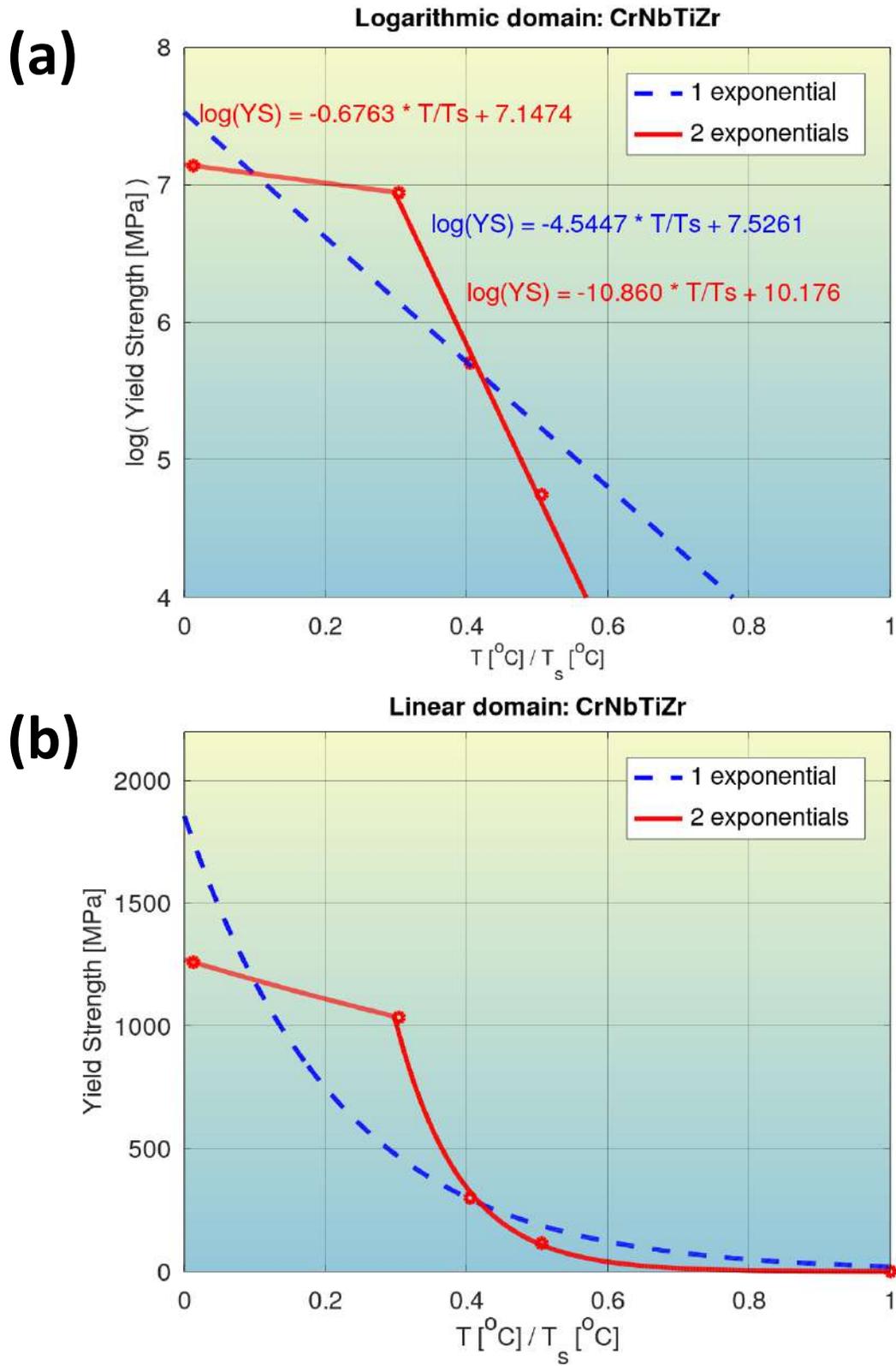

**Fig. S72**: Quantification of modeling accuracy of the bilinear log model, for composition No. 71 from **Tab. S2** (CrNbTiZr, BCC+Laves phases), and comparison to that of a model with a single exponential.

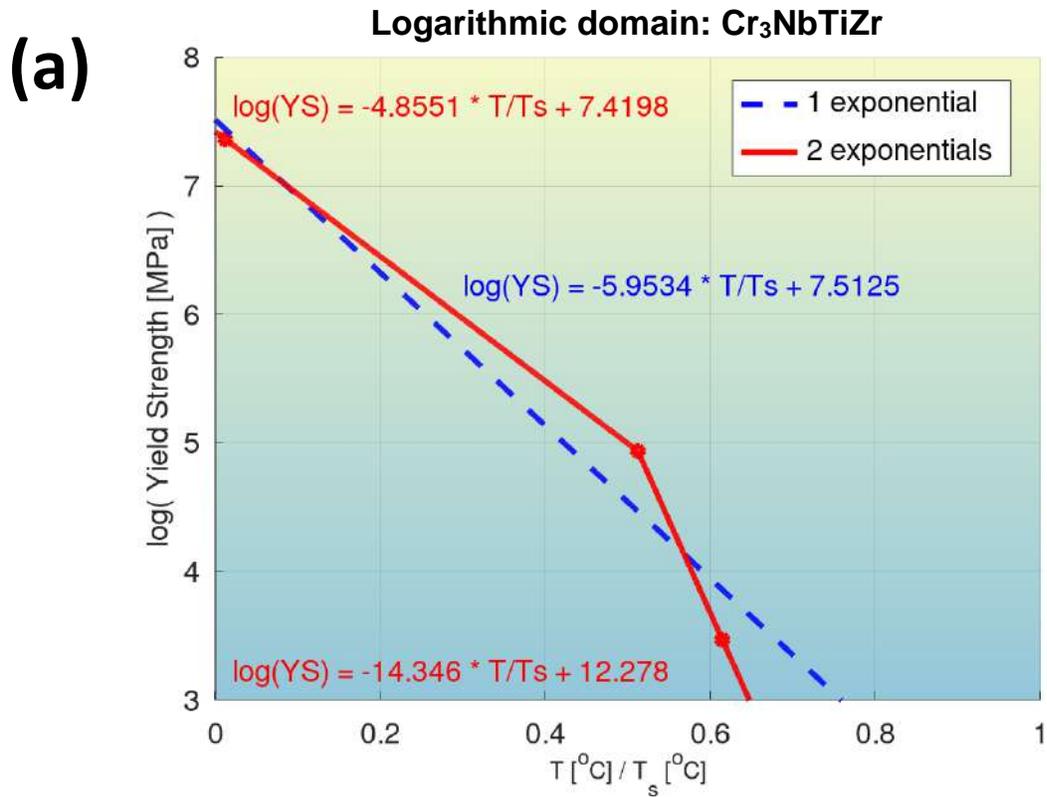

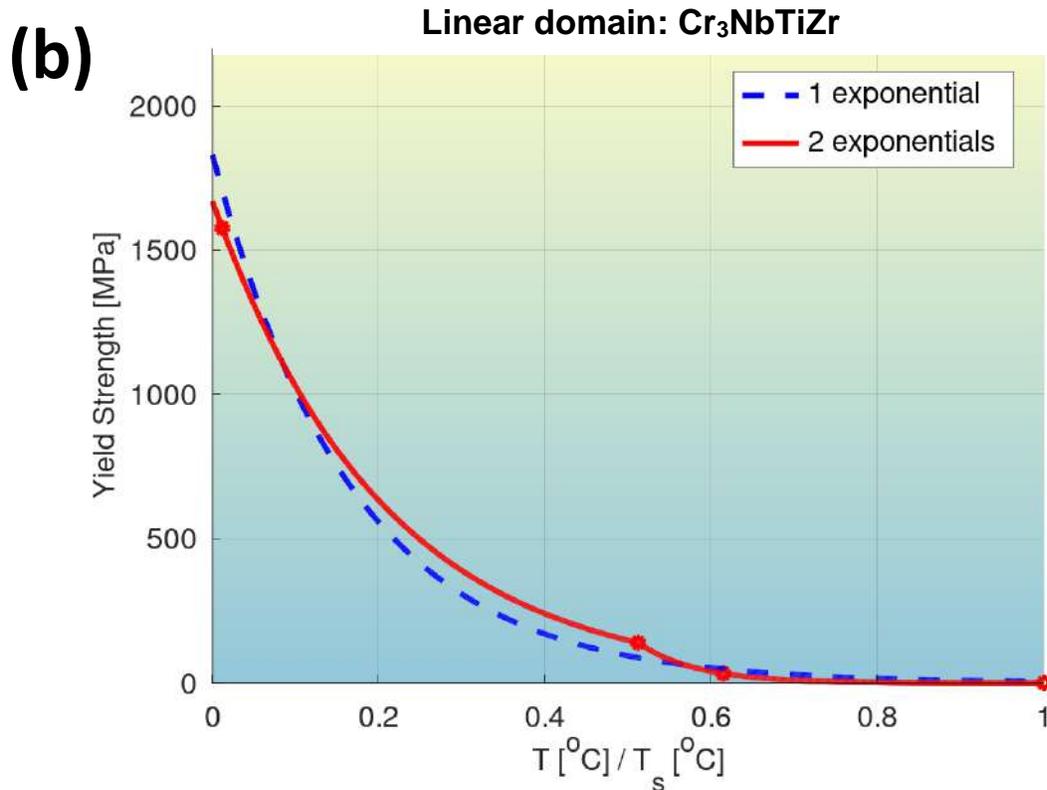

**Fig. S73**: Quantification of modeling accuracy of the bilinear log model, for composition No. 72 from **Tab. S2** ($Cr_3NbTiZr$, BCC+Laves phases), and comparison to that of a model with a single exponential.

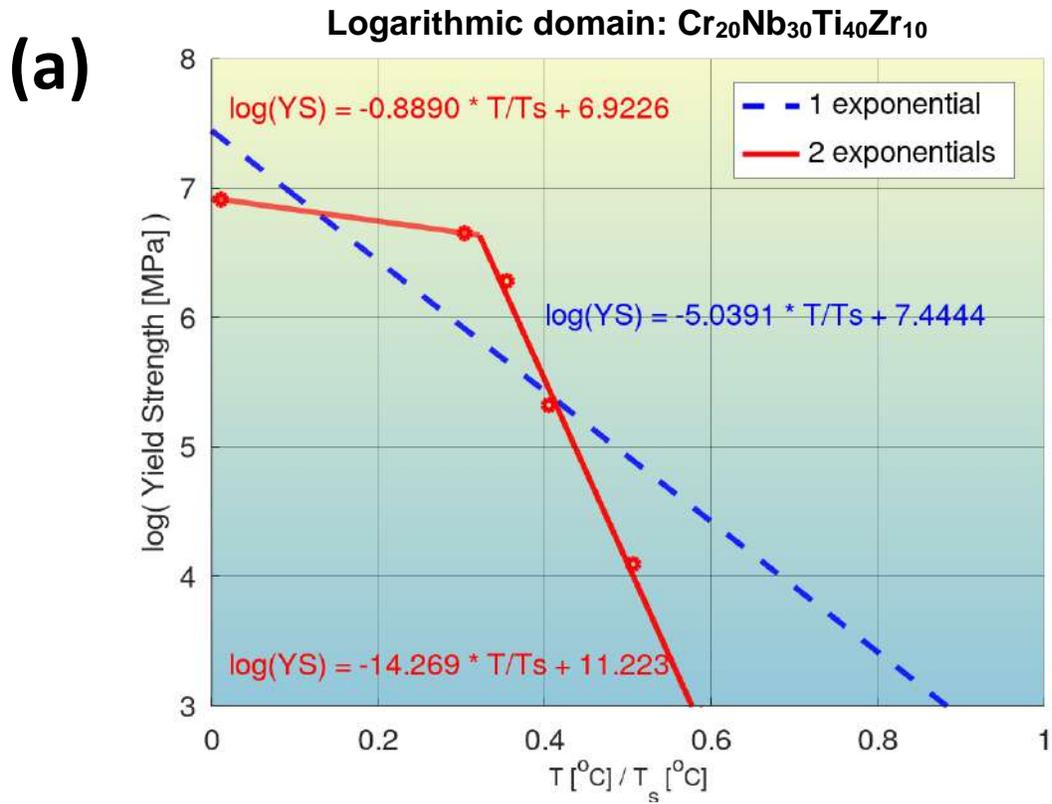

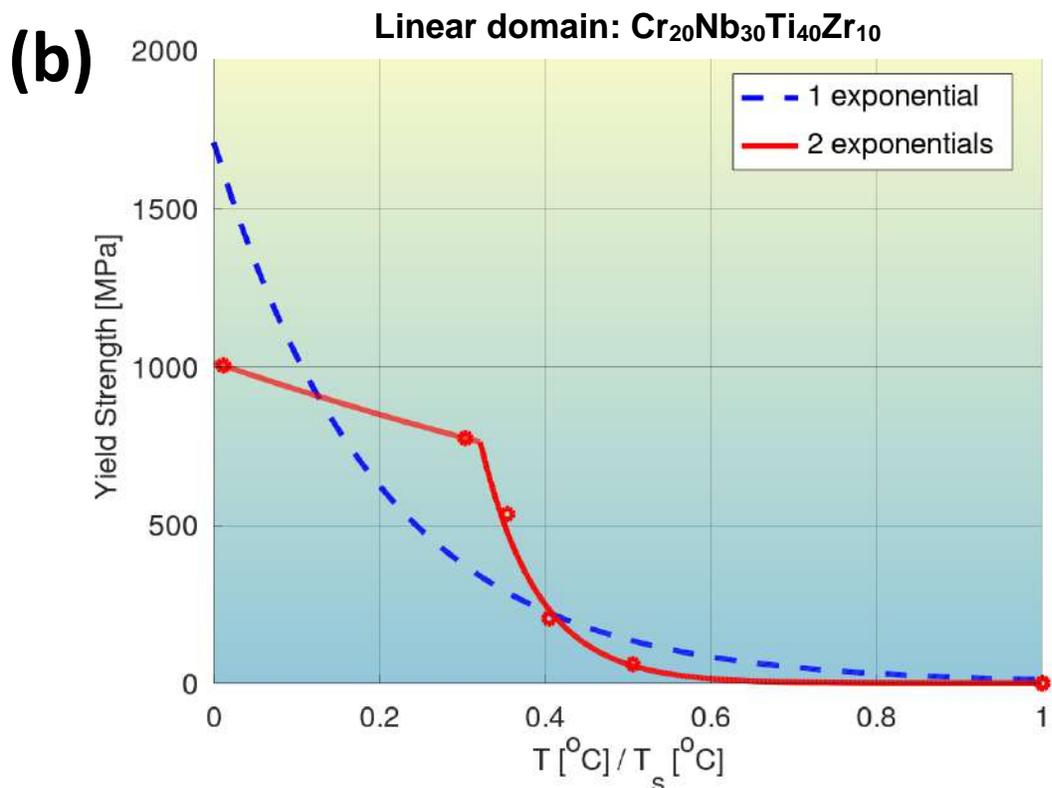

**Fig. S74**: Quantification of modeling accuracy of the bilinear log model, for composition No. 73 from **Tab. S2** ($Cr_{20}Nb_{30}Ti_{40}Zr_{10}$, BCC+Laves phases), and comparison to that of a model with a single exponential.

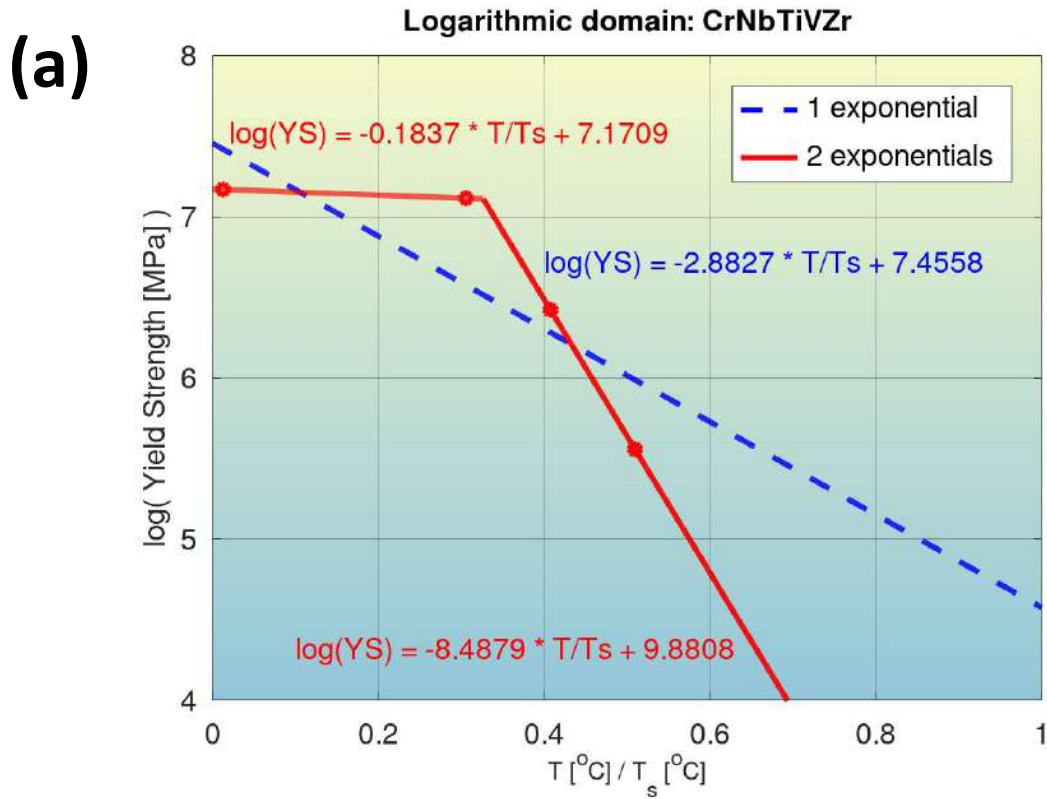

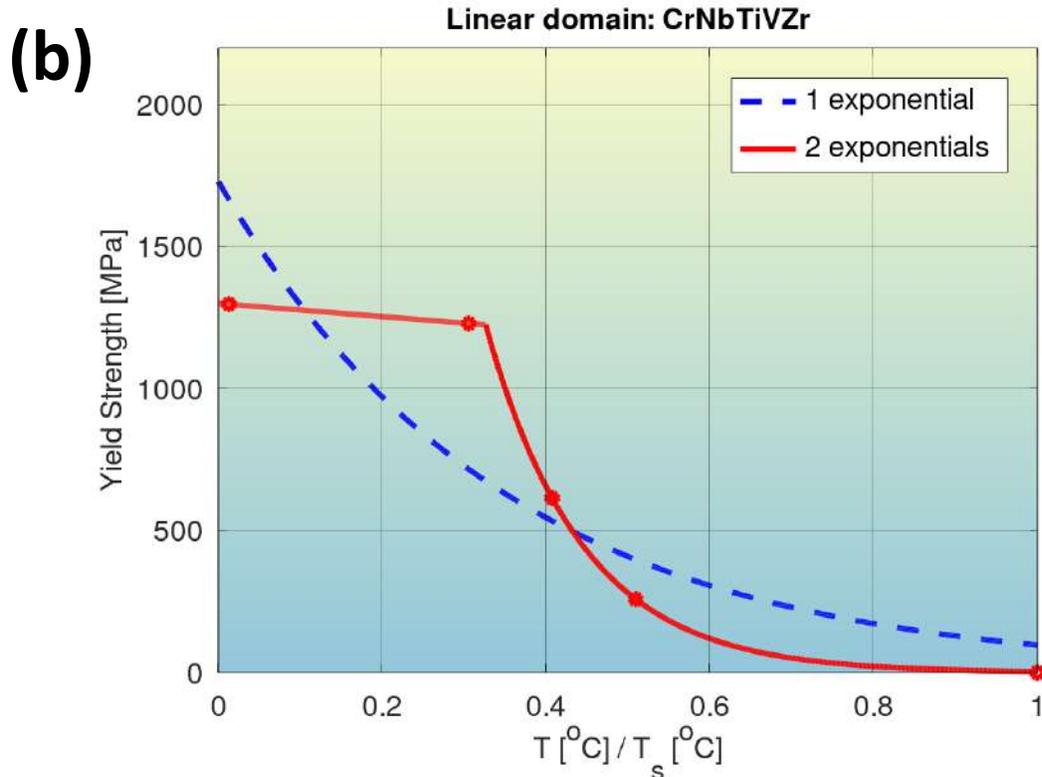

**Fig. S75**: Quantification of modeling accuracy of the bilinear log model, for composition No. 74 from **Tab. S2** (CrNbTiVZr, BCC+Laves phases), and comparison to that of a model with a single exponential.

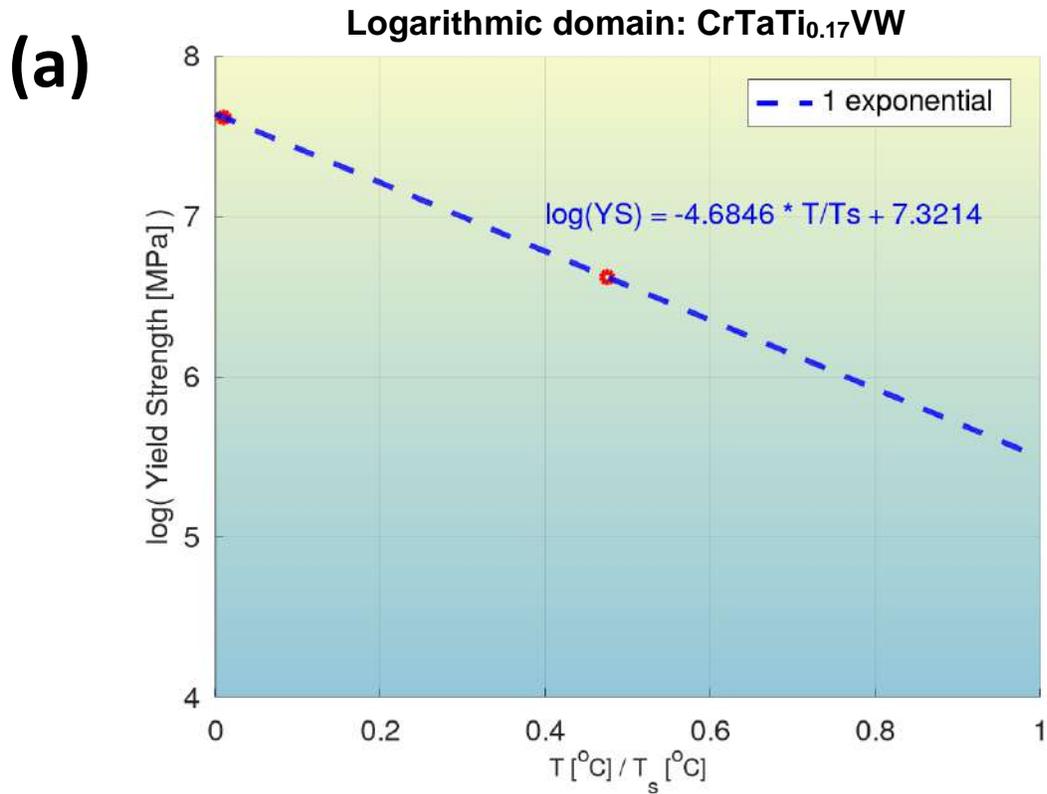

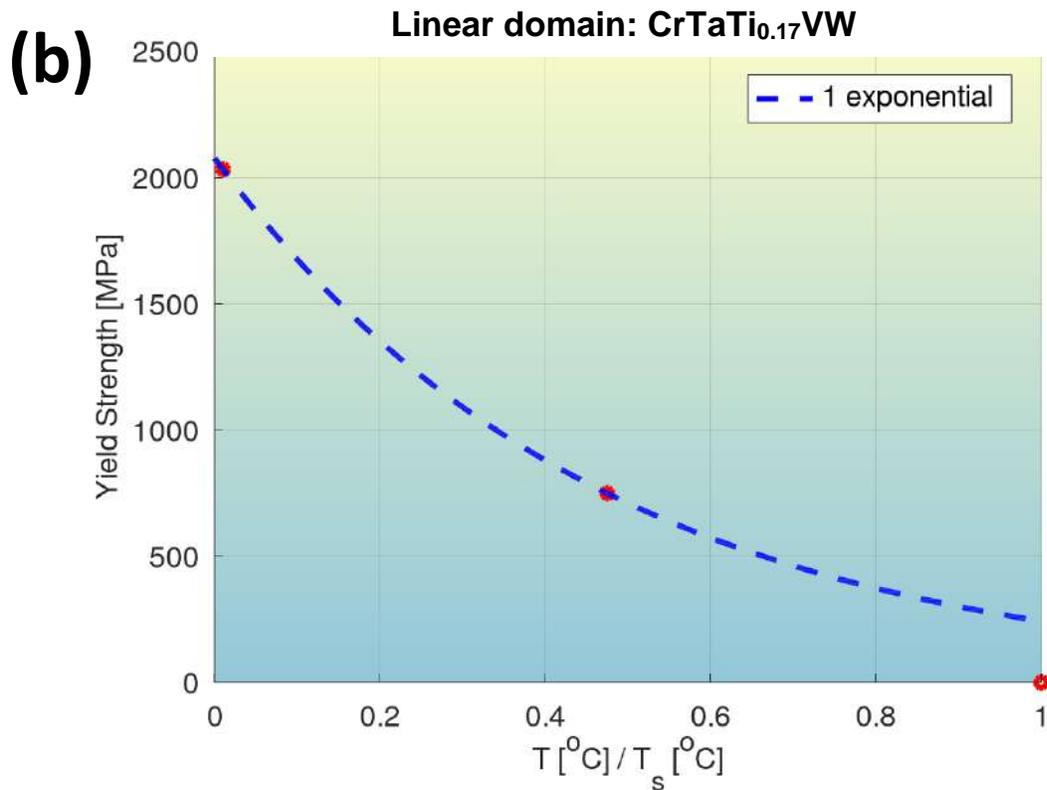

**Fig. S76**: Quantification of modeling accuracy of the bilinear log model, for composition No. 75 from **Tab. S2** ($CrTaTi_{0.17}VW$, BCC+Laves phases), and comparison to that of a model with a single exponential.

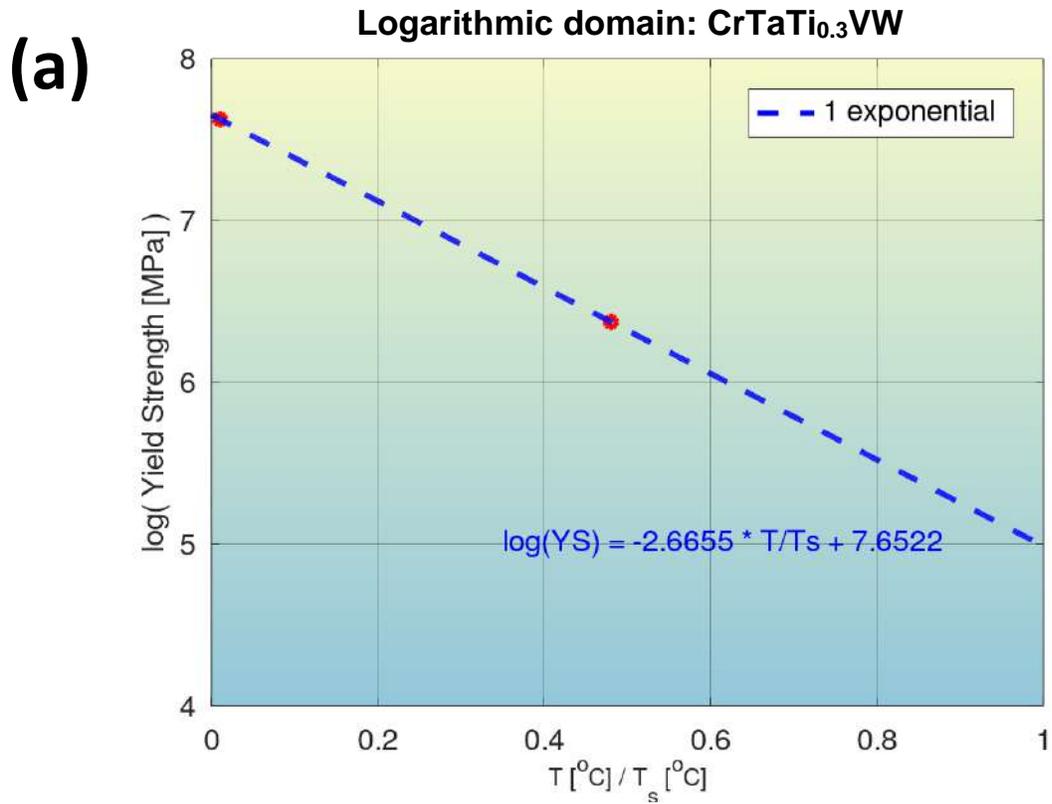

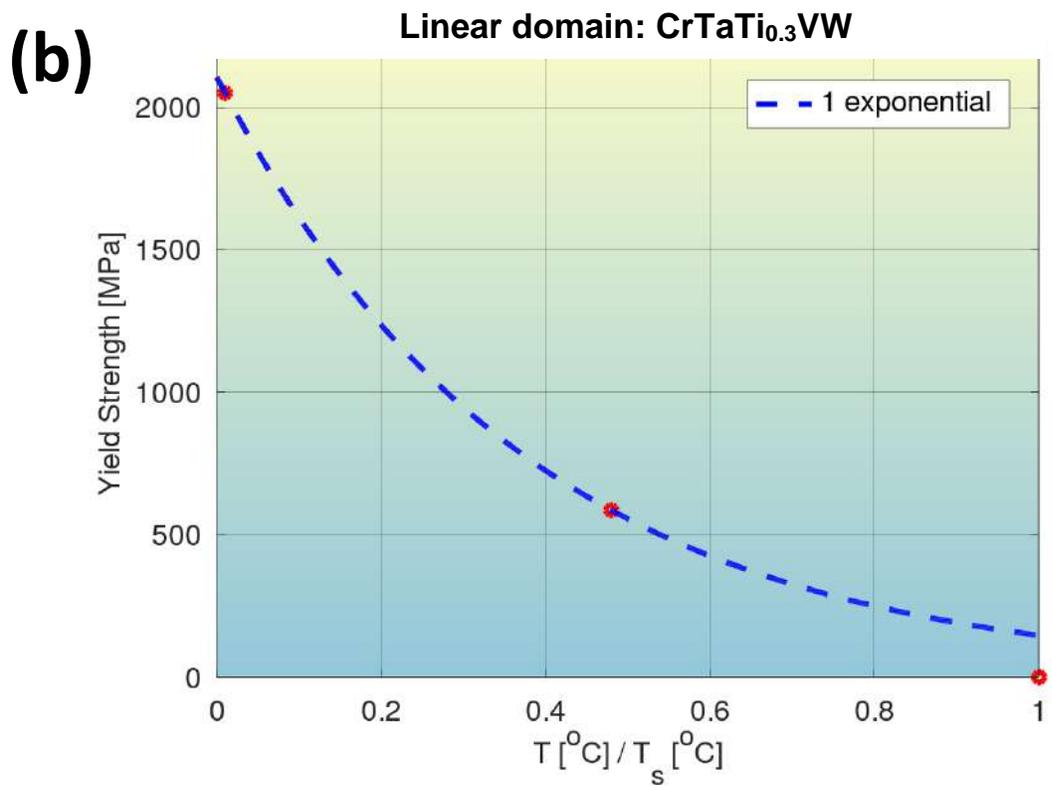

**Fig. S77**: Quantification of modeling accuracy of the bilinear log model, for composition No. 76 from **Tab. S2** (CrTaTi$_{0.3}$VW, BCC+Laves phases), and comparison to that of a model with a single exponential.

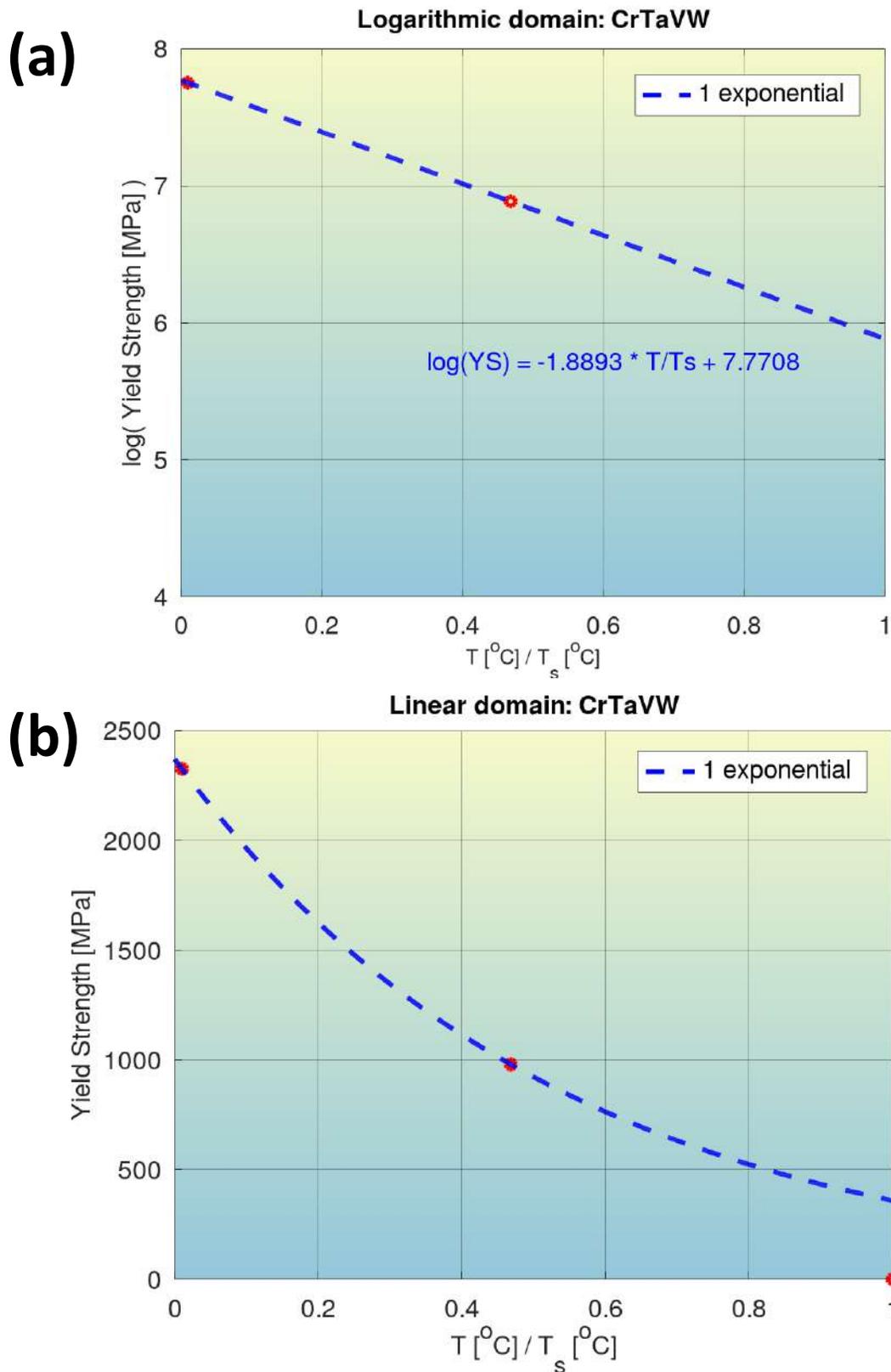

**Fig. S78**: Quantification of modeling accuracy of the bilinear log model, for composition No. 77 from **Tab. S2** (CrTaVW, BCC+Laves phases), and comparison to that of a model with a single exponential.

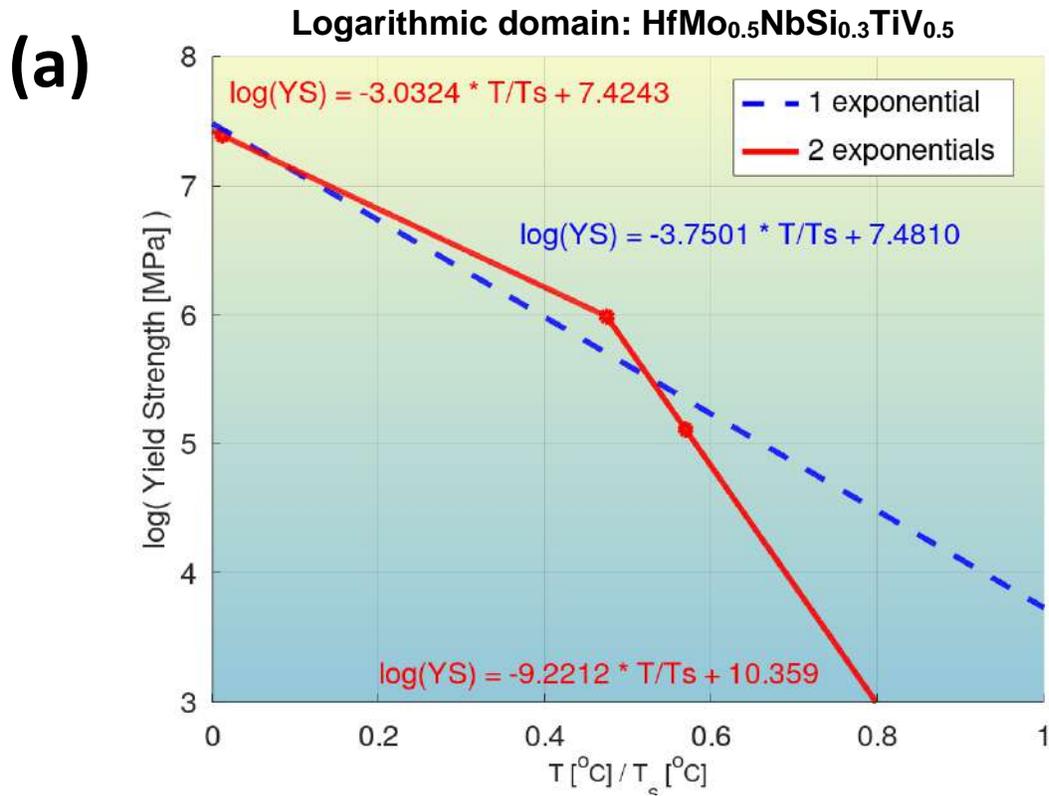

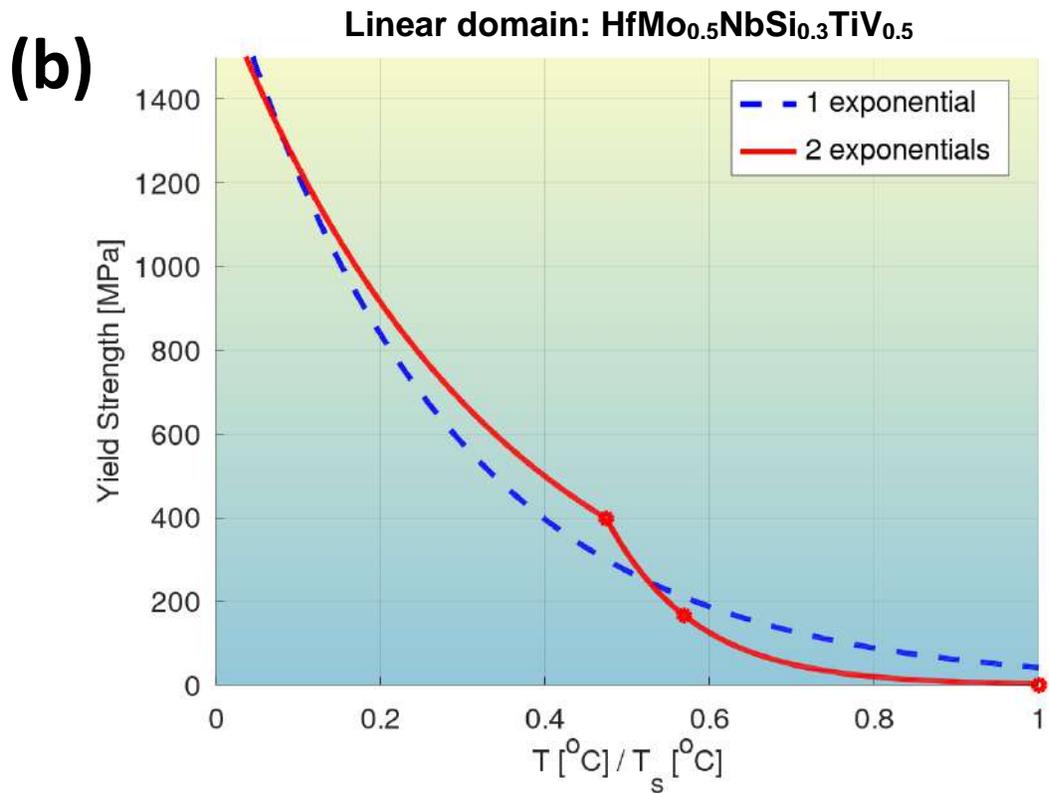

**Fig. S79**: Quantification of modeling accuracy of the bilinear log model, for composition No. 78 from **Tab. S2** (HfMo$_{0.5}$NbSi$_{0.3}$TiV$_{0.5}$, BCC+M5Si3 phases), and comparison to that of a model with a single exponential.

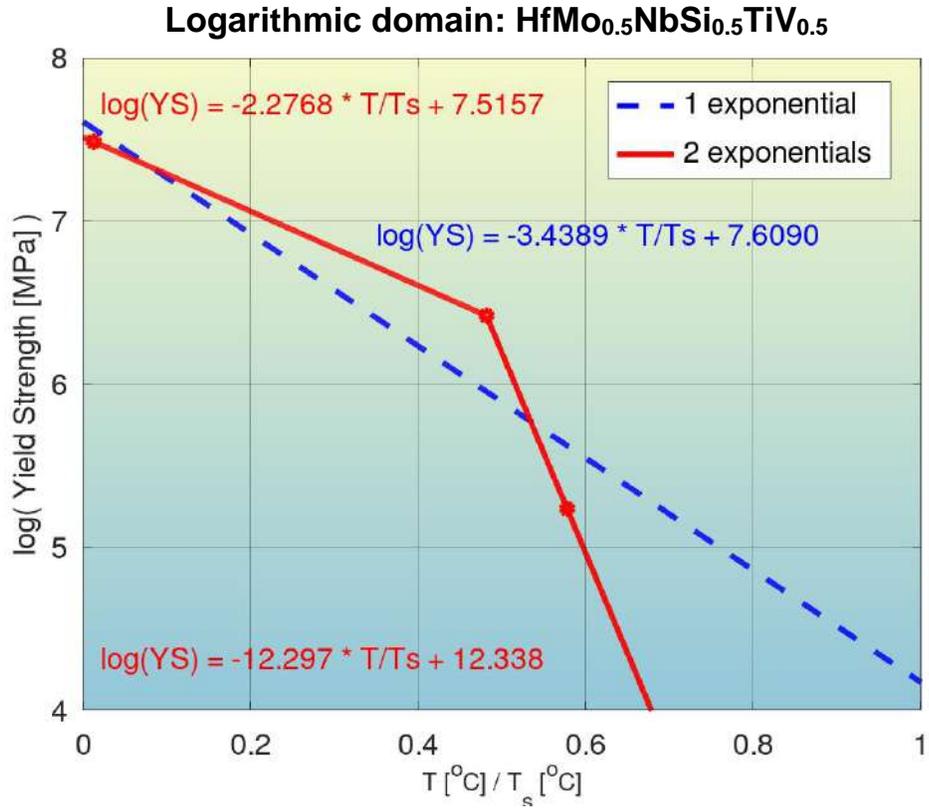
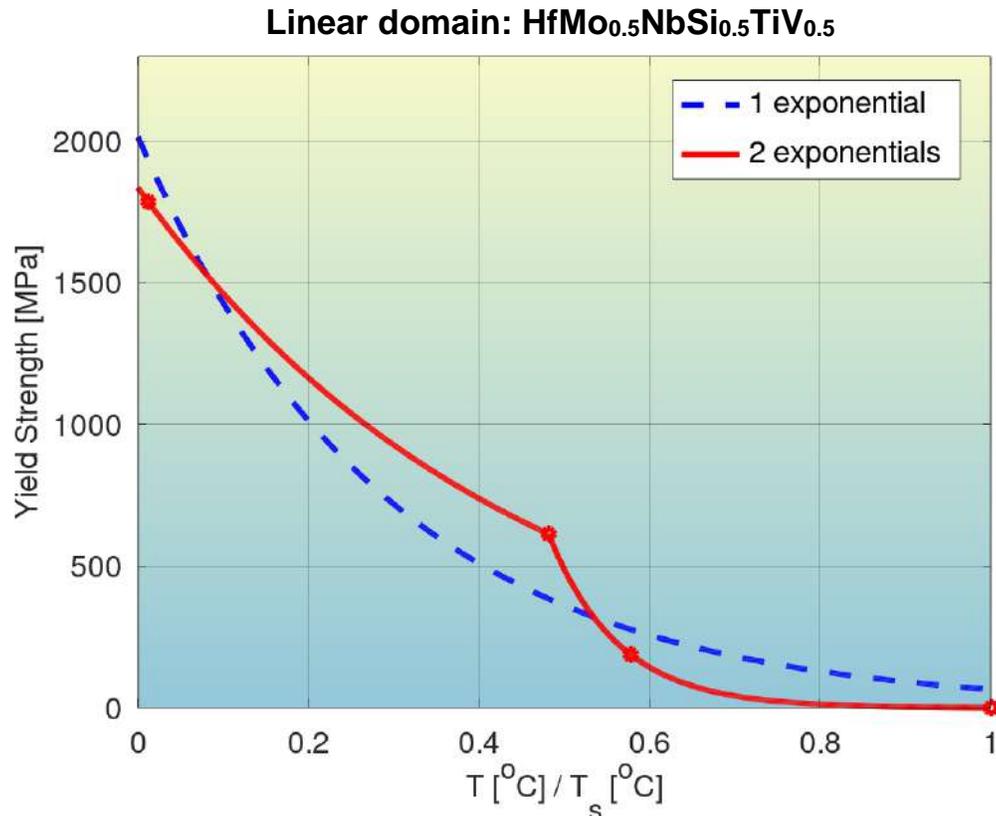

**Fig. S80**: Quantification of modeling accuracy of the bilinear log model, for composition No. 79 from **Tab. S2** (HfMo$_{0.5}$NbSi$_{0.5}$TiV$_{0.5}$, BCC+M5Si3 phases), and comparison to that of a model with a single exponential.

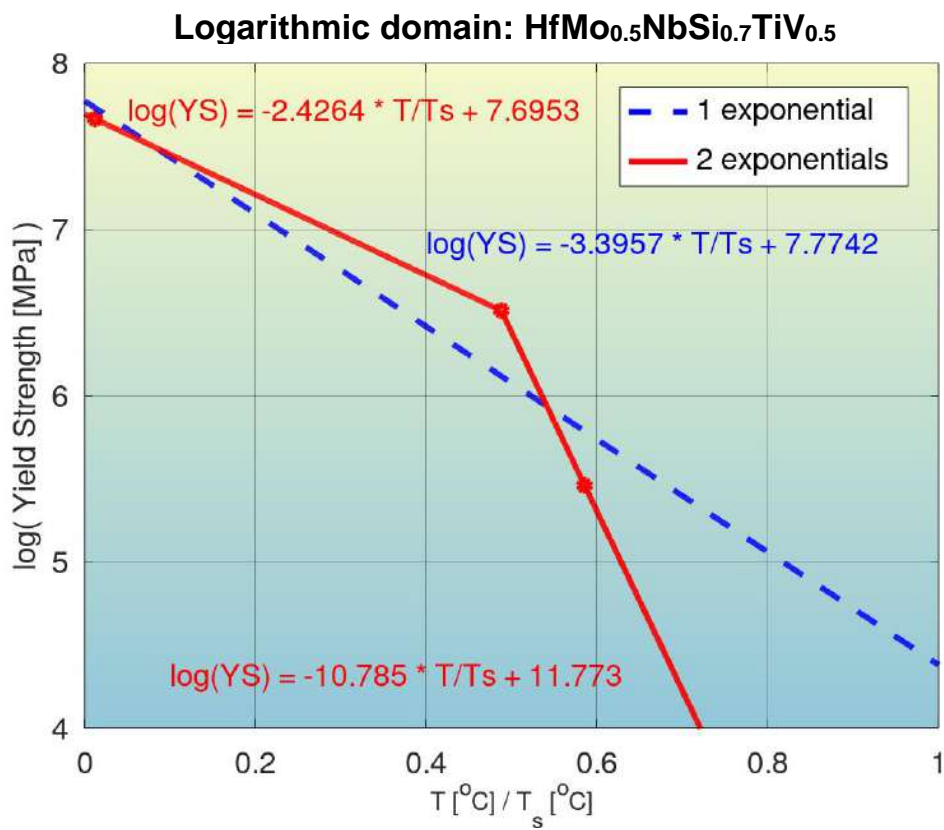

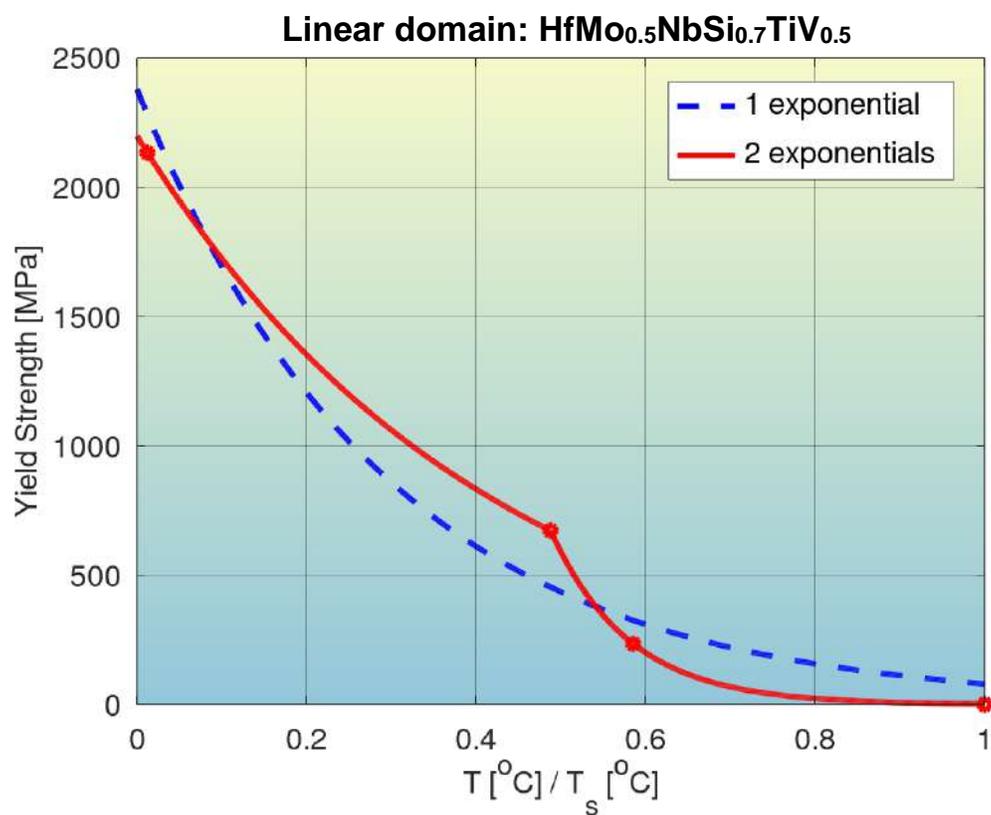

**Fig. S81**: Quantification of modeling accuracy of the bilinear log model, for composition No. 80 from **Tab. S2** (HfMo$_{0.5}$NbSi$_{0.7}$TiV$_{0.5}$, BCC+M5Si3 phases), and comparison to that of a model with a single exponential.

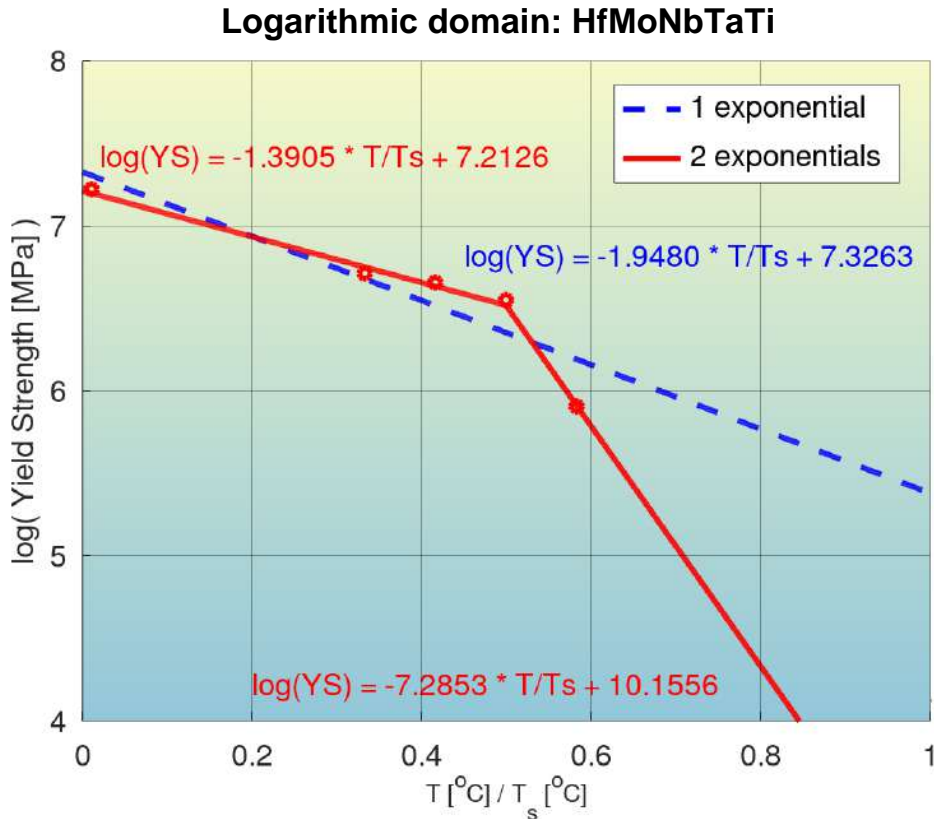

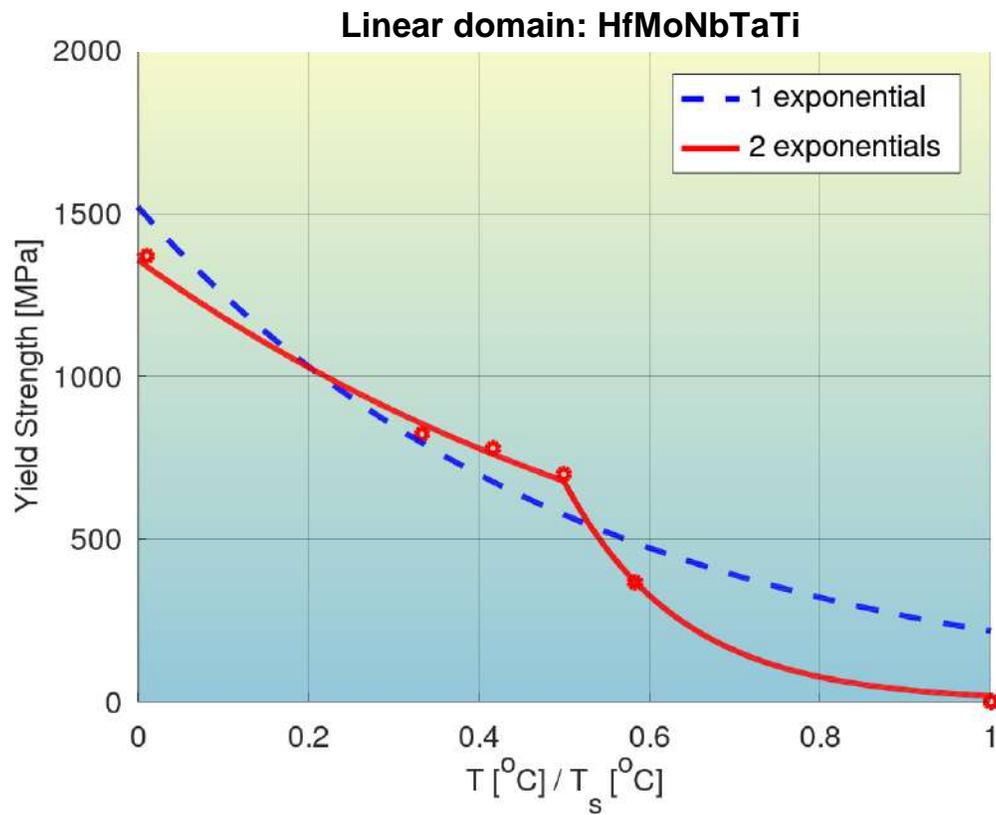

**Fig. S82**: Quantification of modeling accuracy of the bilinear log model, for composition No. 81 from **Tab. S2** (HfMoNbTaTi, BCC phase), and comparison to that of a model with a single exponential.

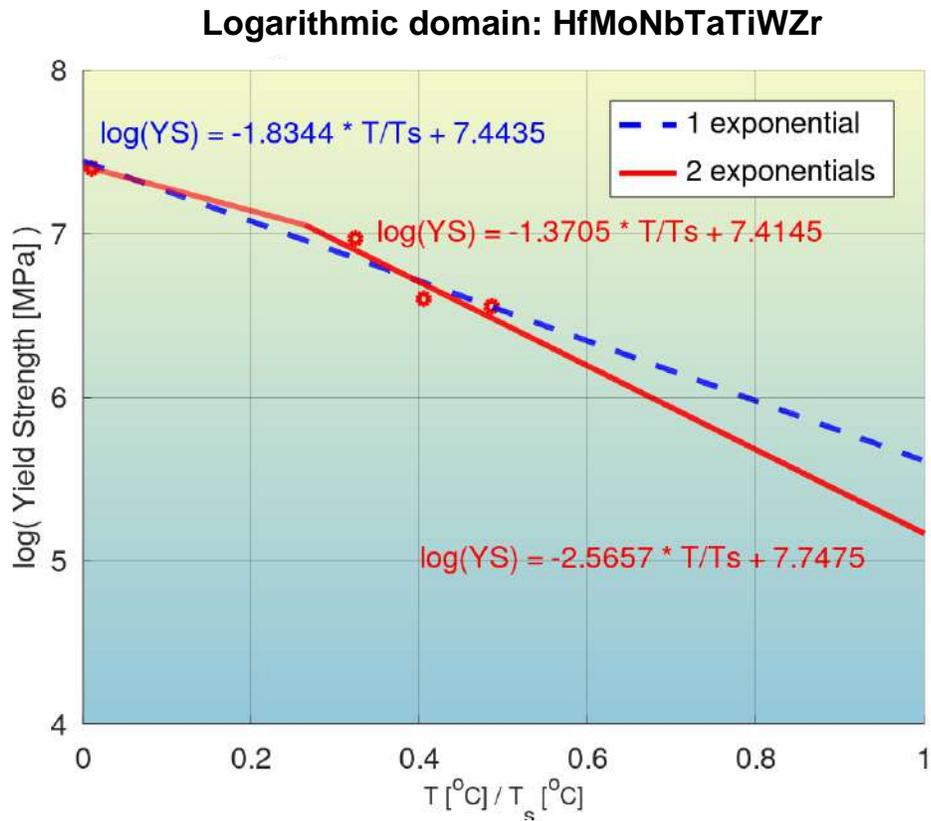

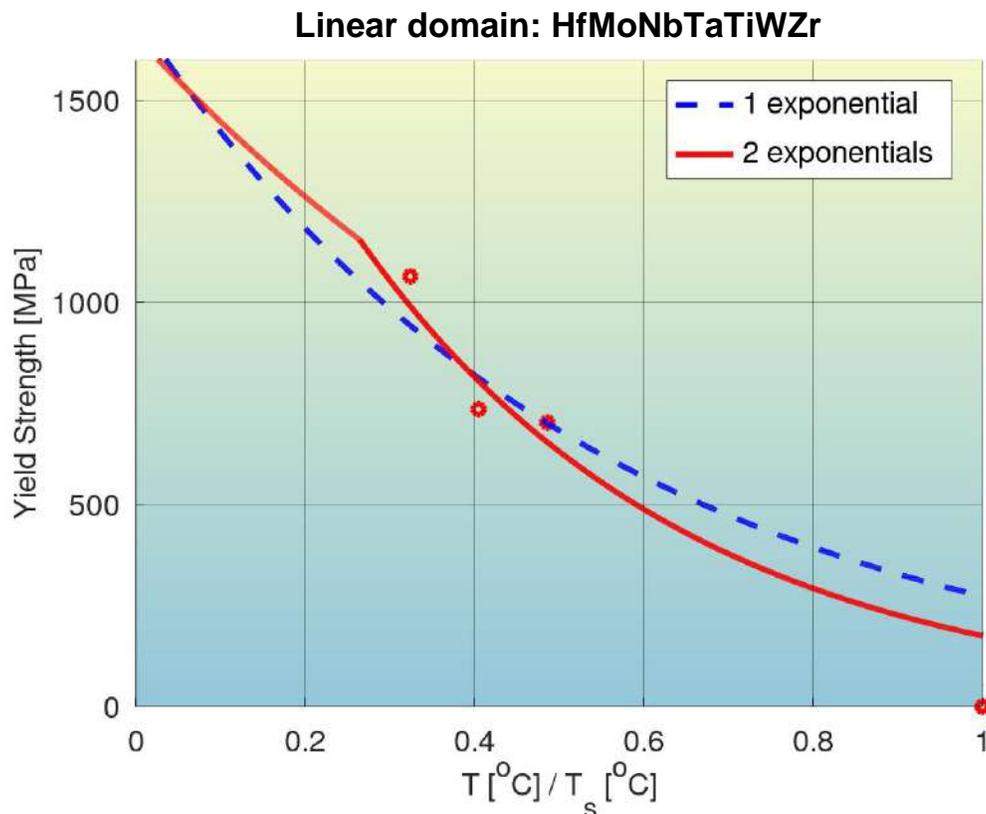

**Fig. S83**: Quantification of modeling accuracy of the bilinear log model, for composition No. 82 from **Tab. S2** (HfMoNbTaTiWZr, BCC+Laves phases), and comparison to that of a model with a single exponential.

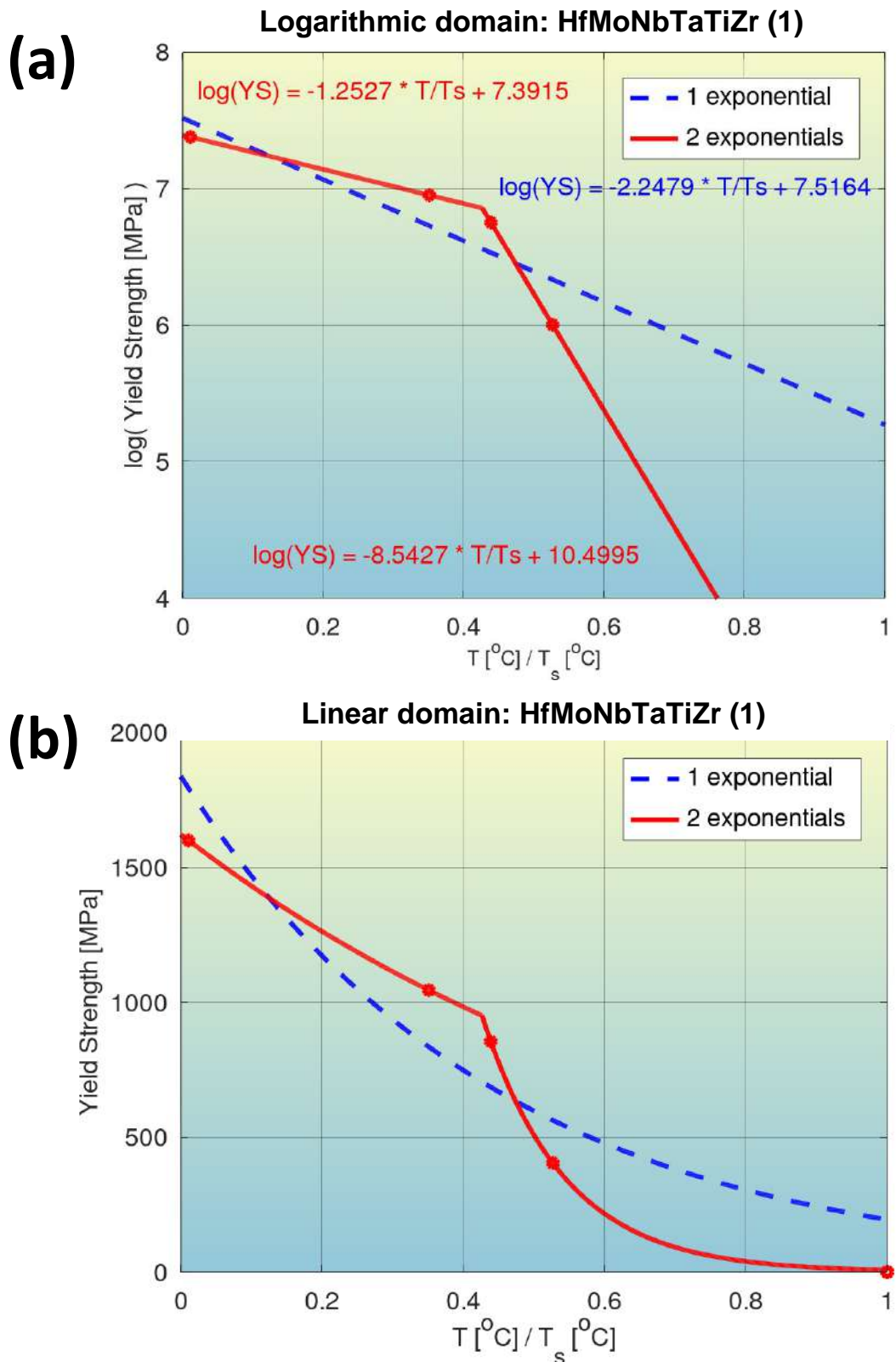

**Fig. S84**: Quantification of modeling accuracy of the bilinear log model, for composition No. 83 from **Tab. S2** (HfMoNbTaTiZr, BCC phase), and comparison to that of a model with a single exponential.

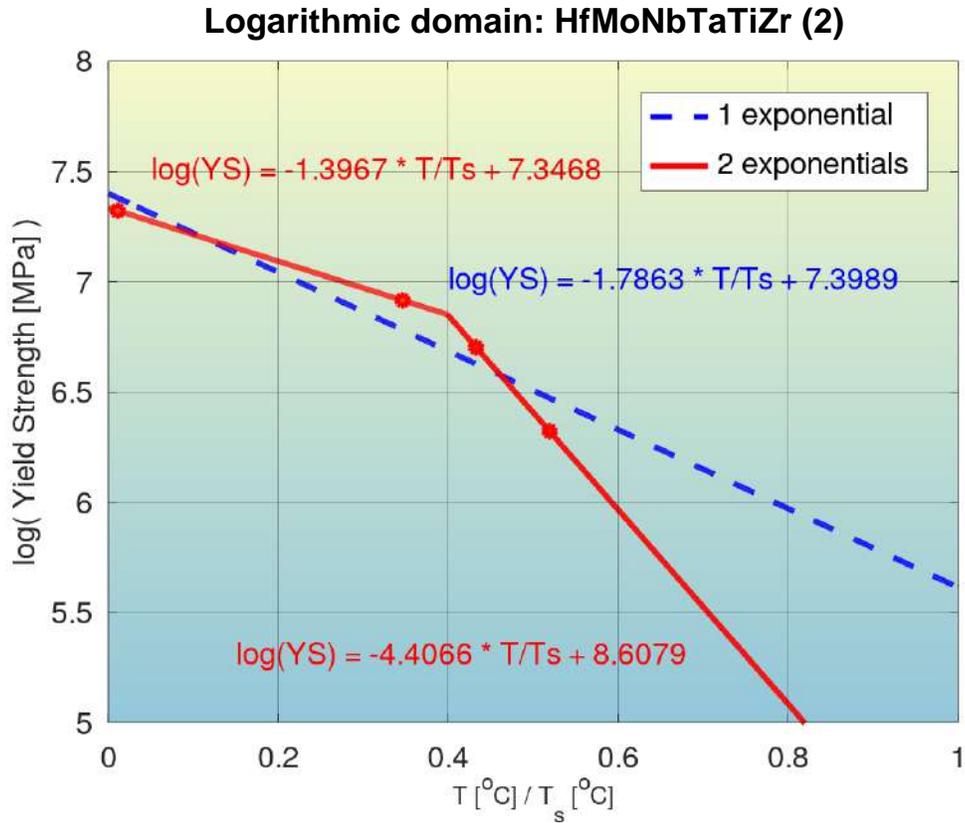

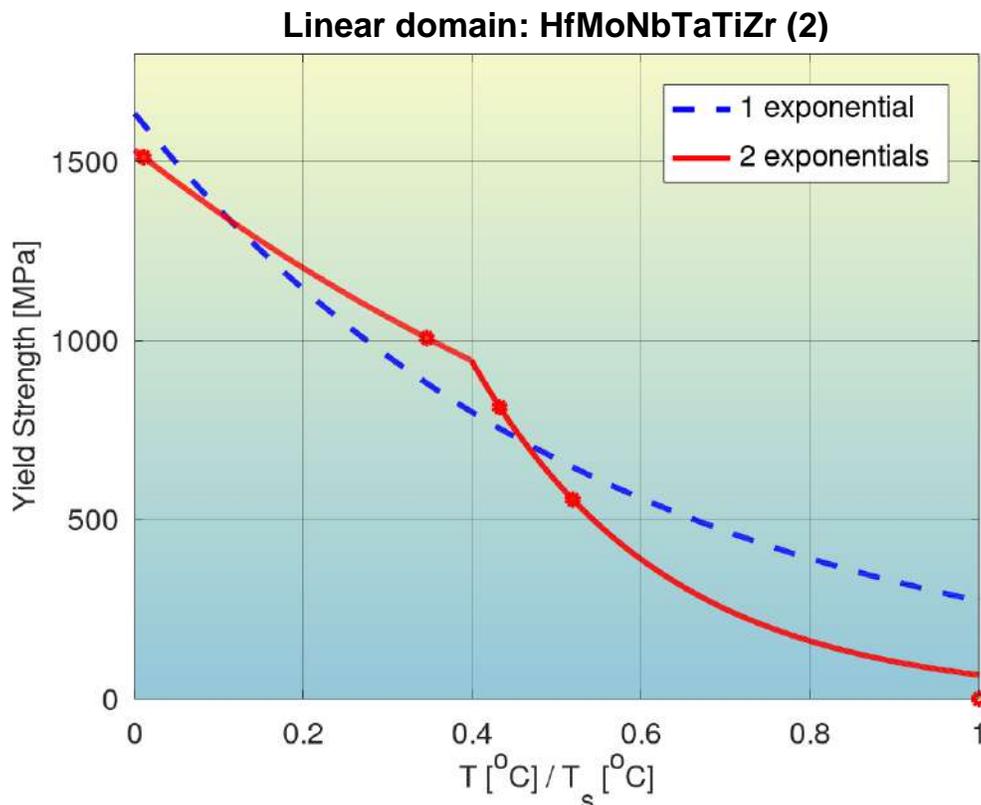

**Fig. S85**: Quantification of modeling accuracy of the bilinear log model, for composition No. 84 from **Tab. S2** (HfMoNbTaTiZr, BCC phase), and comparison to that of a model with a single exponential.

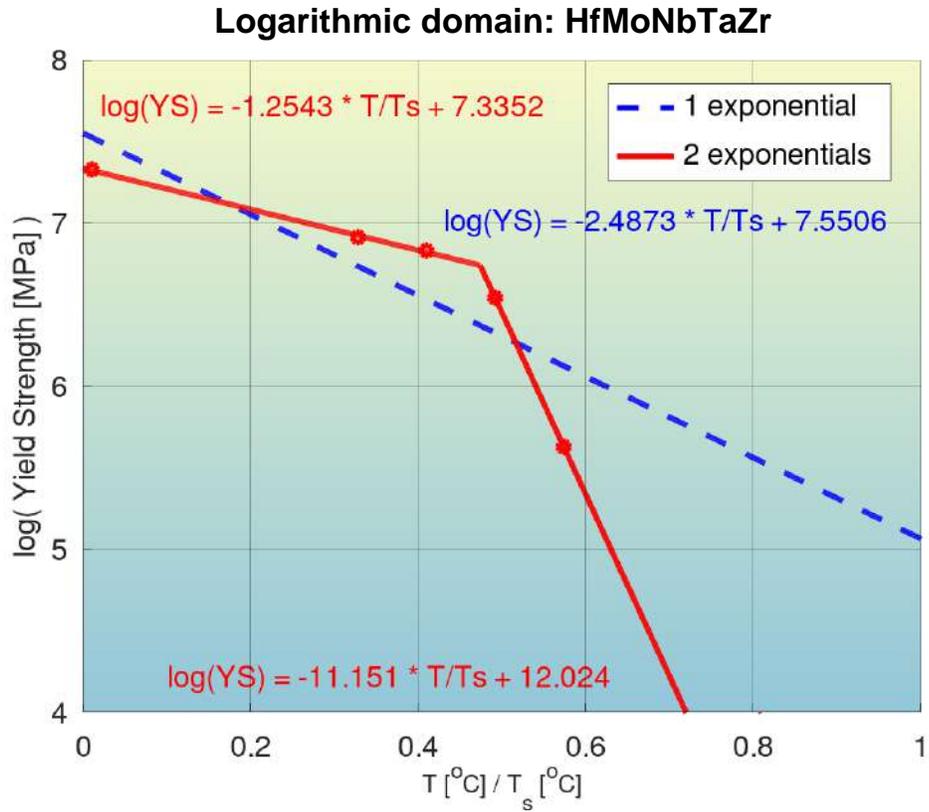
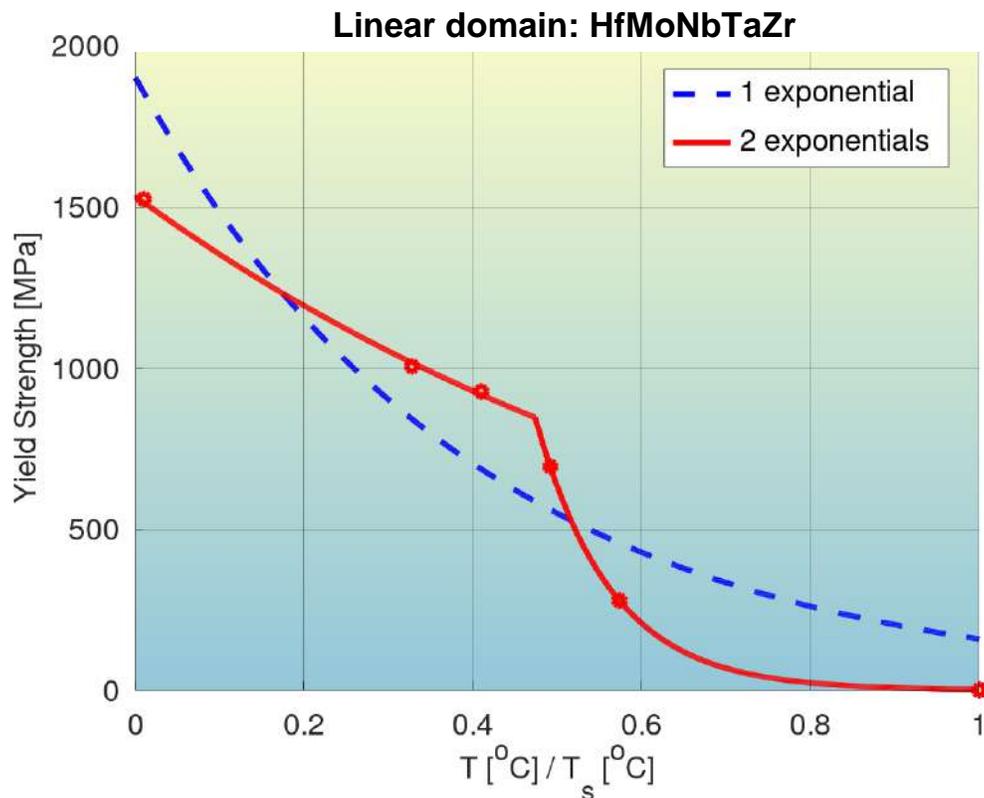

**Fig. S86**: Quantification of modeling accuracy of the bilinear log model, for composition No. 85 from **Tab. S2** (HfMoNbTaZr, BCC phase), and comparison to that of a model with a single exponential.

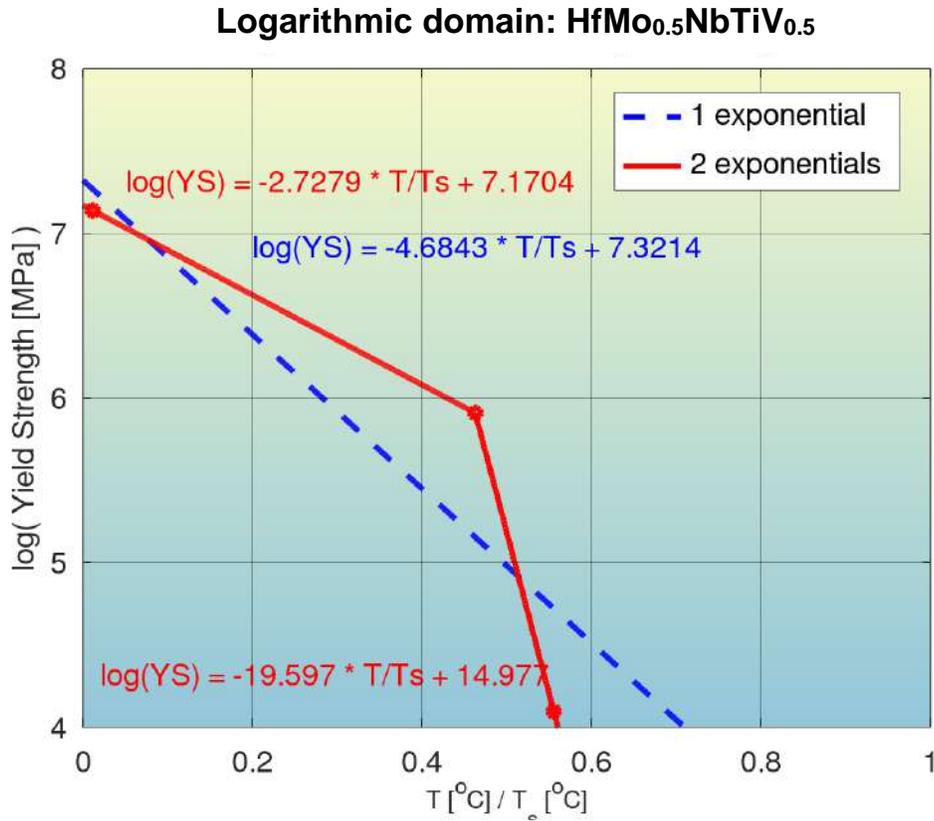

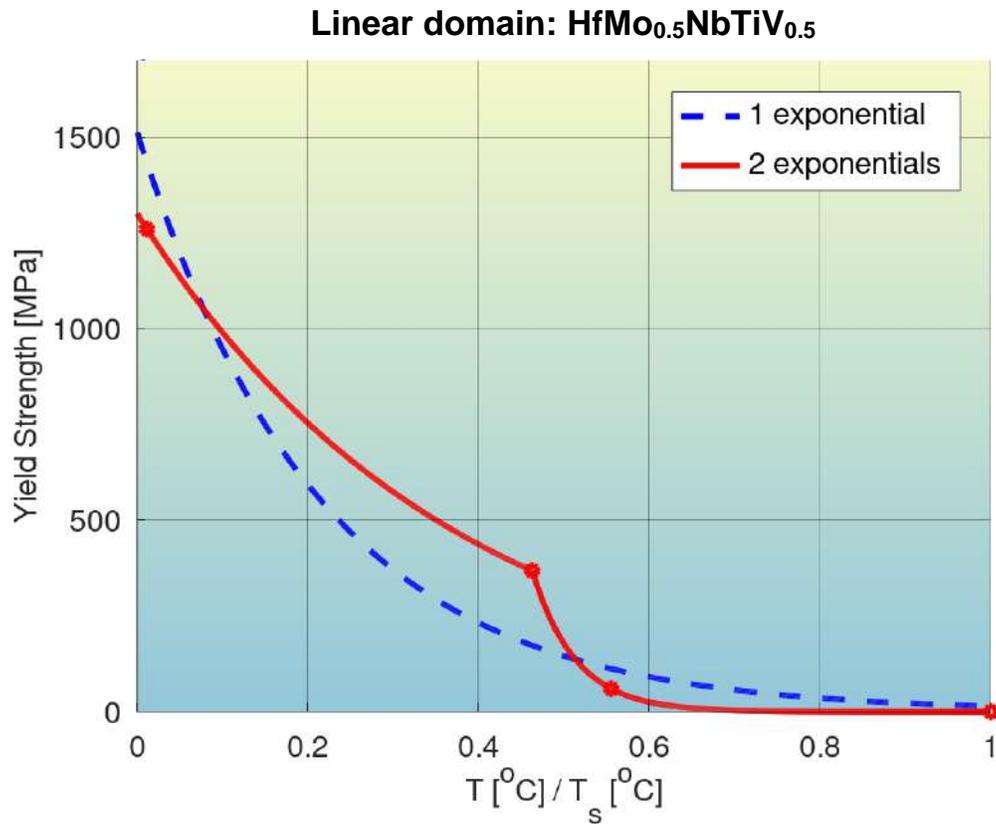

**Fig. S87**: Quantification of modeling accuracy of the bilinear log model, for composition No. 86 from **Tab. S2** (HfMo$_{0.5}$NbTiV$_{0.5}$, BCC phase), and comparison to that of a model with a single exponential.

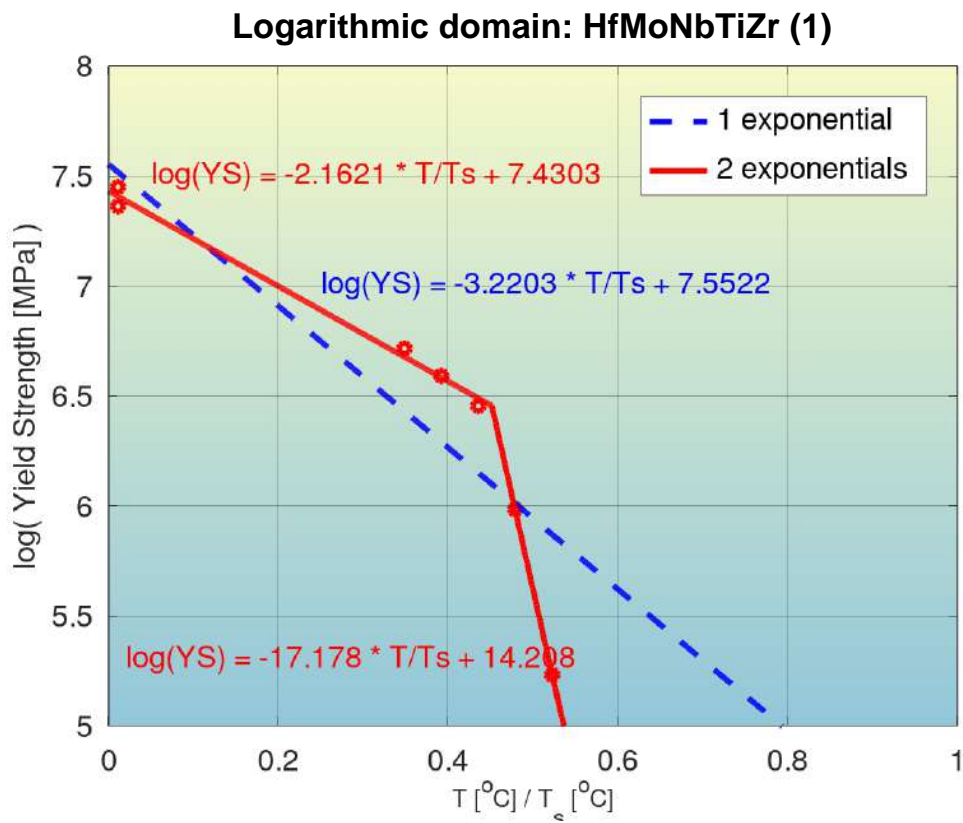

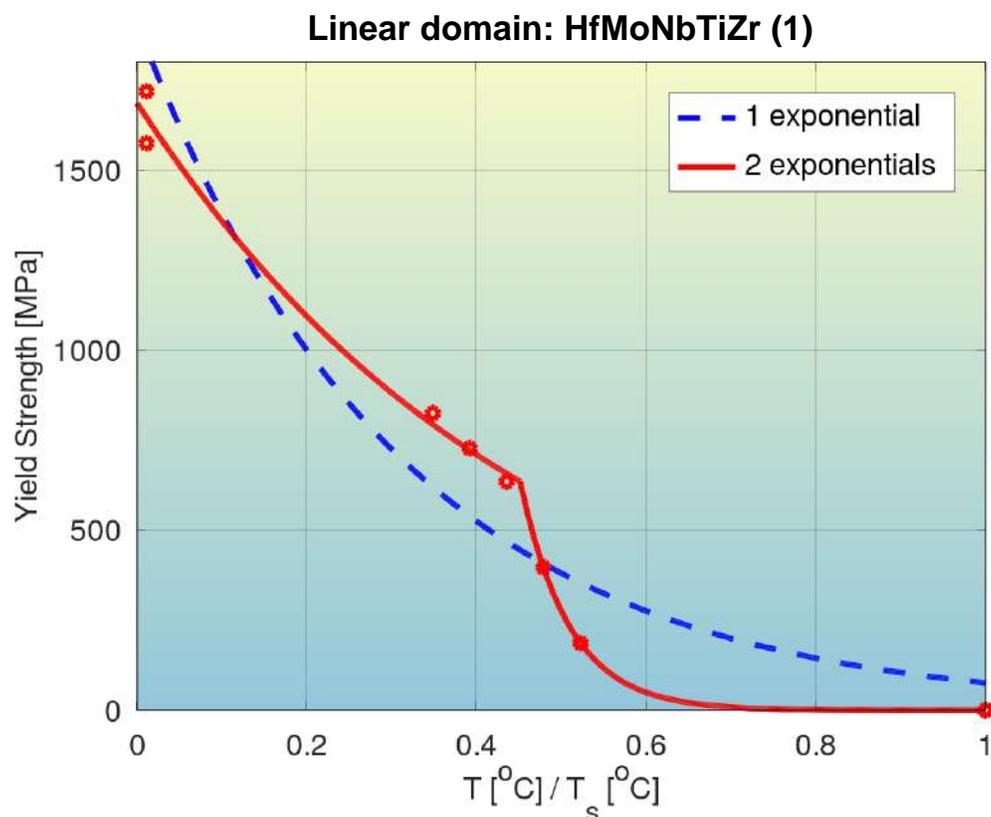

**Fig. S88**: Quantification of modeling accuracy of the bilinear log model, for composition No. 87 from **Tab. S2** (HfMoNbTiZr (1), BCC phase), and comparison to that of a model with a single exponential.

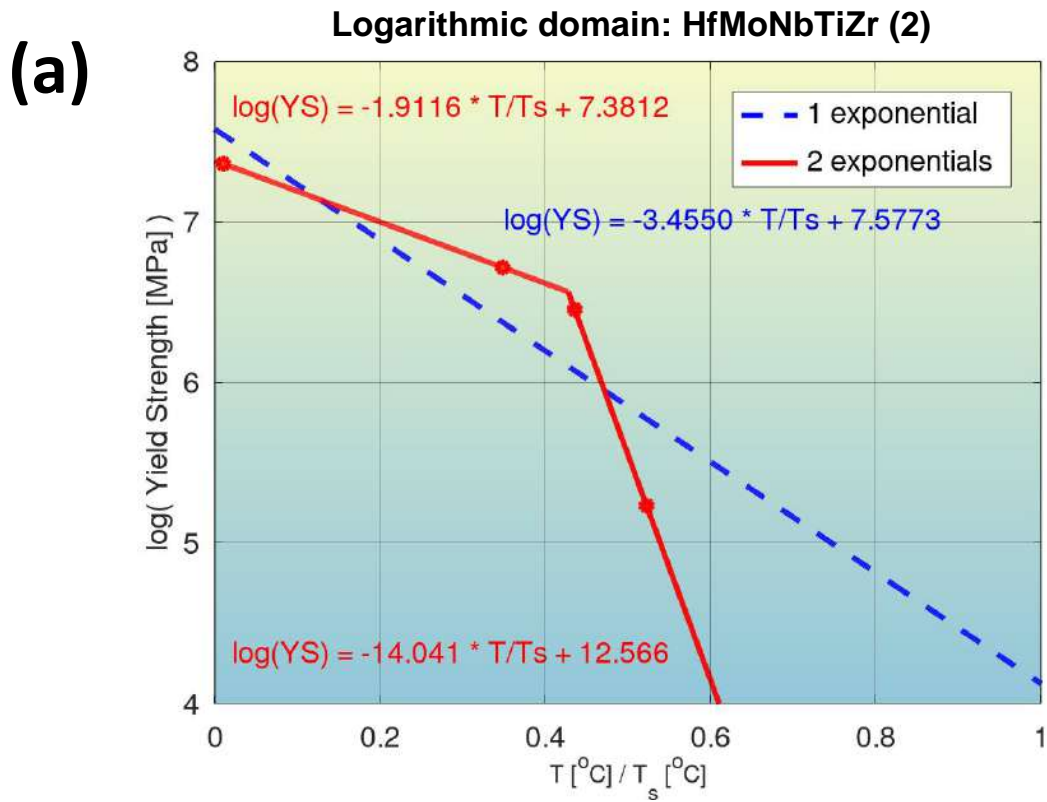

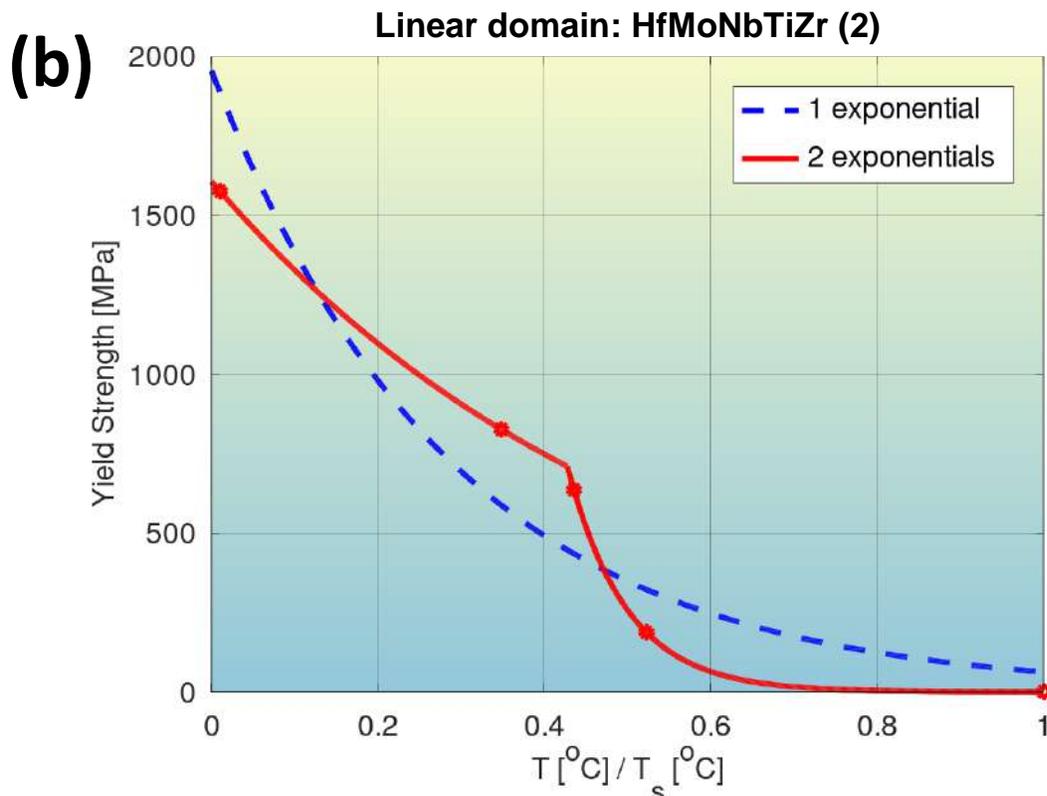

**Fig. S89**: Quantification of modeling accuracy of the bilinear log model, for composition No. 88 from **Tab. S2** (HfMoNbTiZr (2), BCC phase), and comparison to that of a model with a single exponential.

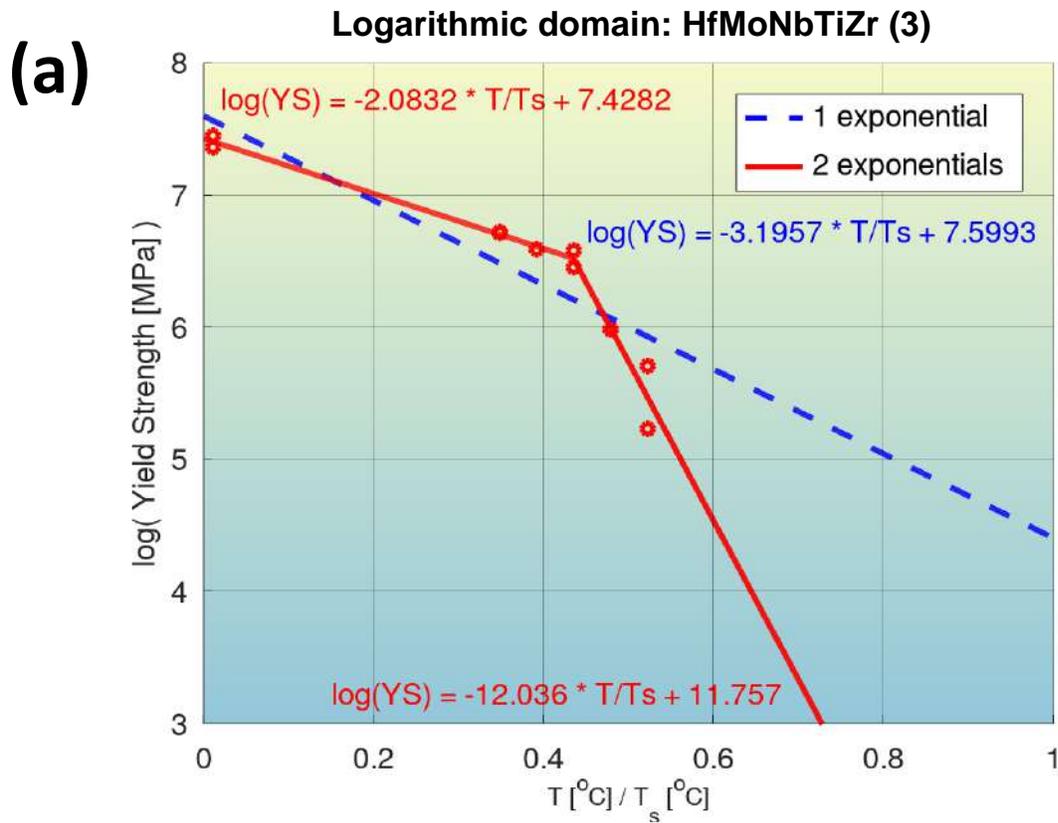
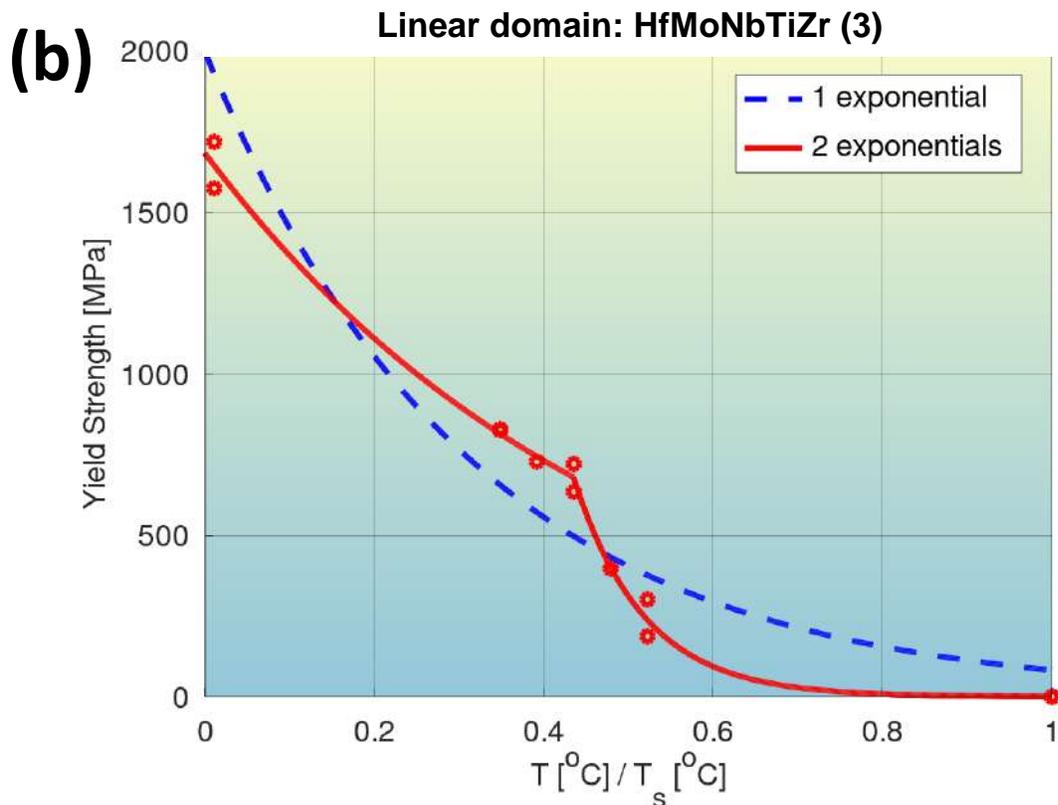

**Fig. S90**: Quantification of modeling accuracy of the bilinear log model, for composition No. 89 from **Tab. S2** (HfMoNbTiZr (3), BCC phase), and comparison to that of a model with a single exponential.

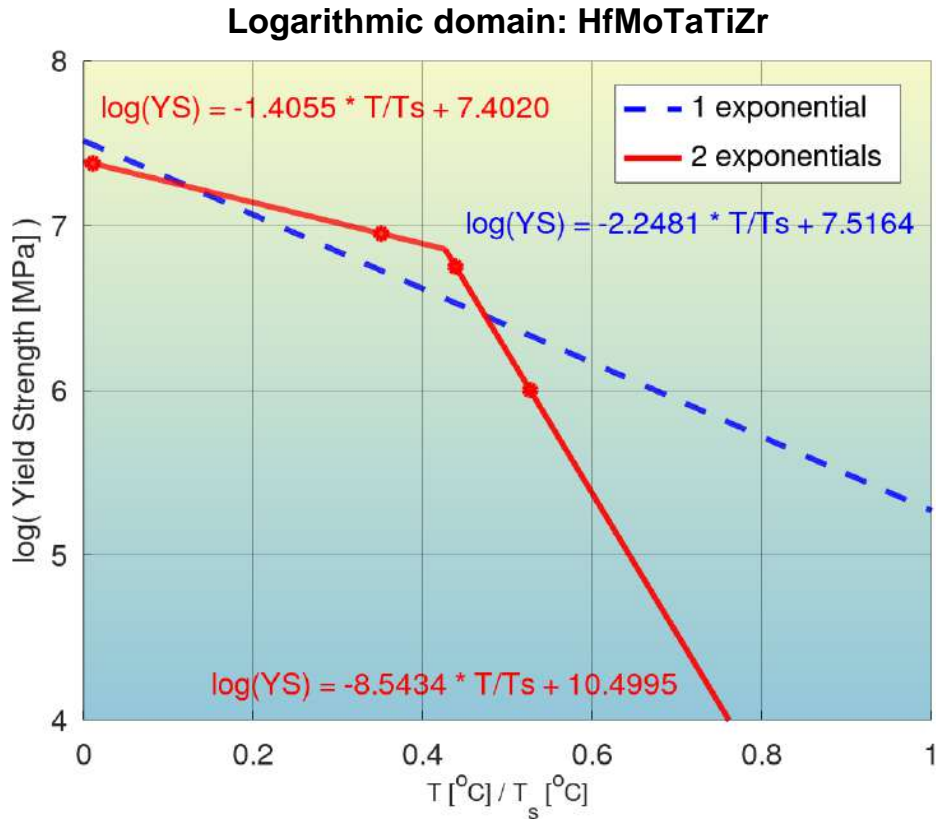

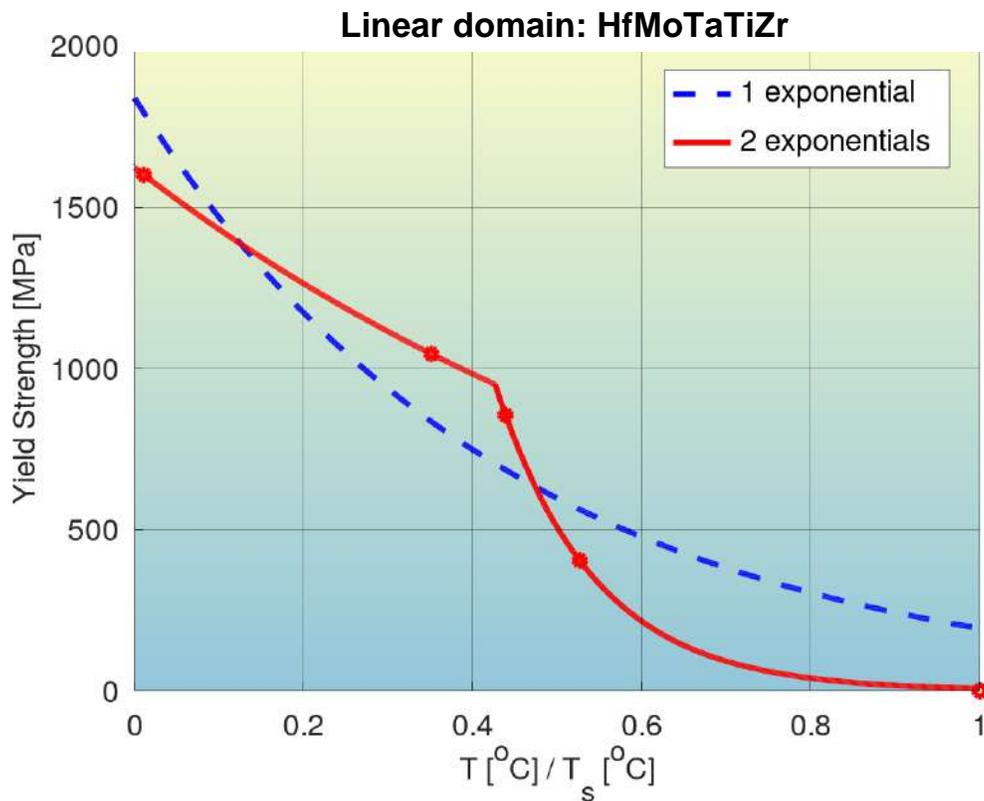

**Fig. S91**: Quantification of modeling accuracy of the bilinear log model, for composition No. 90 from **Tab. S2** (HfMoTaTiZr, BCC phase), and comparison to that of a model with a single exponential.

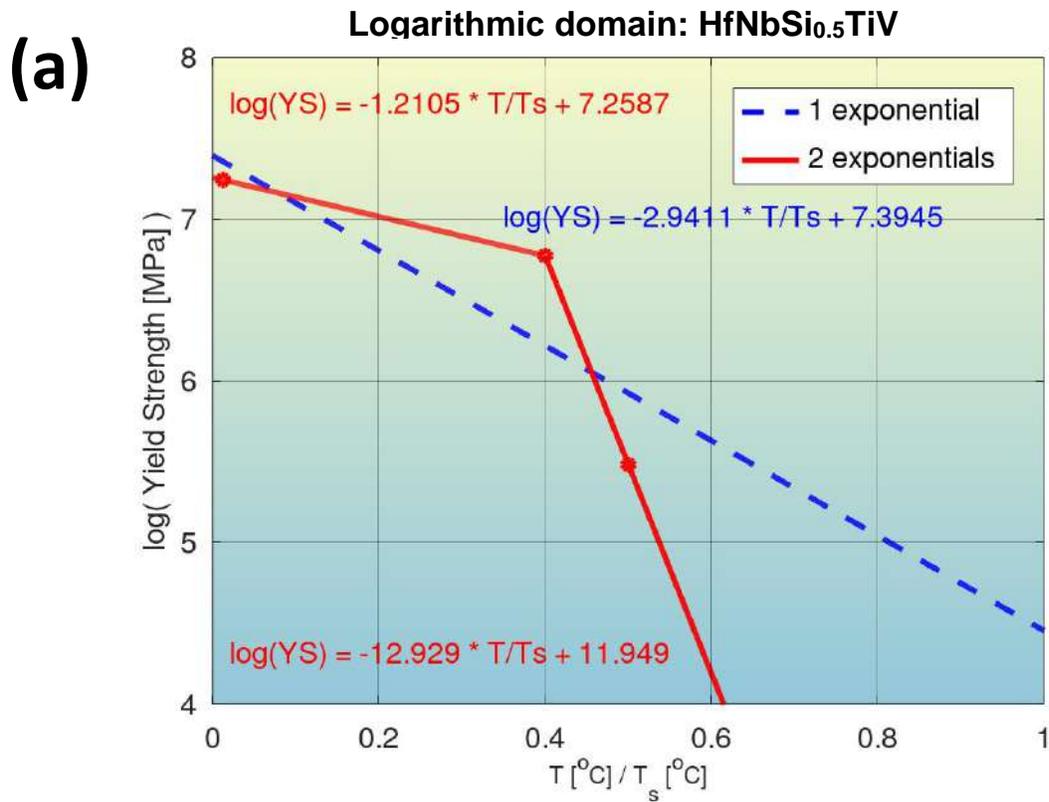

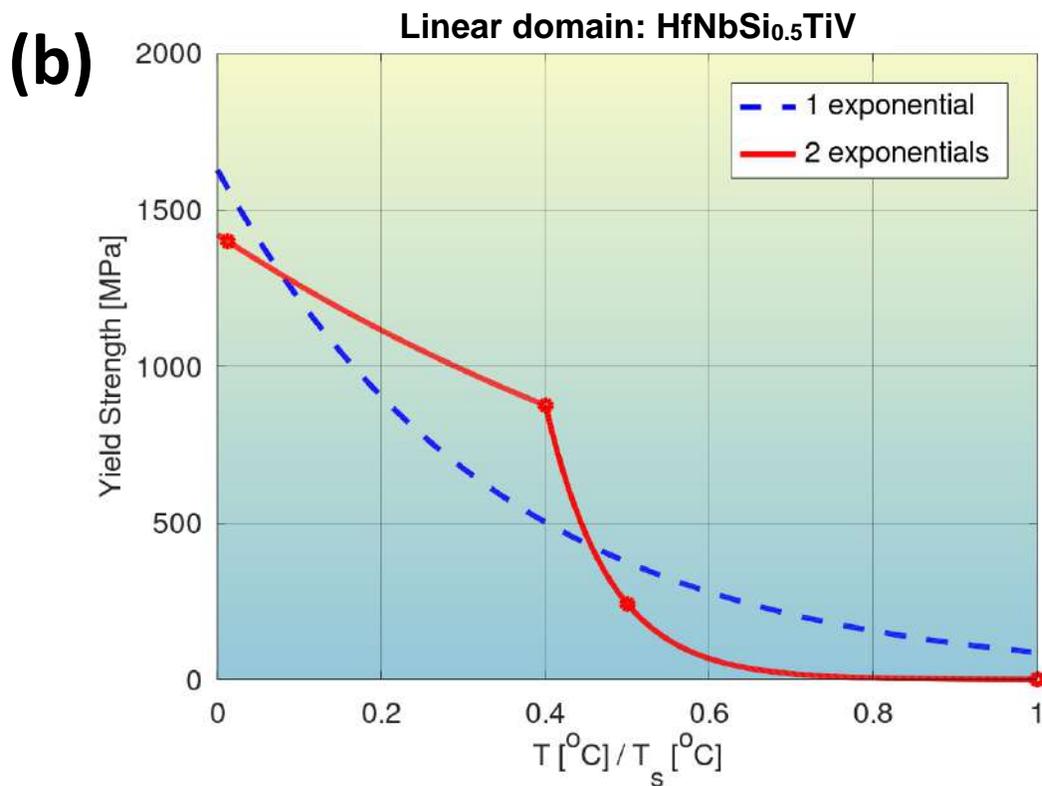

**Fig. S92**: Quantification of modeling accuracy of the bilinear log model, for composition No. 91 from **Tab. S2** (HfNbSi$_{0.5}$TiV, BCC+M5Si3 phases), and comparison to that of a model with a single exponential.

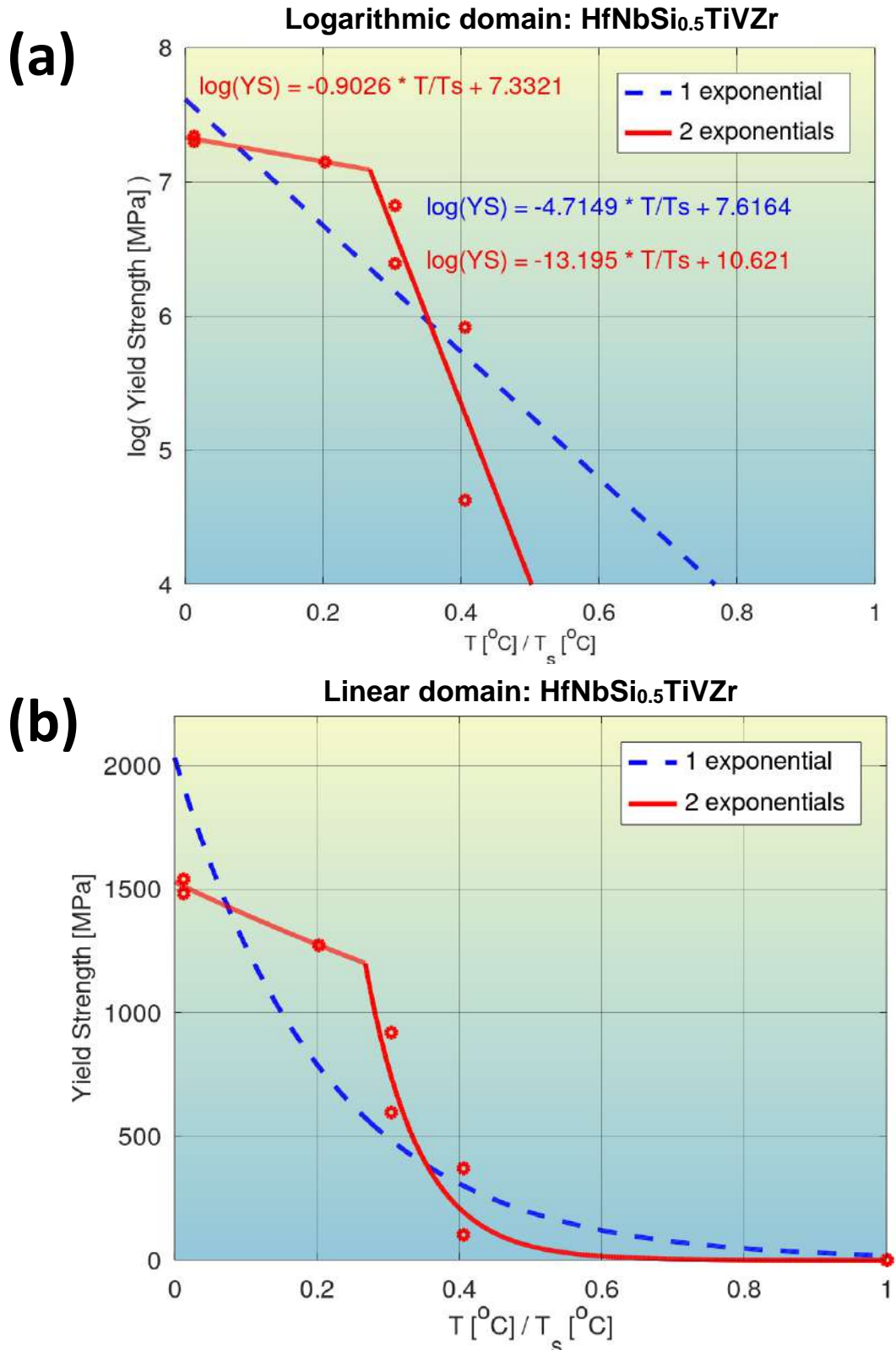

**Fig. S93**: Quantification of modeling accuracy of the bilinear log model, for composition No. 92 from **Tab. S2** (HfNbSi$_{0.5}$TiVZr, BCC+Laves+M5Si3 phases), and comparison to that of a model with a single exponential.

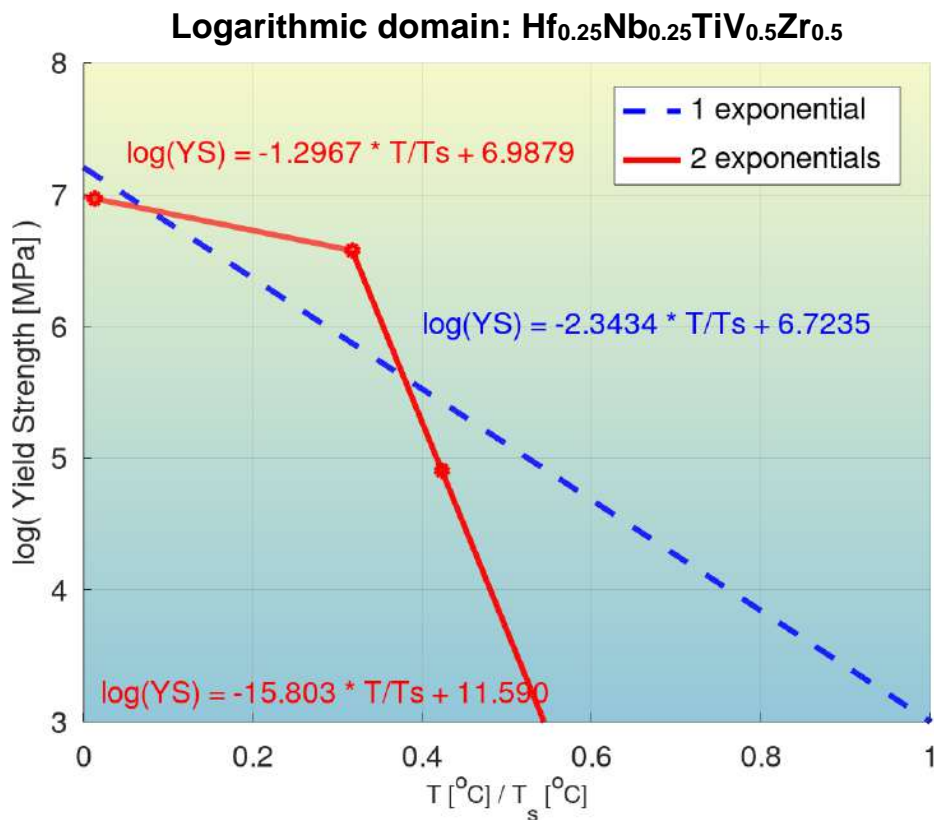

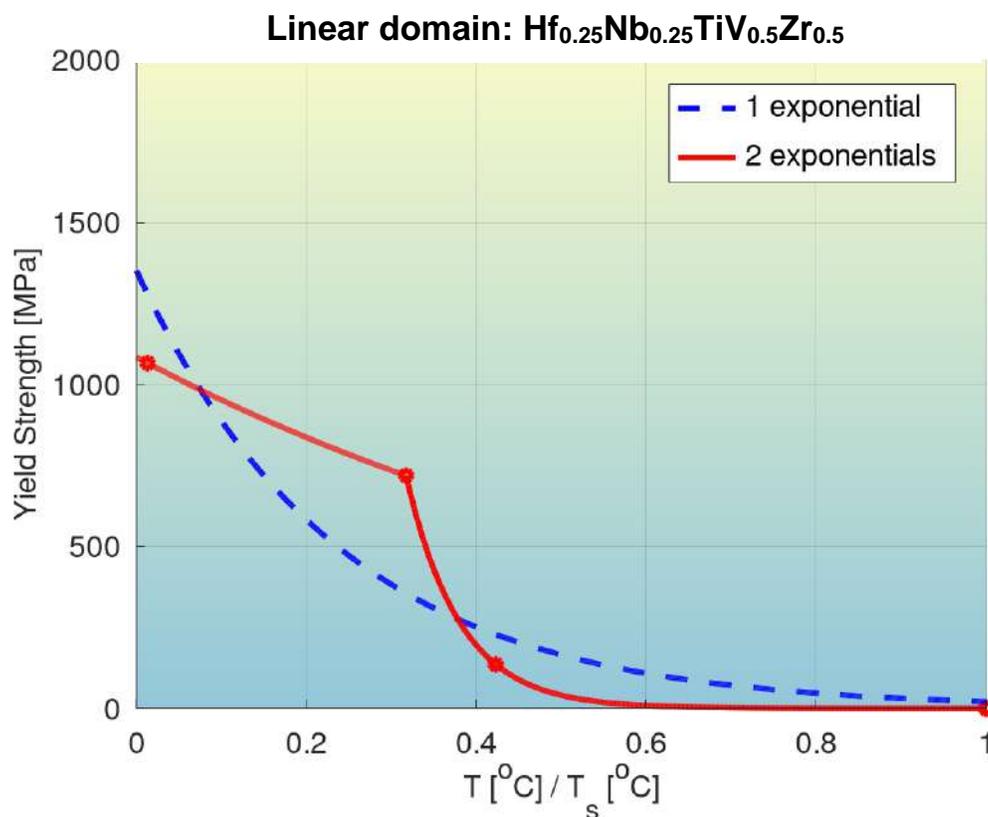

**Fig. S94**: Quantification of modeling accuracy of the bilinear log model, for composition No. 93 from **Tab. S2** (Hf$_{0.25}$Nb$_{0.25}$TiV$_{0.5}$Zr$_{0.5}$, BCC phase), and comparison to that of a model with a single exponential.

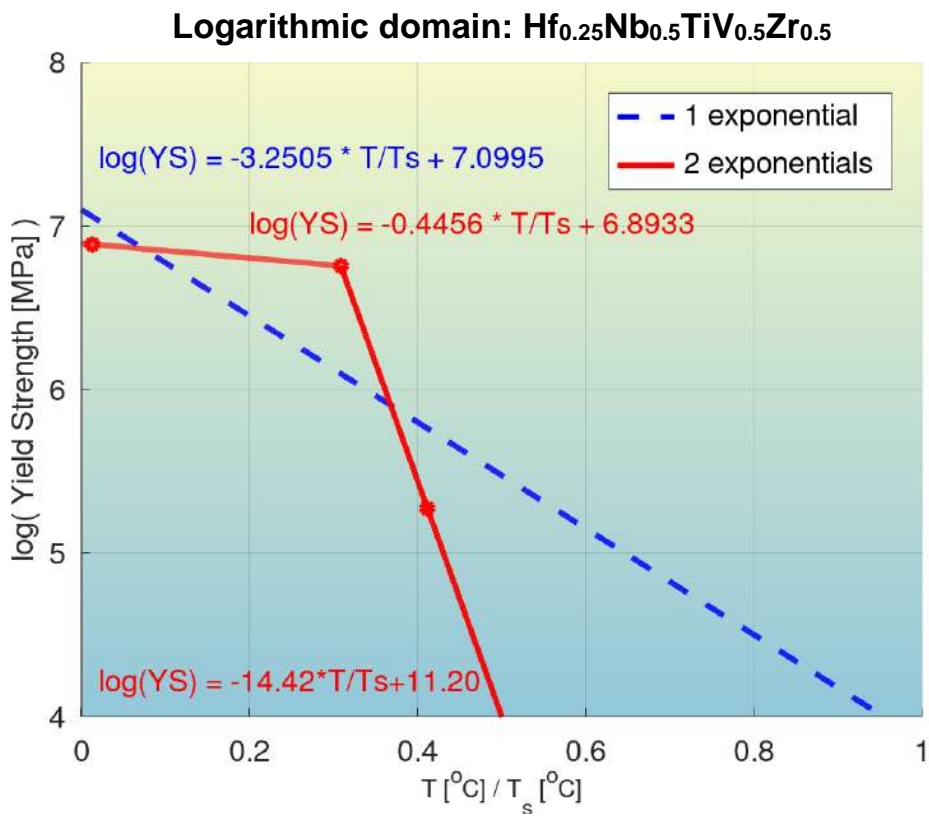

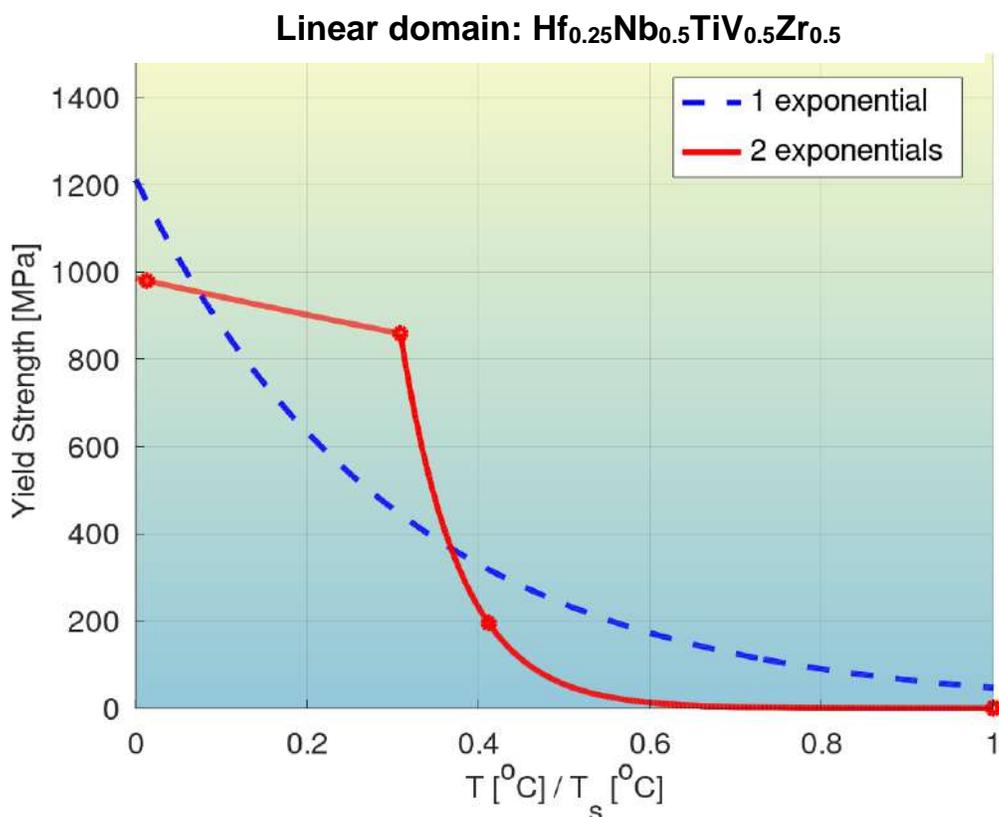

**Fig. S95**: Quantification of modeling accuracy of the bilinear log model, for composition No. 94 from **Tab. S2** ($Hf_{0.25}Nb_{0.5}TiV_{0.5}Zr_{0.5}$, BCC phase), and comparison to that of a model with a single exponential.

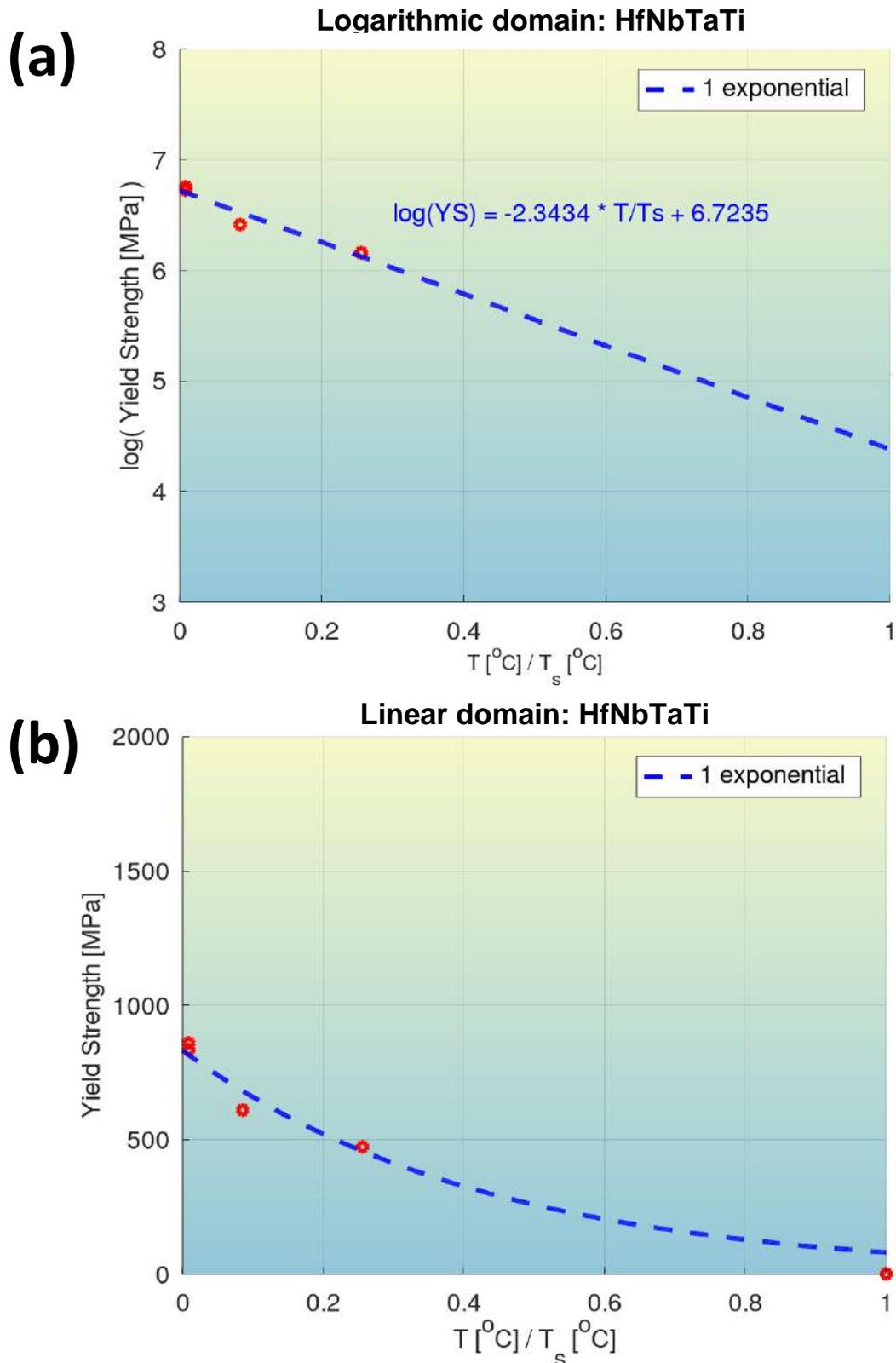

**Fig. S96**: Quantification of modeling accuracy of the bilinear log model, for composition No. 95 from **Tab. S2** (HfNbTaTi, BCC phase), and comparison to that of a model with a single exponential.

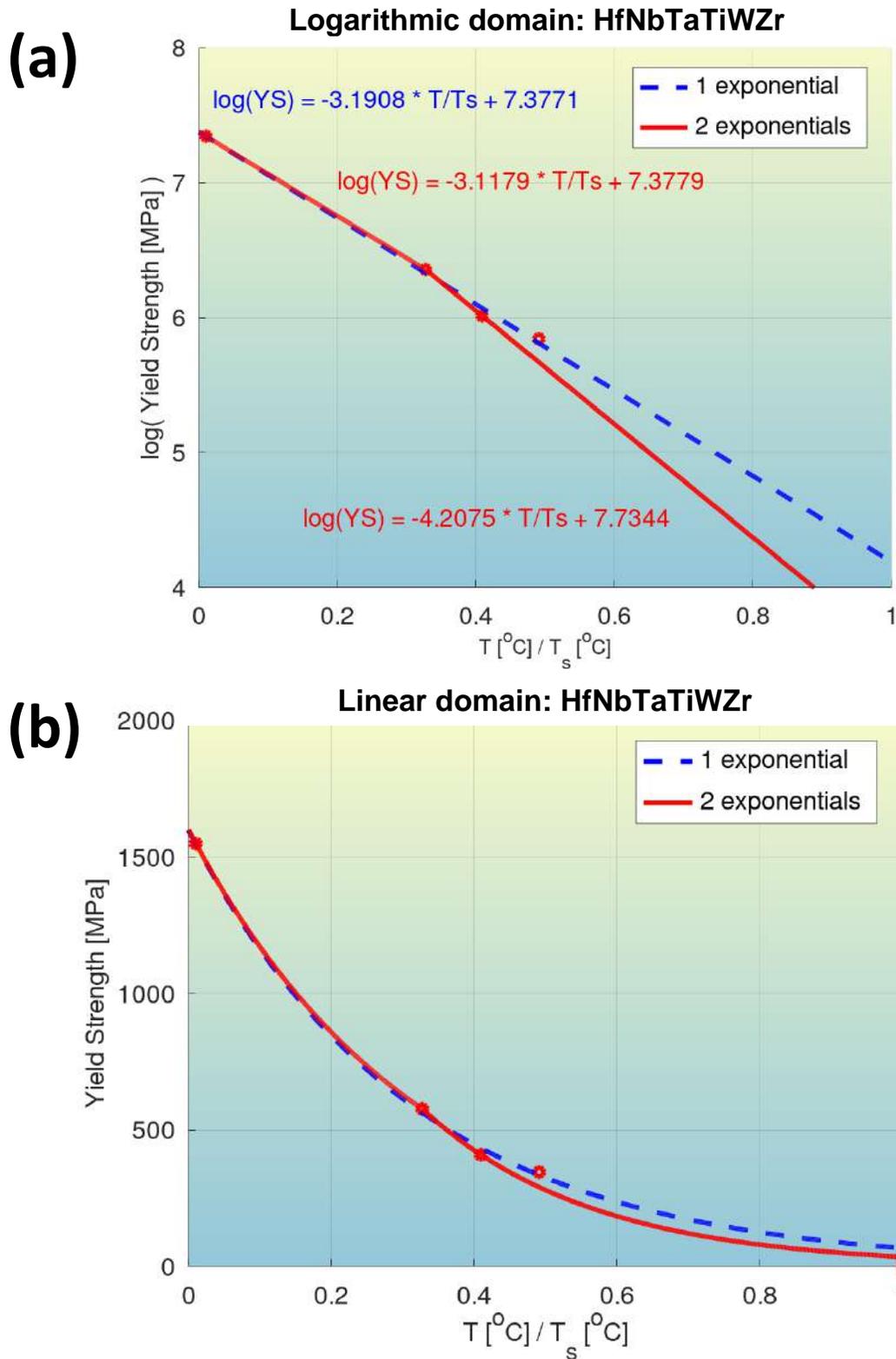

**Fig. S97**: Quantification of modeling accuracy of the bilinear log model, for composition No. 96 from **Tab. S2** (HfNbTaTiWZr, BCC+Laves phases), and comparison to that of a model with a single exponential.

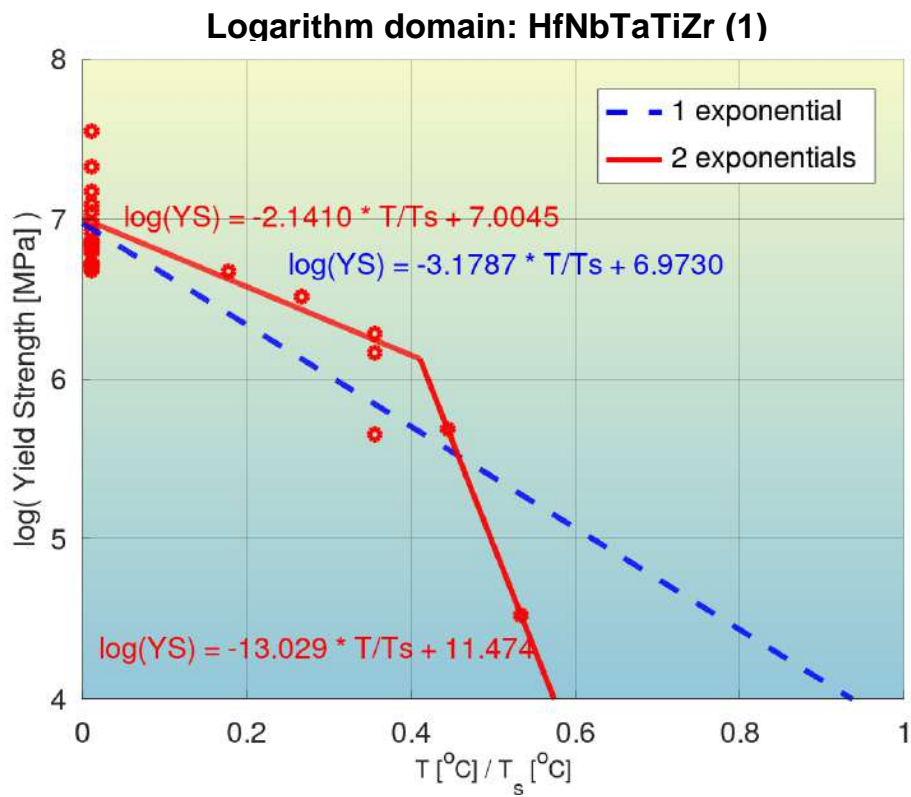

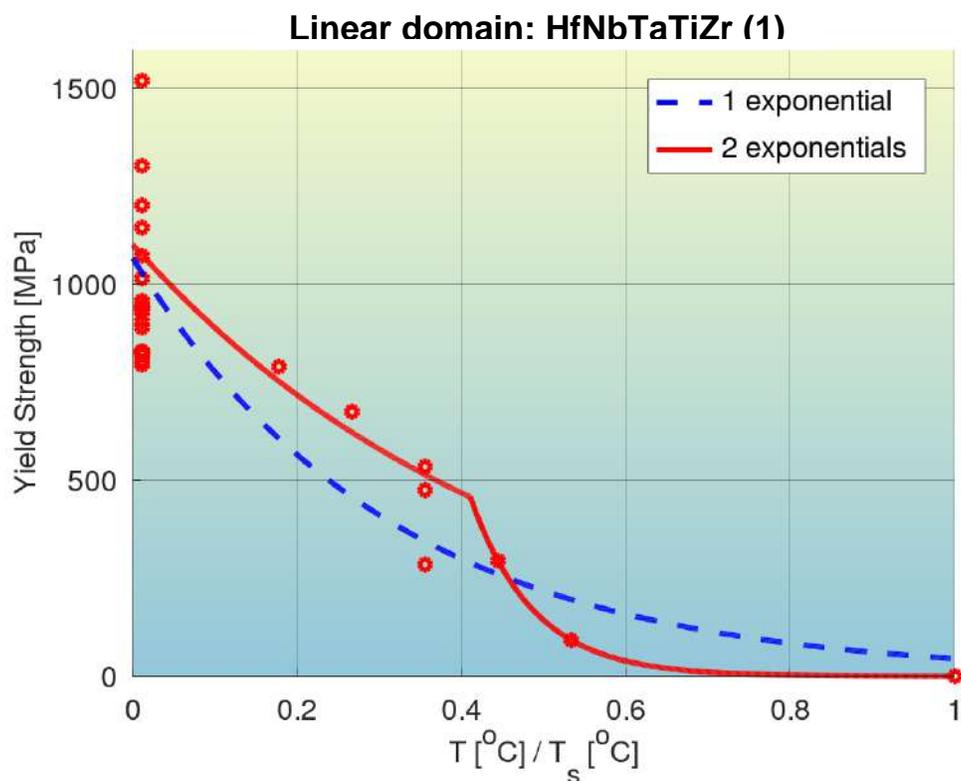

**Fig. S98**: Quantification of modeling accuracy of the bilinear log model, for composition No. 97 from **Tab. S3** (HfNbTaTiZr (1), BCC phase), and comparison to that of a model with a single exponential. The strength measurements at room temperature correspond to different heat treatment and post-processing applied

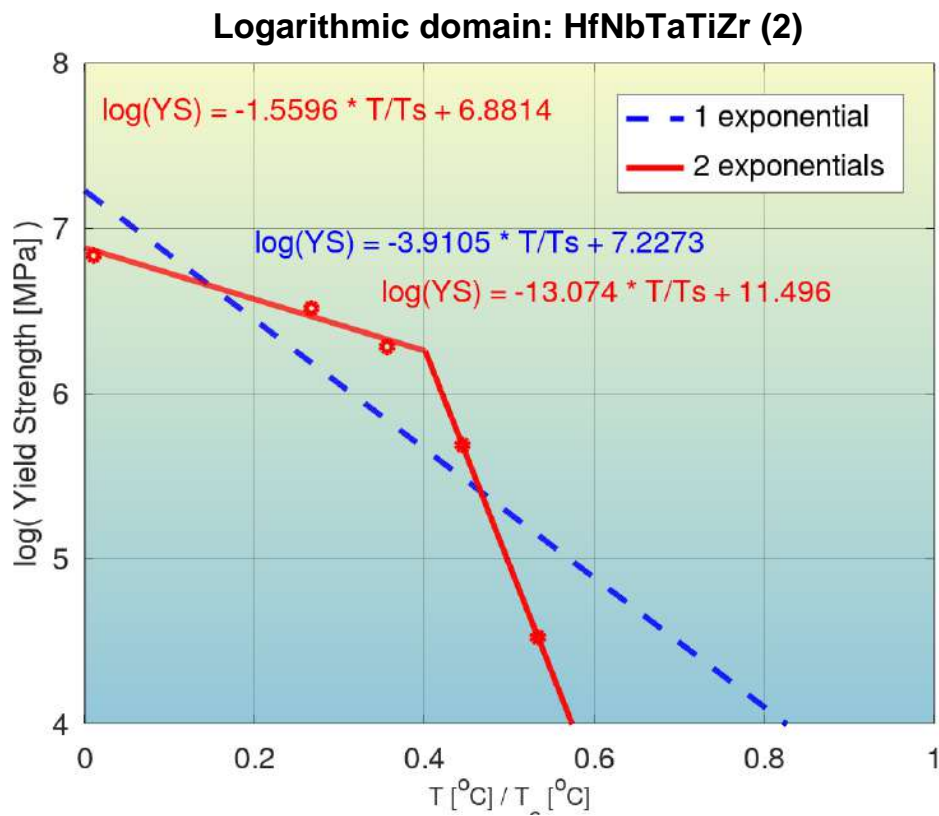

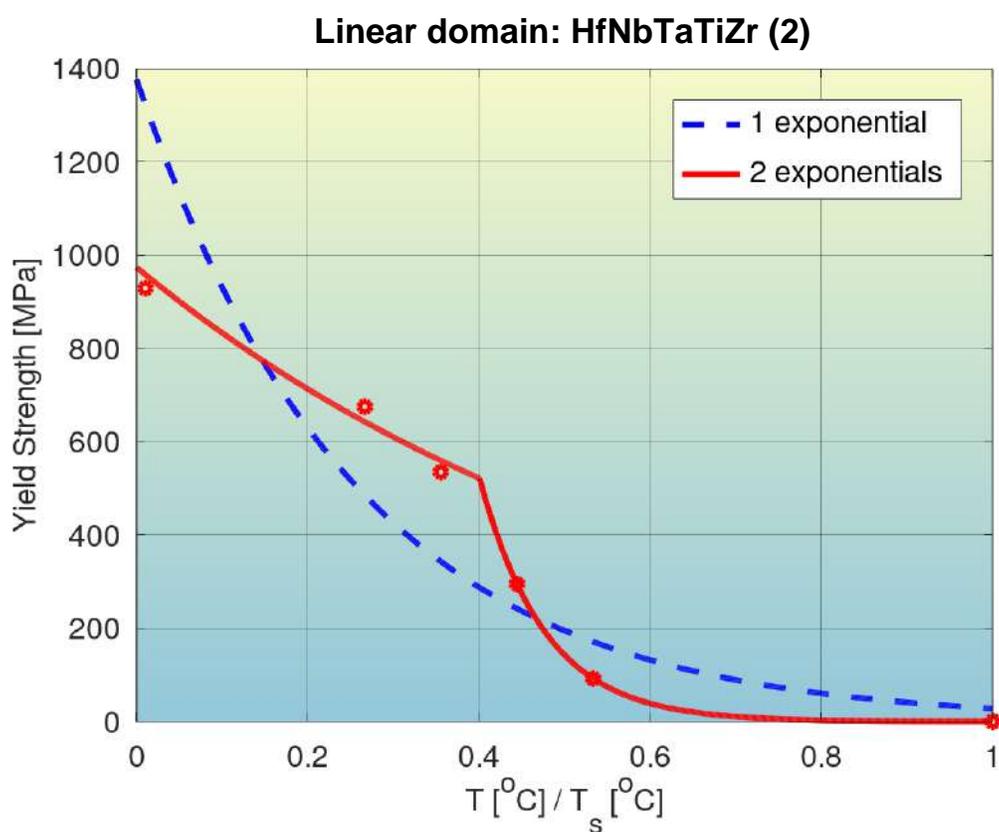

**Fig. S99**: Quantification of modeling accuracy of the bilinear log model, for composition No. 98 from **Tab. S3** (HfNbTaTiZr (2), BCC phase), and comparison to that of a model with a single exponential.

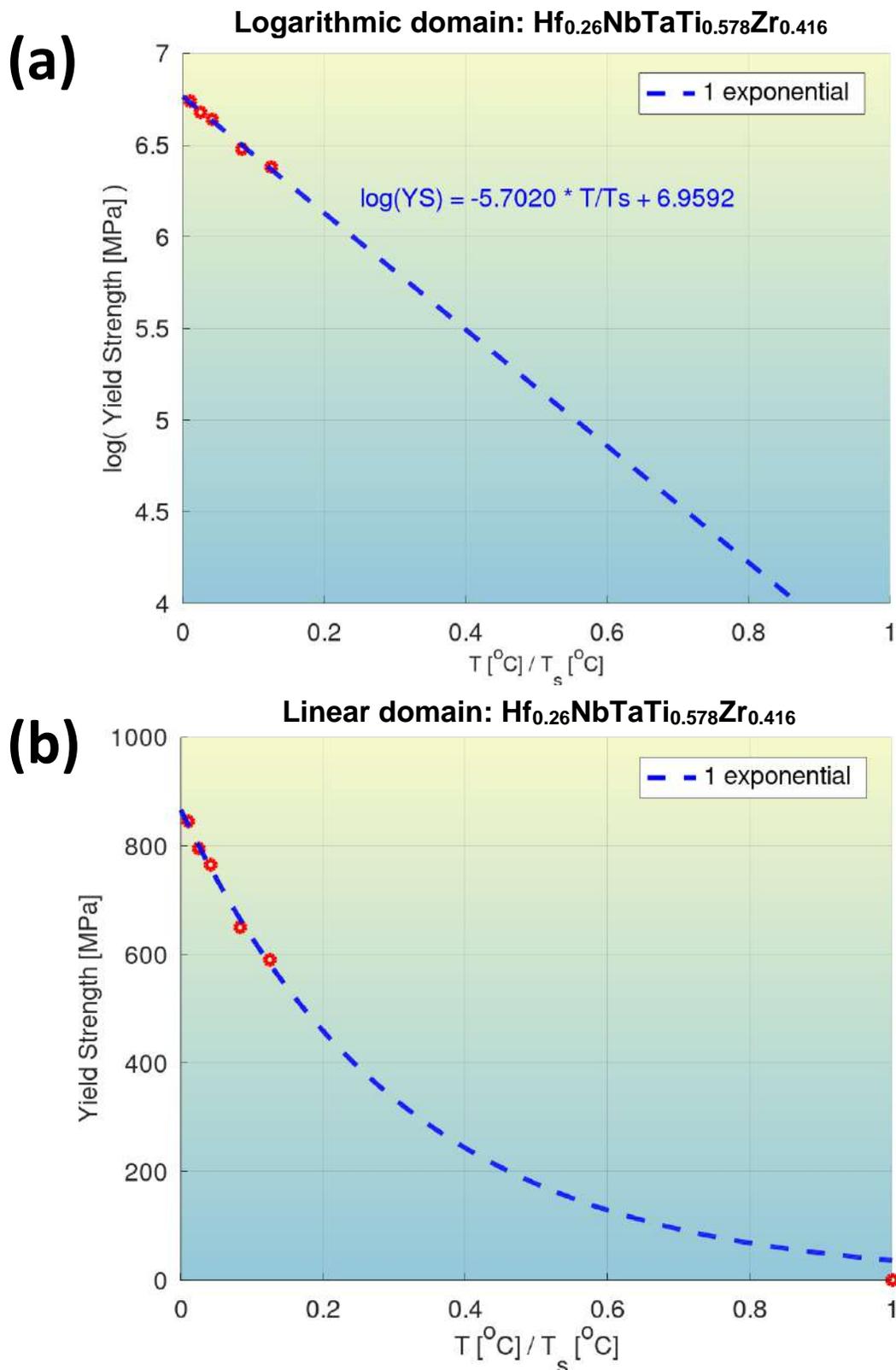

**Fig. S100**: Quantification of modeling accuracy of the bilinear log model, for composition No. 99 from **Tab. S3** (Hf$_{0.26}$NbTaTi$_{0.578}$Zr$_{0.416}$, BCC phase), and comparison to that of a model with a single exponential.

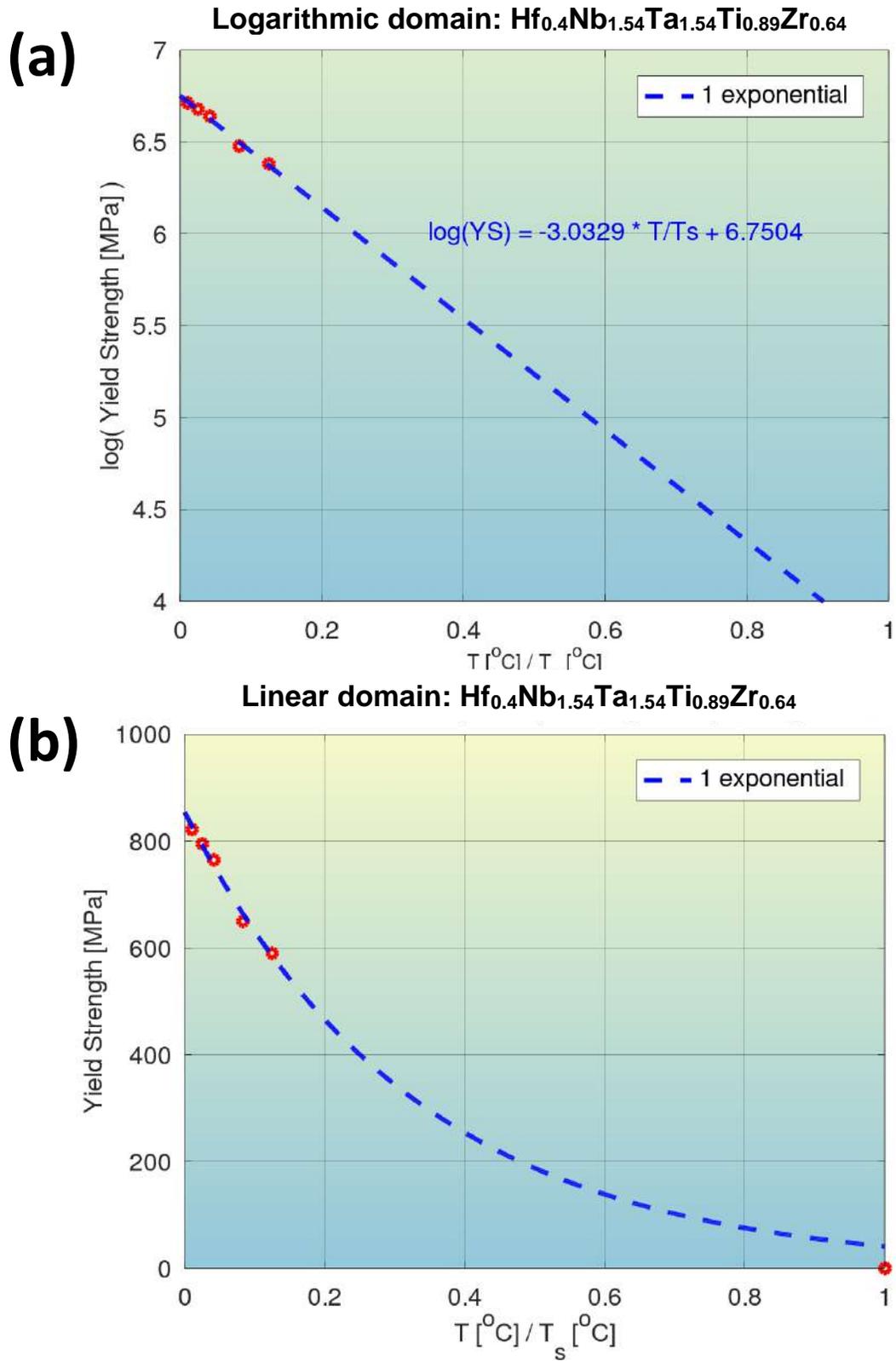

**Fig. S101**: Quantification of modeling accuracy of the bilinear log model, for composition No. 100 from **Tab. S3** (Hf$_{0.4}$Nb$_{1.54}$Ta$_{1.54}$Ti$_{0.89}$Zr$_{0.64}$, BCC phase), and comparison to that of a model with a single exponential.

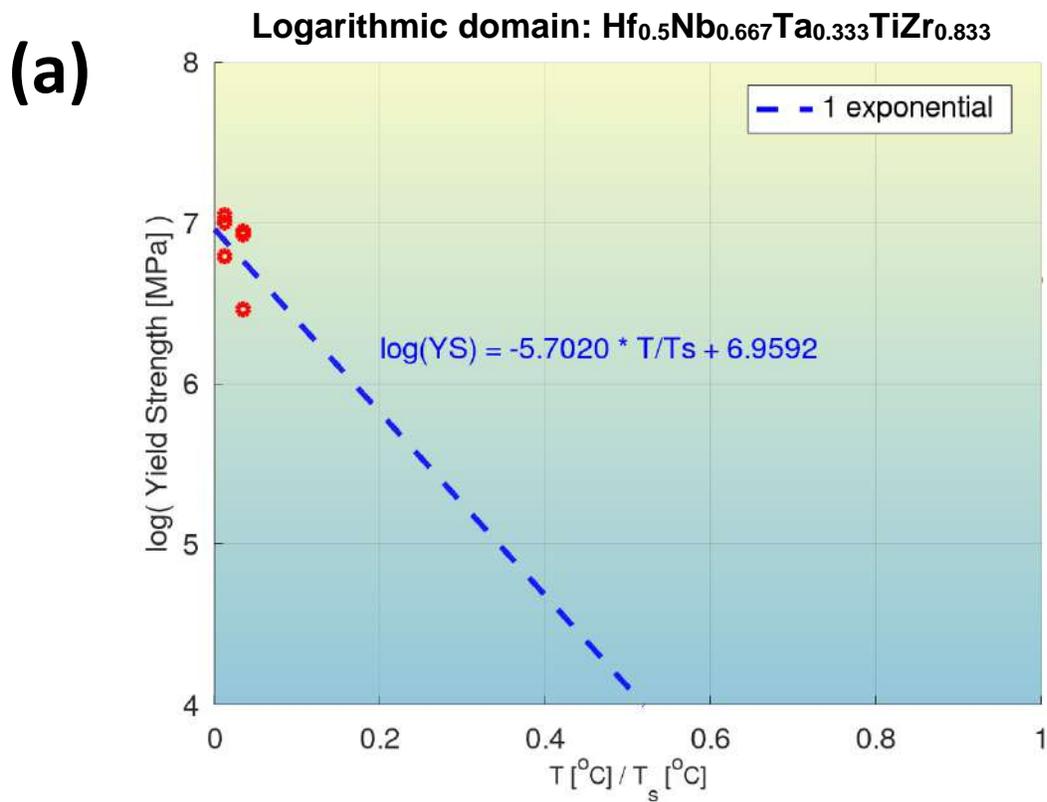

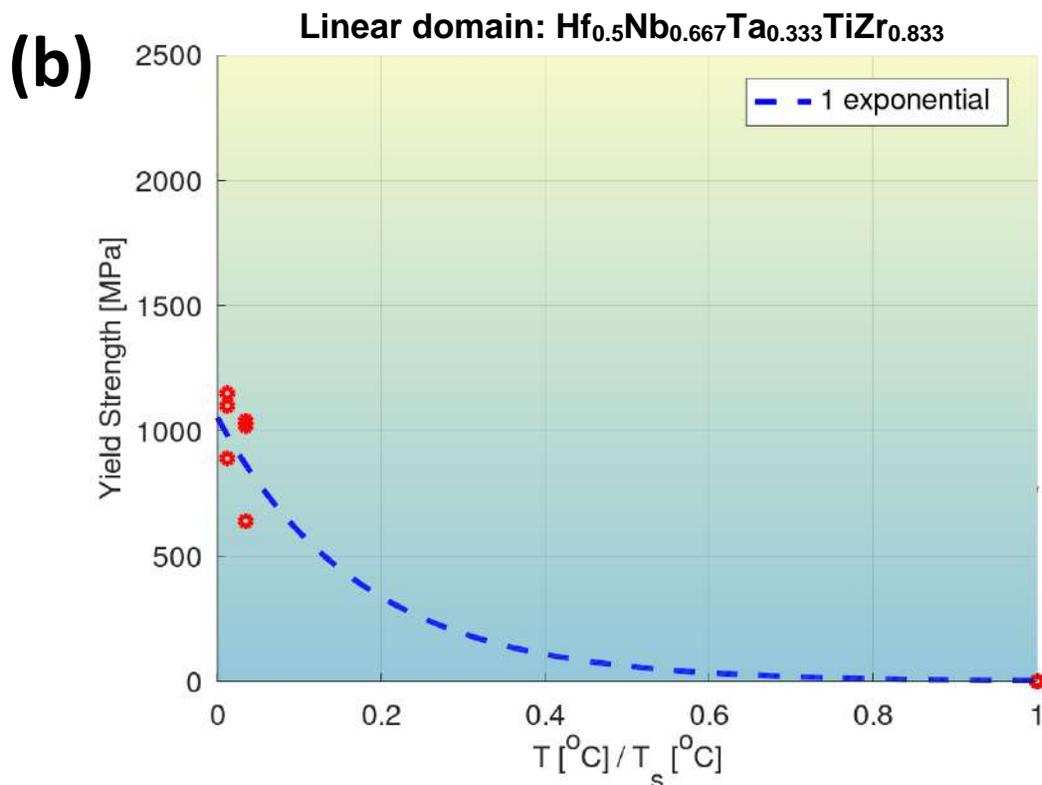

**Fig. S102**: Quantification of modeling accuracy of the bilinear log model, for composition No. 101 from **Tab. S3** ($Hf_{0.5}Nb_{0.667}Ta_{0.333}TiZr_{0.833}$, BCC phase), and comparison to that of a model with a single exponential.

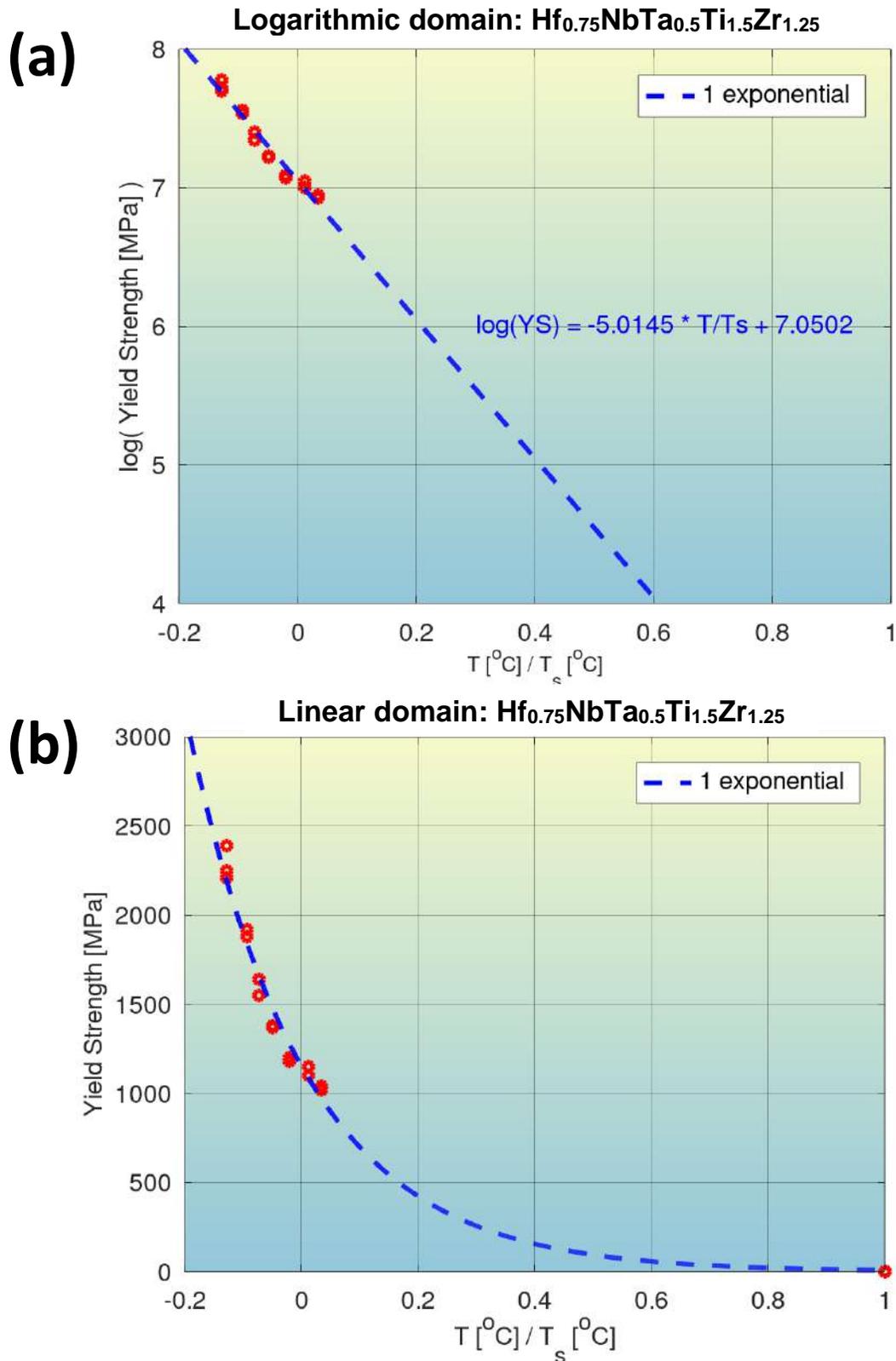

**Fig. S103**: Quantification of modeling accuracy of the bilinear log model, for composition No. 102 from **Tab. S3** (Hf$_{0.75}$NbTa$_{0.5}$Ti$_{1.5}$Zr$_{1.25}$, BCC phase), and comparison to that of a model with a single exponential.

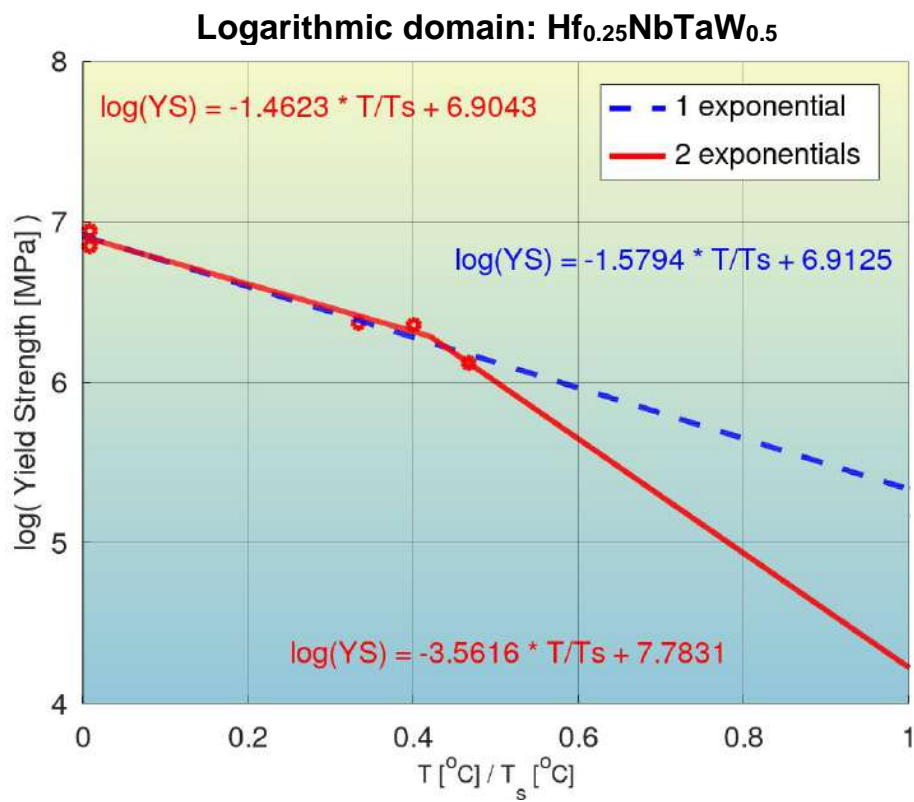

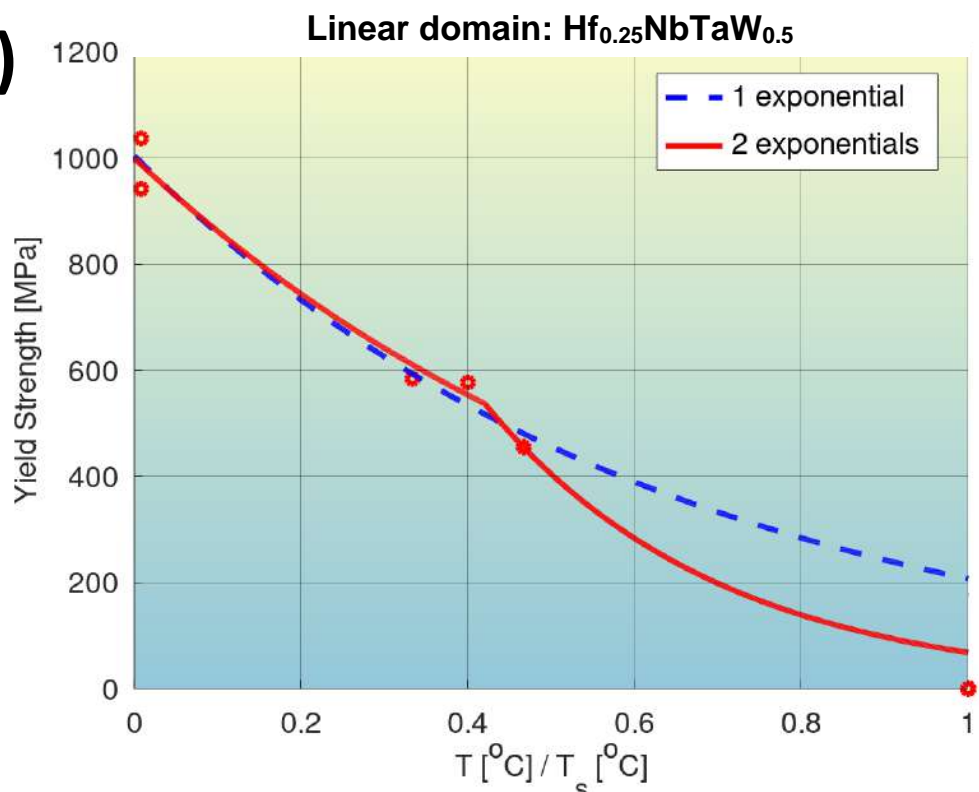

**Fig. S104**: Quantification of modeling accuracy of the bilinear log model, for composition No. 103 from **Tab. S3** ($Hf_{0.25}NbTaW_{0.5}$, BCC phase), and comparison to that of a model with a single exponential.

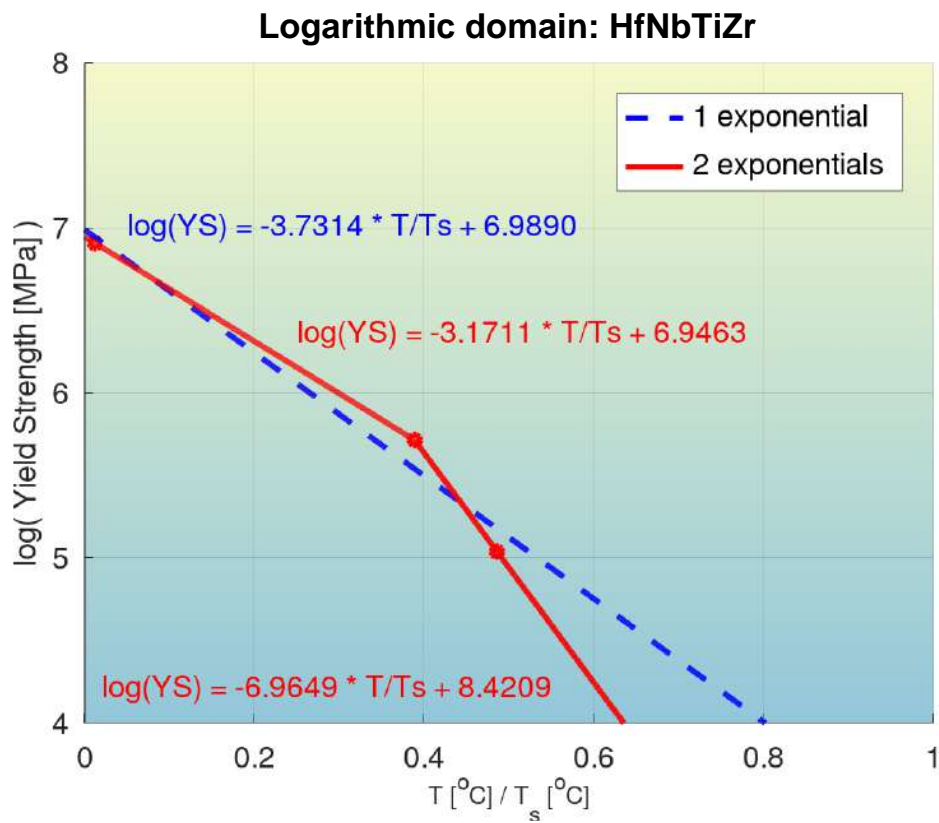

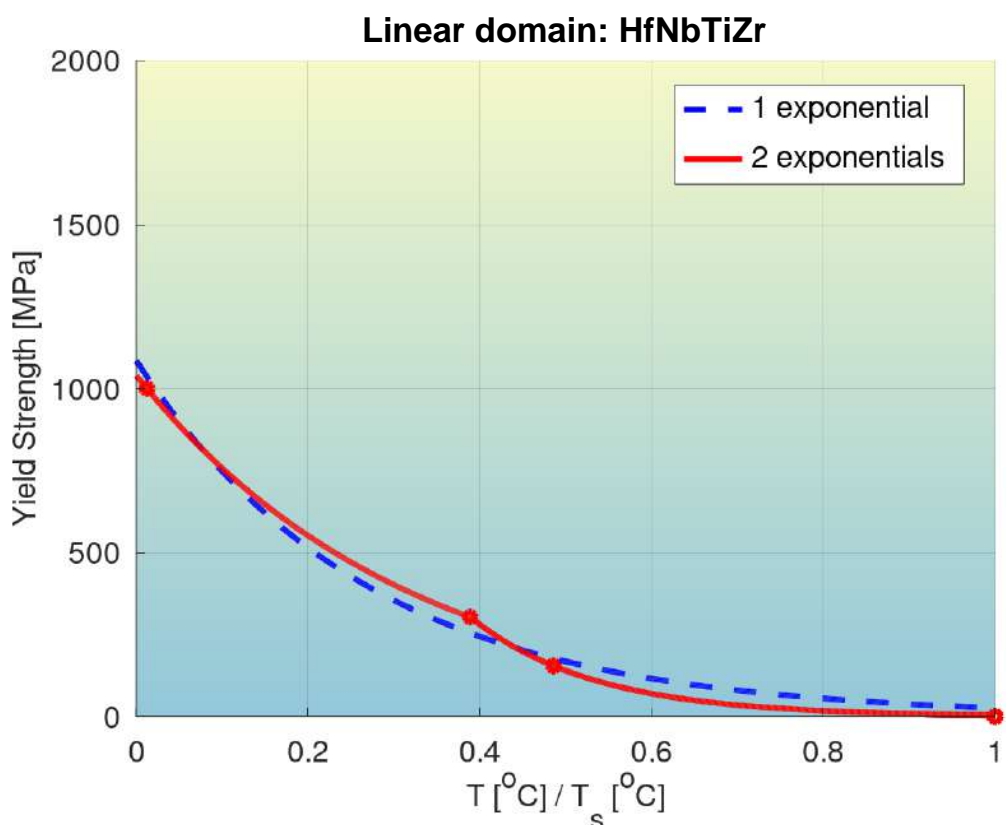

**Fig. S105**: Quantification of modeling accuracy of the bilinear log model, for composition No. 104 from **Tab. S3** (HfNbTiZr, BCC phase), and comparison to that of a model with a single exponential.

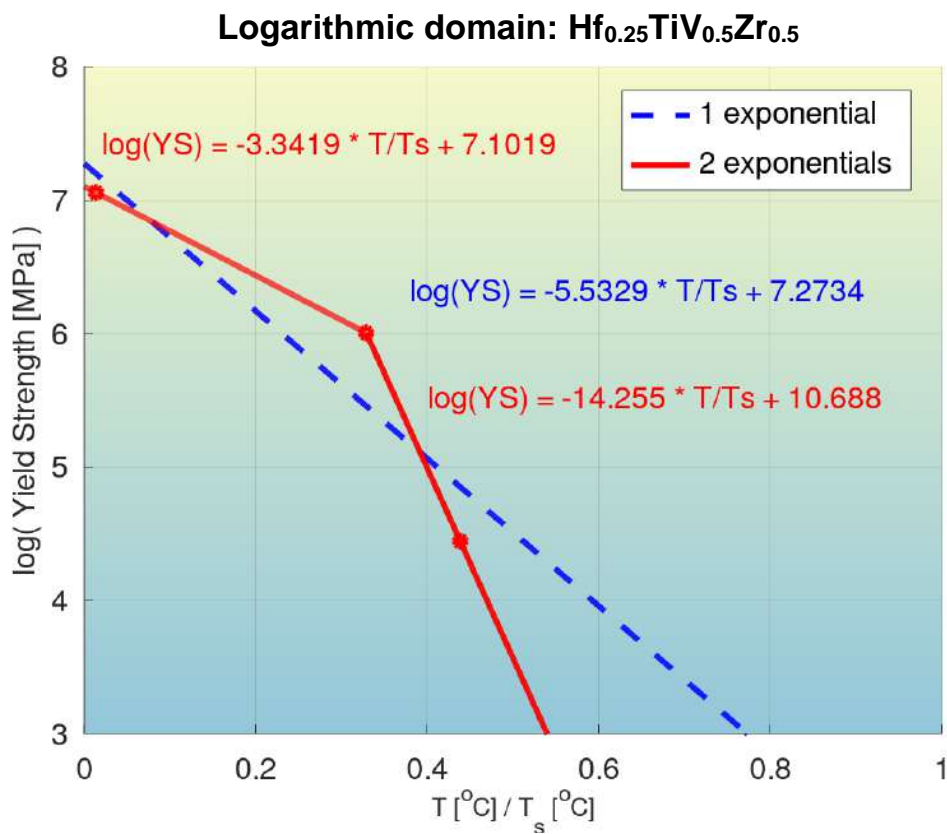

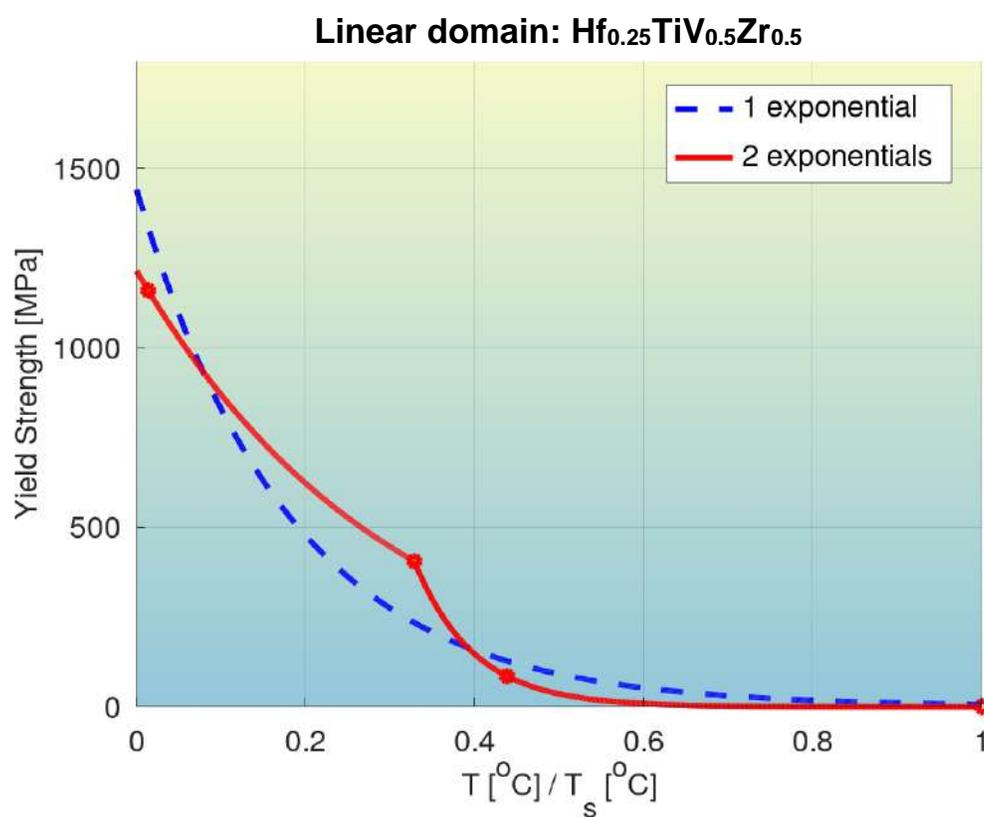

**Fig. S106**: Quantification of modeling accuracy of the bilinear log model, for composition No. 105 from **Tab. S3** ($Hf_{0.25}TiV_{0.5}Zr_{0.5}$, BCC phase), and comparison to that of a model with a single exponential.

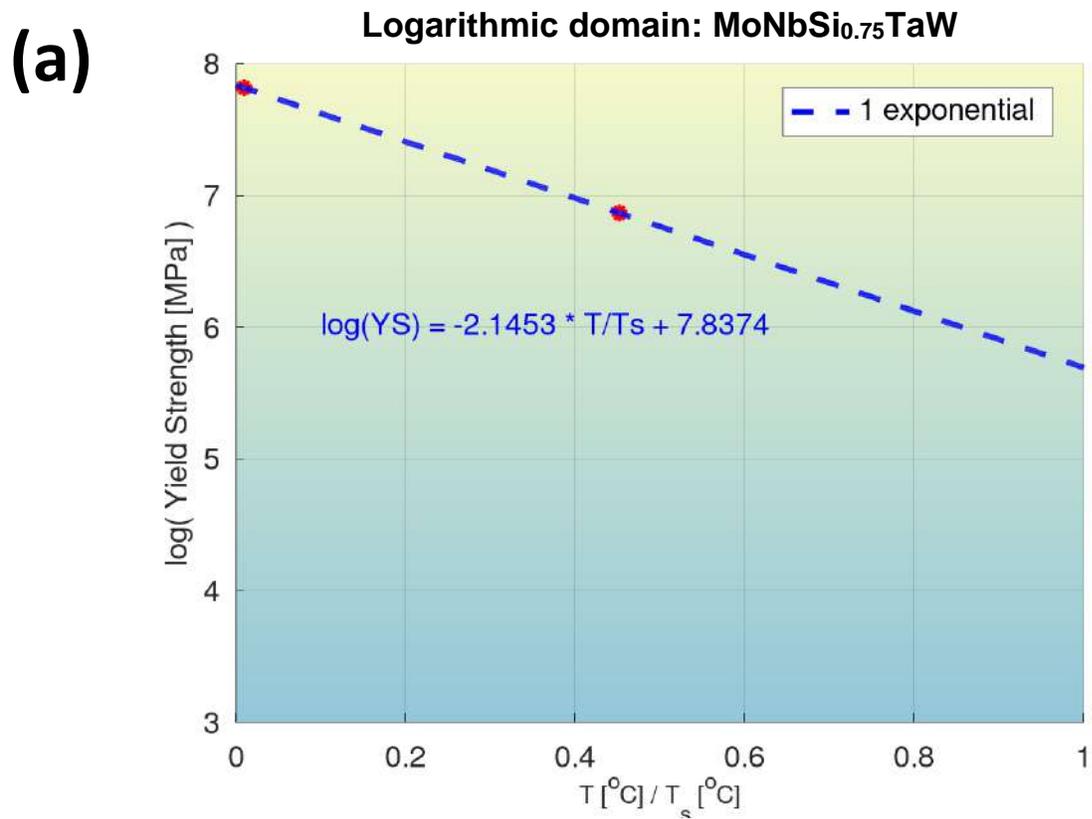

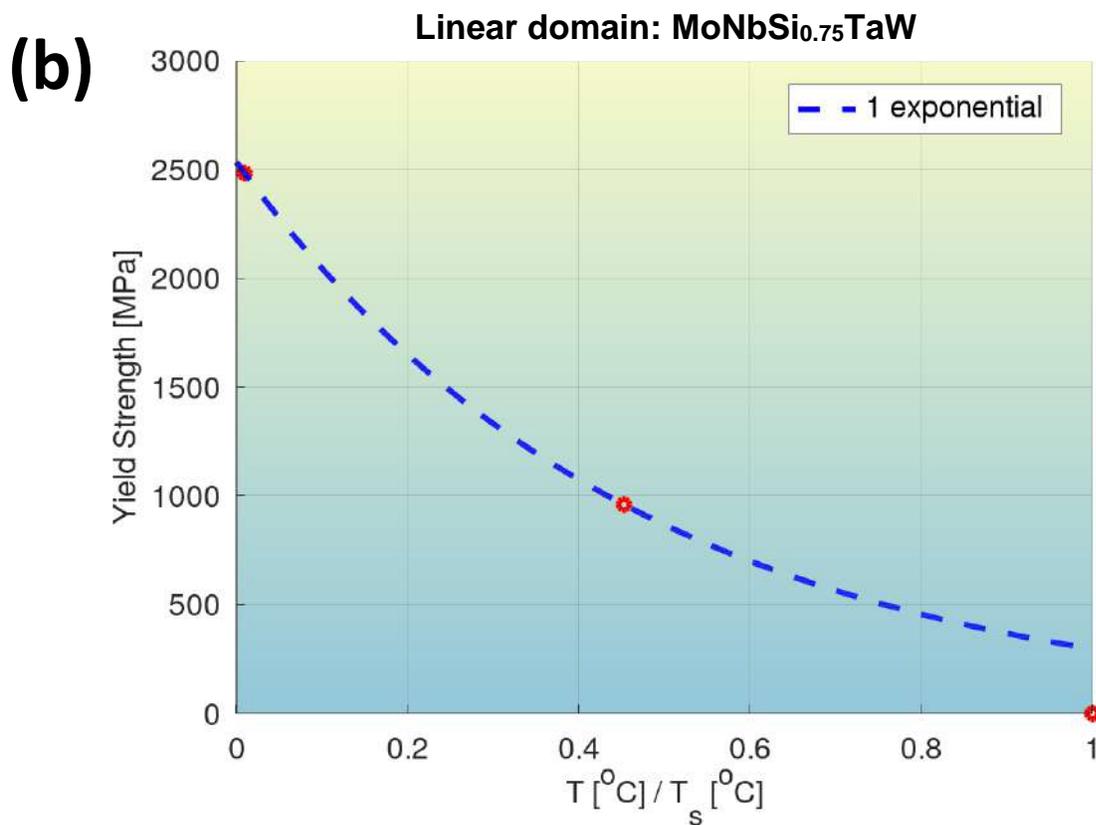

**Fig. S107**: Quantification of modeling accuracy of the bilinear log model, for composition No. 106 from **Tab. S3** (MoNbSi$_{0.75}$TaW, BCC+silicide phases), and comparison to that of a model with a single exponential.

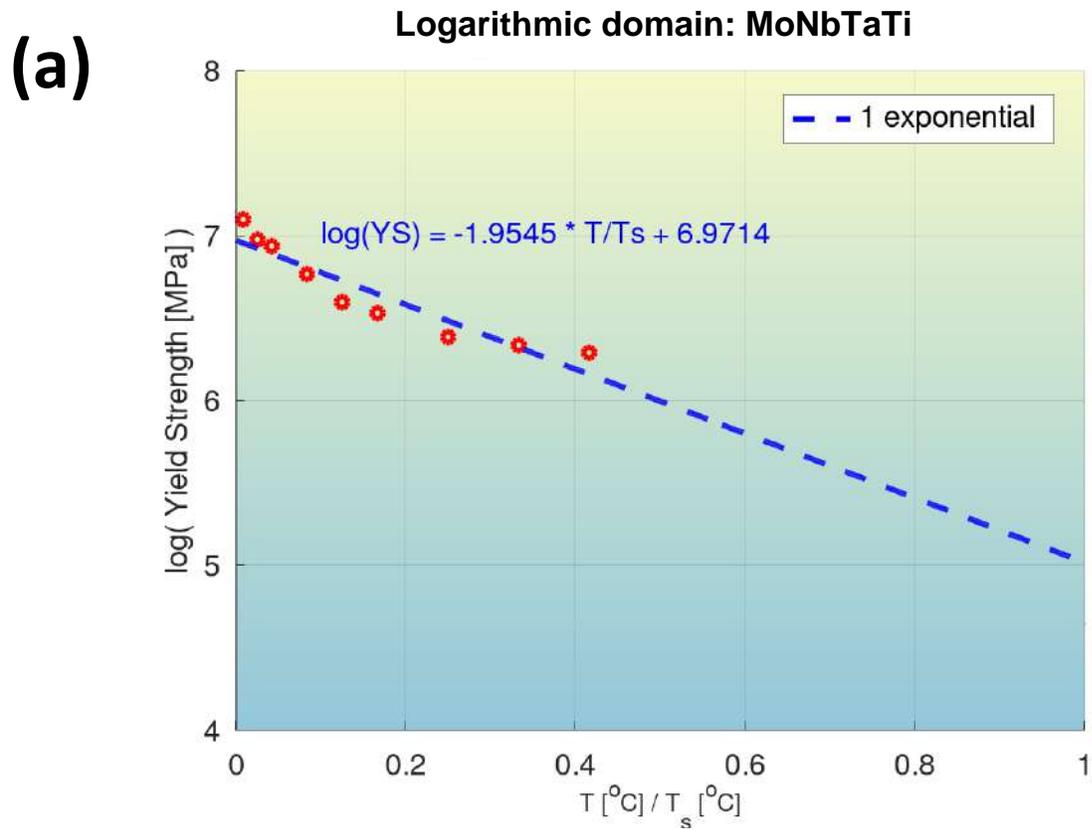
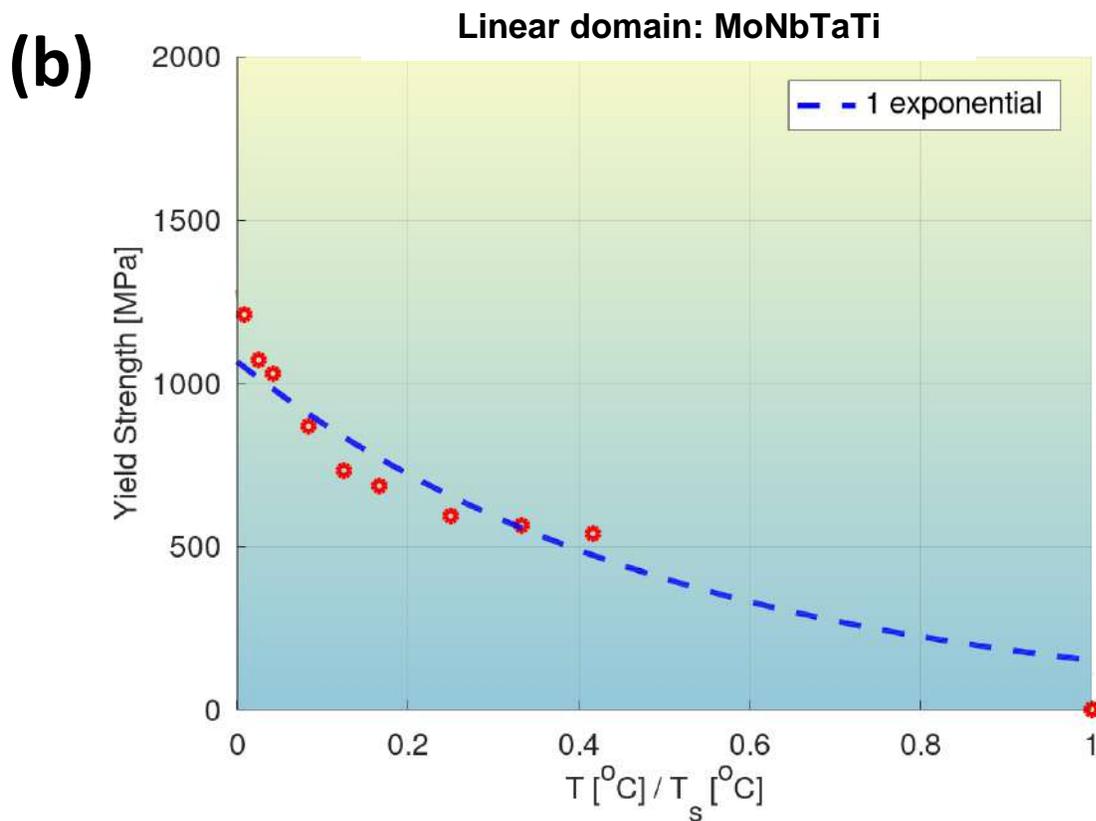

**Fig. S108**: Quantification of modeling accuracy of the bilinear log model, for composition No. 107 from **Tab. S3** (MoNbTaTi, BCC phase), and comparison to that of a model with a single exponential.

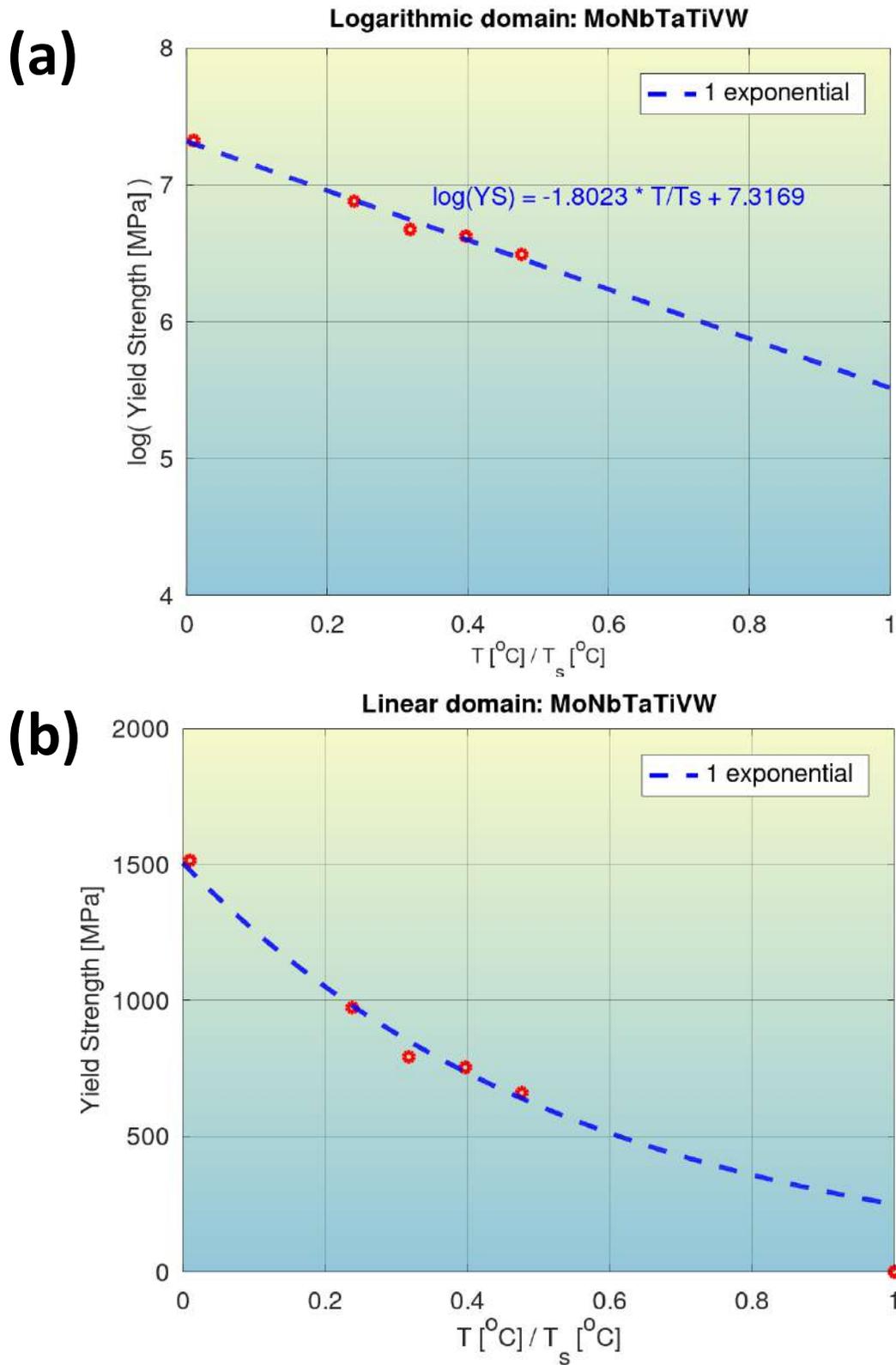

**Fig. S109**: Quantification of modeling accuracy of the bilinear log model, for composition No. 108 from **Tab. S3** (MoNbTaTiVW, BCC phase), and comparison to that of a model with a single exponential.

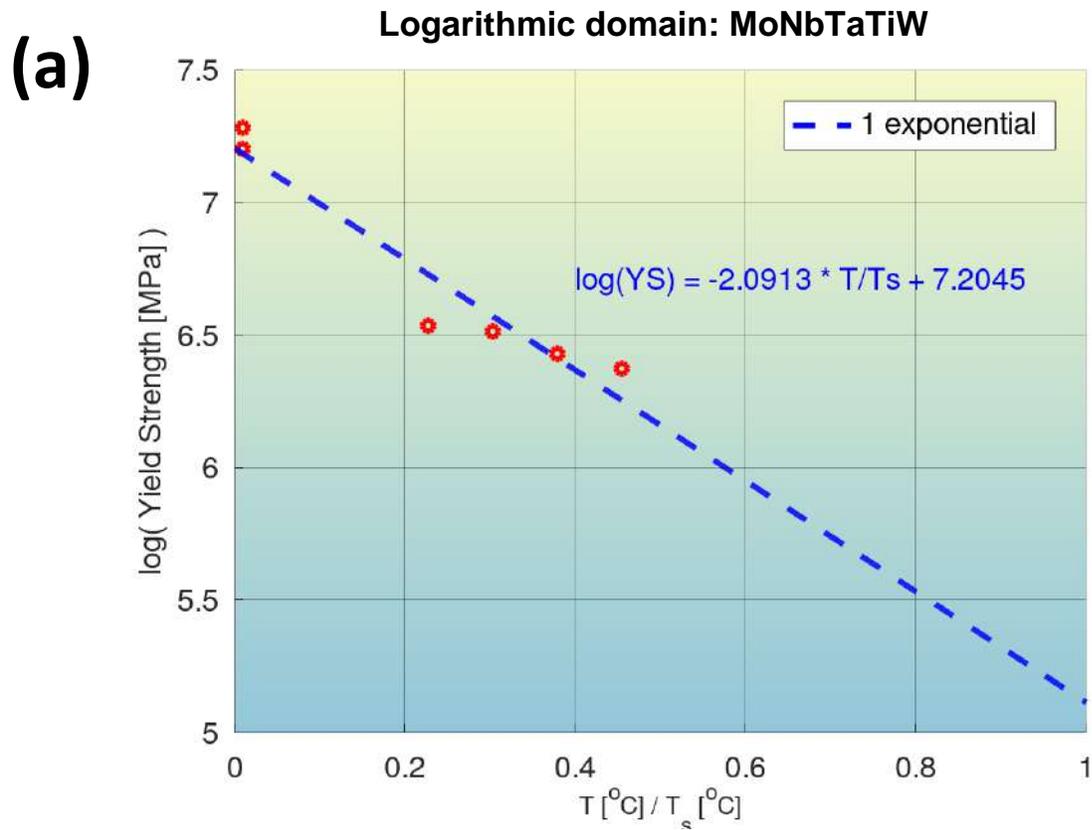
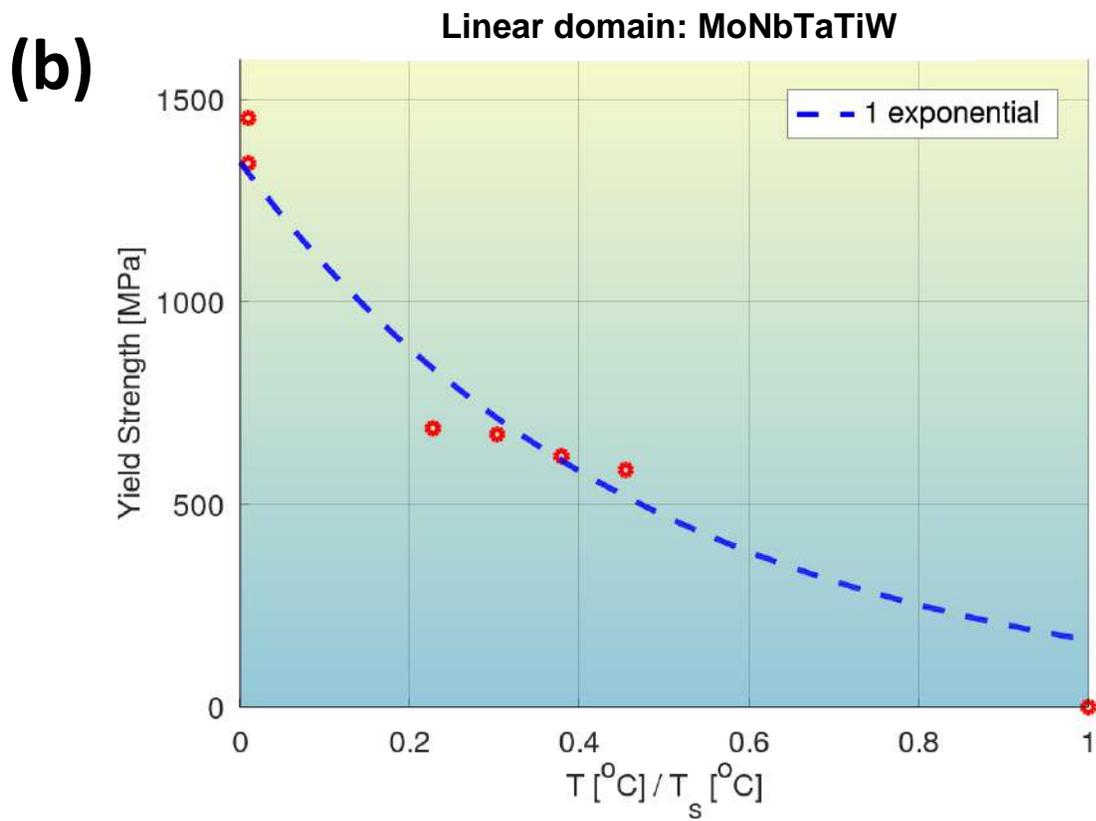

**Fig. S110**: Quantification of modeling accuracy of the bilinear log model, for composition No. 109 from **Tab. S3** (MoNbTaTiW, BCC phase), and comparison to that of a model with a single exponential.

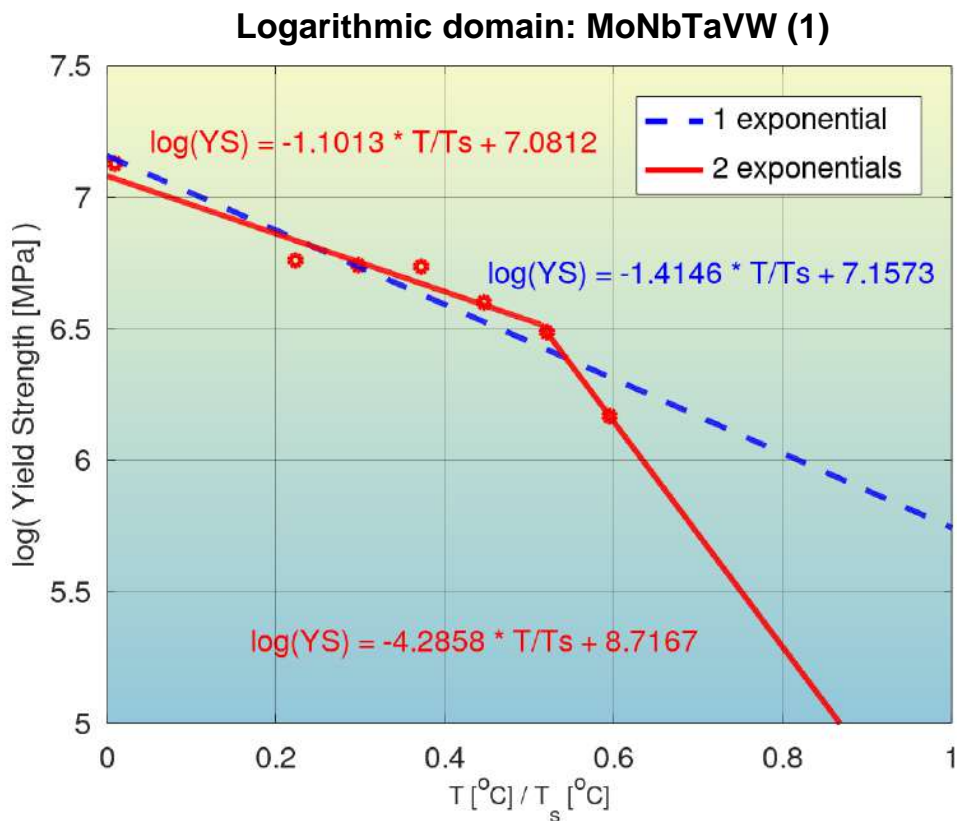

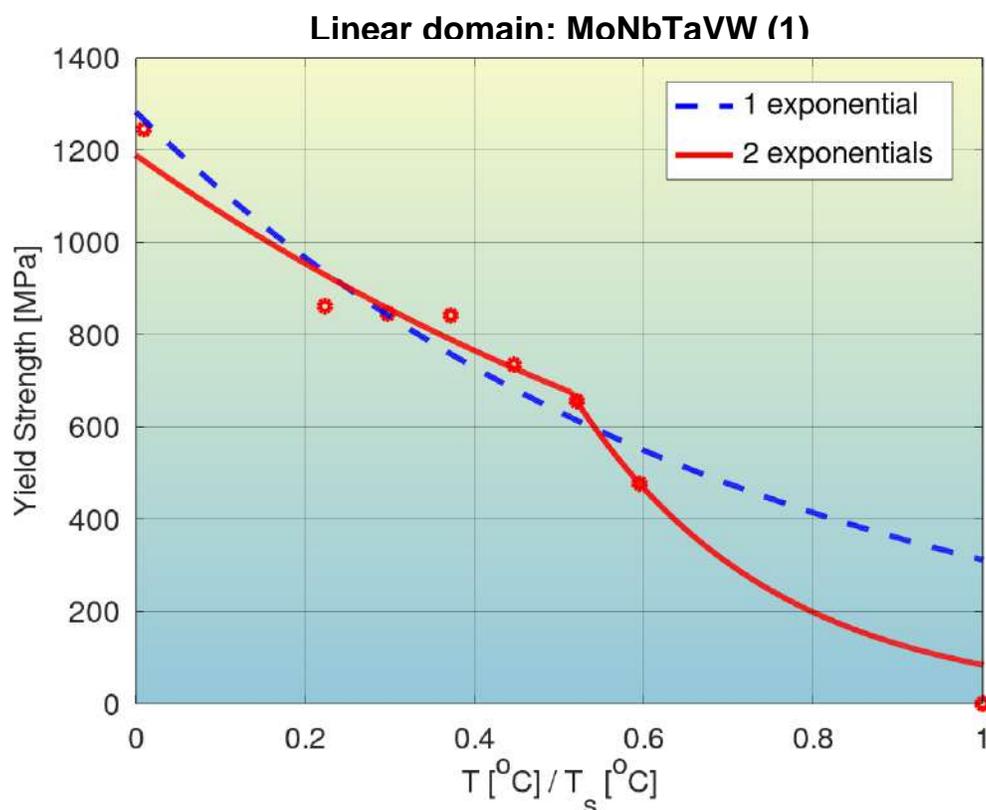

**Fig. S111**: Quantification of modeling accuracy of the bilinear log model, for composition No. 110 from **Tab. S3** (MoNbTaVW (1), BCC phase), and comparison to that of a model with a single exponential.

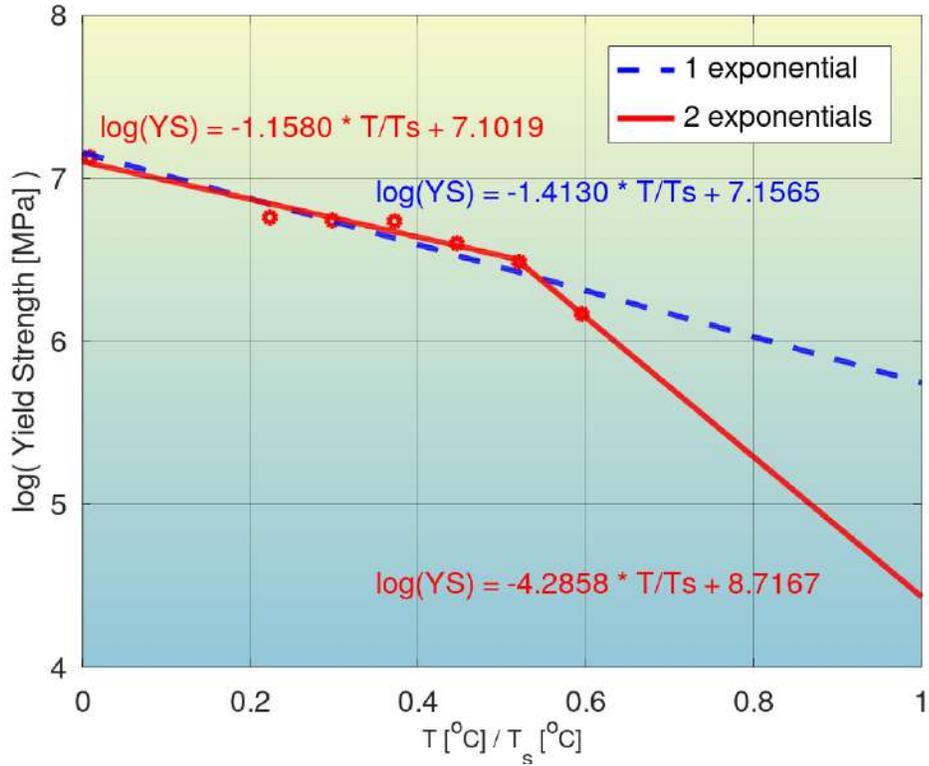
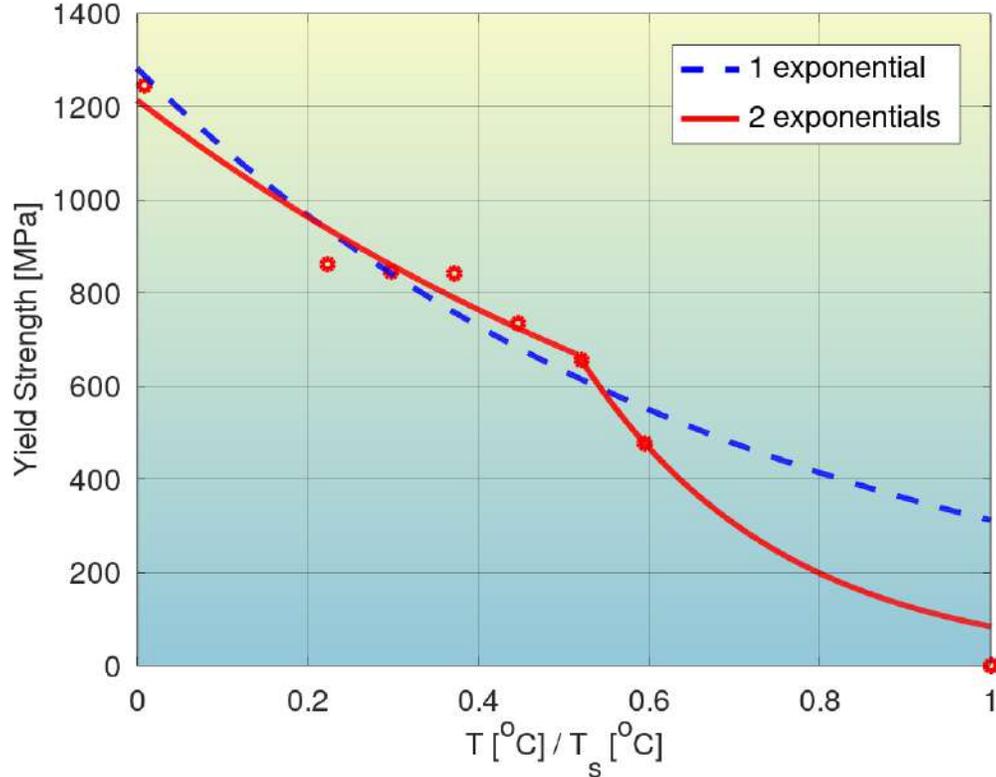

**Fig. S112**: Quantification of modeling accuracy of the bilinear log model, for composition No. 111 from **Tab. S3** (MoNbTaVW (2), BCC phase), and comparison to that of a model with a single exponential.

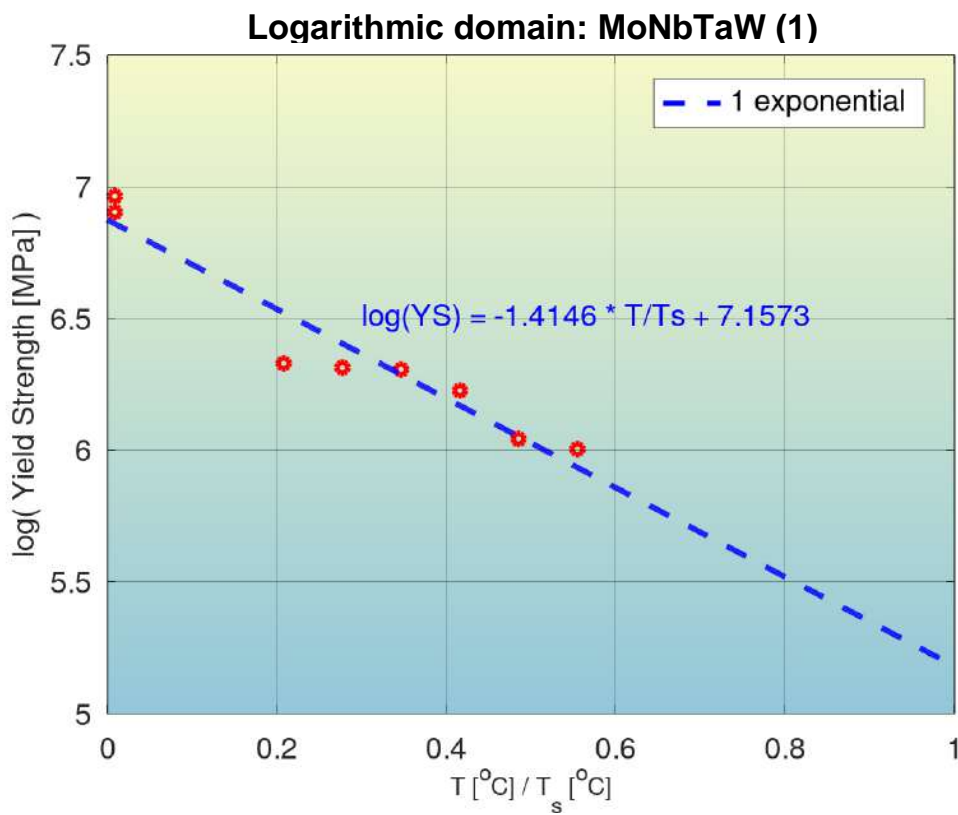

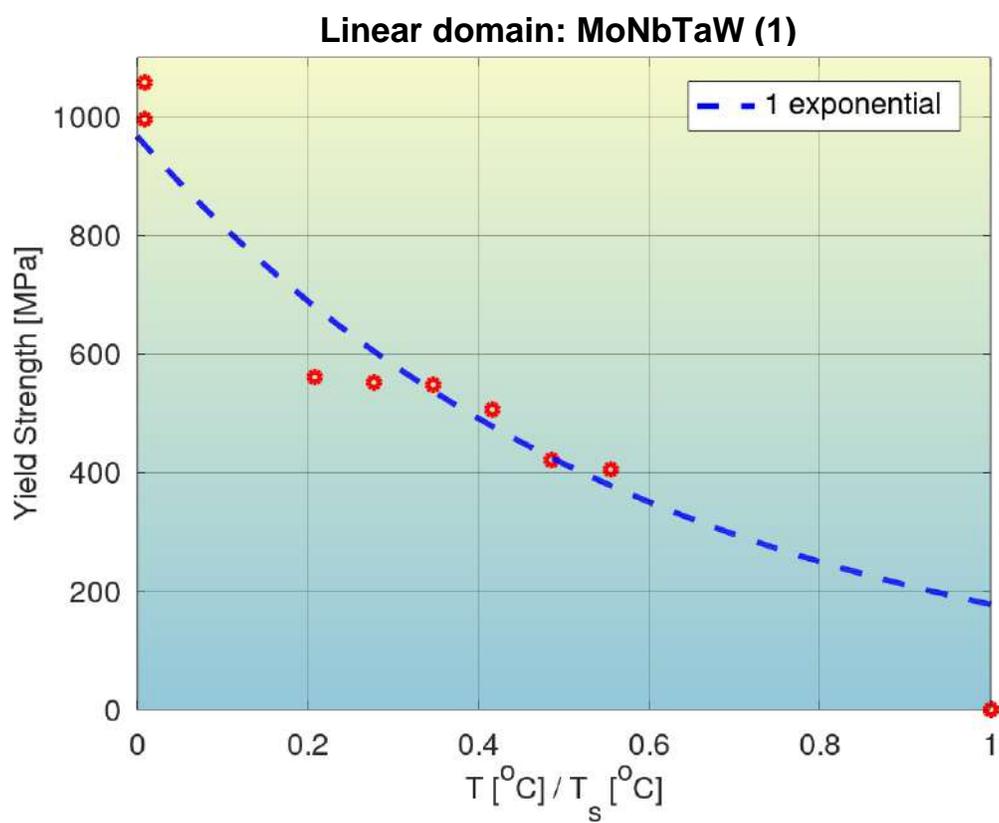

**Fig. S113**: Quantification of modeling accuracy of the bilinear log model, for composition No. 112 from **Tab. S3** (MoNbTaW (1), BCC phase), and comparison to that of a model with a single exponential.

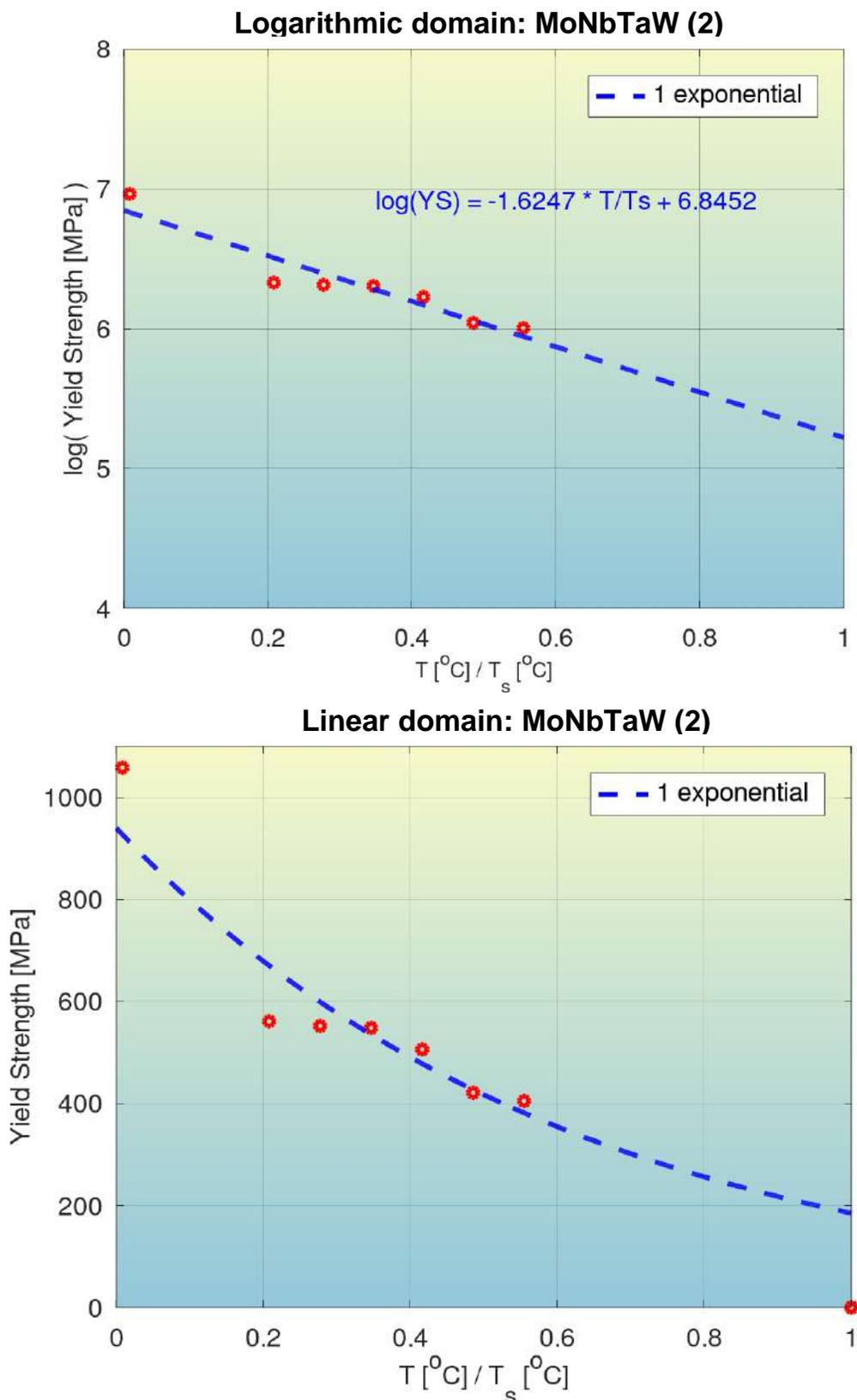

**Fig. S114**: Quantification of modeling accuracy of the bilinear log model, for composition No. 113 from **Tab. S3** (MoNbTaW (2), BCC phase), and comparison to that of a model with a single exponential.

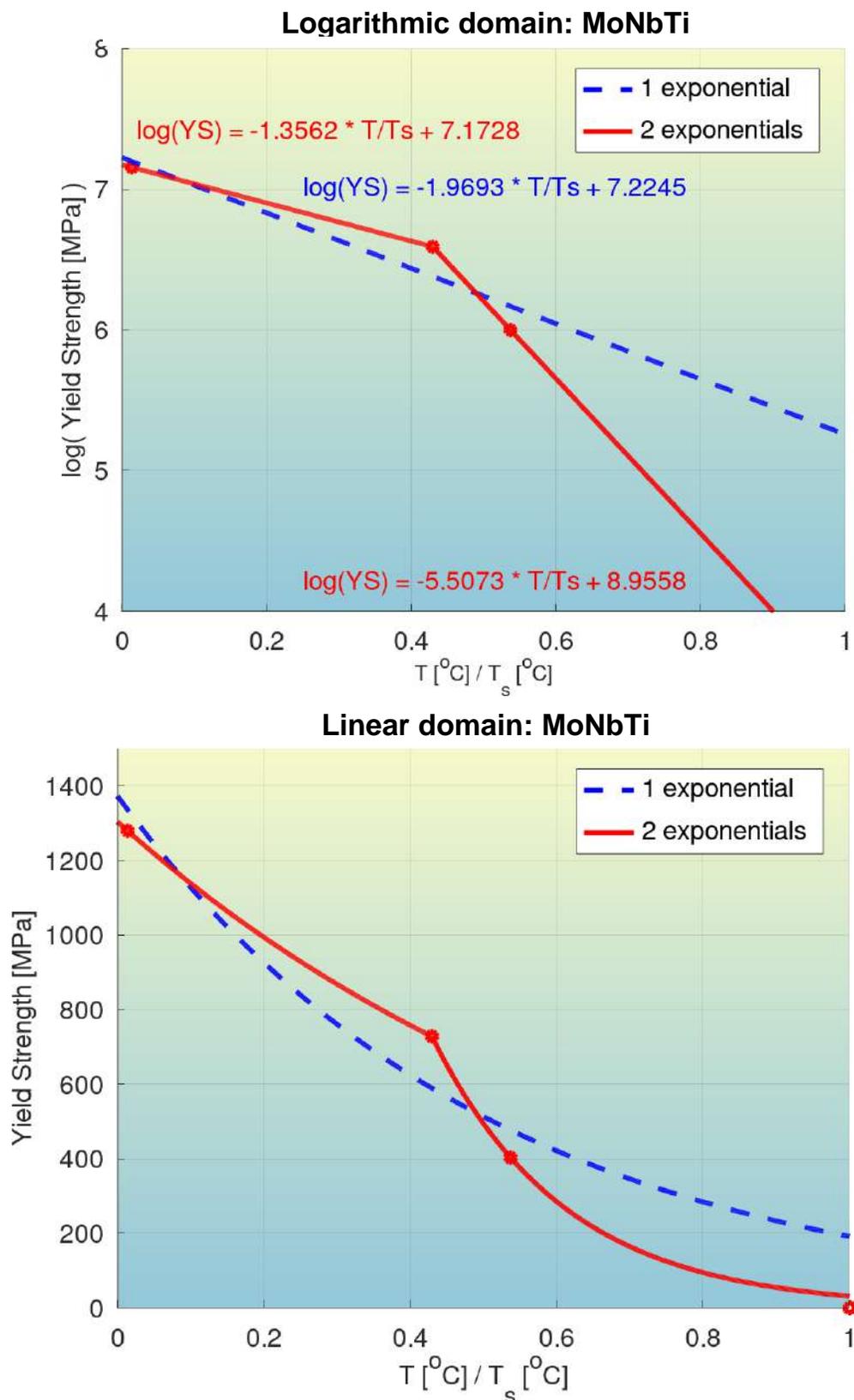

**Fig. S115**: Quantification of modeling accuracy of the bilinear log model, for composition No. 114 from **Tab. S3** (MoNbTi, BCC phase), and comparison to that of a model with a single exponential.

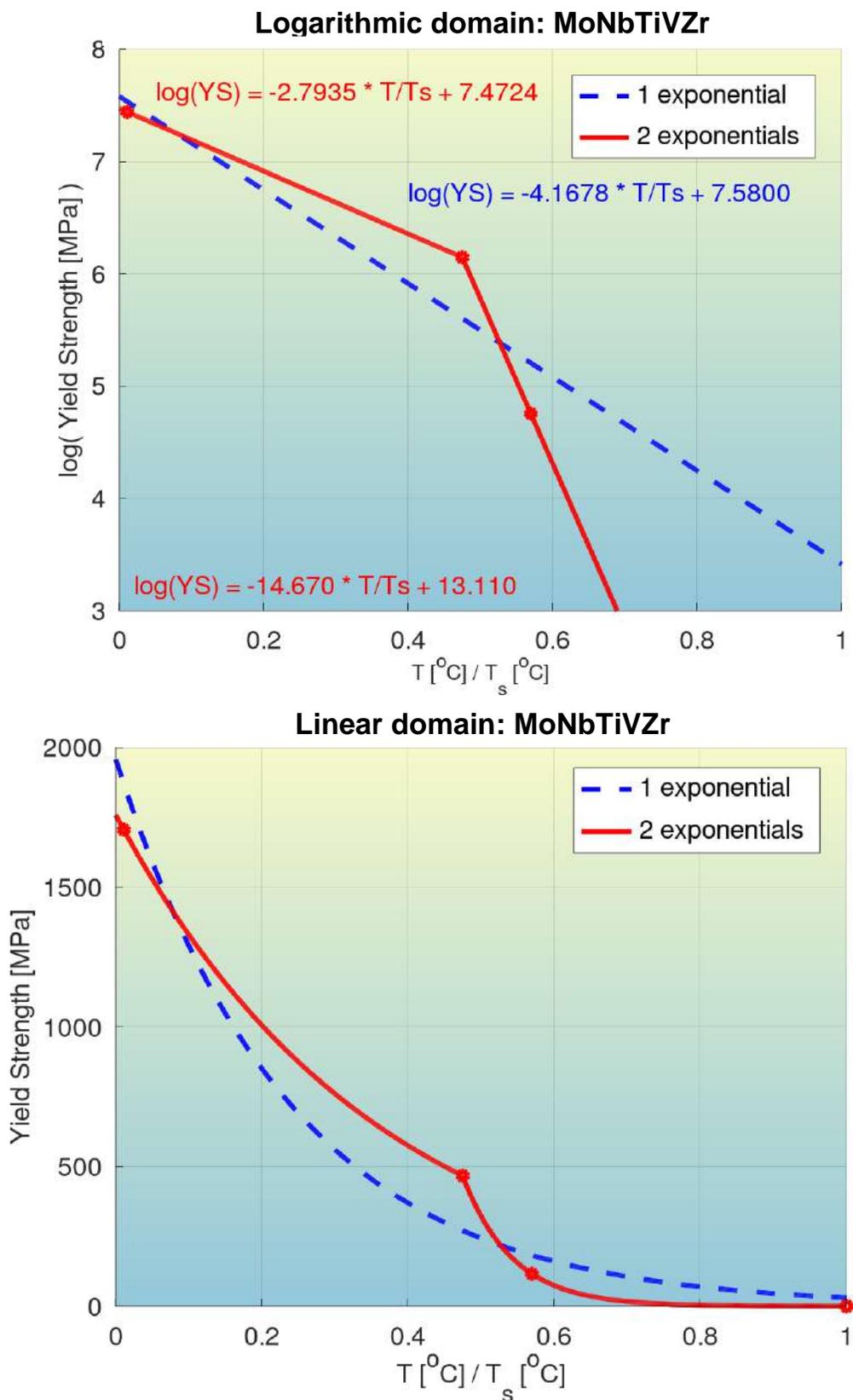

**Fig. S116**: Quantification of modeling accuracy of the bilinear log model, for composition No. 115 from **Tab. S3** (MoNbTiVZr, BCC+Laves phases), and comparison to that of a model with a single exponential.

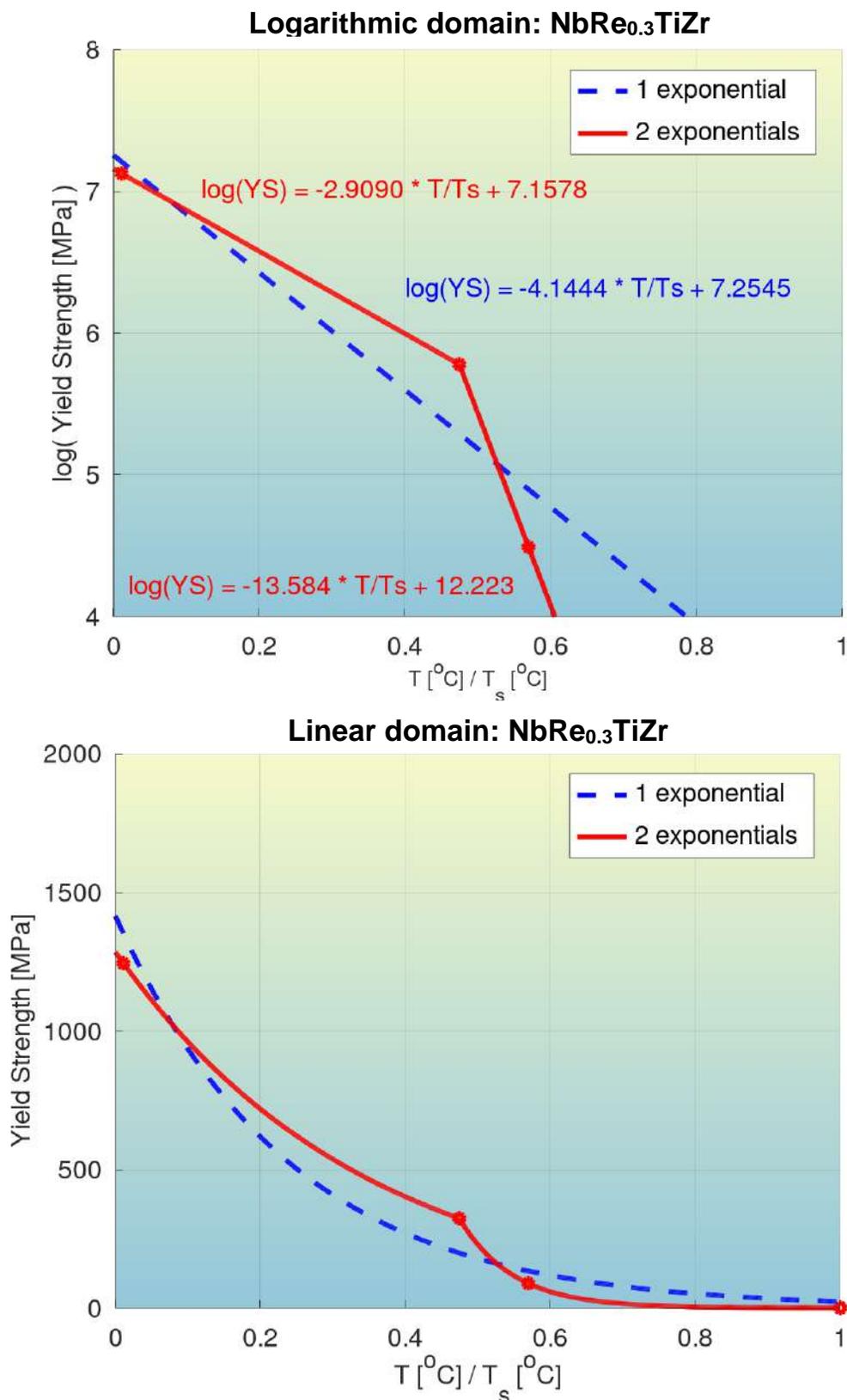

**Fig. S117**: Quantification of modeling accuracy of the bilinear log model, for composition No. 116 from **Tab. S3** (NbRe$_{0.3}$TiZr, BCC+Laves phases), and comparison to that of a model with a single exponential.

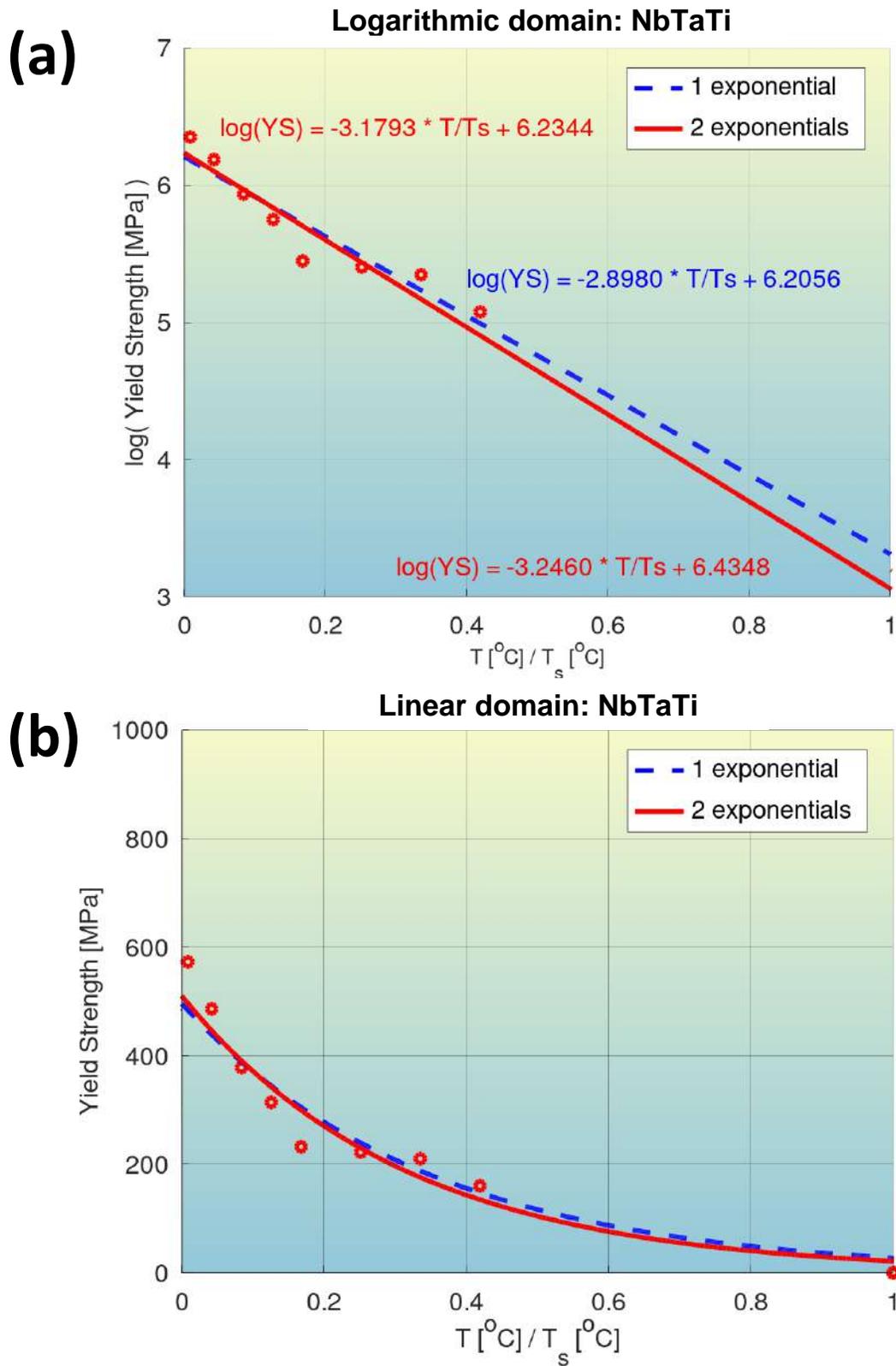

**Fig. S118**: Quantification of modeling accuracy of the bilinear log model, for composition No. 117 from **Tab. S3** (NbTaTi, BCC phase), and comparison to that of a model with a single exponential.

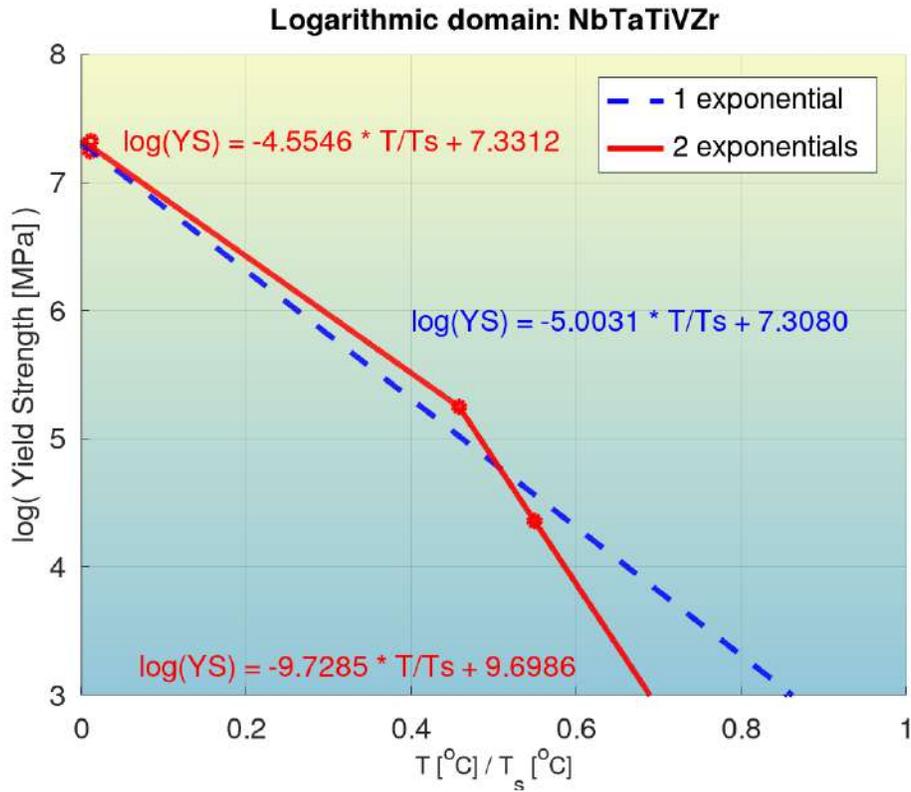

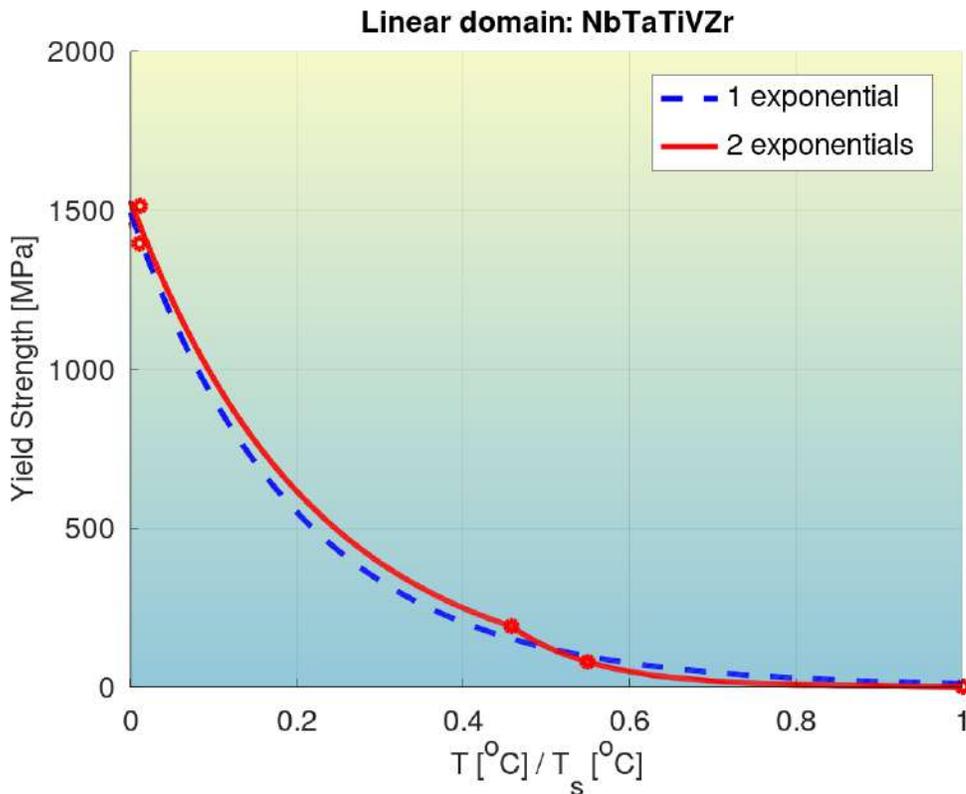

**Fig. S119**: Quantification of modeling accuracy of the bilinear log model, for composition No. 118 from **Tab. S3** (NbTaTiVZr, 2 BCC phases), and comparison to that of a model with a single exponential.

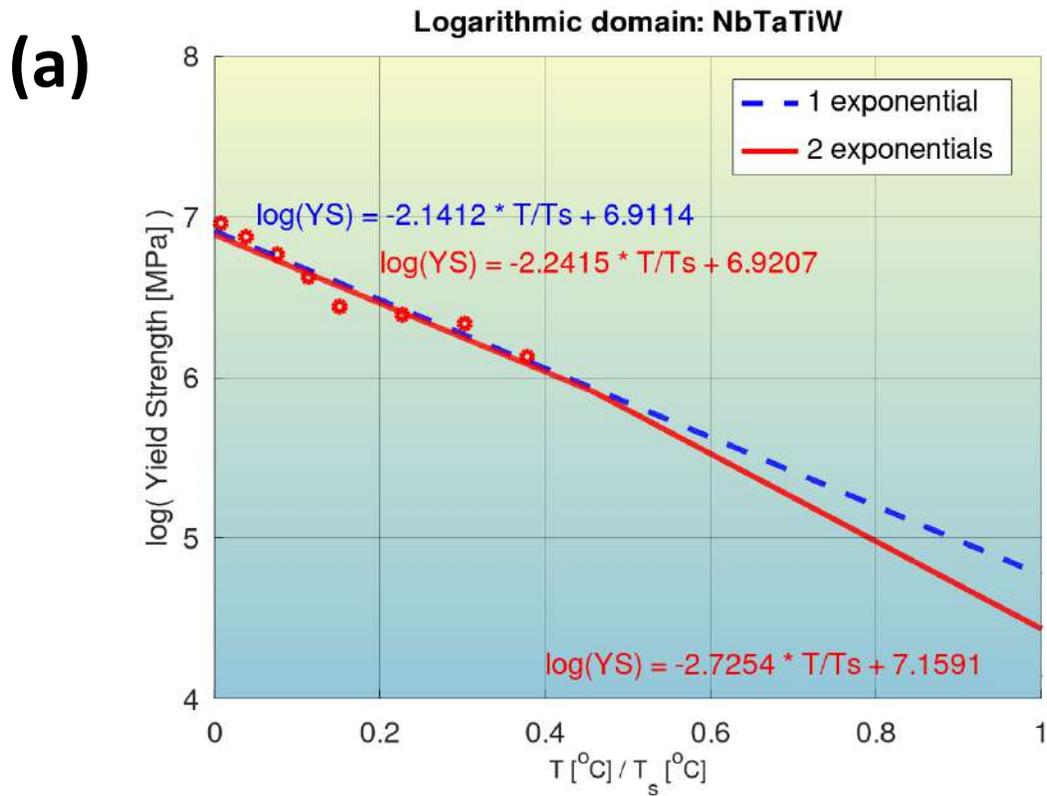

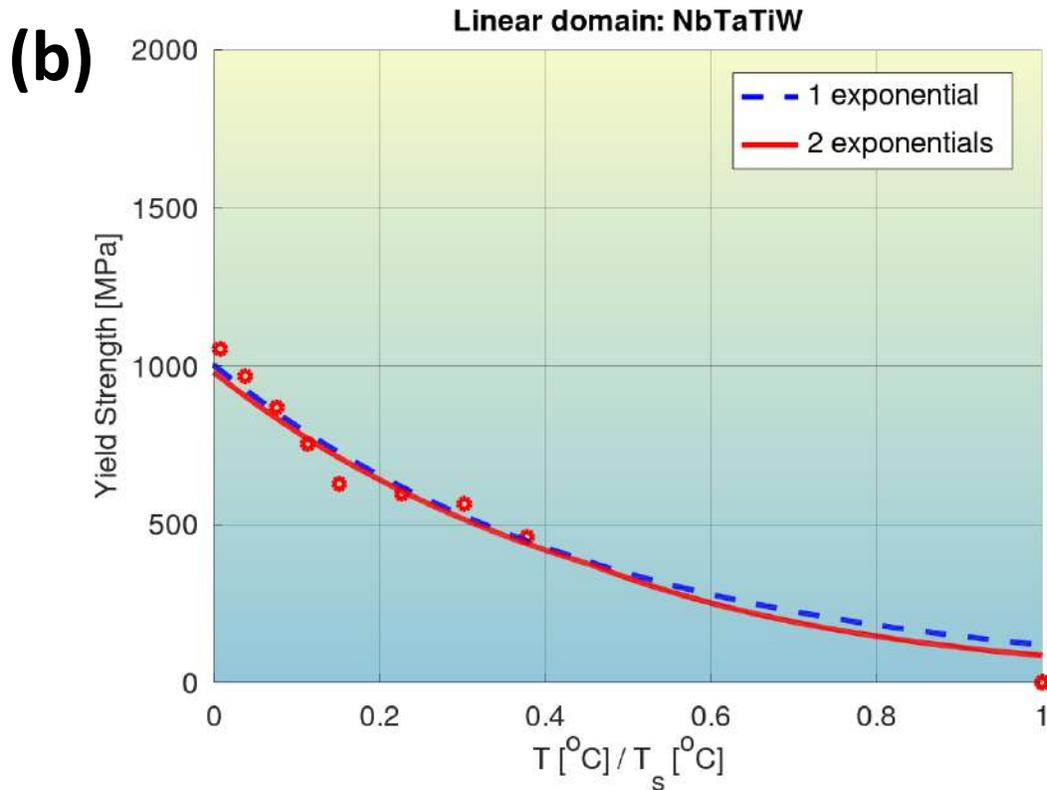

**Fig. S120**: Quantification of modeling accuracy of the bilinear log model, for composition No. 119 from **Tab. S3** (NbTaTiW, BCC phase), and comparison to that of a model with a single exponential.

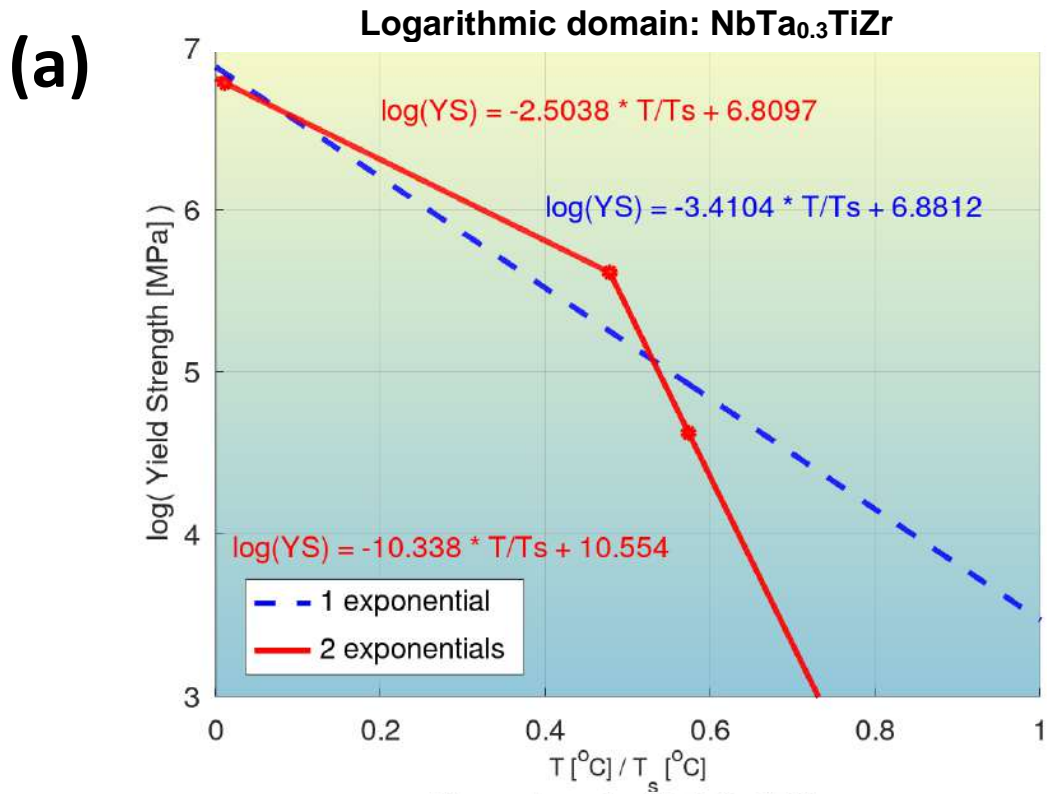
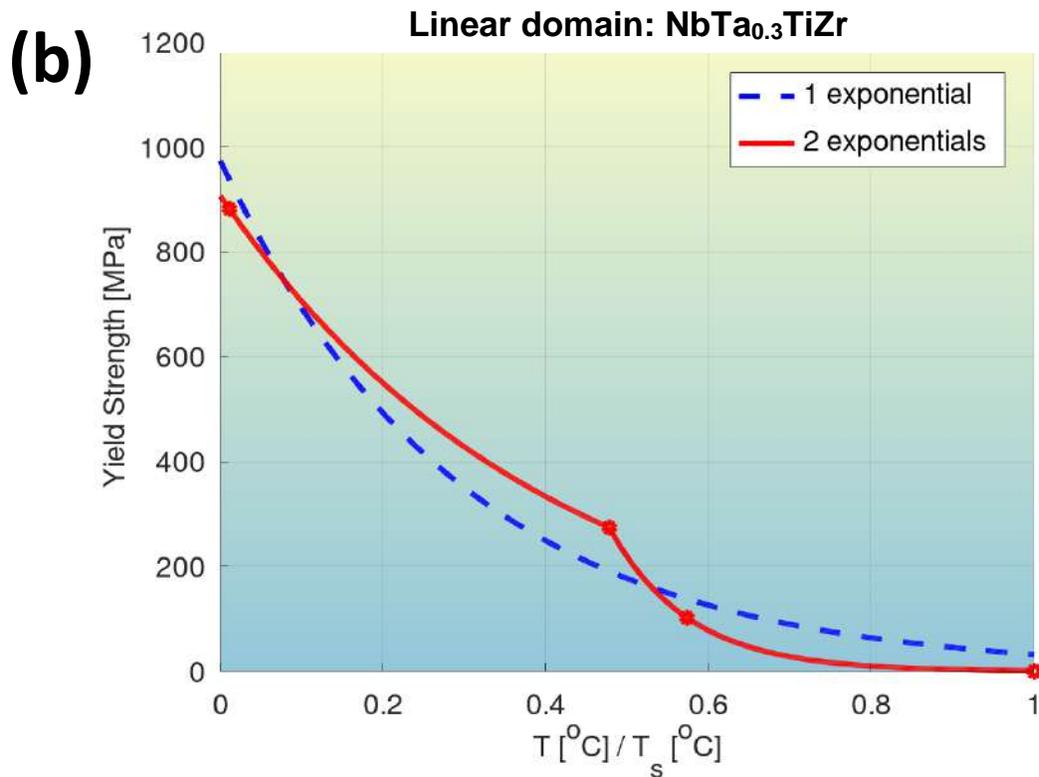

**Fig. S121**: Quantification of modeling accuracy of the bilinear log model, for composition No. 120 from **Tab. S3** (NbTa$_{0.3}$TiZr, BCC phase), and comparison to that of a model with a single exponential.

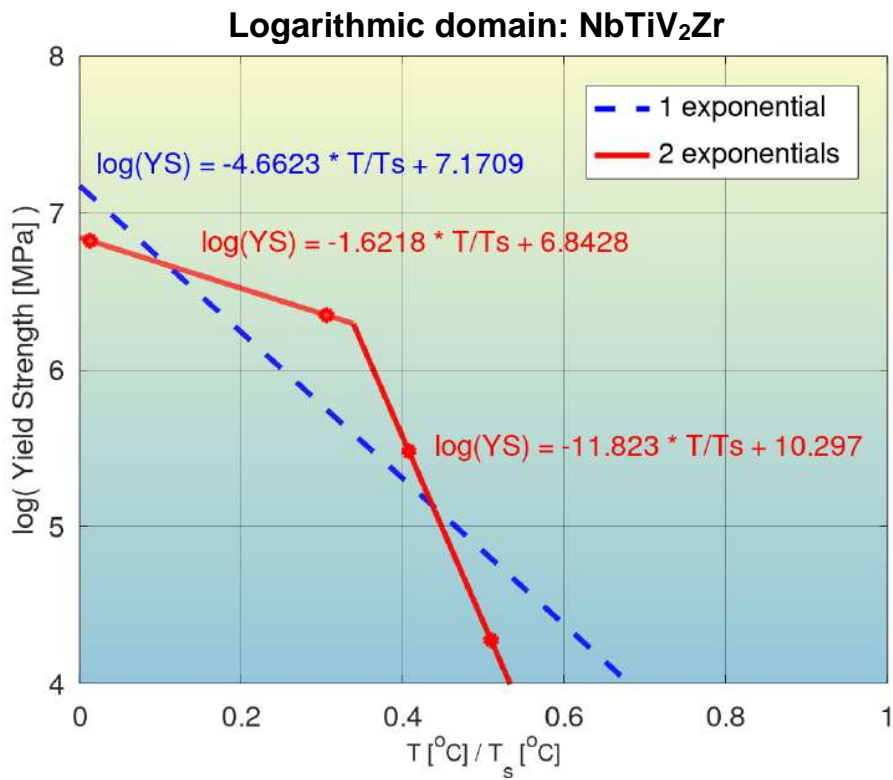

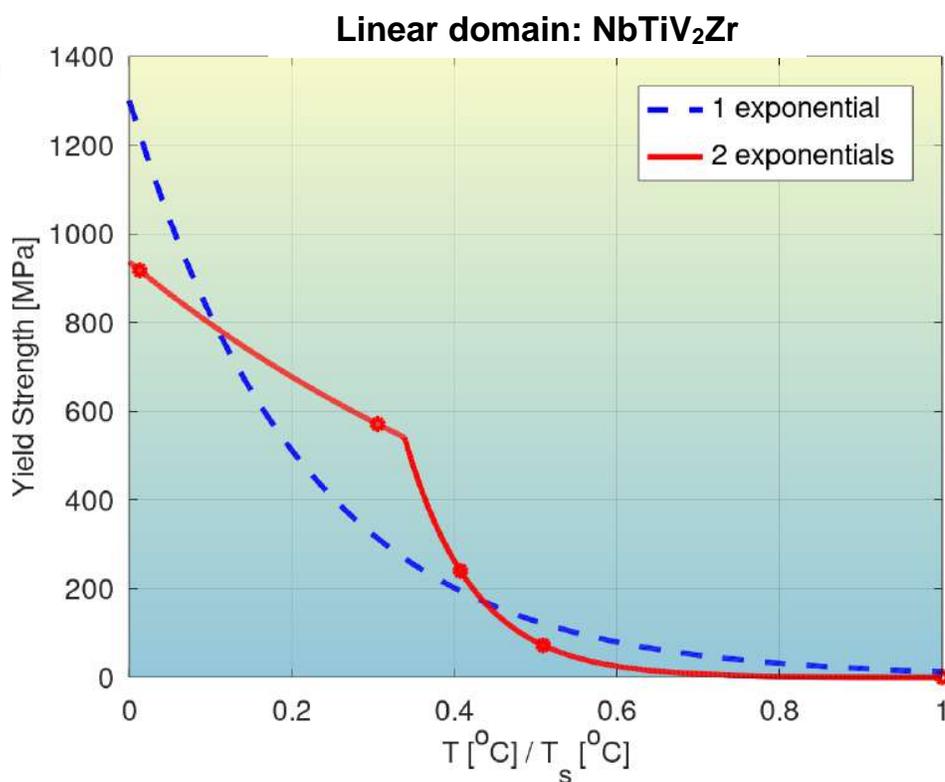

**Fig. S122**: Quantification of modeling accuracy of the bilinear log model, for composition No. 121 from **Tab. S3** (NbTiV$_2$Zr, 3 BCC phases), and comparison to that of a model with a single exponential.

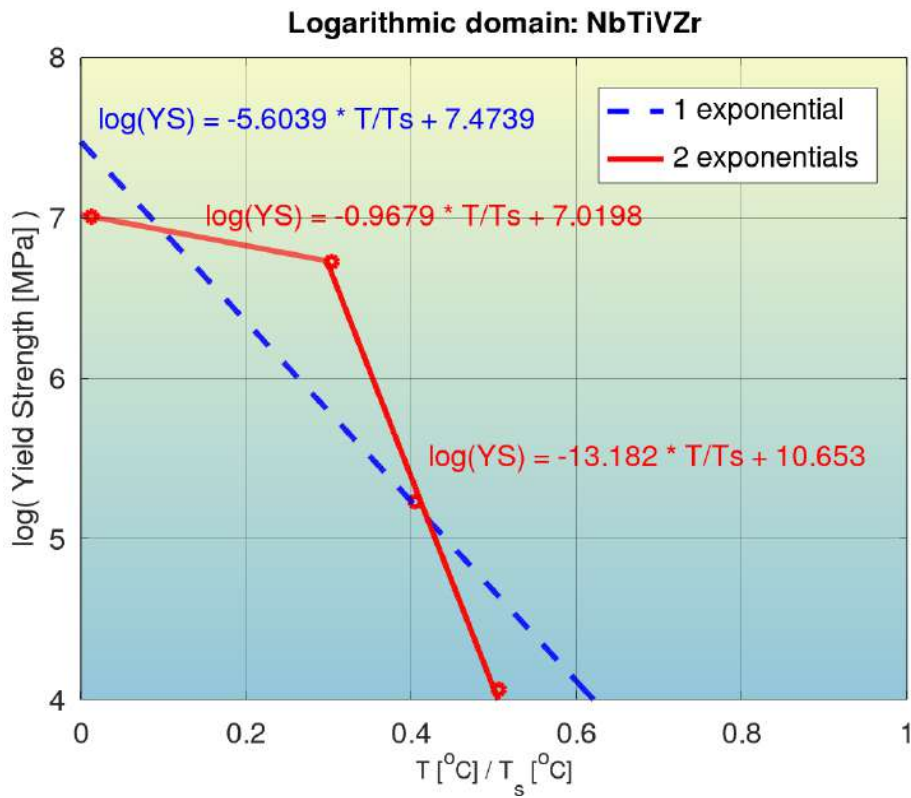
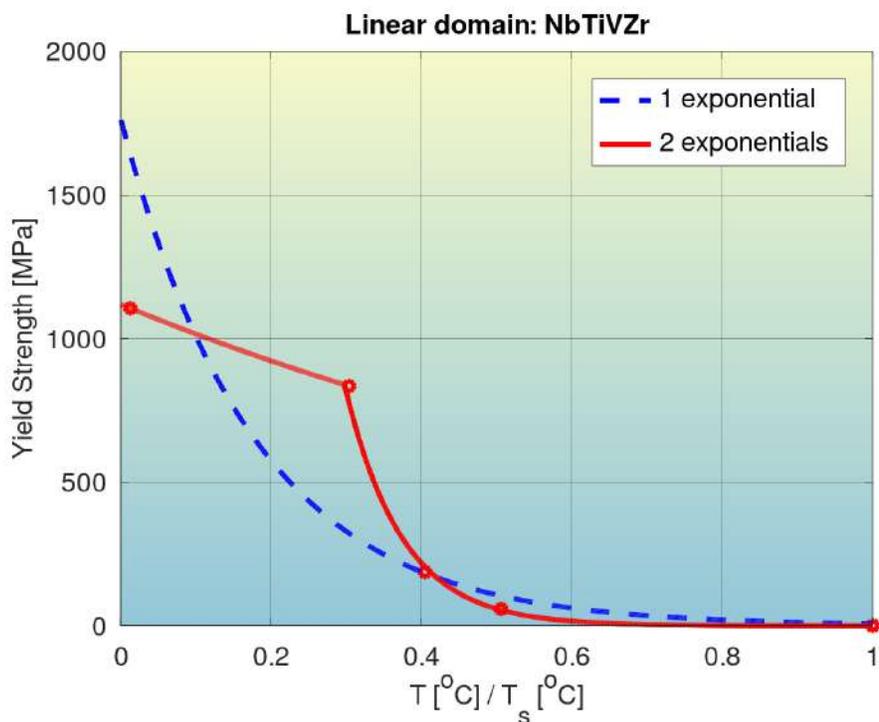

**Fig. S123**: Quantification of modeling accuracy of the bilinear log model, for composition No. 122 from **Tab. S3** (NbTiVZr, 2 BCC phases), and comparison to that of a model with a single exponential.

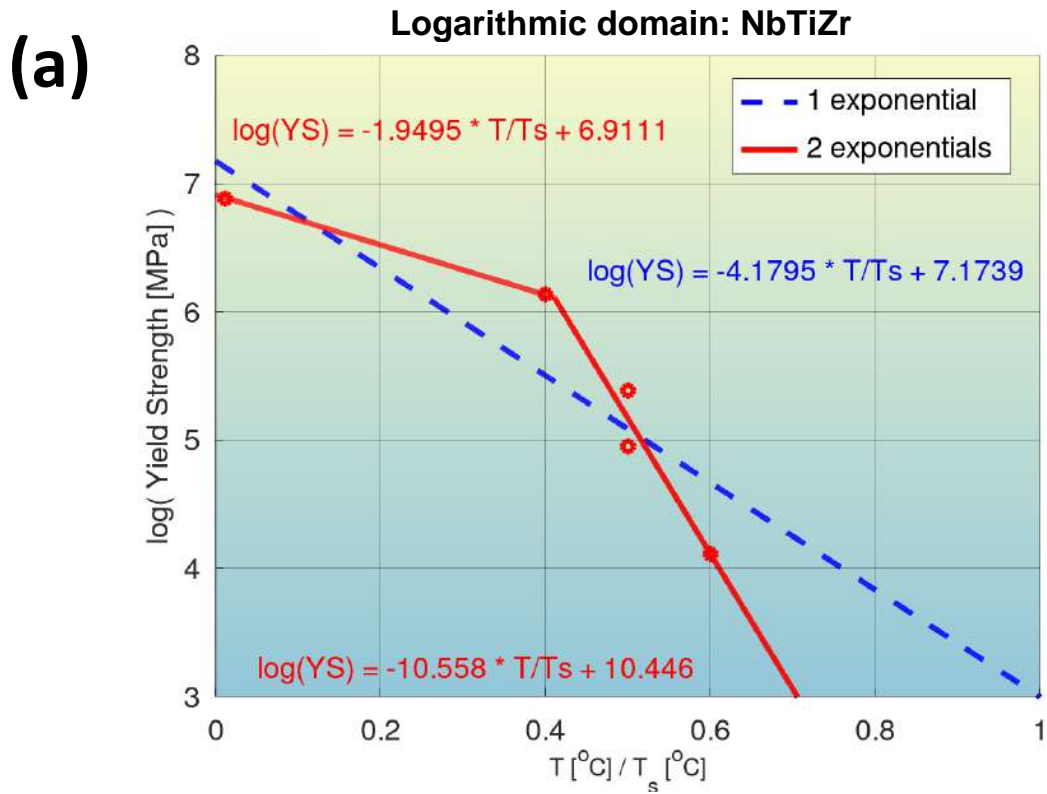

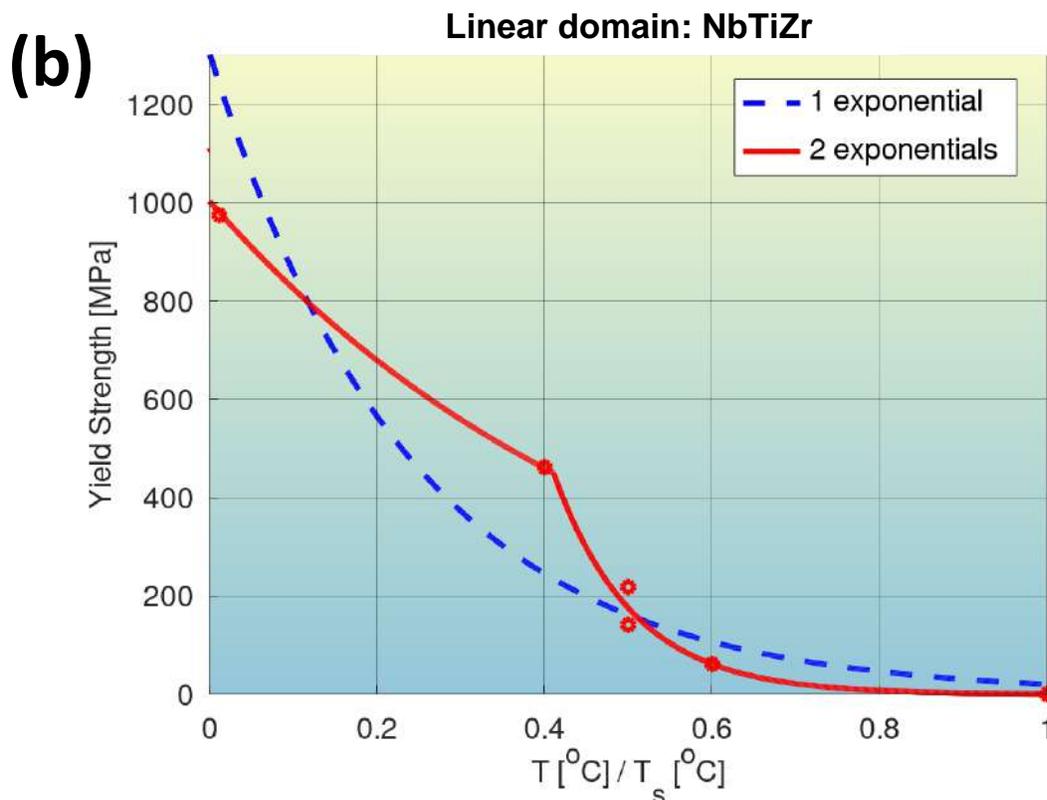

**Fig. S124**: Quantification of modeling accuracy of the bilinear log model, for composition No. 123 from **Tab. S3** (NbTiZr, BCC phase), and comparison to that of a model with a single exponential.

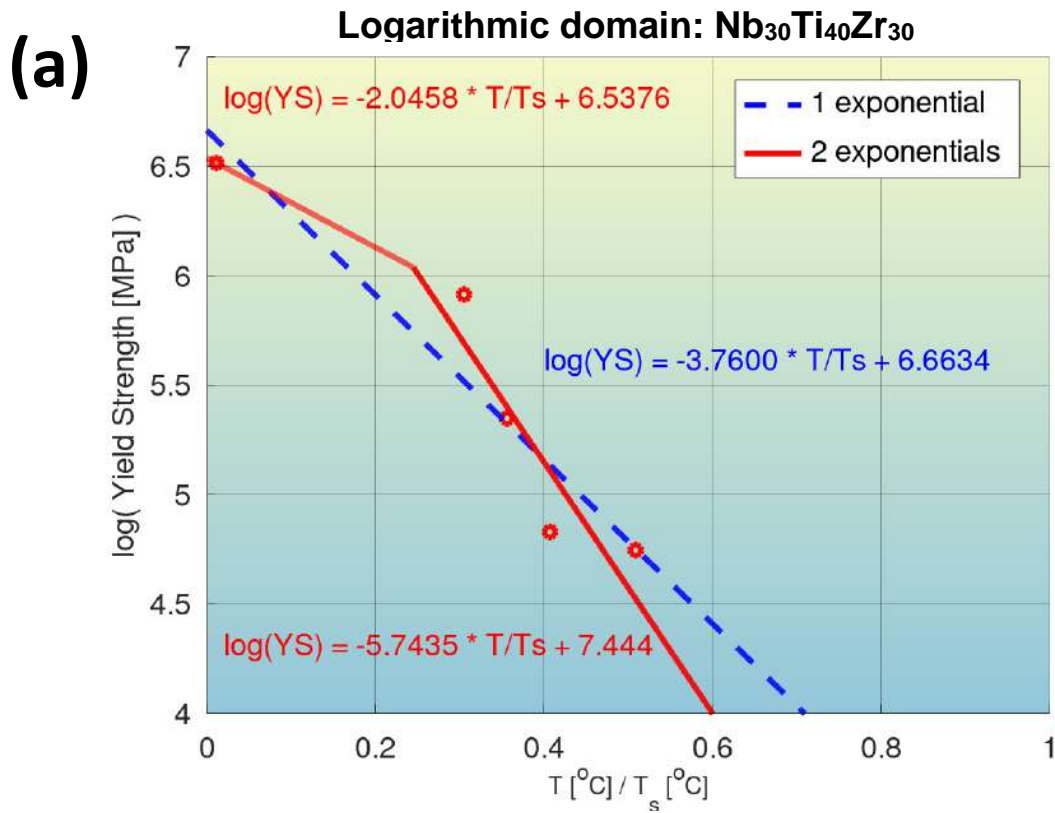

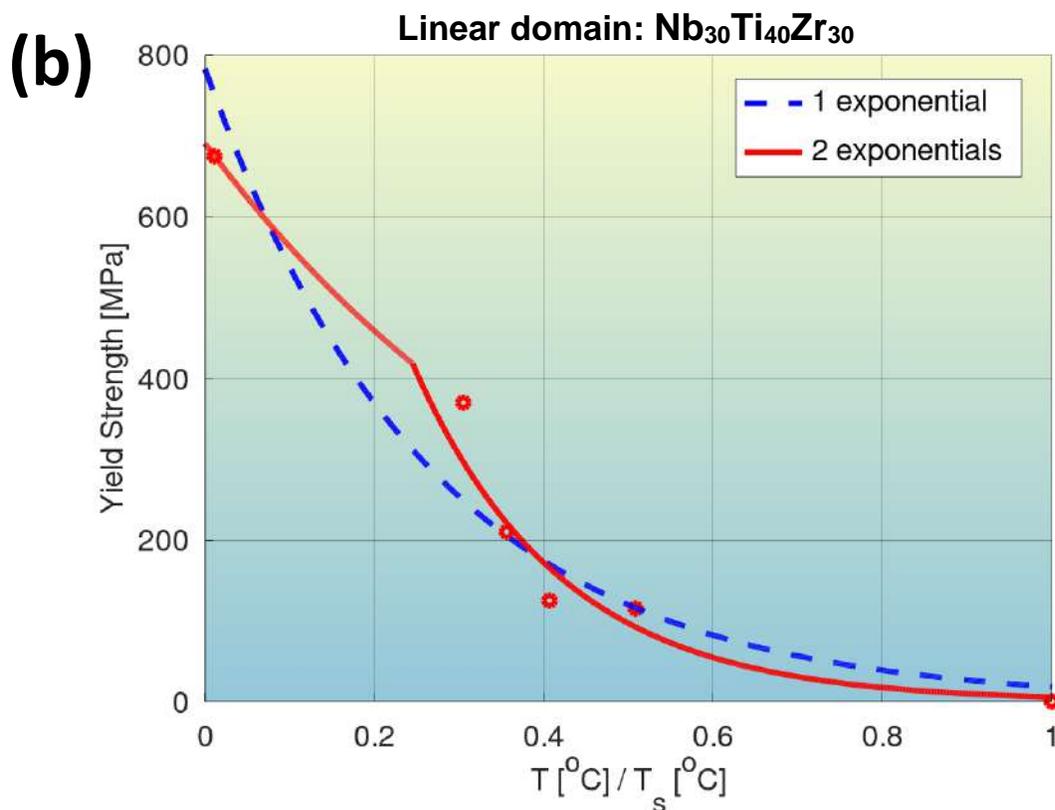

**Fig. S125**: Quantification of modeling accuracy of the bilinear log model, for composition No. 124 from **Tab. S3** ($Nb_{30}Ti_{40}Zr_{30}$, BCC phase), and comparison to that of a model with a single exponential.

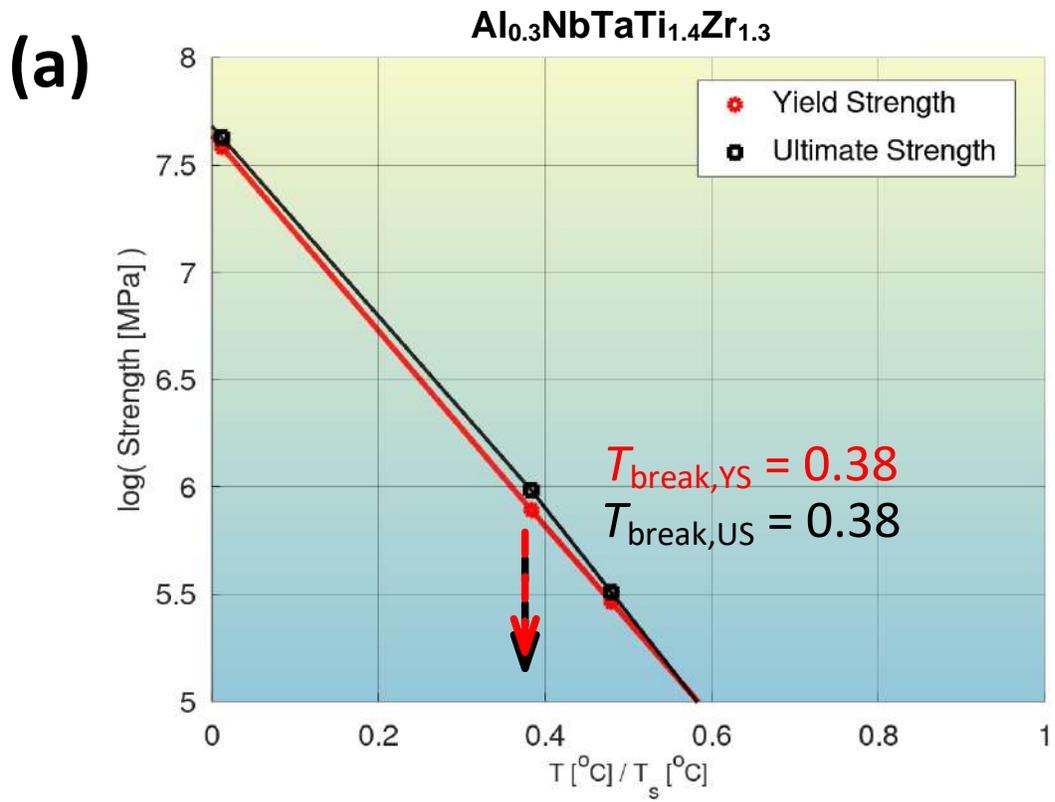

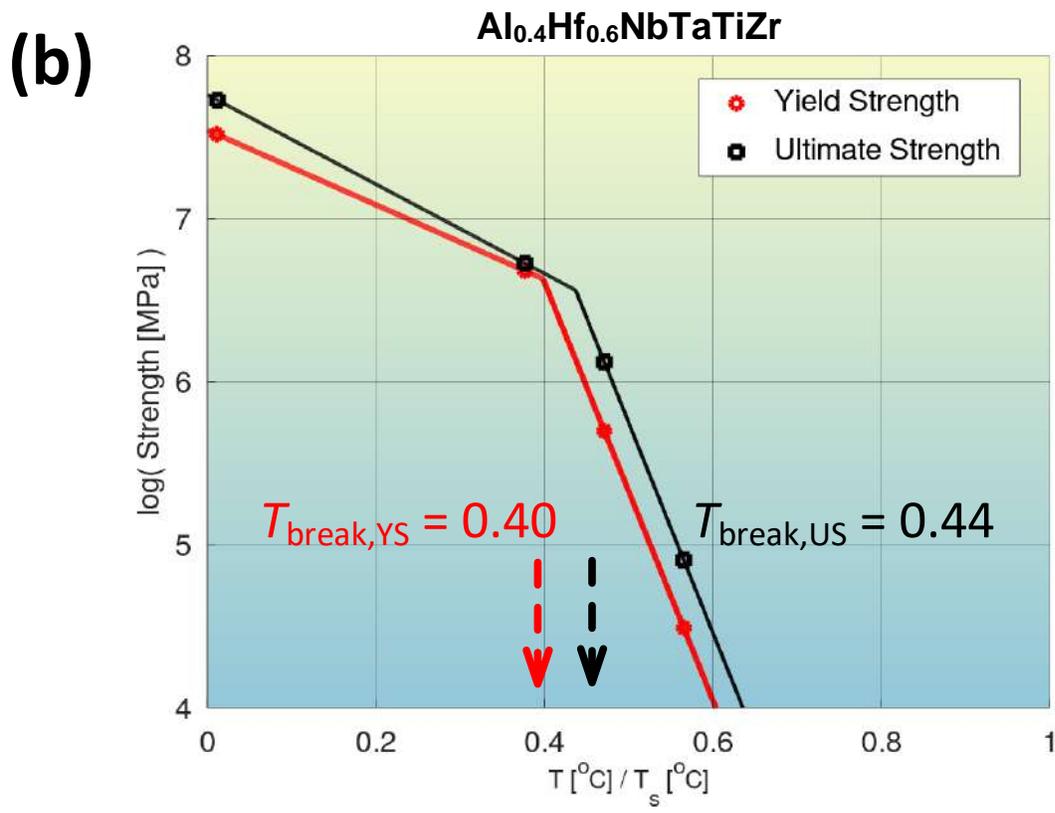

**Fig. S126**: Comparison between break temperatures for the yield strength and ultimate strength for the compositions $Al_{0.3}NbTaTi_{1.4}Zr_{1.3}$ (above) and $Al_{0.4}Hf_{0.6}NbTaTiZr$ (below).

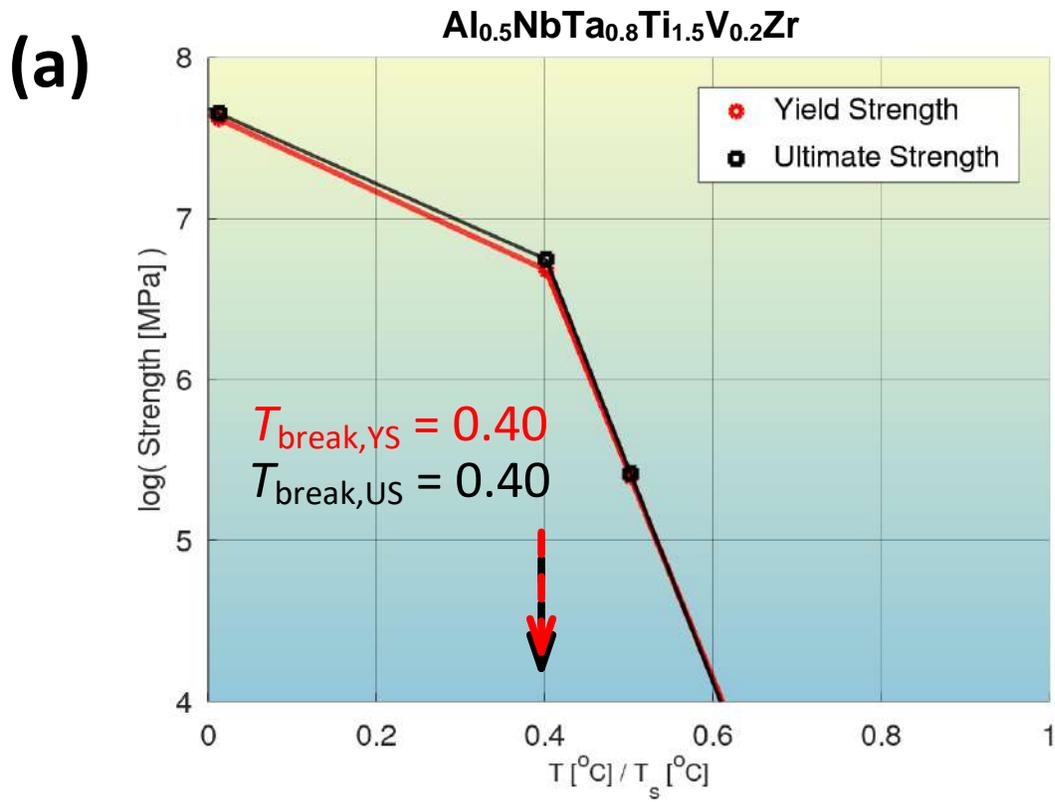

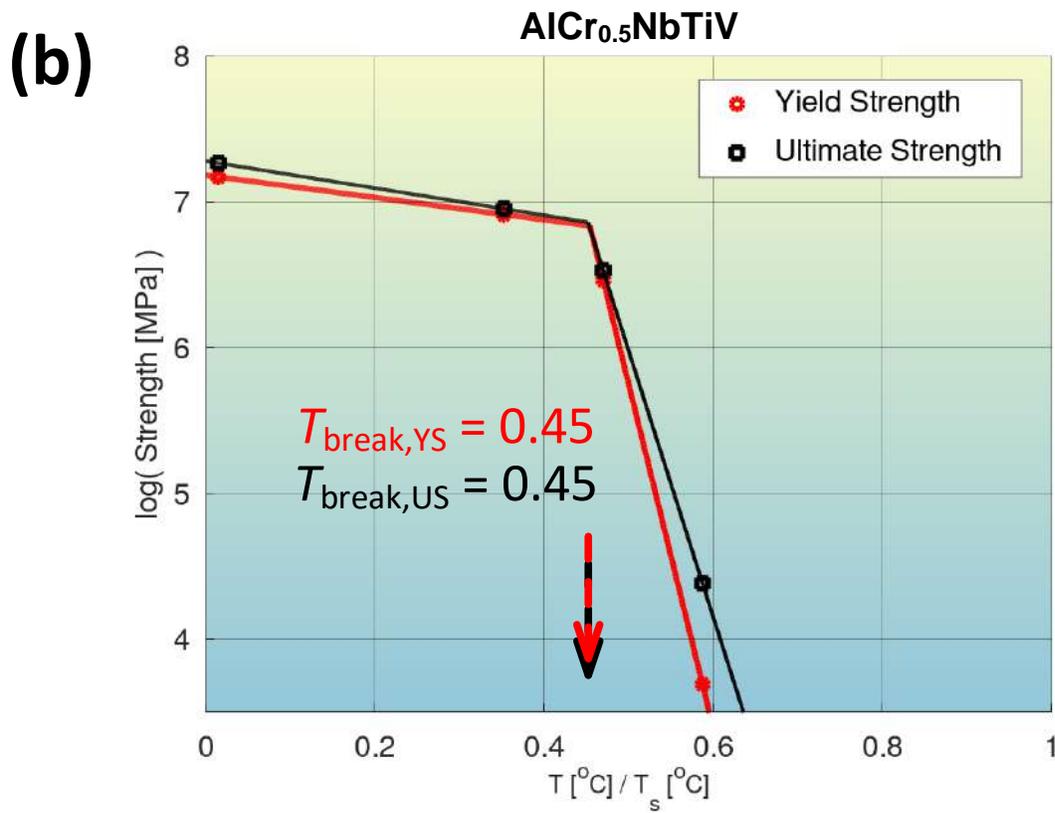

**Fig. S127**: Comparison between break temperatures for the yield strength and ultimate strength for the compositions $Al_{0.5}NbTa_{0.8}Ti_{1.5}V_{0.2}Zr$ (above) and $AlCr_{0.5}NbTiV$ (below).

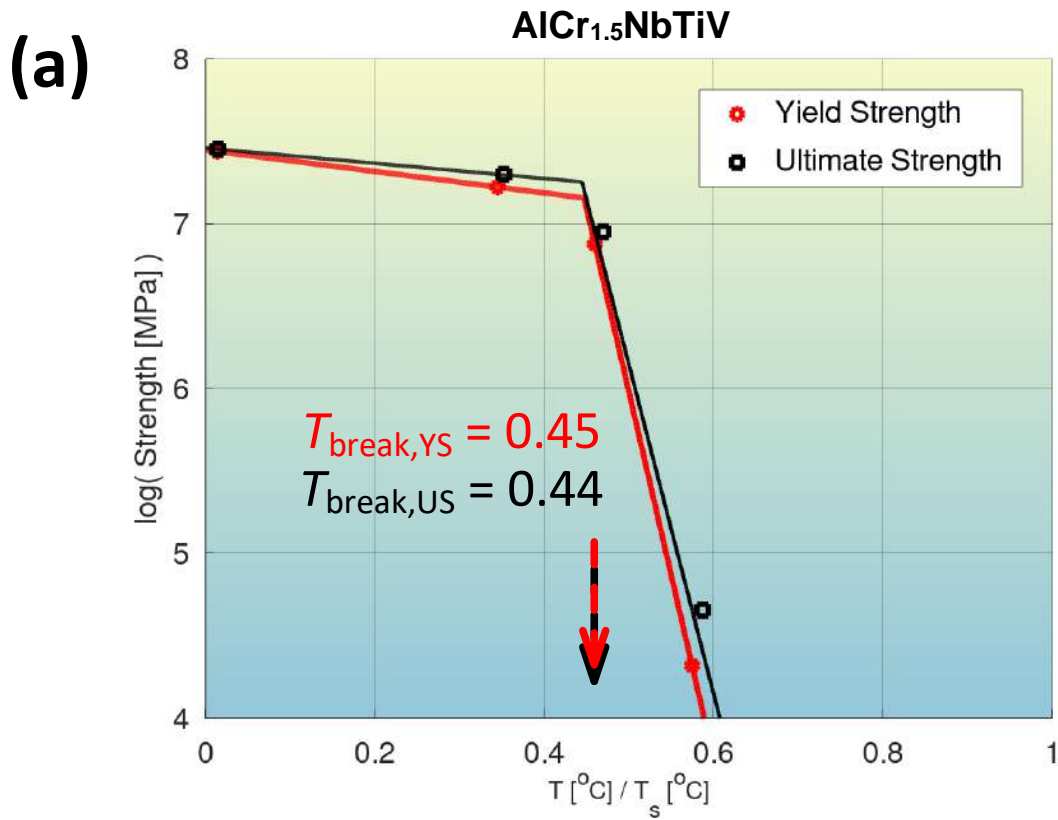

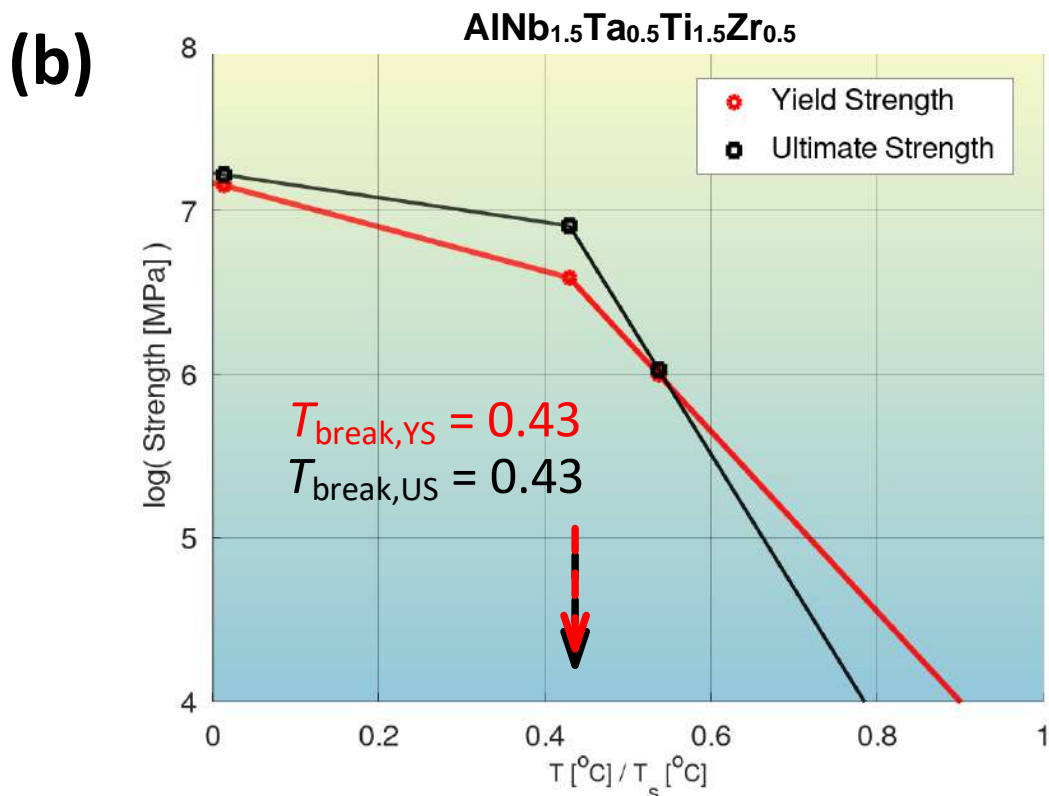

**Fig. S128**: Comparison between break temperatures for the yield strength and ultimate strength for the compositions AlCr$_{1.5}$NbTiV (above) and AlNb$_{1.5}$Ta$_{0.5}$Ti$_{1.5}$Zr$_{0.5}$ (below).

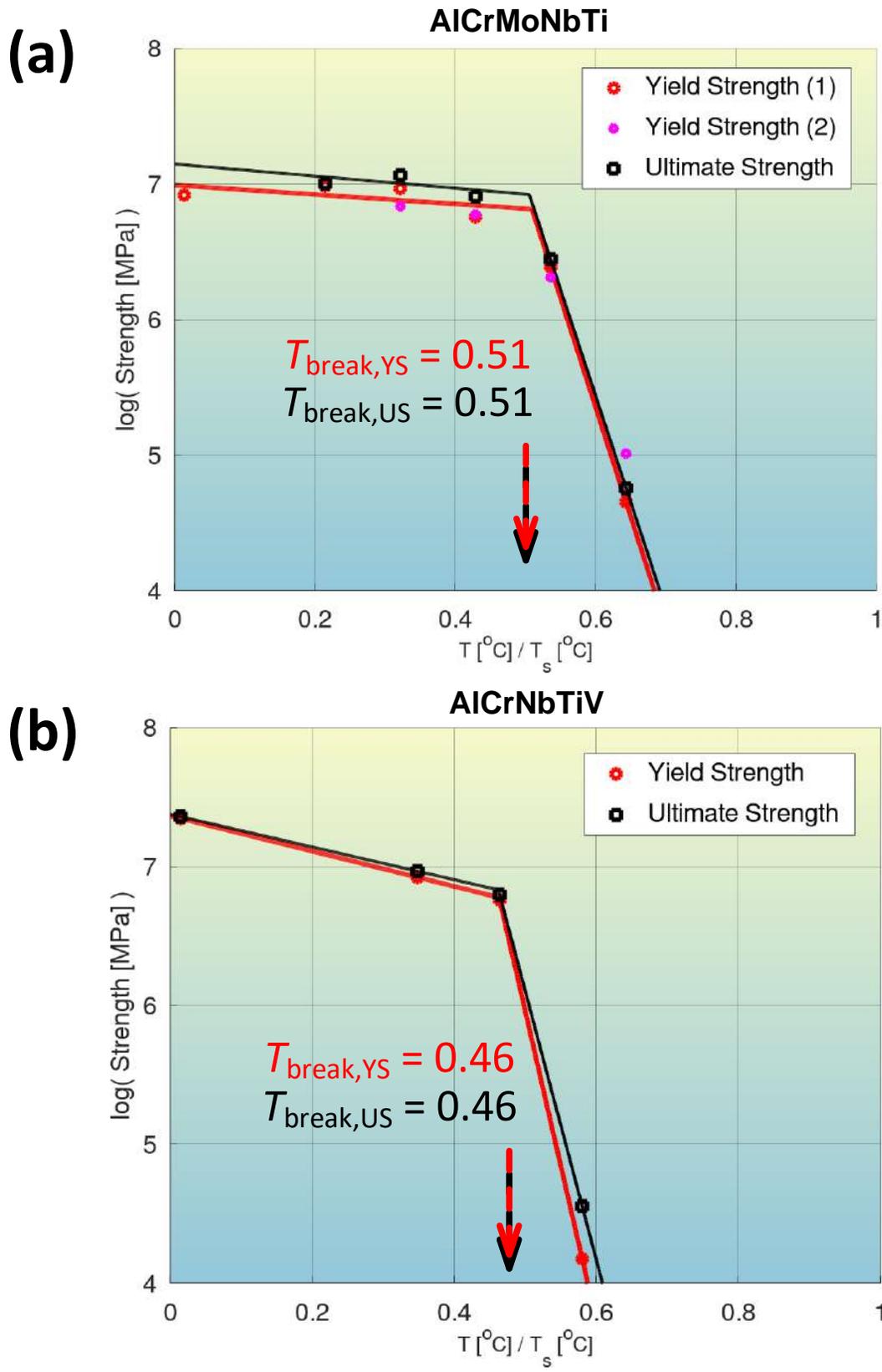

**Fig. S129**: Comparison between break temperatures for the yield strength and ultimate strength for the compositions AlCrMoNbTi (above) and AlCrNbTiV (below).

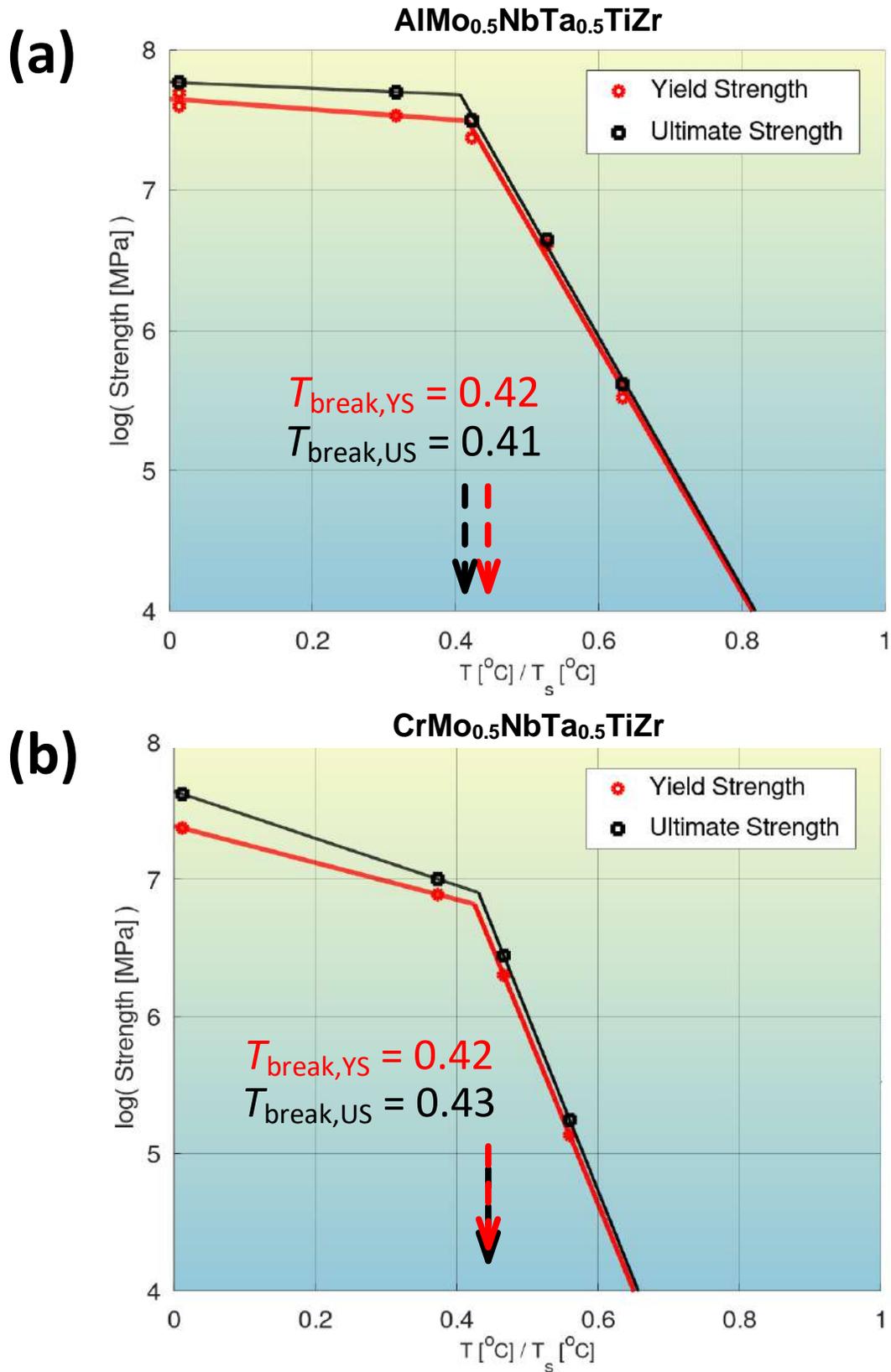

**Fig. S130**: Comparison between break temperatures for the yield strength and ultimate strength for the compositions AlMo$_{0.5}$NbTa$_{0.5}$TiZr (above) and CrMo$_{0.5}$NbTa$_{0.5}$TiZr (below).

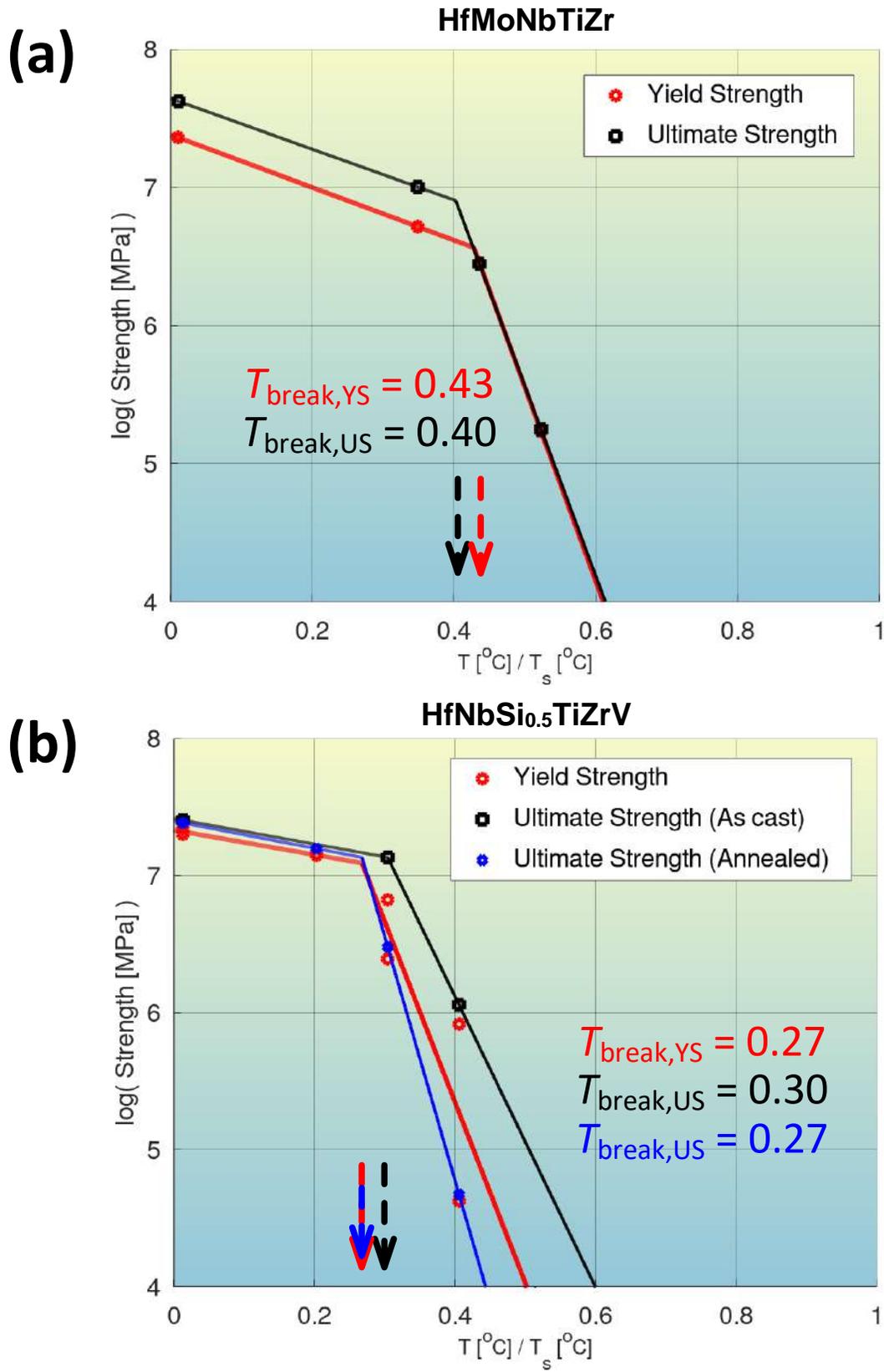

**Fig. S131**: Comparison between break temperatures for the yield strength and ultimate strength for the compositions HfMoNbTiZr (above) and HfNbSi$_{0.5}$TiZrV (below).

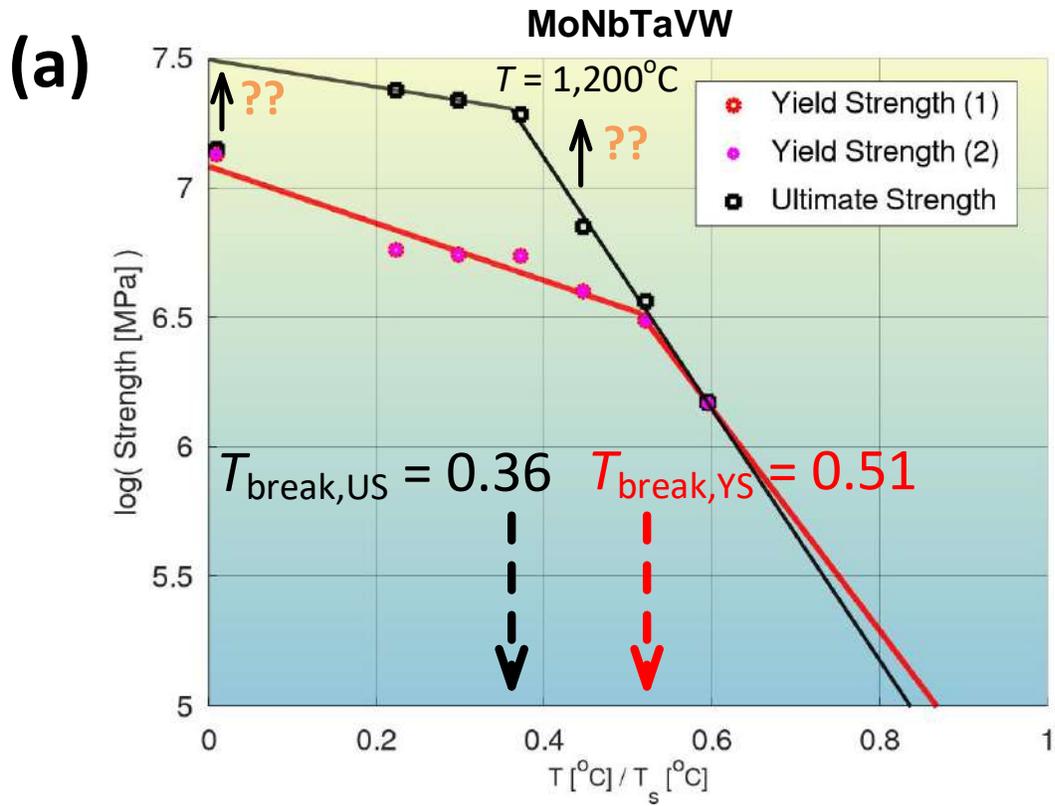
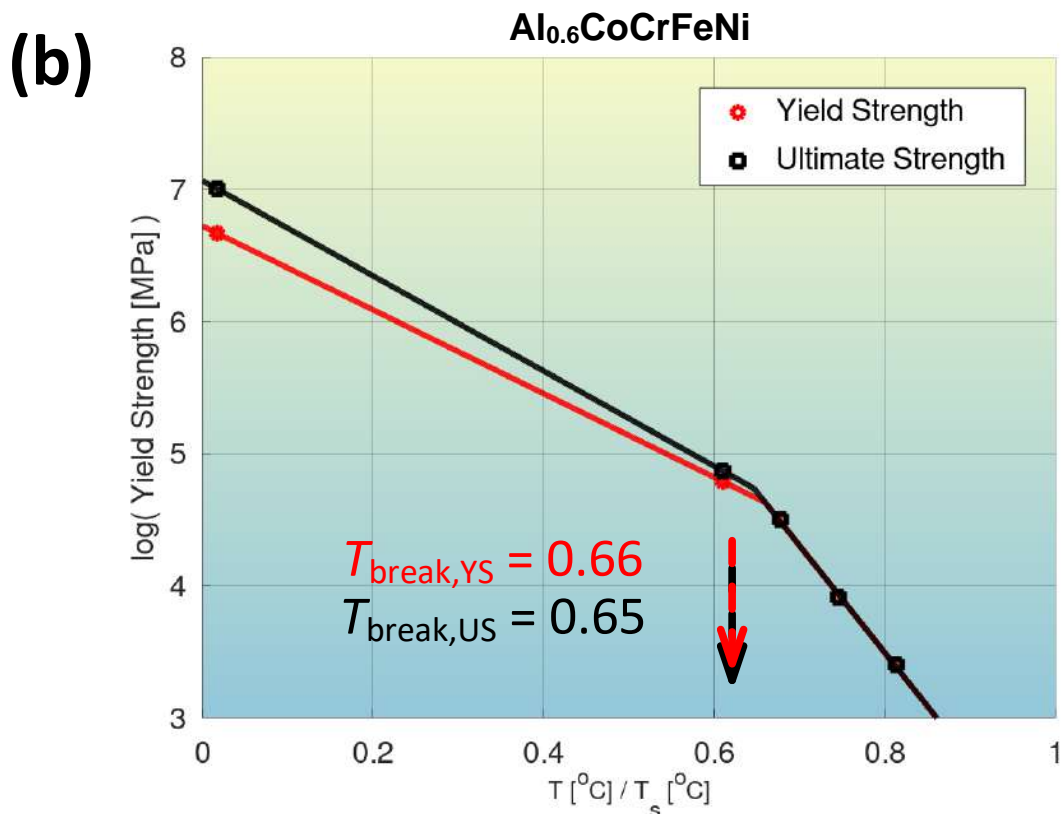

**Fig. S132**: Comparison between break temperatures for the yield strength and ultimate strength for the compositions MoNbTaVW (above) and Al$_{0.6}$CoCrFeNi (below).

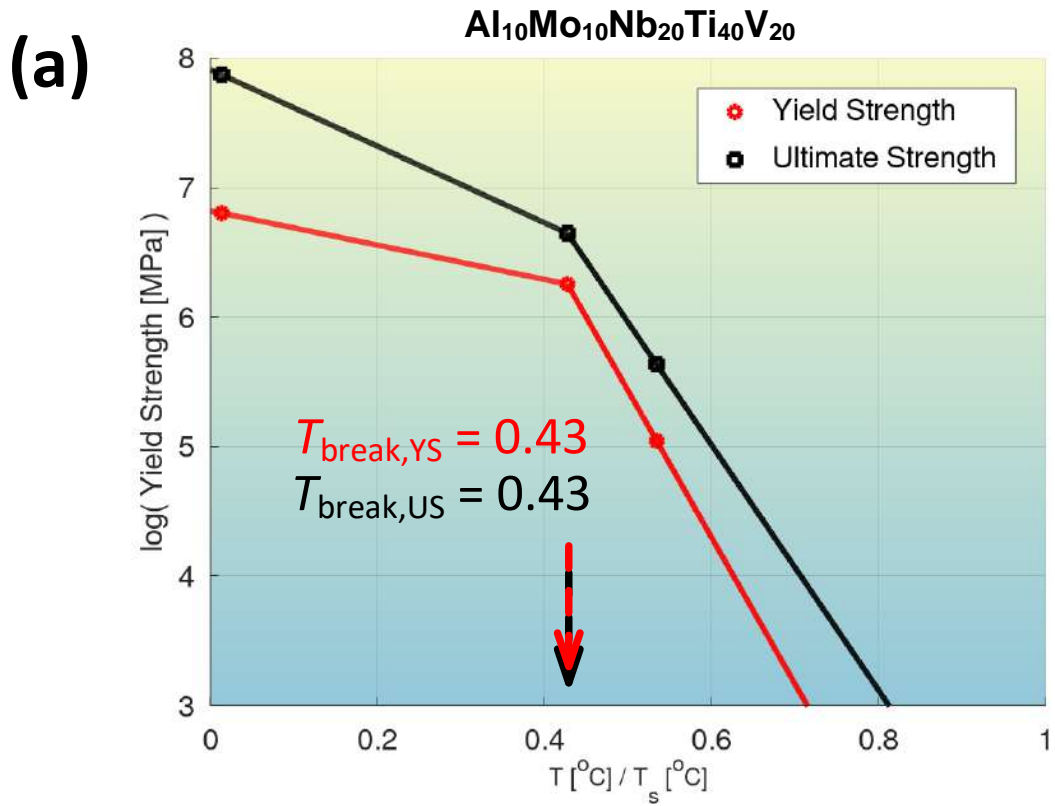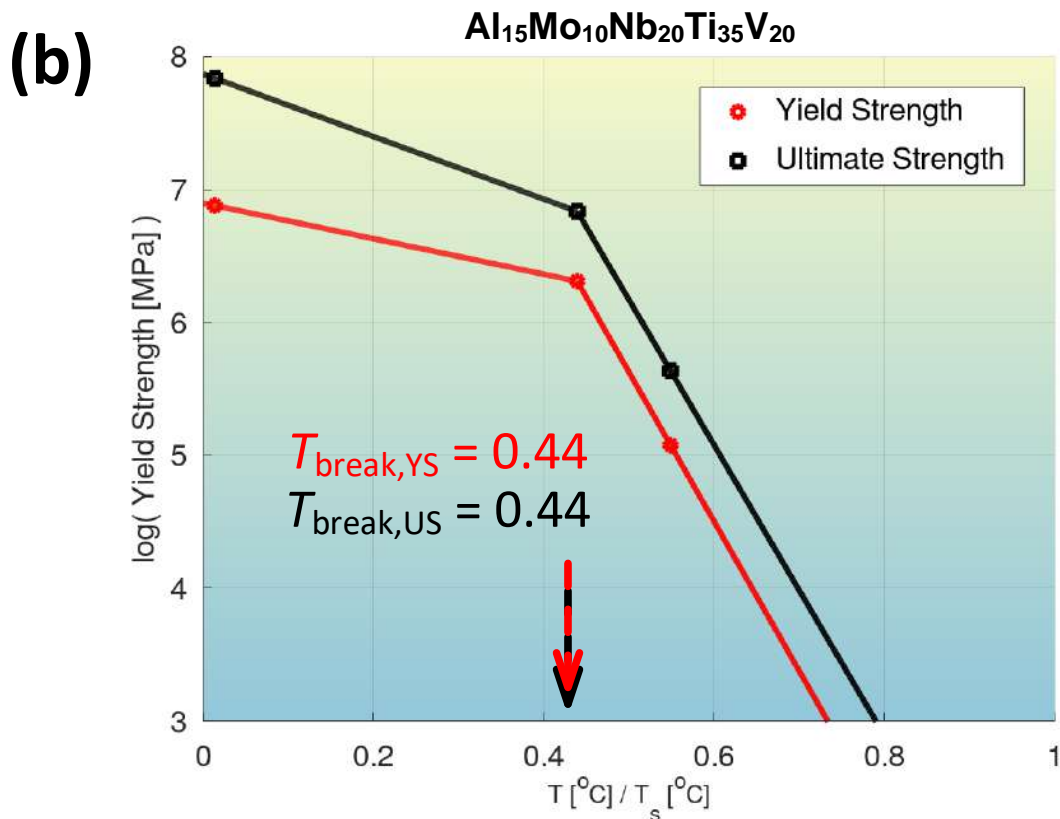

**Fig. S133**: Comparison between break temperatures for the yield strength and ultimate strength for the compositions MoNbTaVW (above) and Al$_{0.6}$CoCrFeNi (below).

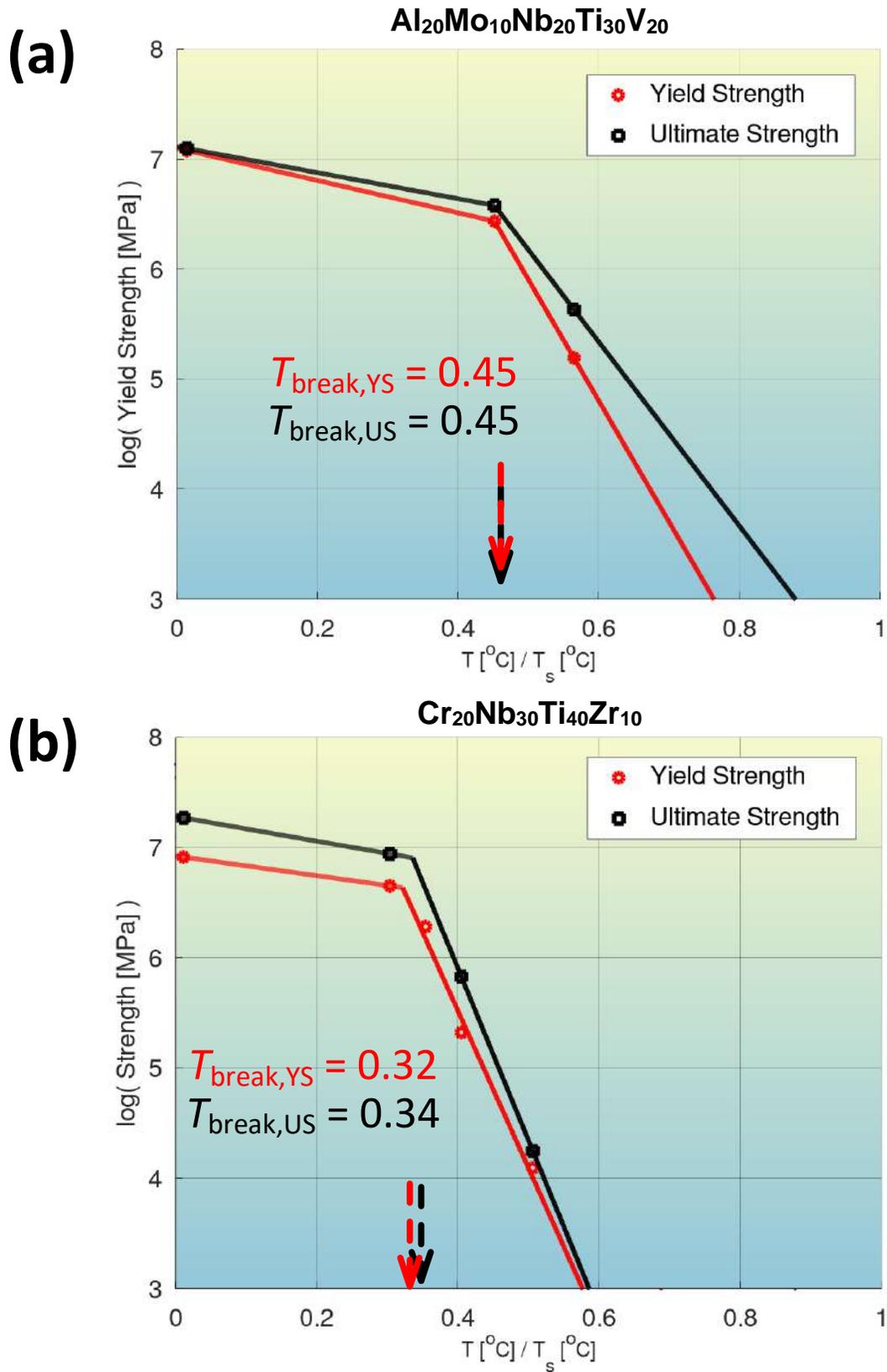

**Fig. S134**: Comparison between break temperatures for the yield strength and ultimate strength for the compositions $Al_{20}Mo_{10}Nb_{20}Ti_{30}V_{20}$ (above) and $Cr_{20}Nb_{30}Ti_{40}Zr_{10}$ (below).

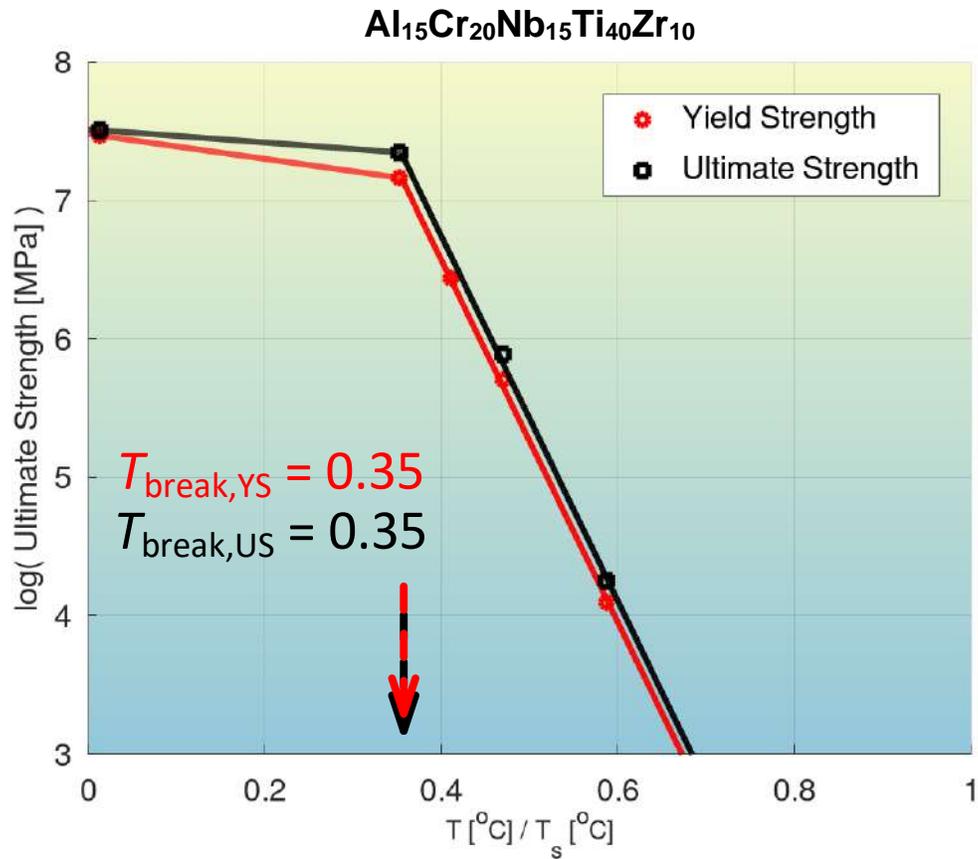

**Fig. S135**: Comparison between break temperatures for the yield strength and ultimate strength for the compositions $Al_{15}Cr_{20}Nb_{15}Ti_{40}Zr_{10}$.

**Supplementary Notes**

*1. Primary Strengthening Mechanisms in Pure Metals and Alloys*

The classical strengthening mechanisms in pure metals and alloys include [1], [2]:

1. Work hardening (also referred to as strain hardening or cold working)

2. Grain-boundary strengthening

3. Solid-solution strengthening

4. Precipitation hardening (also referred to as age hardening)

5. Dispersion strengthening

6. Transformation hardening (also referred to as phase-transformation hardening, martensitic-transformation hardening or simply as martensitic hardening)

7. Surface hardening (also referred to as-case hardening)

In addition, short-range ordering may contribute to certain alloys retaining strengths at elevated temperatures [3], [4].

1.1 Work (or Strain) Hardening

As noted by Hertzberg [5], the origin of work hardening, also referred to as strain hardening or cold working, can be traced back to the Bronze Age. Work hardening is probably the first strengthening mechanism that has been widely used for metals [5]. Skilled craftsmen hammered and bent metals to desired shapes and achieved superior strength in the process. Common work-hardened (cold-worked) commercial products include cold-drawn piano wire or cold-rolled sheet metal. The superior strength of work (strain) hardening results from a dramatic increase in the number of dislocation-dislocation interactions and reduced dislocation mobility. Figure 4.1 of [5] illustrates that the strengths of metals tend to approach very high levels, when there are no dislocations present or when the number of dislocations is extremely high ($\geq 10^{10}$ cm$^{-2}$). Figure 4.1 of [5], further, shows that low- strength levels tend to correspond to the presence of moderate numbers of dislocations (~ $10^3 - 10^5$ cm$^{-2}$).

1.2 Grain-Boundary Strengthening

Grain boundary strengthening refers to hindrance provided by grain boundaries, during deformation of materials. Such hindrance (barrier) tends to make it difficult for dislocations, esp. glide dislocations, to move, and tends to improve the strength. As outlined in Section 4.2 of [5], the Hall-Petch relation describes the impact of grain size on the yield strength of polycrystalline materials [6]:

$$\sigma_{ys} = \sigma_i + \Delta\sigma_{Hall-Petch} = \sigma_i + k_y \ d^{-1/2} \qquad (1)$$

Here, $\sigma_{ys}$ refers to the yield strength of a polycrystalline sample, $\sigma_i$ to the overall resistance of lattice to dislocation movement, $k_y$ to a "locking parameter" which represents relative hardening contribution of grain boundaries, and $d$ to the grain size [5].

1.3 Solid-Solution Strengthening[1]

The work hardening and grain boundary strengthening effects are found in pure metals and alloys alike. However, when two or more elements are combined such as to form a single-phase microstructure, various elastic, electrical and chemical interactions take place between the force (strain) fields of the solute atoms and the dislocations present in the lattice. Such interactions, captured in part in

Fig. **S1**, result in strengthening mechanism, referred to as solid-solution strengthening, resembling the strain hardening outlined above. As a result, solid solutions tend to have higher yield strengths than the pure elements.

In

---

[1] Much of the explanations in this section are adapted from a tutorial video by Prof. Rajesh Prasad from the Department of Applied Mechanics at IIT Delphi (https://www.youtube.com/watch?v=14WeQp_UfNo).

Fig. **S1** (b), we have substituted a larger substitutional solute into the lattice of the pure elements. Because the size of the substitutional solute is larger than of the atoms comprising the lattice, the substitutional solute needs to displace the atoms around, to make room for itself. These displacements cause a deformation or a force (strain) field around the solute atom. Similarly, if the interstitial solute in

Fig. **S1** (c) happens to be of larger size than the void size, then the atoms comprising the lattice will need to move from their original position, and there is a force (strain) field created.

The strain field around a solute or interstitial atom can interact with the strain field of a dislocation and impede movement of the dislocation. For an edge dislocation, there are tensile stresses below the slip plane and compressive stresses above the slip plane. The strain field of the dislocation can interact with the strain field of the solute or interstitial atom and result in increase in the strain energy of the material. This can result in the dislocation line being repelled by the solute or interstitial atom.

But if the tensile of the solute atom overlaps with compressive of the dislocation line, then there will be an attraction and the edge dislocation will likely get attracted to the solute atom. But in this case, if the dislocation continues to move, to create further plastic deformation, it will likely not move away, given the attraction with the solute atom. But in either case, the interaction of the strain field of the solute atom with the strain field of the dislocation impedes the motion of the dislocation.

An important factor affecting the magnitude of the solid-solution strengthening pertains to the difference in size of the atoms comprising the lattice of the pure element vs. the size of the substitutional solute atom. Larger size difference results in larger force (strain) fields and more effective hardening.

Ref. [7] claims to prove pertinence of lattice distortion for solid-solution strengthening in BCC HEAs.

1.4    Precipitation Hardening

Precipitation hardening refers to a hardening mechanism, resulting from dislocations interacting with precipitates in the microstructure, and causing the precipitates to obstruct the movement of the dislocations and improve strength. Precipitation hardening can be achieved by heat treating (aging) an alloy at a desired temperature and for a desired duration. This can result in the precipitates desired, such as $\gamma'$ or $\gamma''$, coming out. For additional information, refer to Section 4.4 of [5].

1.5  Dispersion Strengthening

Alloys can also be strengthened through addition of oxide particles that can obstruct the motion of dislocations. By adding $Al_2O_3$ flakes and $ThO_2$ particles to the lattice of aluminum and nickel, respectively, the strength properties of these metals at elevated temperature can be improved. For additional information, refer to Section 4.5 of [5].

1.6 Transformation Hardening

Transformation hardening, also known as martensitic transformation hardening or simply as martensitic hardening, refers to one of the most common hardening methods used for iron-carbon alloys (steels), both carbon steels and stainless steels [8]. The martensitic transformation refers to a phase transformation of steels from face-centered cubic austenitic ($\gamma$) phase into the stronger martensite phase. The martensitic transformation occurs if carbon or low-alloy steels are quenched sufficiently fast below the martesite starting temperature, $T_{Ms}$.

1.7 Surface Hardening

Surface hardening (or case hardening) refers to a process, where the hardness of a surface (case) of an object is enhanced, while the inner core of the object is kept elastic and tough. This can be accomplished through utilization of several processes, such as a carburizing or nitriding process, where a component is exposed to a carbonaceous or nitrogenous atmosphere at elevated temperature. As the result of such a process, the process surface hardness, wear-resistance and fatigue life are usually enhanced [9].

*2. Analytical Modeling of the Yield Strength of Alloys*

2.1 The work of Hall from 1951 [10] and Petch from 1953 [11]

The Hall-Petch relation, summarized in Eq. (1), describes the impact of grain size on the yield strength of polycrystalline materials. This impact is referred to as grain-boundary strengthening.

2.2 The work of Brown and Ham from 1971 [12]

There are three types of interphase boundaries (IPB) in precipitation hardening:

1. A coherent or ordered IPB

2. A fully-disordered IPB

3. A partially -rdered IPB

In case of coherent or ordered precipitates (IPB), the atoms match up one by one along the IPB. Due to the difference in lattice parameters between the alloy matrix and the precipitate, a coherency strain energy is associated with this type of boundary. In case of fully-disordered IPB, there are no coherency strains. The precipitate particles, however, tend to be non-deforming to dislocations. In case of the partially-ordered IPB, coherency strains are partially relieved by the periodic introduction of dislocations along the interphase boundary.

Brown and Ham presented the theory of strengthening by coherent ordered precipitates, i.e., precipitates with the coherent or ordered IPB. Their theory predicts that coherent ordered precipitates should strengthen the alloy matrix by an amount, $\Delta\sigma$, given by

$$\Delta \sigma = \begin{cases} \dfrac{\gamma_{afb}}{2\mathbf{b}} \left[ \left( \dfrac{3\pi^2 \gamma_{afb} f <r>}{32\, T} \right)^{1/2} - f \right]; \dfrac{f\, T}{\gamma_{afb}} < <r> < \dfrac{4\, T}{\pi\, \gamma_{afb}} \\ \dfrac{\gamma}{2\mathbf{b}} \left[ \left( \dfrac{3\pi f}{8} \right)^{1/2} - f \right]; <r> >> \dfrac{4\, T}{\pi\, \gamma} \end{cases} \quad (2)$$

Here $\gamma_{afb}$ represents the antiphase boundary (APB) energy on the (111) slip plane, $T$ denotes the line tension of dislocation, **b** is a Burgers vector, $f$ is volume fraction of the precipitation particles, and $<r>$ average particle radius [12].

2.3 The work of Ardell et al. from 1976 [13]

Ardell et al. looked at precipitation hardening in Ni-Al alloys containing large volume fractions of $\gamma'$ phase. The authors showed that conventional theory of strengthening by coherent ordered precipitates, referred to as theory of order-strengthening, which was introduced by Brown and Ham [12], and which applied to alloys for which the volume fraction $f$ is small, could not accurately predict the magnitude of the measured values for $\Delta\sigma$. To address this, the authors present amended theory applicable to larger values of $f$.

2.4 The work of Norstrom et al. from 1976 [14]

Norstrom et al. proposed a comprehensive superposition relation describing the yield strength of lath martensite (steel):

$$\sigma_{ys} = \underset{\text{Peierls Stress}}{\sigma_i} + \underset{\text{Solid-solution strengthening}}{k\sqrt{c}} + \underset{\text{Boundary hardening}}{\dfrac{k_y}{\sqrt{d}}} + \underset{\text{Work (or strain) hardening}}{\alpha\, G\, b\, \sqrt{\rho}} \quad (3)$$

Here $c$ represents the concentration of alloying element, $G$ shear modulus, $b$ magnitude of Burgers vector, $\alpha$ a proportionality factor specific to a given material, and $\rho$ dislocation density. Peierls stress is the force needed to move a dislocation within a plane of atoms in the unit cell [15]. A martensite steel refers to steel with the martensite microstructure (steel that has undergone a martensite phase transformation). The martensitic phase transformation refers to a phase transformation of steels from the face-centered-cubic (FCC) austenitic (γ) phase into a stronger martensite phase.

The work (or strain) hardening exhibits square root dependency on the number of dislocations. Material strength tends to be high, if there either is a high concentration of dislocations in the material (say, greater than $10^{14}$ dislocations per m$^2$) or no dislocations.

First E.P. George, W.A. Curtin and C.C. Tasan [16], and then B. Cantor [17], have described the deformation in multi-component high-entropy alloys, in a similar fashion, in terms of flow stress $\sigma_f$, that depends on plastic strain $\epsilon$, strain rate $\dot{\epsilon}$ and temperature $T$, with an initial dislocation-free, large-grain-size yield stress $\sigma_y$, a Taylor work-hardening term $\sigma_{wh}$ (Taylor work hardening rate with increasing plastic strain). and a Hall-Petch grain-boundary-hardening term $\sigma_{gb}$ as:

$$\sigma_{ys} = \sigma_y(T,\dot{\epsilon}) + \sigma_{wh}(\epsilon,T,\dot{\epsilon}) + \sigma_{gb}(d) \tag{4}$$

$$= \frac{KG(\sum_i x_i \delta_i^6)^{2/3}}{b^2}\left[1 - \frac{RT}{\Delta E}\ln\left(\frac{\dot{\epsilon}_o}{\dot{\epsilon}}\right)\right] + \alpha M G b \rho^{1/2} + k d^{-1/2} \tag{5}$$

Here $K$, $\alpha$ and $k$ represent constants, $G$ denotes the shear modulus, $b$ represents the Burgers vector, $x_i$ and $\delta_i$ are, respectively, the molar fraction and atomic misfit of the $i$th component, $\Delta E$ is the activation barrier for dislocation motion between pinning points, $\dot{\epsilon}_o$ is a reference strain rate, $M$ is the Taylor factor converting applied tensile stress resolved shear stress on the primary slip plane, $\rho$ is the dislocation density, and $d$ the gain size.

2.5 The work of Langford et al. from 1977 [18]

Langford et al. have shown that the strength of pearlitic steels can be described as

$$\sigma_{ys} = \sigma_i + k_1 S^{-1/2} + k_2 S^{-1}, \tag{6}$$

where $S$ represents the interlamellar spacing, $k_1$ and $k_2$ are constants, and $\sigma_i$ denotes the resistance of the lattice to dislocation movement. A pearlitic steel, or pearlite, refers to a two-phased, layered (called lamellar) structure composed of alternating layers of ferrites (87.5 wt%) and cementites (12.5 wt%) that occurs in some steels and cast irons. During slow cooling of an iron-carbon alloy (a steel), pearlite can form through an eutectoid reaction, when austenitic steel cools below the eutectoid temperature of 723 °C (1,333 °F).

2.6 The work of Reppich from 1982 [19], [20]

Reppich presented a theoretical concept founded on existing hardening models for estimating the increase in the yield stress due to the pairwise-particle cutting and anti-phase domain-boundary formation in γ′-precipitating Ni-base alloys. The author presents a sound background on conventional APB-hardening theories, covering both small particles [21], [12], and large particles [22]. The author, further, reviews line-tension models for anisotropic, two-phase material [12], the Orowan stress of an anisotropic, two-phase material [23], Labush hardening [24], as well as Fleischer-Friedel hardening.

2.7 The work of Ardell from 1985 [25]

In [25], Ardell provides quite a comprehensive and thorough review of precipitation hardening, in particular on the influence of precipitates on critical resolved shear stress (CRSS) and yield strength of heat treated (aged) alloys. According to Ardell, precipitate particles can impede the motion of dislocations through a variety of interaction mechanisms. Ardell presents the theoretical exposition of the following interaction mechanisms (case of shearing for deforming particles):

1. Chemical strengthening
2. Stacking-fault strengthening
3. Modulus hardening
4. Coherency strengthening

5. Order strengthening

Chemical strengthening results from additional matrix-precipitate interfaces that are created by a dislocation as it shears through a coherent particle. Chemical strengthening is associated with the surface energy of the precipitate-matrix interface, as the particle is sheared by dislocations. Here, the analysis of the interfacial area can be complicated by the distortion of the dislocation line. Stacking-fault strengthening occurs when stacking-fault energies of the precipitate, and the alloy matrix differ.

Modulus hardening occurs when the shear moduli of the alloy matrix and the precipitate differ. This difference can result in an energy change of dislocation-line tension when the dislocation line cuts the precipitate. The dislocation line may bend when entering the precipitate, leading to an increase in the length of the dislocation line affected. According to Eq. (72) of [25], the modulus strengthening, $\Delta\sigma_{MS}$, can be modeled as

$$\Delta\sigma_{MS} = M * 0.0055 * (\Delta G)^{\frac{3}{2}} * \left(\frac{2f}{G}\right)^{\frac{1}{2}} * \left(\frac{r}{b}\right)^{\frac{3m}{2}-1} \tag{7}$$

Similar to Eq. (2), $r$ represents an average particle radius, $b$ denotes the length of a Burgers vector for a dislocation, $f$ represents the volume fraction of the precipitation particles, $G$ the shear modulus of the alloy matrix, $G_p$ the shear modulus of the precipitate, and

$$\Delta G \equiv |G_p - G|. \tag{8}$$

Coherency strengthening arises by virtue of elastic interaction between the strain fields of a coherent misfitting precipitate and a dislocation. Coherency strengthening takes place when the interface between the particles and the matrix is coherent, which depends on parameters like particle size, and the way that particles are introduced. When small particles precipitate from the supersaturated solid solution, coherent interfaces with the alloy matrix is usually formed. Coherency-hardening results from the atomic-volume difference between the precipitate and the matrix, which leads to a coherency strain. The associated stress field interacts with dislocations leading to an increase in yield strength, through a mechanism analogous to the size effect in solid-solution strengthening. Coherency strengthening, $\Delta\sigma_{CS}$, can be modeled as

$$\Delta\sigma_{CS} = M * \alpha_S * (G\ \varepsilon_c)^{\frac{3}{2}} * \left(\frac{r\ f}{0.5\ Gb}\right)^{1/2} \tag{9}$$

The most thoroughly-modeled case for coherency strengthening is that of a pure edge dislocation interacting elastically with a spherical coherent precipitate. In this case, $r$ represents the radius of the precipitate, but $\varepsilon_c$ a misfit parameter given by

$$\varepsilon_c \equiv |\delta|[1 + 2\ G\ (1 - 2\ v_p)/G_p(1 + v_p)]. \tag{10}$$

As before, $G$ denotes the shear modulus of the alloy matrix but $G_p$ the shear modulus of the precipitate. Similarly, $v_p$ represents a Poisson ratio for the precipitate. The quantity, $\delta$, is defined in terms of the difference between lattice parameters of the precipitate of the alloy matrix, $a_p$ and $a$:

$$\delta = \frac{a_p - a}{a}. \tag{11}$$

Ordered strengthening comes into play, when the crystal structure of a coherent precipitate is a superlattice, and the alloy matrix is a disordered solid solution. Ordered strengthening takes place when the precipitate is an ordered structure. This feature causes the bond energy before and after shearing to be different. The resulting increase in the energy per unit area represents the anti-phase-boundary energy, which accumulates gradually as a dislocation passes through a particle. Further drawing upon the work to Ardell, the ordering strengthening can be modeled as

$$\Delta\sigma_{OS} = M * 0.81 * \frac{\gamma_{apb}}{2\ b} * \left(\frac{3\pi\ f}{8}\right)^{1/2} \tag{12}$$

For definition of the terms, refer to Eq. (2), Eq. (7) and Eq. (9).

2.8 The work of Ardell et al. from 1995 [26]

Ardell formulated equations for predicting critical resolved shear stress, when the size of the precipitates for polycrystalline multimodal Nickel-base superalloys fall into the following, three different stages:

$$\tau_c = \begin{cases} \dfrac{0.94\, \gamma_{APB}\, B^{1/2}}{2\, b} \left[1 + 0.7 \left(\dfrac{3\, f\, T}{8\, \gamma_{APB}\, r}\right)^{1/2}\right] - 0.45\, f\, \dfrac{\gamma_{APB}}{2\, b} & \text{for underaged} \\ 0.81\, \dfrac{\gamma_{APB}}{2\, b} \left(\dfrac{3\, \pi\, f}{8}\right)^{1/2} & \text{for peak-aged} \\ \dfrac{1.273\, T}{b\, r} \dfrac{f^{1/2}}{1.137 - f^{1/2}} & \text{for overaged} \end{cases} \quad (13)$$

Here, $\gamma_{AFB}$ refers to APB energy and $f$ to the volume fraction of $\gamma'$ precipitates [26]. The term, $T$, represents the dislocation line tension, $G$ shear modulus, $b$ magnitude of Burgers vector, and $r$ average radius of $\gamma'$ precipitates [26]. The term $B$ is defined through the following relation [13]:

$$B = \dfrac{3\, \pi^2\, \gamma_{APB}\, f\, r}{32\, T} \quad (14)$$

2.9 The work of Reed from 2006 [27]

Depending on the relationship between the particle spacing and precipitate diameter, Reed showed that the critical resolved shear stress can be described by the following equation, accounting for weak-pair coupling, strong-pair coupling as well as the so-called Orowan looping for non-deforming particles:

$$\begin{cases} \tau_{weak} = \dfrac{\gamma_{APB}}{2\, b} \left[\left(\dfrac{6\, \gamma_{APB}\, r\, f}{\pi\, T}\right)^{1/2} - f\right] \\ \tau_{strong} = \sqrt{\dfrac{3}{2}} \left(\dfrac{G\, b}{r}\right) \dfrac{f^{1/2}}{\pi^{3/2}} \left(\dfrac{2\pi\, \gamma_{APB}\, r}{G\, b^2} - 1\right)^{1/2} \\ \tau_{orowan} = \dfrac{G\, b}{L} \end{cases} \quad (15)$$

Here, $L$ denotes the average precipitates spacing that can be expressed in terms of the average size and volume fraction of $\gamma'$ precipitates [27]:

$$L = \left(\dfrac{2\, \pi}{3\, f}\right)^{1/2} r. \quad (16)$$

The so-called Orowan looping describes the ability of dislocations to bow (loop) around a precipitate particle. In case of non-deforming particles, where the spacing is small enough or the precipitate-matrix interface is disordered, dislocations can bow around a precipitate instead of shearing. The Orowan strengthening mechanism is related to the effective spacing between particles considering a finite particle size, but not related to particle strength, because once a particle is strong enough for the dislocations to bow rather than cut, additional increase in the dislocation penetration resistance won't improve strengthening. Precipitation strengthening through Orowan looping can be modeled as

$$\Delta\sigma_{orowan} = \frac{\varphi G b}{d_{Si}} \left(\frac{6\ V_{Si}}{\pi}\right)^{1/3}. \tag{17}$$

Here $d_{Si}$ represents particle size (diameter) and $V_{Si}$ volume fraction.

2.10 The work of Kozar et al. from 2009 [28]

Kozar et al. present a model that accounts for solid-solution strengthening, the Hall-Petch effect, precipitate shearing in strong and weak pair-coupled modes as well as dislocation bowing between precipitates in polycrystalline multimodal Nickel-base superalloys:

$$\begin{aligned}\sigma_{ys} = \ &\text{grain size hardening of the } \gamma \text{ matrix}\\ &+ \text{solid} - \text{solution hardening of the } \gamma \text{ matrix}\\ &+ \text{grain} - \text{size hardening of a primary } \gamma' \text{ phase}\\ &+ \text{anomalous hardening of a primary } \gamma' \text{ phase}\\ &\quad\text{and large secondary } \gamma' \text{ phase}\\ &+ (\text{shearing / bowing of secondary } \gamma' - \text{strong}\\ &\quad\text{pair coupling}\\ &+ \text{shearing / bowing of tertiary } \gamma' - \text{weak}\\ &\quad\text{coupling}) f(T)\end{aligned} \tag{18}$$

Here the grain-size hardening of the γ matrix and the primary γ′ phase is modeling using equations resembling Eq. (1) [28]. The term, $f(T)$, denotes the temperature dependence of dislocation shearing. The solid-solution hardening of the γ matrix is modeled as

$$\Delta\sigma_{ss} = \sum_i \left(\frac{d\sigma}{\sqrt{dC_i}}\sqrt{C_i}\right), \tag{19}$$

where $C_i$ represents the concentration of the $i$th alloying element, but $d\sigma/\sqrt{dC_i}$ is a strengthening coefficient that reflects the strengthening potency of the alloying element. $i$. The anomalous hardening of the primary $\gamma'$ phase and the large secondary $\gamma'$ phase is modeled as

$$\Delta\sigma = f\left[\sigma(T)_{\text{Ni}_3\text{Al}} + \Sigma_i\left(\frac{d\sigma}{\sqrt{dC_i}}\sqrt{C_i}\right)\right], \tag{20}$$

where $f$ represents the fraction of precipitates subject to cross-slip-induced hardening, and $\sigma(T)_{\text{Ni}_3\text{Al}}$ represents the strength of pure Ni$_3$Al as a function of temperature. For the formulation of the shearing / bowing of the secondary $\gamma'$- strong coupling and the shearing / bowing of the tertiary $\gamma'$- weak coupling, refer to [28].

2.9 The work of Collins et al. from 2014 [29]

Collins et al. presented a computational methodology, one that combines the models of precipitation strengthening, dispersion strengthening with grain growth, and grain-boundary hardening. The combined model offer capabilities for predicting properties of the microstructure and the yield strength of Nickel-base superalloys subjected to arbitrary thermal cycles.

2.10 The work of Galindo-Nava et al. from 2015 [30]

Galindo-Nava et al. provide a good overview over models for precipitate shearing and over multimodal precipitate distribution effects. The overview of the precipitate shearing covers classic approaches, unified approaches and Orowan stresses. The authors model the yield stress of unimodal and multimodal $\gamma'$ Nickel-base superalloys as

$$\sigma_y = \sigma_D + \sigma_{SS} + \sigma_P + \sigma_{Oro}, \tag{21}$$

The term, $\sigma_D$, represents grain-boundary strengthening and is modeled, using the Hall-Petch relationship [Eq. (1)]. The term, $\sigma_{SS}$, denotes the strengthening contribution of the solid solution in $\gamma$, the term, $\sigma_P$, stands for precipitation shearing and the term, $\sigma_{Oro}$, for Orowan bypassing.

By employing the Labusch theory [31], which accounts for the increment in the yield strength resulting from solute atoms acting as frictional obstacles for the dislocation slip in a binary alloy, and accounting an extension by Gypen and Deruytterre [32], [33], the authors model the yield strength due to solid-solution hardening as

$$\sigma_{SS} = (1-f)\left(\sum X_i \beta_i^{3/2}\right)^{2/3}. \tag{22}$$

Here $f$ represents volume fraction, $x_i$ is the atom fraction of substitutional element, $i$, in the γ; and $\beta i$ are constants related to the lattice and modulus misfit of the element, $i$, with $N_i$ in the binary system. The factor, $(1 - f)$, accounts for a solid-solution contribution confined to the $c$, as dislocation slip mostly occurs at the matrix.

2.11 The work of Deng et al. from 2020 [34]

Considering multimodal particle-size distributions, Deng et al. proposed the following model for precipitation strengthening, $\sigma_P$, in polycrystalline Nickel-base superalloys during interrupt cooling:

$$\sigma_P = M \sqrt[n]{(\tau_{P,pri})^n + (\tau_{P,sec})^n + (\tau_{P,ter})^n}, \tag{23}$$

Here $M$ represents a Taylor factor, evaluated as $M = 3.06$ [34], but $\tau_{P,pri}$, $\tau_{P,sec}$ and $\tau_{P,ter}$ represent critical resolve shear stress from the primary, secondary and tertiary γ′ precipitates, respectively.